\documentclass[a4paper,11pt]{article}
\usepackage{latexsym}
\usepackage{amsmath,amssymb,bm,ascmac,color,calc}
\usepackage{slashed}
\usepackage[mathscr]{eucal}
\usepackage{graphicx}
\usepackage{cite}
\usepackage{lineno}
\usepackage{jheppub}
\usepackage{multirow}
\usepackage{subfigure}
\usepackage[normalem]{ulem}
\allowdisplaybreaks

\newcommand{\ignore}[1]{}
\renewcommand{\bf}{\textbf}

\newcommand{\GeV}{{\rm\ GeV}}

\newcommand{\TeV}{{\rm\ TeV}}

\def\beq{\begin{equation}}
\def\eeq{\end{equation}}
\newcommand{\ba}{\begin{array}}
\newcommand{\ea}{\end{array}}
\newcommand{\bea}{\begin{eqnarray}}
\newcommand{\eea}{\end{eqnarray} }
\newcommand{\bal}{\begin{align}}
\newcommand{\eal}{\end{align}}
\def\bi{\begin{itemize}}
\def\ei{\end{itemize}}
\def\ben{\begin{enumerate}}
\def\een{\end{enumerate}}
\def\beq{\begin{equation}}
\def\eeq{\end{equation}}
\def\bc{\begin{center}}
\def\ec{\end{center}}
\def\bt{\begin{table}}
\def\et{\end{table}}
\def\btb{\begin{tabular}}
\def\etb{\end{tabular}}

\title{Loop Induced Single Top Partner Production and Decay at the LHC}

\author[a]{Jeong Han Kim}
\author[a]{Ian M. Lewis}

\affiliation{Department of Physics and Astronomy, University of Kansas, Lawrence, Kansas, 66045 USA}

\emailAdd{jeonghan.kim@ku.edu}
\emailAdd{ian.lewis@ku.edu}

\abstract{Most searches for top partners, $T$, are concerned with top partner pair production.  However, as these bounds become increasingly stringent, the LHC energy will saturate and single top partner production will become more important.  In this paper we study the novel signature of the top partner produced in association with the SM top, $pp\rightarrow T\overline{t}+t\overline{T}$, in a model where the Standard Model (SM) is extended by a vector-like $SU(2)_L$ singlet fermion top partner and a real, SM gauge singlet scalar, $S$. In this model, $pp\rightarrow T\overline{t}+t\overline{T}$ production is possible through loops mediated by the scalar singlet.  We find that, with reasonable coupling strengths, the production rate of this channel can dominate top partner pair production at top partner masses of $m_T\gtrsim 1.5$~TeV.  In addition, this model allows for the exotic decay modes $T\rightarrow tg$, $T\rightarrow t \gamma$, and $T\rightarrow t S$.  In much of the parameter space the loop induced decay $T\rightarrow tg$ dominates and the top partner is quite long lived.  New search strategies are necessary to cover these decay modes. We project the the sensitivity of the high luminosity LHC to $pp\rightarrow T\overline{t}+t\overline{T}$ via a realistic collider study.  We find with 3 ab$^{-1}$, the LHC is sensitive to this process for masses $m_T\lesssim2$~TeV.  In addition, we provide appendices detailing the renormalization of this model.  
}

\begin{document}

\maketitle

\section{Introduction}
The Large Hadron Collider (LHC) is quickly accumulating data at the energy frontier of particle physics.  While the the LHC is searching for many types of beyond-the-Standard Model (BSM) physics, of particular interest are searches for partners of the SM top quark.   In many models that solve the naturalness problem, top quark partners are postulated to exist and cancel the quadratic corrections to the Higgs mass, stabilizing the Higgs at the electroweak (EW) scale.   However, as BSM physics remains elusive, it is necessary to go beyond the typical search strategies. In this paper, we will consider a simple model with new, exotic signals of top partners at the LHC.  These novel signatures will help fill in gaps in the coverage of BSM searches.

The focus of the paper will be on a fermionic top partner, $T$.  These top partners are ubiquitous in composite Higgs~\cite{Agashe:2004rs,Agashe:2005dk,Agashe:2006at,Contino:2006qr,Giudice:2007fh,Azatov:2011qy,Serra:2015xfa} and little Higgs models~\cite{ArkaniHamed:2002qy,ArkaniHamed:2002pa,Low:2002ws,Chang:2003un,Csaki:2003si,Perelstein:2003wd,Chen:2003fm,Berger:2012ec}.  Most searches for these top partners are concerned with double production, $T\overline{T}$.  The utility of this mode is that the production rate only depends on the strong force coupling, and, hence, is fairly model independent.  However, as bounds on the top partner mass, $m_T$, become multi-TeV, the LHC energy will be saturated and the utility of this channel greatly diminished.  In such cases, single production of a top partner in association with another quark or $W$ boson may be promising~\cite{Willenbrock:1986cr,Han:2003wu,Han:2005ru,DeSimone:2012fs,Backovic:2015bca,Liu:2016jho,Aguilar-Saavedra:2013qpa,Ortiz:2014iza,Matsedonskyi:2014mna,Liu:2015kmo,Backovic:2015lfa,Zhang:2017nsn,Liu:2017sdg}, since there is more available phase space. 

Typically, single top partner production is mediated by $W$ or $Z$ bosons and the relevant top partner-$W/Z$ couplings are usually proportional to the mixing angle between the $T$ and SM top quark  $t$.  This mixing angle is constrained by EW precision measurements to be quite small~\cite{Lavoura:1992np,Maekawa:1995ha,He:2001tp,Dawson:2012di,Aguilar-Saavedra:2013qpa,Ellis:2014dza,Chen:2017hak}, suppressing the single top partner production rate.  In this paper, we consider a model with a SM gauge singlet scalar~\cite{Barger:2007im,OConnell:2006rsp,Pruna:2013bma,Chen:2014ask,Buttazzo:2015bka,Robens:2015gla,Dawson:2015haa,Costa:2015llh,Kanemura:2015fra,Kanemura:2016lkz,Robens:2016xkb,Lewis:2017dme,Kanemura:2017gbi,Dawson:2017jja}, $S$, in addition to a top partner~\cite{Fox:2011qc,Ellis:2015oso,McDermott:2015sck,Falkowski:2015swt,Anandakrishnan:2015yfa,Serra:2015xfa,Gupta:2015zzs,Han:2015dlp,Knapen:2015dap,Craig:2015lra,Dolan:2016eki,Nakamura:2017irk}.  Besides being a simple addition to the SM, singlet scalars can help provide a strong first order EW phase transition necessary for EW baryogenesis~\cite{Ham:2004cf,Profumo:2007wc,Espinosa:2011ax,No:2013wsa,Curtin:2014jma,Huang:2015tdv,Huang:2016cjm,Chen:2017qcz}.  With this particle content, a new tree-level flavor off-diagonal coupling $S-T-t$ is allowed and it is not suppressed by a mixing angle.  This new coupling introduces new mechanisms for single $T$ production.  First, if the mass of the scalar is greater than the $T$ and top quark masses, it is possible that we can search for resonant production of a top partner in association with a top quark through $S$ decays~\cite{Fichet:2016xpw}.   Even if resonant production is not possible, the new scalar can mediate loop induced $pp\rightarrow T\overline{t}+t\overline{T}$ production ($Tt$).  Although loop suppressed, such a process will become increasing important at the LHC as more data is gained, precision of measurements is increased, and the phase space for pair production of heavy particles is squeezed.  As we will see, the production rate of this mechanism can be larger than pair production for $m_T\gtrsim 1.5$~TeV and reasonable coupling constants.  Additionally, $Tt$ is the dominant single $T$ production mode for small $T-t$ mixing.  

In addition to novel production mechanisms, this model introduces new decay channels for the top partner.  Typically, top partners are searched for in the $T\rightarrow tZ$, $T\rightarrow th$, and $T\rightarrow bW$ with approximate branching ratios of $25\%,25\%$, and $50\%$, respectively~\cite{Aaboud:2017zfn, Aaboud:2017qpr,Sirunyan:2017pks, Sirunyan:2017usq}.  However, with a new scalar boson, the decays of the top partner can be significantly altered from the usual expectations.  If the scalar is light enough, $T\rightarrow tS$ is available at tree level.  The precise signature of this decay depends on how the scalar decays and if it mixes with the Higgs boson~\cite{Dolan:2016eki}.  Nevertheless, new search strategies are necessary.  If the scalar mass $m_S>m_T$, then $T\rightarrow tS$ is forbidden and the traditional decays may be expected to dominate.  However, these decay widths are typically suppressed by the top-partner and top mixing angle, and, as we will show, the loop induced decays $T\rightarrow tg$, $T\rightarrow tZ$ and $T\rightarrow t\gamma$ can dominate.  This is similar to the decay patterns of excited quarks~\cite{DeRujula:1983ak,Kuhn:1984rj,Baur:1987ga,Baur:1989kv,Han:2010rf,Sirunyan:2017yta}, which couple to the SM through dipole operators.  In the model with a top partner and scalar, these decays are completely calculable and give rise to new phenomena.  In particular, the top partner becomes quite long lived, necessitating an update of search strategies.

In this paper we study a simplified model containing a top partner and a real, SM gauge singlet scalar.  We will show that this model has interesting signatures and that LHC is sensitive to new regions of parameter space via $pp\rightarrow T\overline{t}+t\overline{T}$ production.  In Section~\ref{sec:model} we introduce the model and couplings of the new particles.   The production and decay rates of the top partner are studied in Section~\ref{sec:proddec}, and the production and decay rates and scalar are studied in Section~\ref{sec:proddecscalar}.      Current experimental constraints on top partners and scalar singlets are presented in Section~\ref{sec:const}.  In Section~\ref{sec:results}, we perform a realistic collider study for the process $pp\rightarrow T\overline{t}+t\overline{T}\rightarrow t\overline{t}S\rightarrow t\overline{t}gg$.  We conclude in Section~\ref{sec:conc}.  In addition, we attach three appendices with necessary calculation details.  In Appendix~\ref{Appe:renorm} we present the details of the wave-function and mass renormalization of the top sector.  Vertex counterterms for $T-t-g$, $T-t-\gamma$, and $T-t-Z$ are presented in Appendix~\ref{Appe:vertrenorm}.  In Appendix~\ref{app:DR} we give the parameterization of energy smearing for the collider study.

\section{The Model}
\label{sec:model}
We consider a model consisting of a vector-like $SU(2)_L$ singlet top partner, $\mathcal{T}_2$, and a real SM gauge singlet scalar $S$.  A similar model has been consider in Ref.~\cite{Dolan:2016eki}.
For simplicity and to avoid flavor constraints, the top partner is only allowed to couple to the third generation SM quarks:
\begin{eqnarray}
Q_{L}=\begin{pmatrix} t_{1L} \\ b_L \end{pmatrix},\quad \mathcal{T}_{1R},\quad {\rm and}\quad b_R.
\end{eqnarray}
The allowed Yukawa interactions and mass terms are
\begin{eqnarray}
-\mathcal{L}_{Yuk}&=&y_b \overline{Q}_L \Phi b_R+\widetilde{y}_t\overline{Q}_L \widetilde{\Phi} \mathcal{T}_{1R}+\widetilde{\lambda}_t \overline{Q}_L \widetilde{\Phi} \mathcal{T}_{2R}+\widetilde{M}_2 \overline{\mathcal{T}}_{2L} \mathcal{T}_{2R}+\widetilde{M}_{12}\overline{\mathcal{T}}_{2L} \mathcal{T}_{1R}.\nonumber\\
&&+\widetilde{\lambda}_1 S \overline{\mathcal{T}}_{2L} \mathcal{T}_{1R}+\widetilde{\lambda}_2 S\overline{\mathcal{T}}_{2L}\mathcal{T}_{2R}+{\rm h.c.},\label{eq:Lag}
\end{eqnarray}
where $\Phi$ is the SM Higgs doublet, $\widetilde{\Phi}=i\sigma^2 \Phi^*$, and $\sigma^2$ is a Pauli matrix.    The most general renormalizable scalar potential has the form~\cite{Chen:2014ask}
\begin{eqnarray}
\label{eq:pot}
V(\Phi,S)&=&-\mu^2 \Phi^\dagger\Phi +\lambda(\Phi^\dagger\Phi)^2+\frac{a_1}{2} \Phi^\dagger\Phi S +\frac{a_2}{2}\Phi^\dagger \Phi S^2\\
&&+b_1 S+\frac{b_2}{2} S^2 +\frac{b_3}{3!}S^3 +\frac{b_4}{4!}S^4.\nonumber
\end{eqnarray}

After EW symmetry breaking (EWSB), in general both the scalar $S$ and Higgs doublet $\Phi$ can develop vacuum expectation values (vevs): $\langle \Phi \rangle^T=(0,v/\sqrt{2})$ and $\langle S\rangle =x$ where $v=246$~GeV is the SM Higgs doublet vev.  Since $S$ is a gauge singlet and there are no discrete symmetries imposed, shifting to the vacuum $S=x+s$ is a field redefinition that leaves all the symmetries intact.  Hence, it is unphysical and we are free to choose $x=0$~\cite{Chen:2014ask}.  Two possible ways to understand this are: (1) All possible interaction terms of $S$ are already contained in the scalar potential and Yukawa interactions, Eqs.~(\ref{eq:Lag}) and~(\ref{eq:pot}).  Hence, shifting to the vacuum $S=x+s$ does not introduce any new interactions and is unphysical.  (2) After $S$ obtains a vev, any discrete symmetry that $S$ has is broken and all interactions in Eqs.~ (\ref{eq:Lag}) and ~(\ref{eq:pot}) are possible.  Hence, the scalar $S$ can be interpreted as the field after already shifting to the vacuum with $x=0$.

Also after EWSB, it is possible for the scalar $S$ and Higgs boson $h$ to mix.  However, since the focus of this paper is the production and decay of the top partner, for simplicity we set the scalar mixing angle to zero.  This is equivalent to setting $a_1=0$ in Eq.~(\ref{eq:pot}).  Hence, $h$ and $S$ are mass eigenstates with masses $m_h=125$~GeV~\cite{Aad:2015zhl,ATLAS-CONF-2017-046,Sirunyan:2017exp} and $m_S$, respectively; such that $h$ is the observed Higgs boson~\cite{Chatrchyan:2012xdj,Aad:2012tfa}.

There is another possible simplification of the Lagrangian.  Since $\mathcal{T}_{2R}$ and $\mathcal{T}_{1R}$ have the same quantum numbers and $\mathcal{T}_{2L}$ and $\mathcal{T}_{2R}$ are two different Weyl-spinors, the off-diagonal vector-like mass-term, $\widetilde{M}_{12}$, can be removed via the field redefinitions~\cite{Dawson:2012mk}
\begin{eqnarray}
\widetilde{M}_2 \mathcal{T}_{2R}&=& M_2 t_{2R}-\widetilde{M}_{12}t_{1R},\quad \mathcal{T}_{2L}=t_{2L},\quad{\rm and}\quad \mathcal{T}_{1R}=t_{1R}
\end{eqnarray}
The Yukawa interactions and mass terms are then
\begin{eqnarray}
-\mathcal{L}_{Yuk}&=&y_b \overline{Q}_L \Phi b_R+y_t \overline{Q}_L \widetilde{\Phi} t_{1R}+\lambda_t \overline{Q}_L \widetilde{\Phi}t_{2R}+M_2 \overline{t}_{2L}t_{2R}\nonumber\\
&&+\lambda_1 S\,\overline{t}_{2L}t_{1R}+\lambda_2S\,\overline{t}_{2L}t_{2R}+{\rm h.c.}.
\end{eqnarray}
For simplicity, we assume all couplings are real.

The relevant kinetic terms are then
\begin{eqnarray}
\mathcal{L}_{kin}=\left|D_\mu\Phi\right|^2+\frac{1}{2}\left(\partial_\mu S\right)^2+\overline{Q}_Li\slashed{D}Q_L+\overline{t}_{1R}i\slashed{D}t_{1R}+\overline{t}_2i\slashed{D}t_2+\overline{b}_Ri\slashed{D}b_R,
\end{eqnarray}
where the covariant derivatives are
\begin{eqnarray}
D_\mu\Phi&=&(\partial_\mu+i\frac{g}{2}\sigma^a W^a_\mu+i\frac{g'}{2} B_\mu)\Phi\\
\slashed{D}Q_L&=&\left(\slashed{\partial}+i\frac{g}{2}\sigma^a \slashed{W}^a+i\frac{g'}{6} \slashed{B}+ig_ST^A \slashed{G}^A\right)Q_L\nonumber\\
\slashed{D}t_{1R}&=&\left(\slashed{\partial}+i\frac{2}{3}g' \slashed{B}+ig_ST^A \slashed{G}^A\right)t_{1R}\nonumber\\
\slashed{D}t_{2}&=&\left(\slashed{\partial}+i\frac{2}{3}g' \slashed{B}+ig_ST^A \slashed{G}^A\right)t_{2}\nonumber\\
\slashed{D}b_{R}&=&\left(\slashed{\partial}-i\frac{g'}{3} \slashed{B}+ig_ST^A \slashed{G}^A\right)b_{R},
\end{eqnarray}
where $\sigma^a$ are Pauli matrices and $T^A$ are the fundamental $SU(3)$ representation matrices.

\subsection{Scalar Couplings to Top Partners}
After EWSB, in the unitary gauge $\Phi=(0,(h+v)/\sqrt{2})^T$ the quark masses and Yukawa interactions are
\bea
\displaystyle-\mathcal{L}_{Yuk} = \overline{\chi}_L M \chi_R + h \,\overline{\chi}_L Y_h \chi_R+S\, \overline{\chi}_L Y_S \chi_R+\frac{h+v}{\sqrt{2}} y_b \overline{b}_Lb_R+{\rm h.c.},  \label{eq:Lmass}
\eea
where the top quark and partner are
\bea
\chi_\tau = \begin{pmatrix} t_{1\tau} \\ t_{2\tau}\end{pmatrix}
\eea 
with $\tau=L,R$, and the mass and Yukawa matrices are
\bea 
M=\frac{1}{\sqrt{2}}\begin{pmatrix} y_t v & \lambda_t v\\0& \sqrt{2}M_2 \end{pmatrix}, \quad Y_h = \frac{1}{\sqrt{2}}\begin{pmatrix} y_t & \lambda_t \\0 & 0\end{pmatrix},\quad{\rm and}\quad Y_S = \begin{pmatrix} 0 & 0 \\ \lambda_1 & \lambda_2\end{pmatrix}.
\eea
The top-quark mass matrix can be diagonalized via the bi-unitary transformation
\bea 
\begin{pmatrix} t_{1\tau} \\ t_{2\tau}\end{pmatrix}=\begin{pmatrix} \cos\theta_\tau & \sin\theta_\tau \\ -\sin\theta_\tau & \cos\theta_\tau \end{pmatrix}\begin{pmatrix} t_\tau \\ T_\tau \end{pmatrix}.\label{eq:TransLaw}
\eea
  The mass eigenstates are $t$ and $T$ with masses $m_t=173$~GeV~\cite{Patrignani:2016xqp} and $m_T$, respectively, such that $t$ is the observed SM-like top quark.  Upon diagonalization, the Higgs Yukawa coupling, $y_t,\lambda_t$, and the vector like mass $M_2$ can be expressed in terms of the mixing angle $\theta_L$ and masses $m_t,m_T$:
\begin{eqnarray}
M_2^2 &=& m_T^2\,\cos^2\theta_L+m_t^2\,\sin^2\theta_L\nonumber\\
y_t &=& \sqrt{2}\frac{m_t m_T}{v\,M_2}\nonumber\\
\lambda_t&=&\frac{m_T^2-m_t^2}{\sqrt{2}\,v\,M_2}\sin2\theta_L.
\end{eqnarray}
Additionally, only one of the mixing angles $\theta_L$ and $\theta_R$ is free:
\begin{eqnarray}
m_T\,\tan\theta_R = m_t\,\tan\theta_L.
\end{eqnarray}
The independent parameters of this theory are then
\begin{eqnarray}
\theta_L,\, m_T,\, m_S,\, \lambda_1,\,{\rm and}\, \lambda_2.
\end{eqnarray}

After rotating to the mass eigenbasis, the quark masses and scalar couplings are
\begin{eqnarray}
-\mathcal{L}_{Yuk}&=&h\left[\lambda_{tt}^h \overline{t}t+\lambda_{TT}^h\overline{T}T+\overline{t}\left(\lambda_{tT}^hP_R+\lambda_{Tt}^h P_L\right)T+\overline{T}\left(\lambda_{Tt}^h P_R+\lambda_{tT}^h P_L\right)t\right]\nonumber\\
&+&S\left[\lambda_{tt}^S \overline{t}t+\lambda_{TT}^S\overline{T}T+\overline{t}\left(\lambda_{tT}^SP_R+\lambda_{Tt}^S P_L\right)T+\overline{T}\left(\lambda_{Tt}^S P_R+\lambda_{tT}^S P_L\right)t\right]\nonumber\\
&+&m_t\,\overline{t}t+m_T\overline{T}T+m_b\left(1+\frac{h}{v}\right)\overline{b}b,\label{eq:Yuk}
\end{eqnarray}
where $m_b=y_b v/\sqrt{2}$ is the bottom quark mass, the Higgs boson couplings are

\bea
\lambda_{tt}^h &=& \displaystyle\frac{1}{\sqrt{2}}\cos\theta_L\left(y_t\cos\theta_R-\lambda_t\sin\theta_R\right),\quad \lambda_{tT}^h=\displaystyle\frac{1}{\sqrt{2}}\cos\theta_L\left(y_t\sin\theta_R+\lambda_t\cos\theta_R\right),\label{eq:htt}\\
\lambda_{Tt}^h&=&\displaystyle\frac{1}{\sqrt{2}}\sin\theta_L\left(y_t\cos\theta_R-\lambda_t\sin\theta_R\right),\quad\lambda_{TT}^h=\displaystyle\frac{1}{\sqrt{2}}\sin\theta_L\left(y_t\sin\theta_R+\lambda_t\cos\theta_R\right),\nonumber
\eea
and the scalar $S$ couplings are
\bea
\lambda_{tt}^S&=& -\sin\theta_L\left(\lambda_1\cos\theta_R-\lambda_2\sin\theta_R\right),\quad\lambda_{tT}^S=-\sin\theta_L\left(\lambda_1\sin\theta_R+\lambda_2\cos\theta_R\right),\nonumber\\
\lambda_{Tt}^S&=&\cos\theta_L\left(\lambda_1\cos\theta_R-\lambda_2\sin\theta_R\right),\quad\lambda_{TT}^S=\cos\theta_L\left(\lambda_1\sin\theta_R+\lambda_2\cos\theta_R\right).\label{eq:Stt}
\eea
\subsection{$Z$ and $W^\pm$ Couplings to Top Partners}
After diagonalizing the top quark mass matrix, the $Z$ and $W$ couplings to the third generation and top partner are altered as well as introducing the flavor off diagonal coupling $t-T-Z$.  The interactions relevant for our analysis are
\begin{eqnarray}
\mathcal{L}&\supset&-\frac{g}{\sqrt{2}}\left\{ W^{+\mu}\left[\cos\theta_L \overline{t} \gamma_\mu P_L b  +  \sin\theta_L \overline{T} \gamma_\mu P_L b\right]+{\rm h.c.}\right\}\label{eq:gauge}\\
&-&\frac{g}{c_W}Z^\mu\left\{\overline{t}\gamma_\mu \left[\left(g_L^Z-\frac{1}{2}\sin^2\theta_L\right)P_L+g_R^Z P_R\right] t  +\frac{1}{4}\sin2\theta_L\overline{t}\gamma_\mu P_L T\right.\nonumber\\
&&\left.\quad\quad\quad\quad+\frac{1}{4}\sin2\theta_L \overline{T}\gamma_\mu P_L t+\overline{T}\gamma_\mu\left[g_R^Z+\frac{1}{2}\sin^2\theta_L P_L\right]T\right\},\nonumber
\end{eqnarray}
where $c_W=\cos\theta_W$, $s_W=\sin\theta_W$, $\theta_W$ is the weak mixing angle, $g$ is the weak coupling constant, $g_L^Z=\frac{1}{2}-\frac{2}{3}s_W^2$, and $g_R^Z=-\frac{2}{3}s^2_W$.  Since electromagnetism and $SU(3)$ are unbroken, the top quark and partner just couple to photons and gluons according to their electric and color charges.  We use the $Z$-mass, the Fermi decay constant, and the electric coupling at the $Z$-pole as input parameters~\cite{Patrignani:2016xqp}:
\bea
m_Z = 91.1876~{\rm GeV},\quad G_F=1.16637\times 10^{-5}~{\rm GeV}^{-2},\quad \alpha(m_Z)^{-1} = 127.9.
\eea
The other EW parameters ($g,\theta_W,v,m_W$) are calculated using the tree level relations
\bea
G_F = \frac{1}{\sqrt{2}\,v^2},\quad m_Z = \frac{e}{2s_W c_W} v,\quad g=e/s_W,\quad m_W = \frac{1}{2}g\,v,
\eea
where $m_W$ is the $W$-mass.

\subsection{Effective Field Theory}
\label{sec:EFT}
\begin{figure}[tb]
\begin{center}
\subfigure[]{\includegraphics[width=0.28\textwidth,clip]{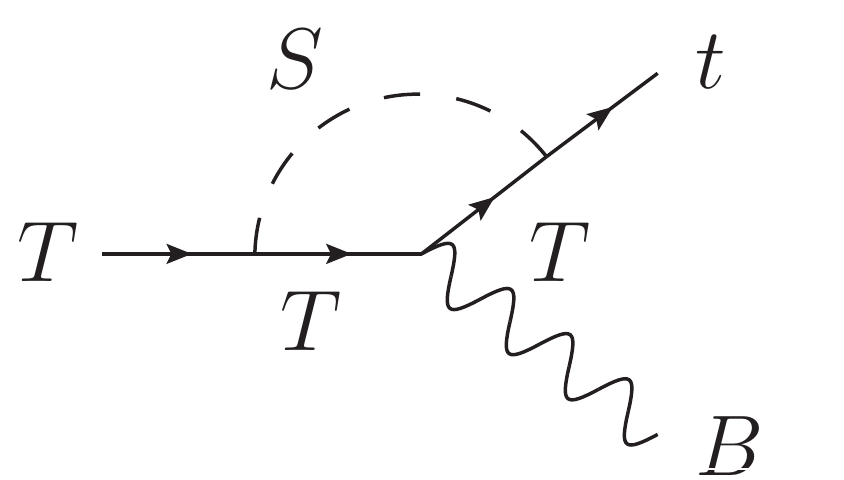}}
\subfigure[]{\includegraphics[width=0.28\textwidth,clip]{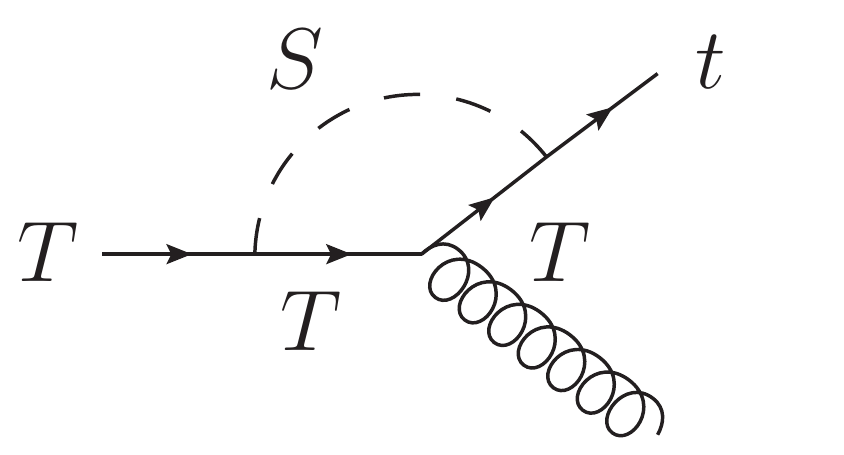}}
\end{center}
\caption{\label{fig:zeromass} Top partner decay diagrams in the limit that $v\rightarrow 0$ and EW symmetry is restored.  Counterterms and external flavor changing self-energies are not shown.}
\end{figure}

In the limit that $m_S\gg m_T,v$, the scalar $S$ can be integrated out.  The lowest dimension operators that contribute to top partner production and decay are the dipole operators:
\begin{eqnarray}
\mathcal{L}_{EFT}=c_B \overline{T_L}\sigma^{\mu\nu}t_R B_{\mu\nu}+c_G \overline{T_L}\sigma^{\mu\nu}T^A t_R G^{A}_{\mu\nu}+{\rm h.c.},\label{eq:EFT}
\end{eqnarray}
where the hypercharge and gluon field strength tensors are
\begin{eqnarray}
B_{\mu\nu}&=&\partial_\mu B_\nu-\partial_\nu B_\mu,\\
G_{\mu\nu}^A&=& \partial_\mu G^A_\nu-\partial_\nu G^A_\mu-g_S f^{ABC}G^B_\mu G^C_\nu,
\end{eqnarray}
and $f^{ABC}$ is the $SU(3)$ structure constant.
These interactions arise from the processes shown in Fig.~\ref{fig:zeromass}.  Taking the limit that $m_S\gg m_T$ and that EW symmetry is restored ($v\rightarrow 0$, $\sin\theta_L\rightarrow 0$), we calculate $T\rightarrow tB$ and $T\rightarrow tg$. The details of the necessary renormalization counterterms can be found in Appendices~\ref{Appe:renorm} and~\ref{Appe:vertrenorm}. Matching onto the EFT, we find the Wilson coefficients:
\begin{eqnarray}
c_B&=&\frac{e}{c_W}\frac{2}{3}\frac{\lambda_1\lambda_2}{24\pi^2}\left(1+\frac{3}{4}\ln\frac{m_T^2}{m_S^2}\right)\frac{m_T}{m_S^2}\\
c_G&=&g_S\frac{\lambda_1\lambda_2}{24\pi^2}\left(1+\frac{3}{4}\ln\frac{m_T^2}{m_S^2}\right)\frac{m_T}{m_S^2}.
\end{eqnarray}
Note that the ratio of the Wilson coefficients $c_G/c_B = 3\,g_s c_W/(2\,e)$ is completely determined by the the ratio of the strong and Hypercharge coupling constants.  This is because the structure of the loop diagrams in Fig.~\ref{fig:zeromass} are essentially the same with the only difference being the external gauge boson and their couplings to the top partner.  Also, although the operators in Eq.~(\ref{eq:EFT}) are dimension five, the Wilson coefficients are suppressed by two powers of $m_S$ ($m_T/m_S^2$) and not one power ($1/m_S$).  The dipole operators couple left- and right-chiral fields.  Hence, the loop diagram needs an odd number of changes in chirality.  From just the couplings, the diagrams in Fig.~\ref{fig:zeromass} have an even number of chiral flips.  An additional mass insertion is needed and one power of $m_T$ in the numerator is necessary.  The operators are then suppressed by $m_T/m_S^2$ and not $1/m_S$.

\section{Production and Decay of Top Partner}
\label{sec:proddec}
We now discuss the production and decay of the top partner, $T$, in the model presented in Sec.~\ref{sec:model}.  To produce the the numerical results we implement the model in \texttt{FeynArts}~\cite{Hahn:2000kx} via \texttt{FeynRules}~\cite{Christensen:2008py,Alloul:2013bka}.  Matrix element squareds are then generated with \texttt{FormCalc}~\cite{Hahn:1998yk}.  We use the \texttt{NNPDF2.3QED}~\cite{Ball:2013hta} parton distribution functions (pdfs) as implemented in \texttt{LHAPDF6}~\cite{Buckley:2014ana}.  We also use the strong coupling constant as implemented in \texttt{LHAPDF6}.  Details on the wave-function renormalization and vertex counterterms needed for the calculations in this section can be found in the Appendices~\ref{Appe:renorm} and~\ref{Appe:vertrenorm}.
\subsection{Top Partner Production Channels}
\label{sec:pro}

\begin{figure}[tb]
\begin{center}
\subfigure[]{\includegraphics[width=0.28\textwidth,clip]{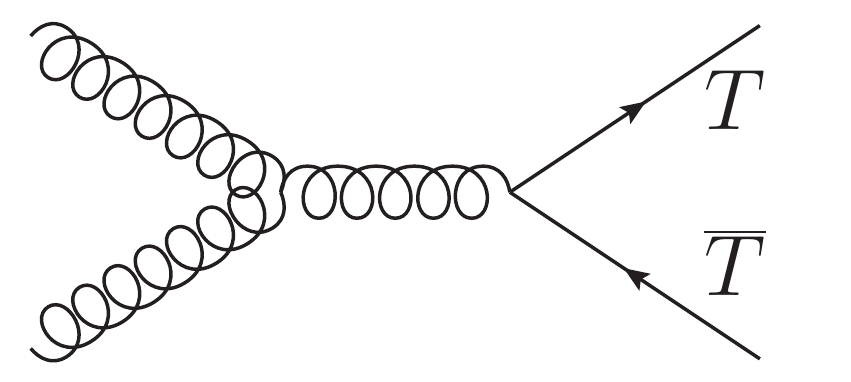}\label{fig:ggTTs}}
\subfigure[]{\includegraphics[width=0.20\textwidth,clip]{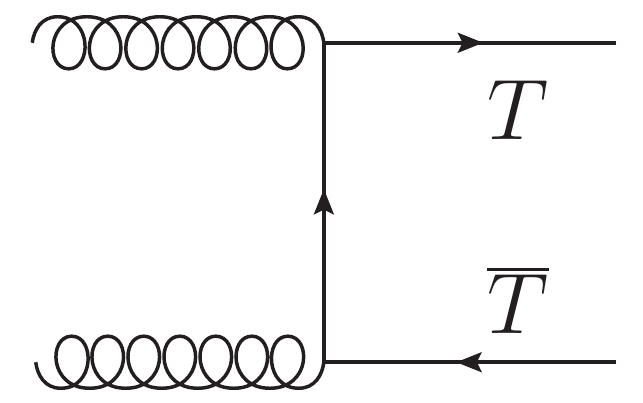}\label{fig:ggTTt}}
\subfigure[]{\includegraphics[width=0.28\textwidth,clip]{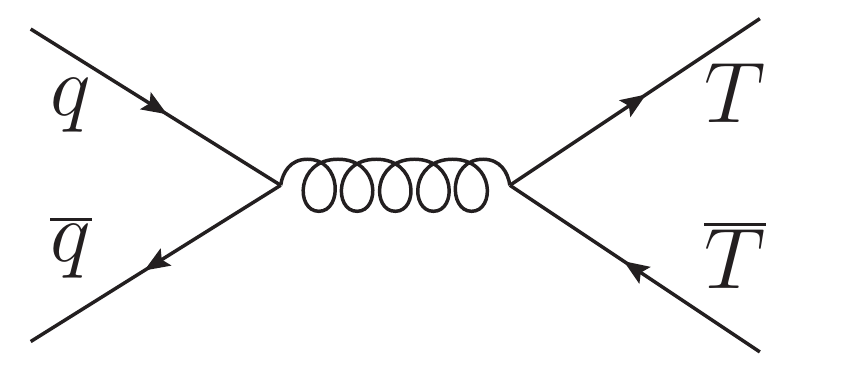}\label{fig:qqTTs}}\\
\subfigure[]{\includegraphics[width=0.23\textwidth,clip]{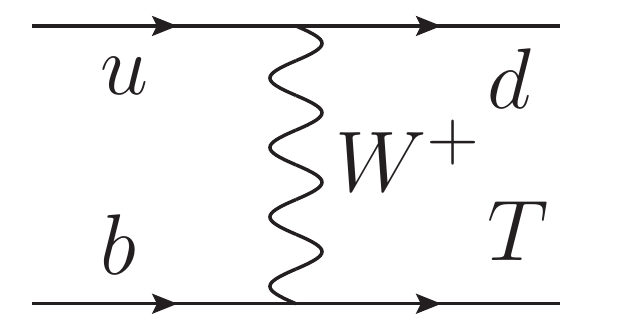}\label{fig:Tjet}}
\subfigure[]{\includegraphics[width=0.23\textwidth,clip]{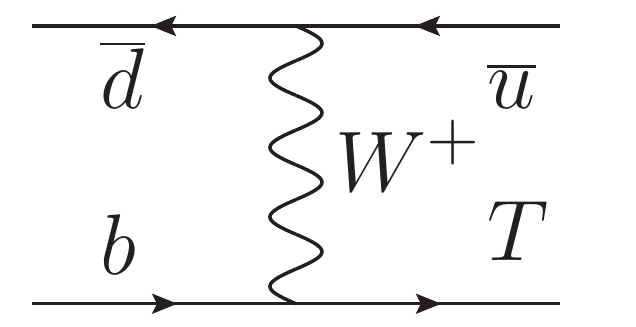}\label{fig:Tjet1}}
\subfigure[]{\includegraphics[width=0.23\textwidth,clip]{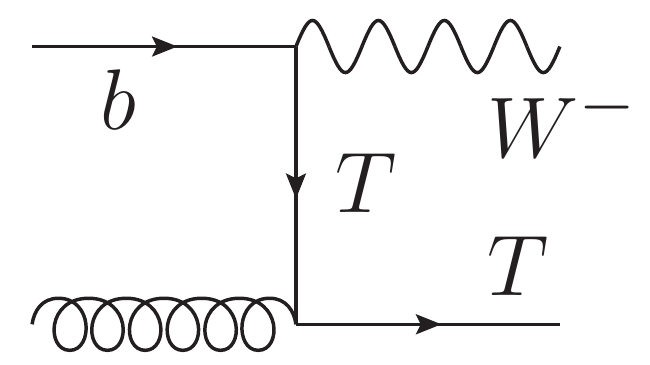}\label{fig:TW}}
\subfigure[]{\includegraphics[width=0.28\textwidth,clip]{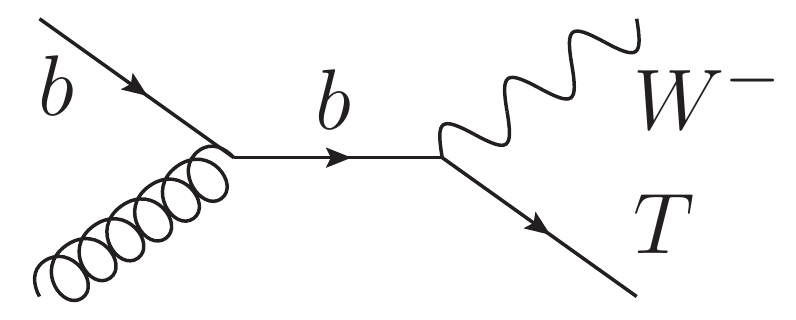}\label{fig:TW1}}
\end{center}
\caption{Standard production modes of top partners at the LHC for (a-c) pair production, (d,e) top partner plus jet production, and (f,g) top partner plus $W^-$ production.  There are conjugate processes for (d-g) that are not shown here.\label{fig:Prod_Feyn1}}
\end{figure}

\begin{figure}[tb]
\begin{center}
\subfigure[]{\includegraphics[width=0.48\textwidth,clip]{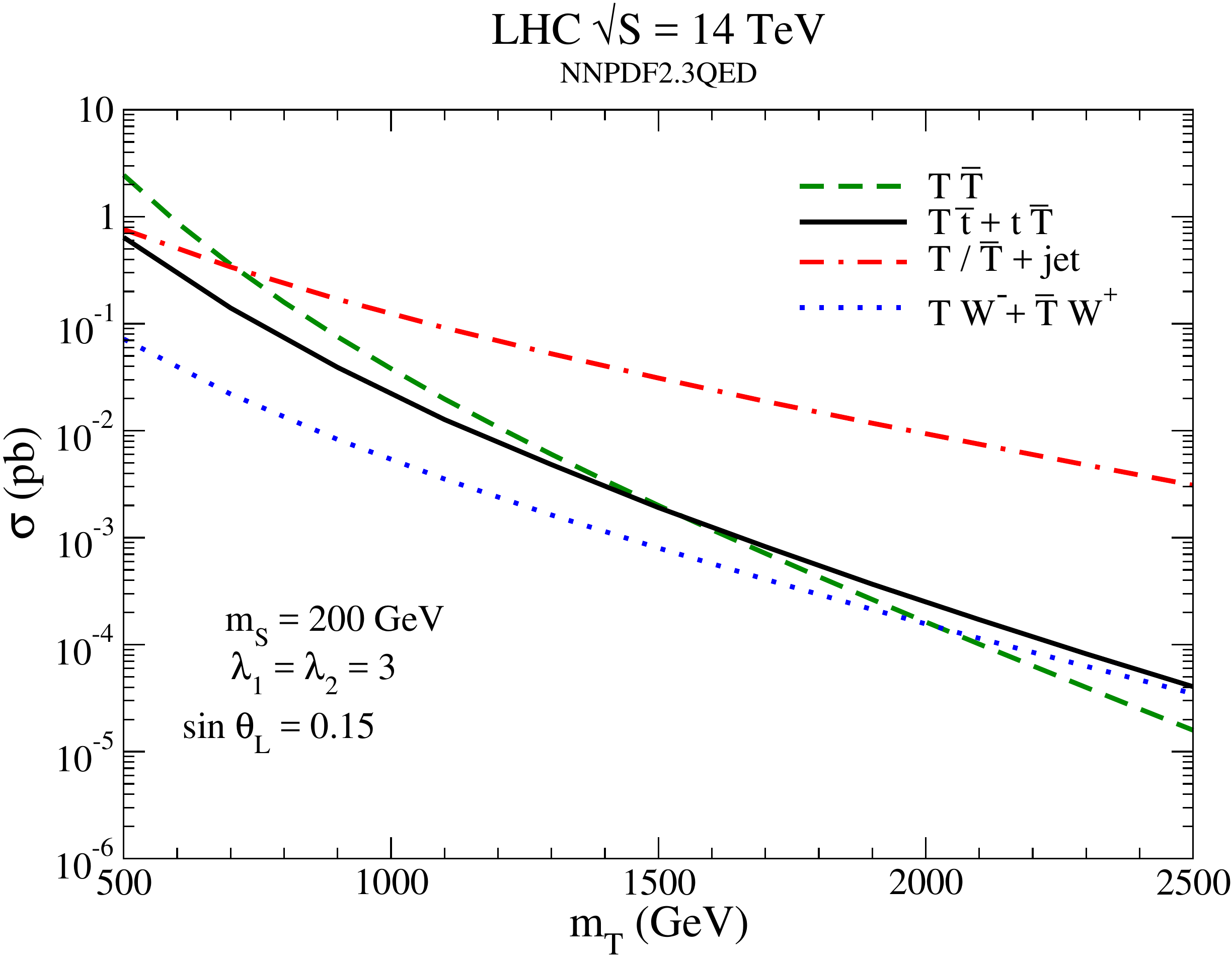}\label{fig:prod_mtp_sth15}}
\subfigure[]{\includegraphics[width=0.48\textwidth,clip]{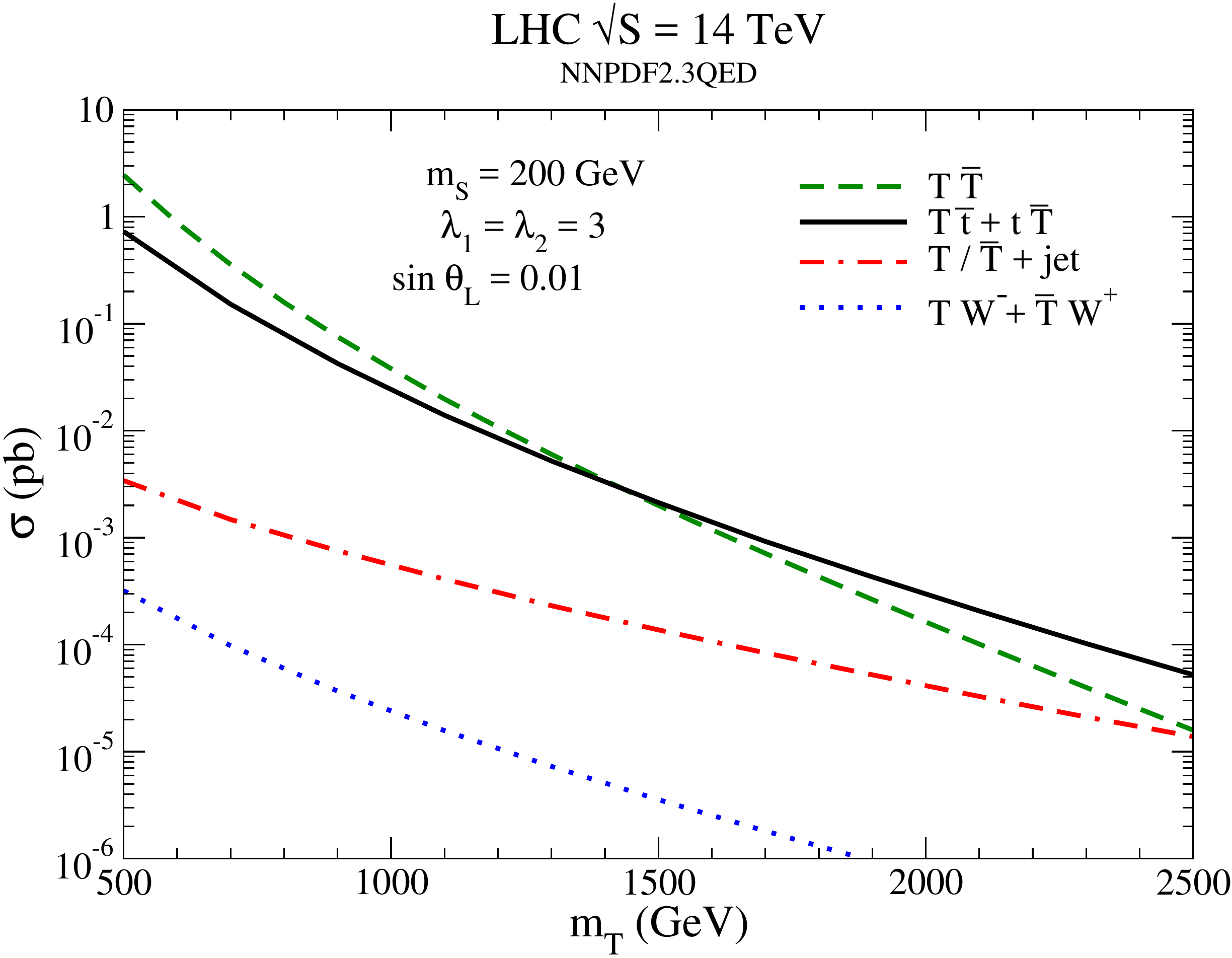}\label{fig:prod_mtp_sth01}}
\end{center}
\caption{\label{fig:prod_mtp} Production cross sections at the $\sqrt{S}=14$~TeV LHC for (green dashed) top partner pair production, (black solid) top partner production in association with a top quark, (red dash-dash-dot) top partner plus jet production, and (blue dotted) top partner plus $W^\pm$ production.  The parameters are set at a scalar mass $m_S=200$~GeV, couplings $\lambda_1=\lambda_2=3$, and mixing angles (a) $\sin\theta_L=0.15$ and (b) $\sin\theta_L=0.01$.  Factorization, $\mu_f$, and renormalization, $\mu_r$, scales are set to the sum of the final state particle masses.}
\end{figure}

\begin{figure}[tb]
\begin{center}
\subfigure[]{\includegraphics[width=.32 \textwidth,clip]{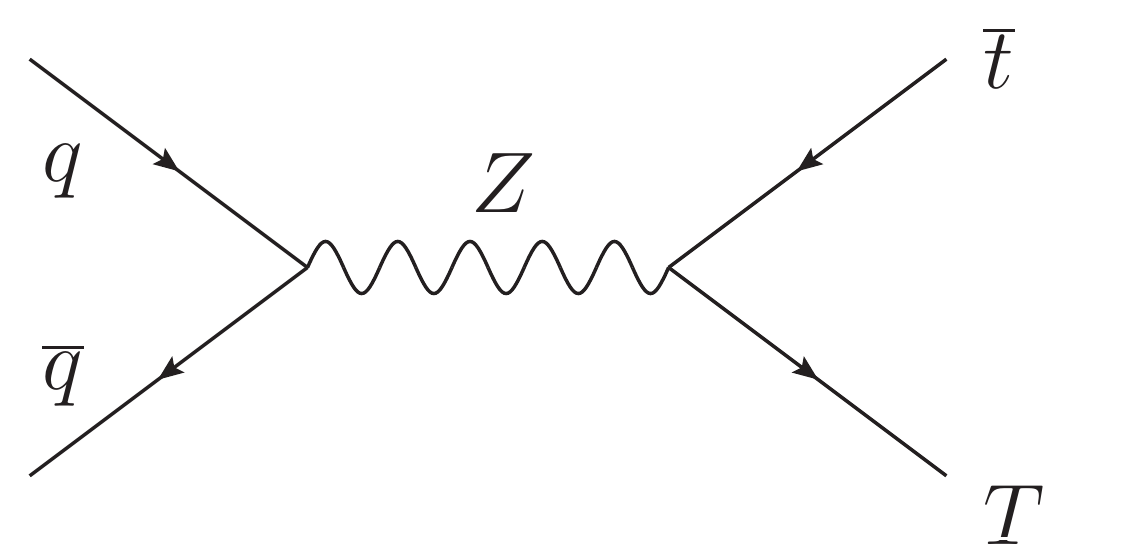}}
\subfigure[]{\includegraphics[width=.32 \textwidth,clip]{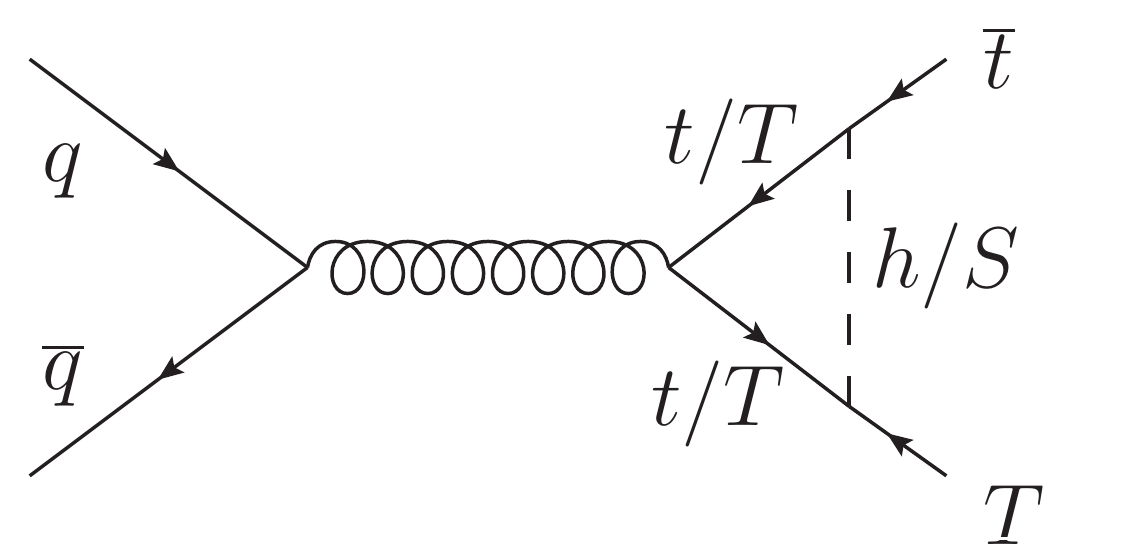}\label{fig:FeynLoop1}}
\subfigure[]{\includegraphics[width=.32 \textwidth,clip]{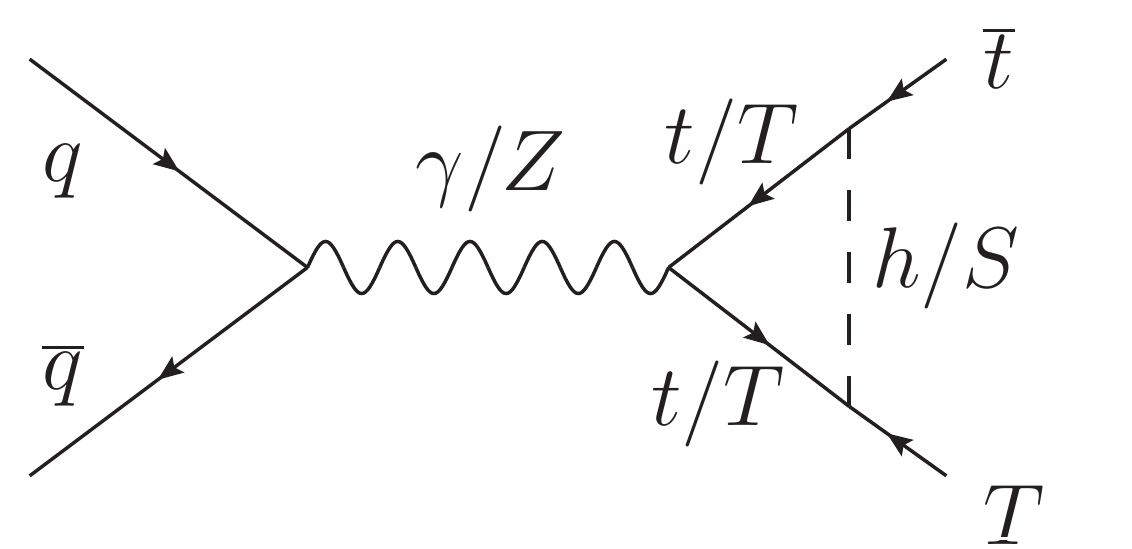}\label{fig:FeynLoopZ}}\\
\subfigure[]{\includegraphics[width=0.32\textwidth,clip]{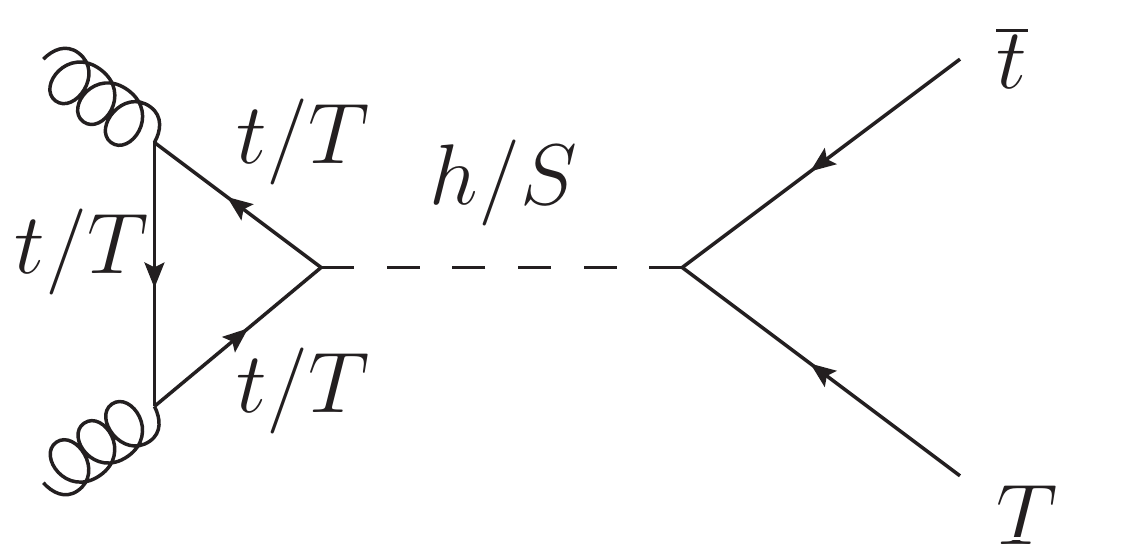}\label{fig:Feyn_SRes}}
\subfigure[]{\includegraphics[width=0.32\textwidth,clip]{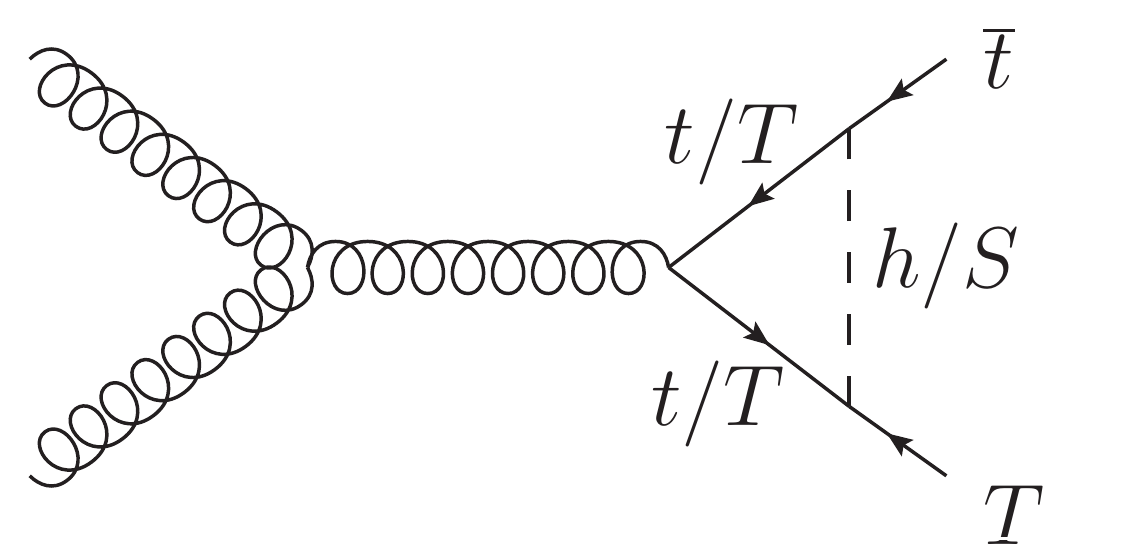}}
\subfigure[]{\includegraphics[width=0.32\textwidth,clip]{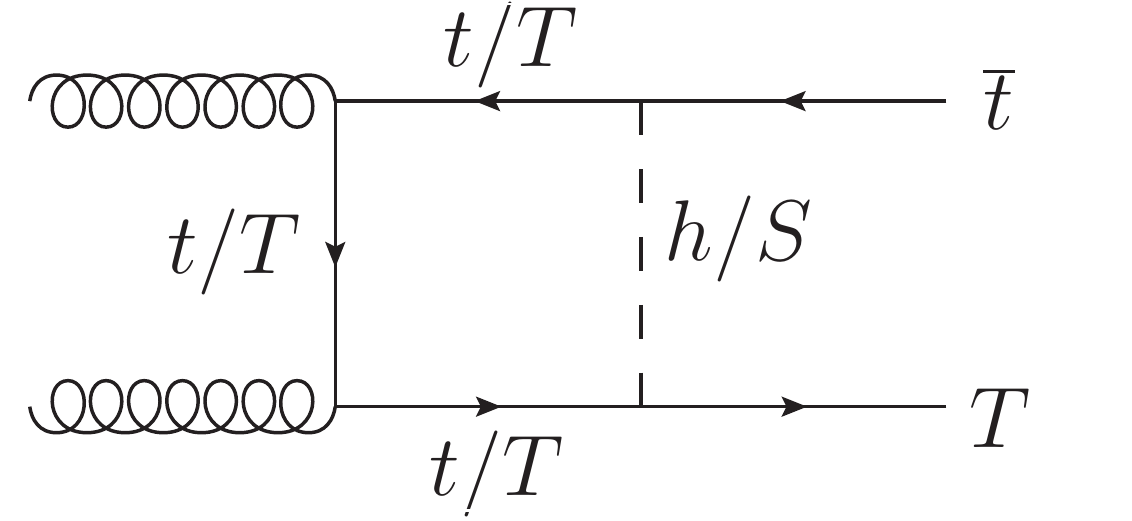}}\\
\subfigure[]{\includegraphics[width=0.22\textwidth,clip]{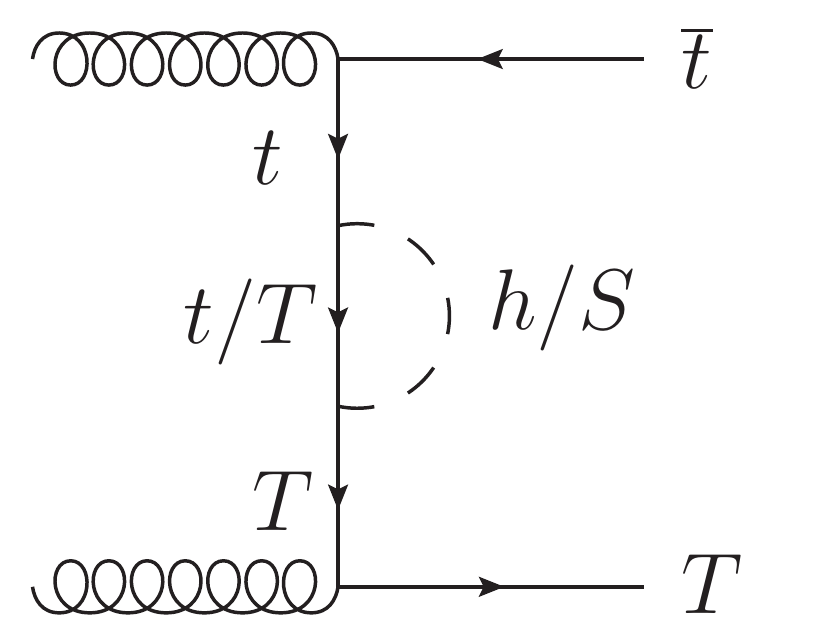}}
\subfigure[]{\includegraphics[width=0.32\textwidth,clip]{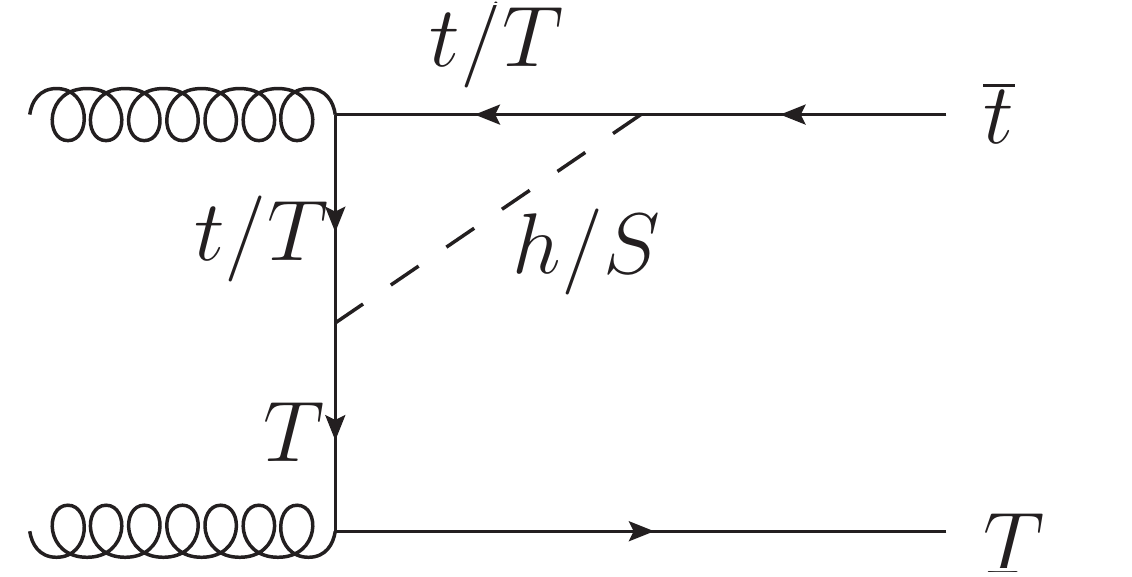}}
\subfigure[]{\includegraphics[width=0.32\textwidth,clip]{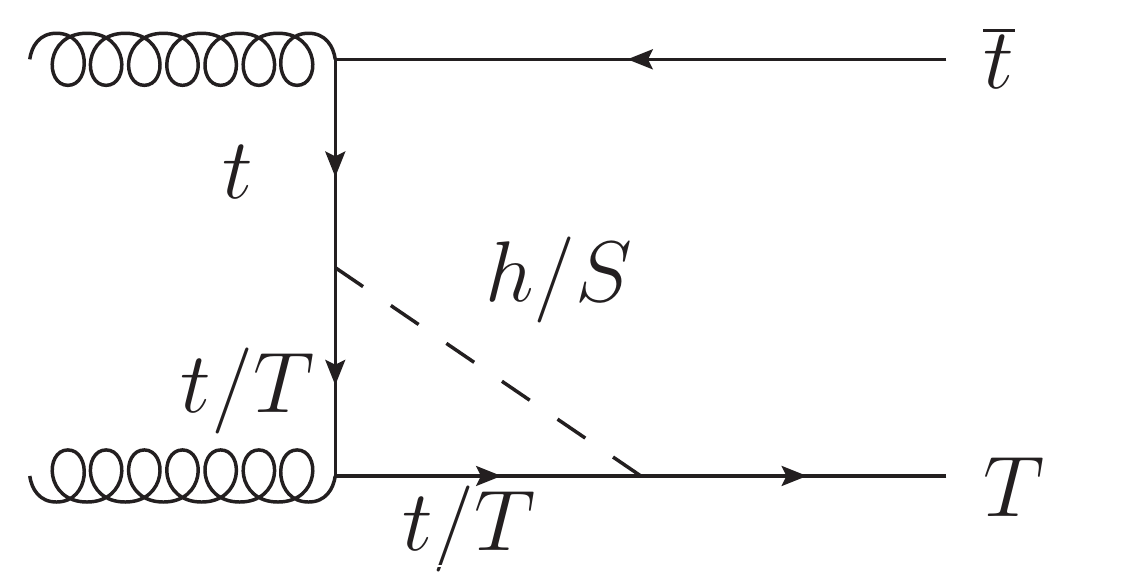}\label{fig:FeynLoopFin}}
\end{center}
\caption{ Representative Feynman diagrams for single production of $T$ in association with a top quark for (a) tree level through an $s$-channel $Z$, (b-c) quark-antiquark initial state, and (d-i) gluon fusion.  We have not shown the conjugate process, counterterms, off-diagonal self energies of the external top and top partner, or any loops with internal Goldstone bosons, $Z$, or $W^\pm$.
\label{fig:Prod_Feyn2}}
\end{figure} 

There are many possible production channels for top partners.  Figure~\ref{fig:Prod_Feyn1} shows the classic tree level mechanisms: (a-c) top partner pair production ($T\overline{T}$), (d,e) top partner plus jet production ($T$+jet), and (f,g) top partner plus $W^\pm$ production ($TW$)\footnote{There is also $q\overline{q}'\rightarrow bT$ through an $s$-channel $W$ boson.  However, due to being $s$-channel, this mode is suppressed relative to the other single top production channels as-well-as still being suppressed by $\sin\theta_L$.}.  We collectively refer to final states with a single $T$ produced in association with a SM particle as single top partner production.  Although top partner pair production is dominant for much of the parameter region, single top partner plus jet production can become important for very massive $T$ despite the $b$-quark pdf suppression~\cite{Dolan:2016eki,Han:2003wu,Han:2005ru,Willenbrock:1986cr,DeSimone:2012fs,Aguilar-Saavedra:2013qpa,Backovic:2015bca,Liu:2016jho}.  This is mainly due to two effects: the gluon pdf drops precipitously at high mass suppressing the $T\overline{T}$ rate and top partner pair production starts saturating the available LHC phase space at high energies. This can be clearly seen in Fig.~\ref{fig:prod_mtp}, which compares the cross sections of various top partner production modes as a function of the top partner mass $m_T$.  At the $\sqrt{S}=14$~TeV LHC and for a mixing angle of $\sin\theta_L=0.15$, Fig.~\ref{fig:prod_mtp_sth15}, the $T$+jet production becomes larger than that of top partner pair production at a mass around $m_T\sim700$~GeV and $TW$ production is comparable to $T\overline{T}$ production for $m_T\sim2.5$~TeV.

 However, for the simplest model where the SM is augmented by a single $SU(2)_L$ singlet top partner, single top partner production relies on the $b-W-T$ coupling.  This coupling is proportional to the to the $T-t$ mixing angle $\sin\theta_L$, as can be seen in Eq.~(\ref{eq:gauge}).  Hence, the production cross section is proportional to $\sin^2\theta_L$ and vanishes as the mixing angle goes to zero.  In fact, as shown in Fig.~\ref{fig:prod_mtp_sth01}, $T\overline{T}$ always dominates $T$+jet and $TW$ for $\sin\theta_L=0.01$ at the $\sqrt{S}=14$~TeV LHC for all masses shown.

In the model presented in Sec.~\ref{sec:model}, in addition to the production modes in Fig.~\ref{fig:Prod_Feyn1}, the flavor-off diagonal couplings between the new scalar, top partner, and top quark introduces new loop level production mode: top partner production in association with a top quark ($Tt$).  Representative Feynman diagrams with flavor off-diagonal scalar couplings for this process are show in Fig.~\ref{fig:Prod_Feyn2}.  We do not show the conjugate process; counterterm diagrams; diagrams with Goldstone bosons, $Z$s, or $W^\pm$s internal to the loop; or external off-diagonal self-energy diagrams between the top quark and top partner.  However, these are included in the calculation.    Although $Tt$ production is allowed at tree level for non-zero $\sin\theta_L$, as with $T$+jet and $TW$ production, the tree level $Tt$ cross section is proportional to $\sin^2\theta_L$.  Hence, it vanishes as $\sin\theta_L$ vanishes.  However, the $S-t-T$ and $S-T-T$ couplings do not vanish for $\sin\theta_L=0$ and the loop level production survives.

For $m_S>m_T+m_t$, it is possible for the scalar to resonantly decay into the top partner and top through the diagram in Fig.~\ref{fig:Feyn_SRes}.  If the scalar is not too heavy, it will be possible to produce it and look for this decay channel at the LHC.  This type of signal has been much studied and searched for~\cite{Fichet:2016xpw,Greco:2014aza,Sirunyan:2017bfa,Dobrescu:2009vz,Barcelo:2011wu,Bini:2011zb}.  However, if the scalar is too heavy it will not be possible to produce it at the LHC.  In this case, the EFT presented in Sec.~\ref{sec:EFT} is relevant.  As can be clearly seen, the production cross section is then suppressed by $1/m_S^{4}$.  For large scalar masses it is always negligible compared to pair production.  Hence, for our discussion of $T$ production we focus on the scenario where $m_S<m_T+m_t$.  However, as we will see, for $m_S\gg m_T$ the decay channels of the top partner are interesting and present a new phenomenology.

The importance of $Tt$ production can be seen in Fig.~\ref{fig:prod_mtp}.  For $m_S=200$~GeV and both $\sin\theta_L=0.15$ and $\sin\theta_L=0.01$, at the $\sqrt{S}=14$~TeV LHC the top partner plus top production rate is greater than that of top partner pair production for $m_T\gtrsim 1.5$~TeV.  While for $\sin\theta_L=0.15$, $T$+jet production is consistently larger than $Tt$ production, the situation changes drastically for smaller mixing angles. As can be seen by comparing Figs.~\ref{fig:prod_mtp_sth15} and~\ref{fig:prod_mtp_sth01}, the $Tt$ rate does not greatly decrease as $\sin\theta_L$ becomes small.  Figure~\ref{fig:prod_mtp_sth01} shows that $Tt$ is the dominant single top partner production mechanism for small mixing angles.

\begin{figure}[tb]
\begin{center}
\subfigure[]{\includegraphics[width=0.48\textwidth,clip]{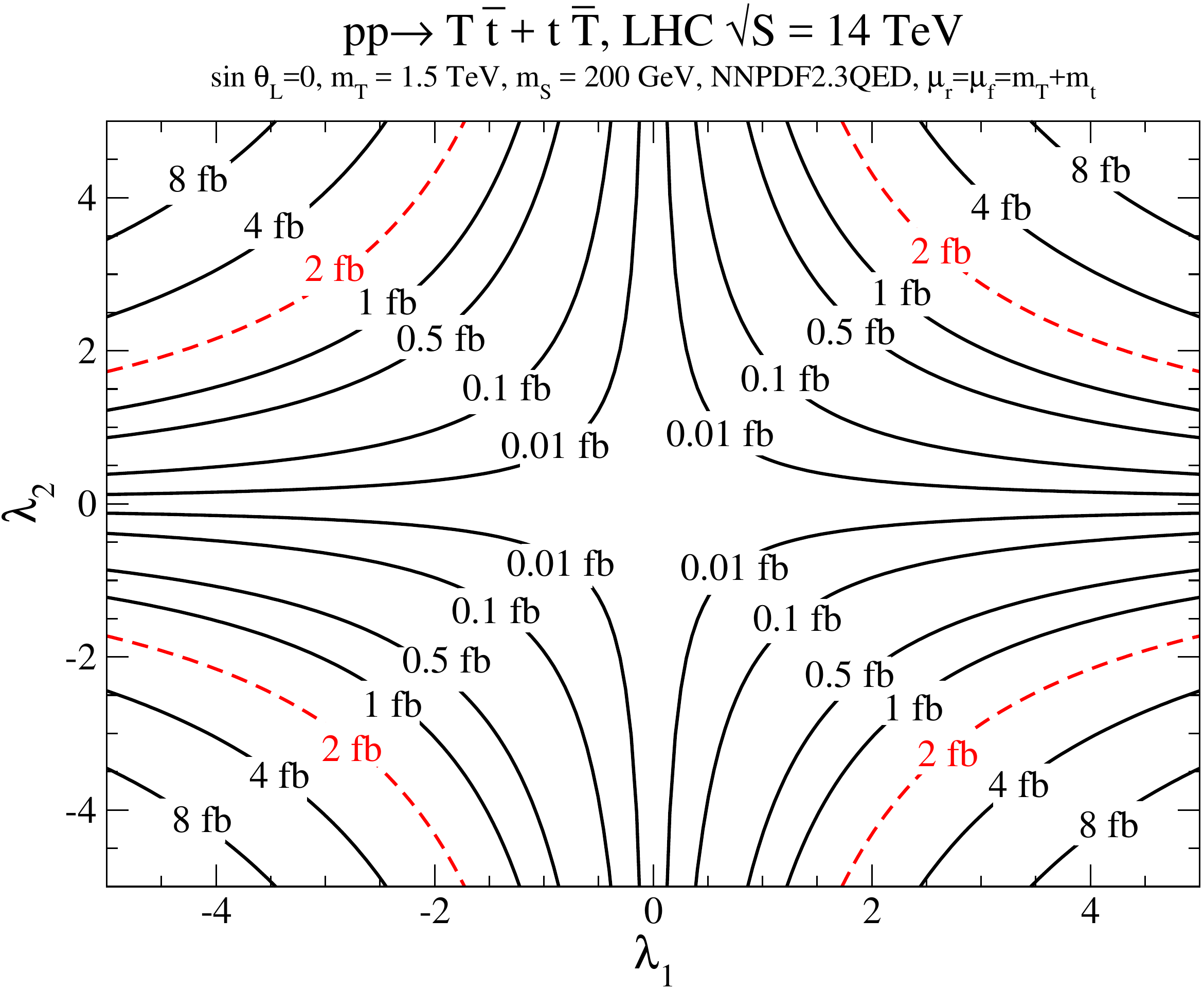}\label{fig:prod_lambda_15TeV}}
\subfigure[]{\includegraphics[width=0.48\textwidth,clip]{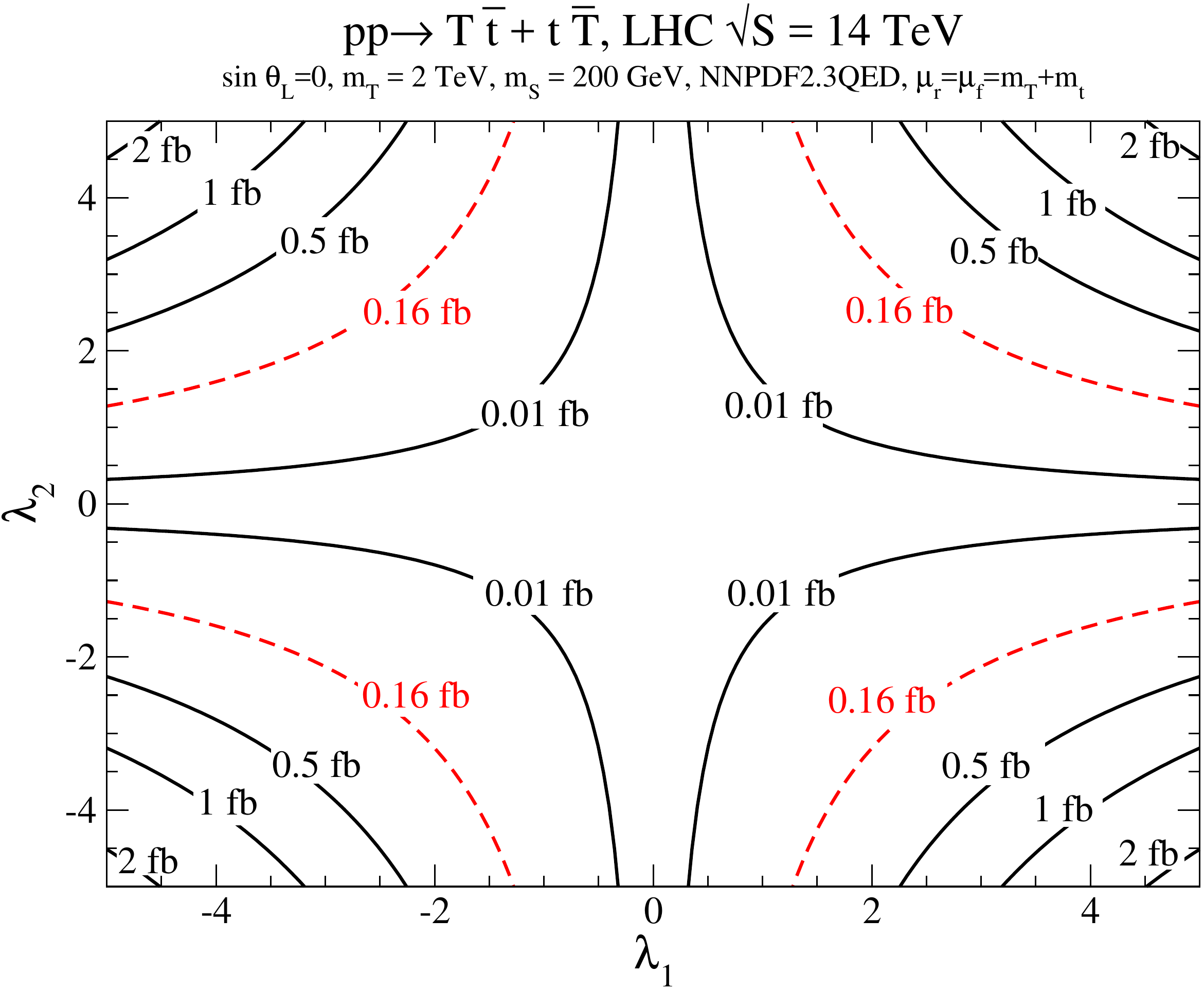}\label{fig:prod_lambda_20TeV}}
\end{center}
\caption{\label{fig:prod_lambda}
Contours of production cross sections at the $\sqrt{S}=14$~TeV LHC for top partner production in association with a top quark in the $\lambda_2-\lambda_1$ coupling constant plane for top partner masses (a) $m_T=1.5$~TeV and (b) $m_T=2$~TeV.  The red dashed lines indicate the $T\overline{T}$ production cross section.  The other parameters are $m_S=200$~GeV and $\sin\theta_L=0$.  The factorization and renormalization scales are set at $\mu_r=\mu_f=m_T+m_t$ for $Tt$ production and $\mu_r=\mu_f=2m_T$ for the $T\overline{T}$ cross section.} 
\end{figure}

In Fig.~\ref{fig:prod_lambda} we show contours of LHC cross sections for top partner plus top production in $\lambda_1-\lambda_2$ plane in the zero mixing $\sin\theta_L=0$ limit.  This is presented for both $m_T=1.5$~TeV, Fig.~\ref{fig:prod_lambda_15TeV}, and $m_T=2$~TeV, Fig.~\ref{fig:prod_lambda_20TeV}.  The shapes of the contours can be understood by noting that in the zero mixing limit the $Tt$ production rate is proportional to the coupling constants squared:
\begin{eqnarray}
\sigma(pp\rightarrow T\overline{t}+t\overline{T})\propto \lambda_1^2\lambda_2^2\label{eq:propto}
\end{eqnarray}
Hence, contours of constant cross section correspond to $|\lambda_1|\propto |\lambda_2|^{-1}$.  For comparison, we also show the top production pair production rate (red dashed lines).  As can be seen, there is a significant amount of parameter space for which the $Tt$ rate dominates $T\overline{T}$.  Using the simple relation in Eq.~(\ref{eq:propto}), for $\sin\theta_L=0$ and $m_S=200$~GeV, we find that at the $\sqrt{S}=14$~TeV LHC the $Tt$ cross section is larger than the $T\overline{T}$ cross section for 
\begin{eqnarray}
\sqrt{|\lambda_1\lambda_2|}&\gtrsim&2.9\quad{\rm for}~m_T=1.5~{\rm GeV~and}\nonumber\\
\sqrt{|\lambda_1\lambda_2|}&\gtrsim&2.5\quad{\rm for}~m_T=2~{\rm TeV.}
\end{eqnarray}

\begin{figure}[tb]
\begin{center}
\subfigure[]{\includegraphics[width=0.48\textwidth,clip]{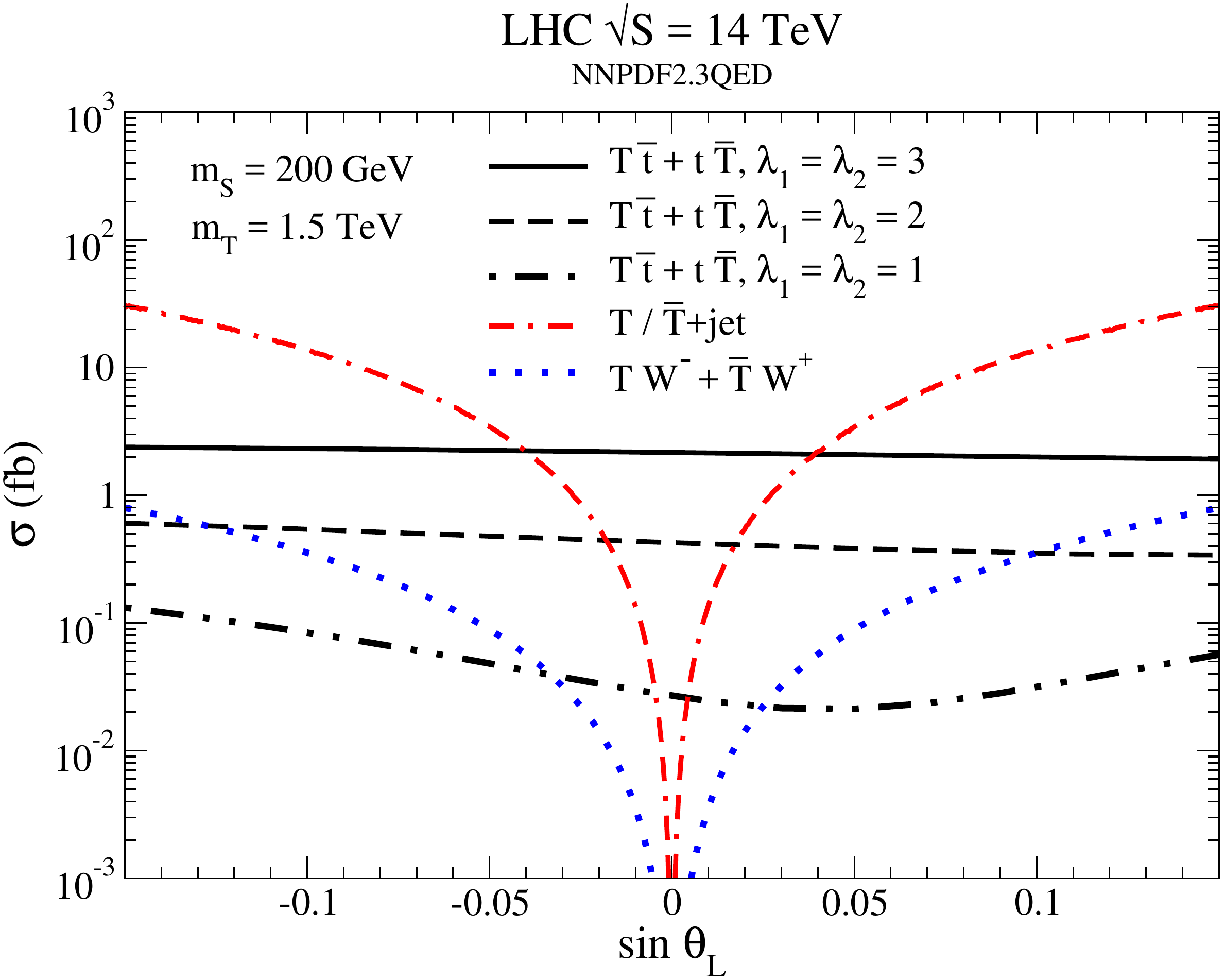}\label{fig:prod_sth}}
\subfigure[]{\includegraphics[width=0.48\textwidth,clip]{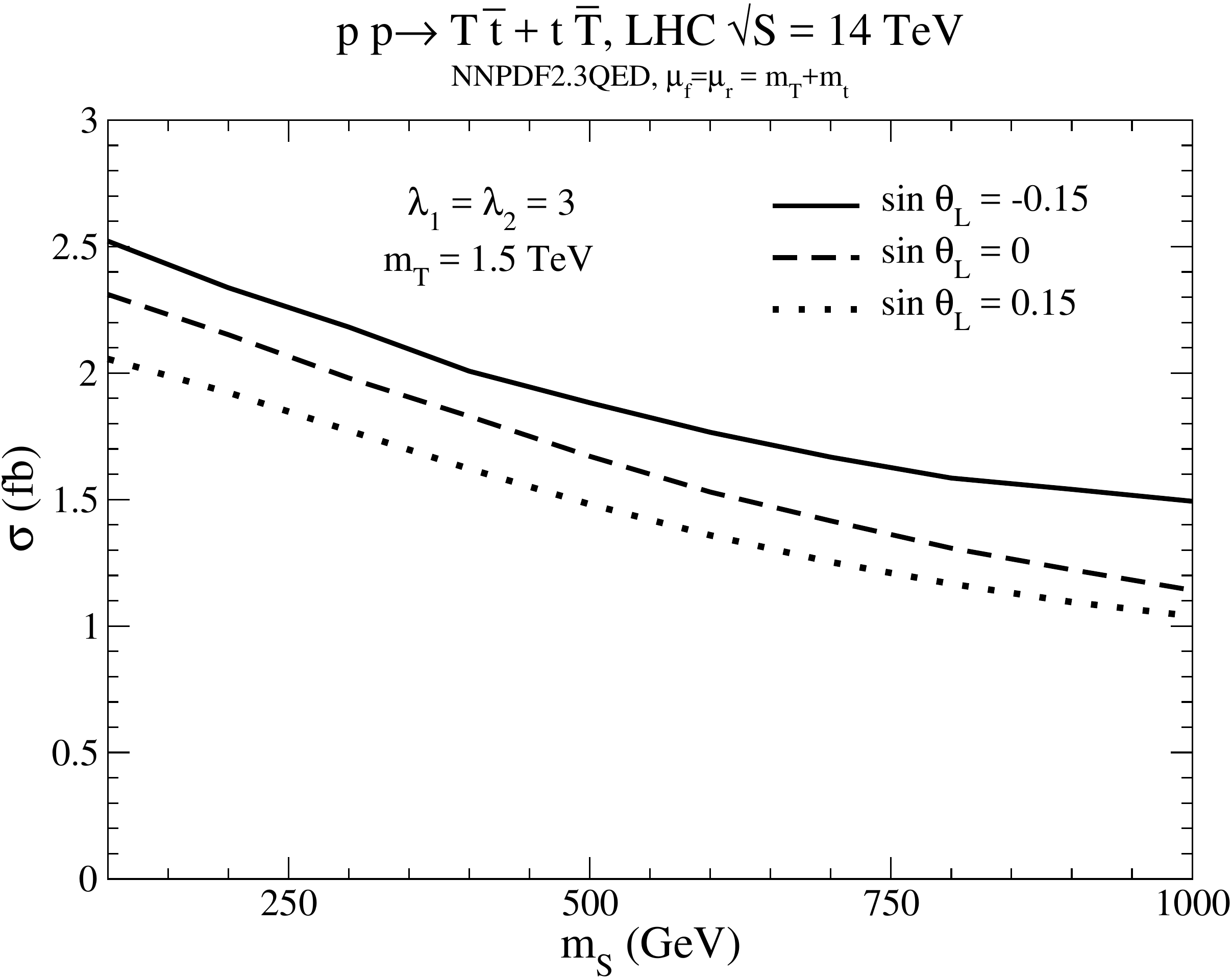}\label{fig:prod_msc}\label{fig:prodmsc}}
\end{center}
\caption{(a) Production cross sections at the $\sqrt{S}=14$~TeV LHC as a function of top partner-top mixing angle $\sin\theta_L$ for top partner production in association with a top quark for (black solid) $\lambda_{1,2}=3$, (black dashed) $\lambda_{1,2}=2$, and (black dash-dot-dot) $\lambda_{1,2}=1$.  Also shown are (red dash-dash-dot) top partner plus jet production and (blue dotted) top partner plus $W^\pm$ production.  The top partner mass is $m_T=1.5$~TeV and the scalar mass $m_S=200$~GeV.  (b) Cross sections of top partner production in association with a top partner as a function of the scalar mass $m_S$ at the $\sqrt{S}=14$~TeV LHC.  The parameters are set at a top partner mass $m_T=1.5$~TeV, coupling constants $\lambda_{1,2}=3$, and mixing angle (solid) $\sin\theta_L=-0.15$, (dashed) $\sin\theta_L=0$, and (dotted) $\sin\theta_L=0.15$.  For both (a) and (b) the factorization, $\mu_f$, and renormalization, $\mu_r$, scales are set to the sum of the final state particle masses.\label{fig:prod_mscsth} }
\end{figure}

In Fig.~\ref{fig:prod_sth} we show various single top partner production rates as a function of $\sin\theta_L$ for $m_T=1.5$~TeV and $m_S=200$~GeV.  At small mixing angles all the single top partner rates vanish except $Tt$.  It is expected that searches for $T$+jet production will limit $\sin\theta_L\lesssim 0.02-0.06$~\cite{Backovic:2015bca}.  Hence, this is the parameter region where top partner plus top production is most important.  Also, for larger coupling constants $\lambda_{1,2}$, the $Tt$ rate has little dependence on $\sin\theta_L$, while for smaller $\lambda_{1,2}$ the dependence is stronger.  This can be understood by noting that for non-zero mixing angles, loop diagrams involving the Higgs, $Z$ boson, $W$ boson, and Goldstone bosons contribute to $Tt$.  For smaller $\lambda_{1,2}$ these contributions can compete with the scalar $S$ contributions, introducing more $\sin\theta_L$ dependence.  For larger $\lambda_{1,2}$, the scalar $S$ loops always dominate and mixing angle dependence is milder.

The dependence of the $Tt$ production rate on the scalar mass is shown in Fig.~\ref{fig:prod_msc} for $\lambda_{1,2}=3$ and $m_T=1.5$~TeV.  For all mixing angles, the cross section is larger for smaller scalar mass.  The dependence of the cross section on $m_S$ does not change greatly for different $\sin\theta_L$.  
 
\subsubsection{Summary}
\label{sec:prodsumm} 

Table~\ref{tab:prodsum} summarizes the results of top partner production with $m_S=200$~GeV.  The left column gives parameter regions for which $Tt$ production is the dominant single top partner production mode.  The right column gives parameter regions for which $Tt$ production dominates $T\overline{T}$ double production.  For small mixing angles, $Tt$ is the dominant single top production mode, while $Tt$ production dominates $T\overline{T}$ production at large $m_T$.  Also, $Tt$ production is maximized for smaller scalar masses.
\begin{table}[tb]
\begin{center}
\makebox[\textwidth][c]{
\begin{tabular}{|c|c|}\hline
Single Production & Double Production\\\hline\hline
$m_T=1.5$~TeV, $\lambda_{1,2}=3$, $|\sin\theta_L|\lesssim 0.04$ & $m_T\gtrsim 1.5$~TeV, $\lambda_{1,2}=3$, $\sin\theta_L=0$\\
$m_T=1.5$~TeV, $\lambda_{1,2}=2$, $|\sin\theta_L|\lesssim 0.02$ & $m_T=1.5$~TeV, $\sqrt{|\lambda_1\lambda_2|}\gtrsim2.9$, $\sin\theta_L=0$\\
$m_T=1.5$~TeV, $\lambda_{1,2}=1$, $|\sin\theta_L|\lesssim 0.005$& $m_T=2$~TeV, $\sqrt{|\lambda_1\lambda_2|}\gtrsim2.5$, $\sin\theta_L=0$ \\\hline 
\end{tabular}}
\end{center}
\caption{Parameter regions for which $Tt$ production in the dominant single production mode (left column) and $Tt$ production is greater than $T\overline{T}$ pair production (right column) with scalar mass $m_S=200$~GeV. \label{tab:prodsum}}
\end{table}

\subsection{Top Partner Decay Channels}
\label{sec:dec} 

\begin{figure}[tb]
\begin{center}
\subfigure[]{\includegraphics[width=0.28\textwidth,clip]{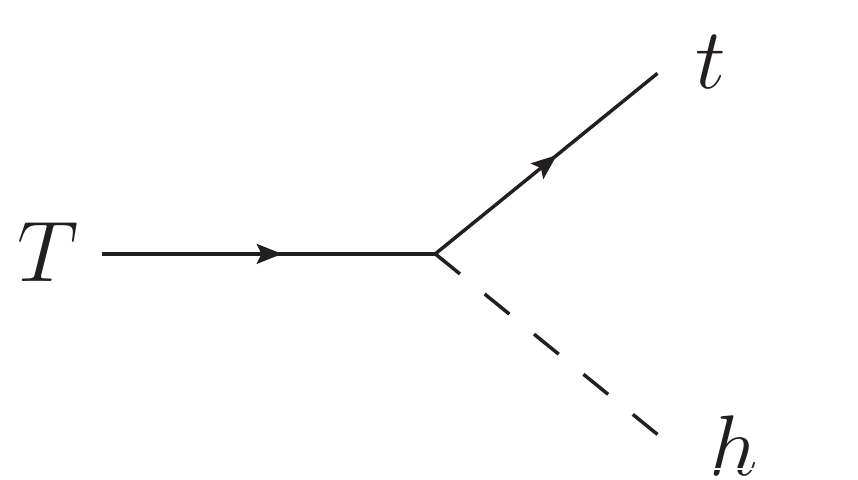}\label{fig:Tth}}
\subfigure[]{\includegraphics[width=0.28\textwidth,clip]{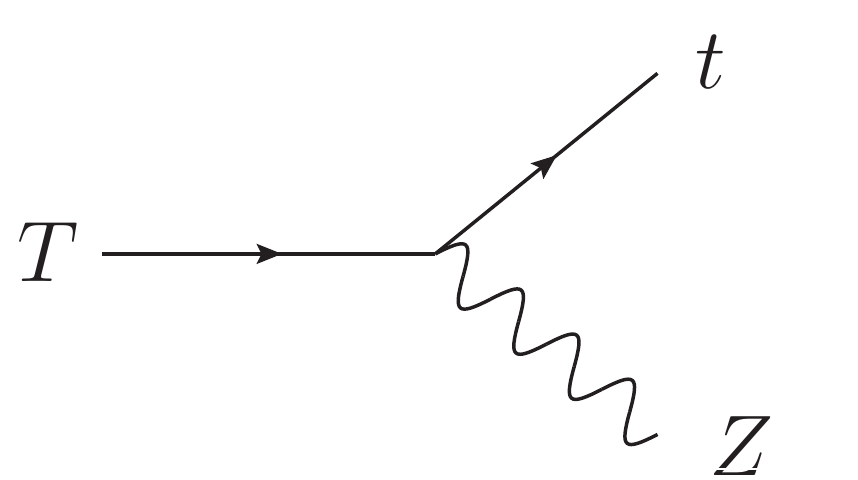}\label{fig:TtZ}}
\subfigure[]{\includegraphics[width=0.28\textwidth,clip]{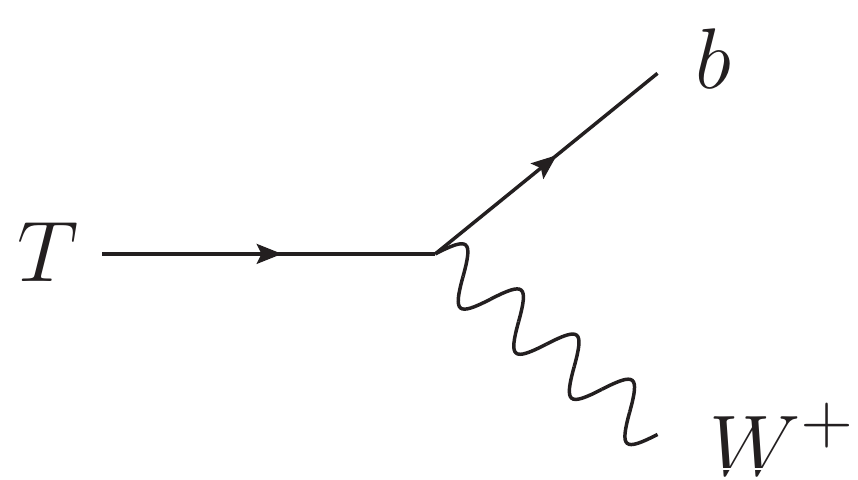}\label{fig:TbW}}
\subfigure[]{\includegraphics[width=0.28\textwidth,clip]{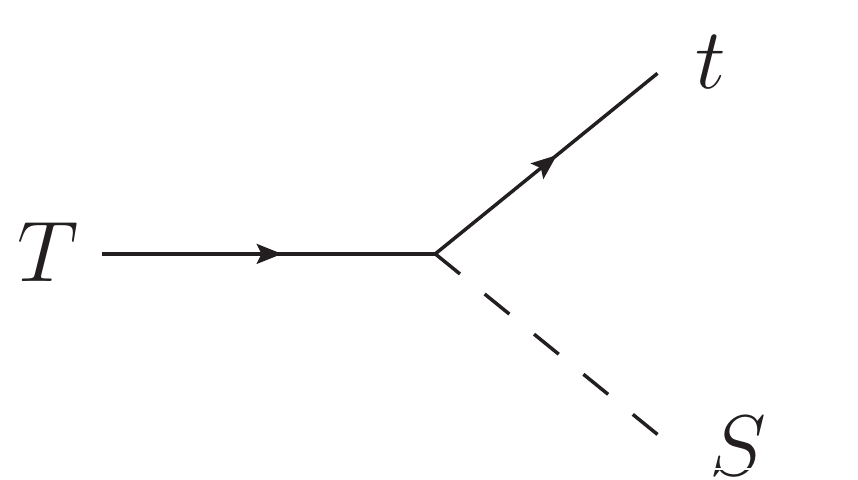}\label{fig:TtS}}
\subfigure[]{\includegraphics[width=0.28\textwidth,clip]{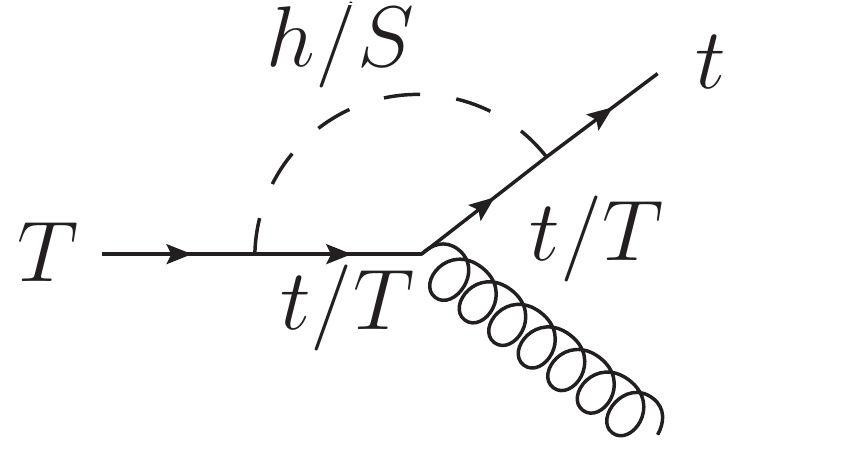}\label{fig:Ttg}}
\subfigure[]{\includegraphics[width=0.28\textwidth,clip]{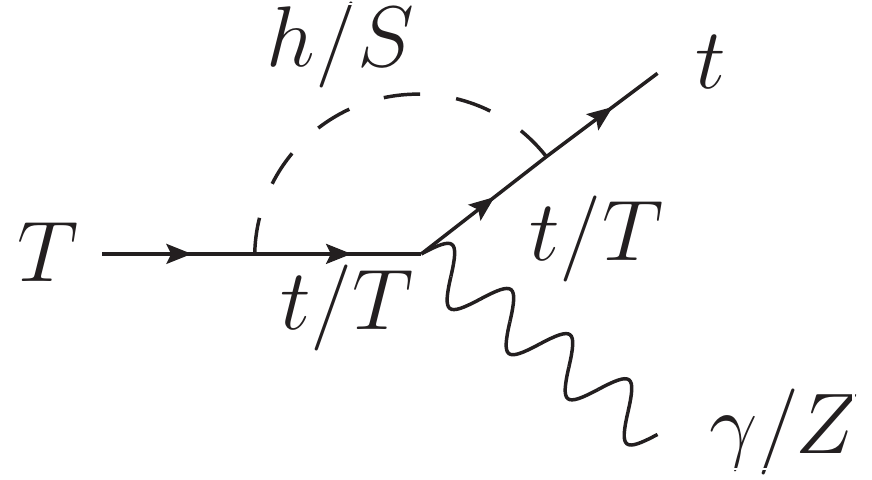}\label{fig:TtZgam}}
\subfigure[]{\includegraphics[width=0.28\textwidth,clip]{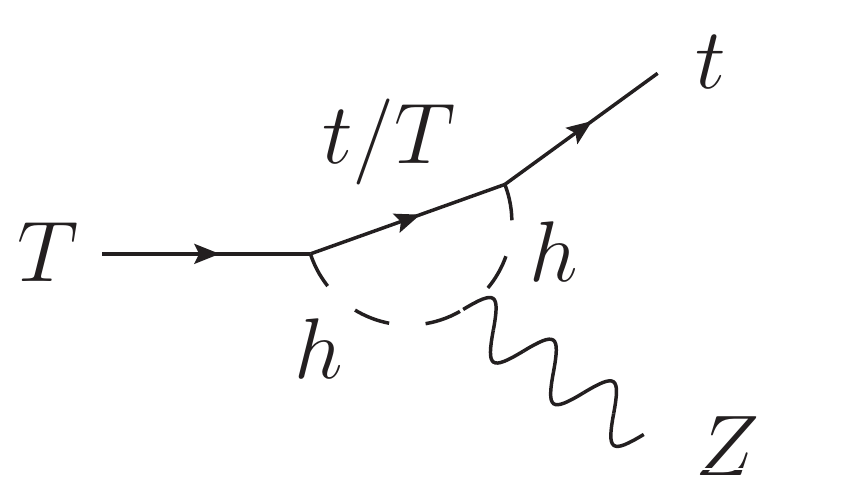}\label{fig:TtZLoop}}
\end{center}
\caption{\label{fig:TdecayAG} Representative Feynman diagrams for top partner decay at (a-d) tree level and (e-g) one-loop level.  Not shown here are external self-energies, external vacuum polarizations, or loops with Goldstone, $W$, or $Z$ bosons.} 
\end{figure}

Figure~\ref{fig:TdecayAG} shows representative Feynman diagrams for top partner decays.  Searches for top partners typically rely on the $T\rightarrow th$, $T\rightarrow tZ$, and $T\rightarrow bW$ decays~\cite{Aaboud:2017zfn, Aaboud:2017qpr,Sirunyan:2017pks, Sirunyan:2017usq} as shown in Fig.~\ref{fig:Tth}-\ref{fig:TbW}.  However, in the model presented in Sec.~\ref{sec:model}, new top partner decay modes are available.  For small enough scalar masses, $m_S+m_t<m_T$, there is a new tree level decay $T\rightarrow tS$, as shown in Fig.~\ref{fig:TtS}.  Additionally, there are possible loop level decays, shown in Figs.~\ref{fig:Ttg}-\ref{fig:TtZLoop}, that are important when the $T\rightarrow tS$ channel is kinematically forbidden and, as we will see, for sufficiently small mixing angle $\sin\theta_L$. These new decay channels can change search strategies for fermionic top partners.   

Again, in the loop diagrams in Fig.~\ref{fig:TdecayAG}, we do not show external self-energies, external vacuum polarizations, loops with $Z$ bosons, loops with $W^\pm$ bosons, or loops with Goldstone bosons, although they are included in the calculations.   Additionally, we only consider the leading contributions to each decay channel.  Hence, $T\rightarrow t\gamma$ and $T\rightarrow tg$ are calculated at one loop.  For $T\rightarrow th$ and $T\rightarrow bW$, we only consider tree level decays.  While there are loop corrections, for $T\rightarrow th$ they will be dependent on $\lambda_{tt}^S$, $\lambda_{TT}^h$, $\lambda_{tT}^h$, $\lambda_{Tt}^h$, $W-b-T$, or the $Z-t-T$ couplings in Eqs.~(\ref{eq:htt}-\ref{eq:gauge}), all of which are proportional to $\sin\theta_L$.  There is also a diagram proportional to the $S-S-h$ coupling, which we have set to zero.  The $S-S-h$ coupling is not technically natural and can be generated through a loop of top quarks and top partners.  However, this would would be a two loop contribution to $T\rightarrow th$ and can be safely ignored.  Similarly, $T\rightarrow bW$ loop level contributions depend on $\lambda_{tT}^h$, $\lambda_{Tt}^h$, $\lambda_{TT}^h$, $Z-t-T$ or $W-T-b$ couplings in Eqs.~(\ref{eq:htt}-\ref{eq:gauge}), which are also proportional to $\sin\theta_L$.  Since both the tree and one loop level contributions to $T\rightarrow th$ and $T\rightarrow Wb$ are always proportional to $\sin\theta_L$, we expect the tree level diagrams to dominate throughout parameter space and we do not calculate the loop contributions to these decays.  The decay $T\rightarrow tZ$ is more complicated.  The tree level component vanishes as $\sin\theta_L\rightarrow 0$.  However, the loop contribution in Fig.~\ref{fig:TtZgam} does not vanish as $\sin\theta_L\rightarrow 0$ since it depends on the $\lambda_{Tt}^S$, $\lambda_{TT}^S$, and the $Z-T-T$ couplings in Eqs.~(\ref{eq:Stt},\ref{eq:gauge}) which are non-zero for $\sin\theta_L= 0$.  Hence, we calculate tree and loop level diagrams to $T\rightarrow tZ$ so that the dominant contributions are included for all $\sin\theta_L$.  

\subsubsection{$m_S<m_T-m_t$}
\label{sec:declowms}

\begin{figure}[tb]
\begin{center}
\subfigure[]{\includegraphics[width=0.45\textwidth,clip]{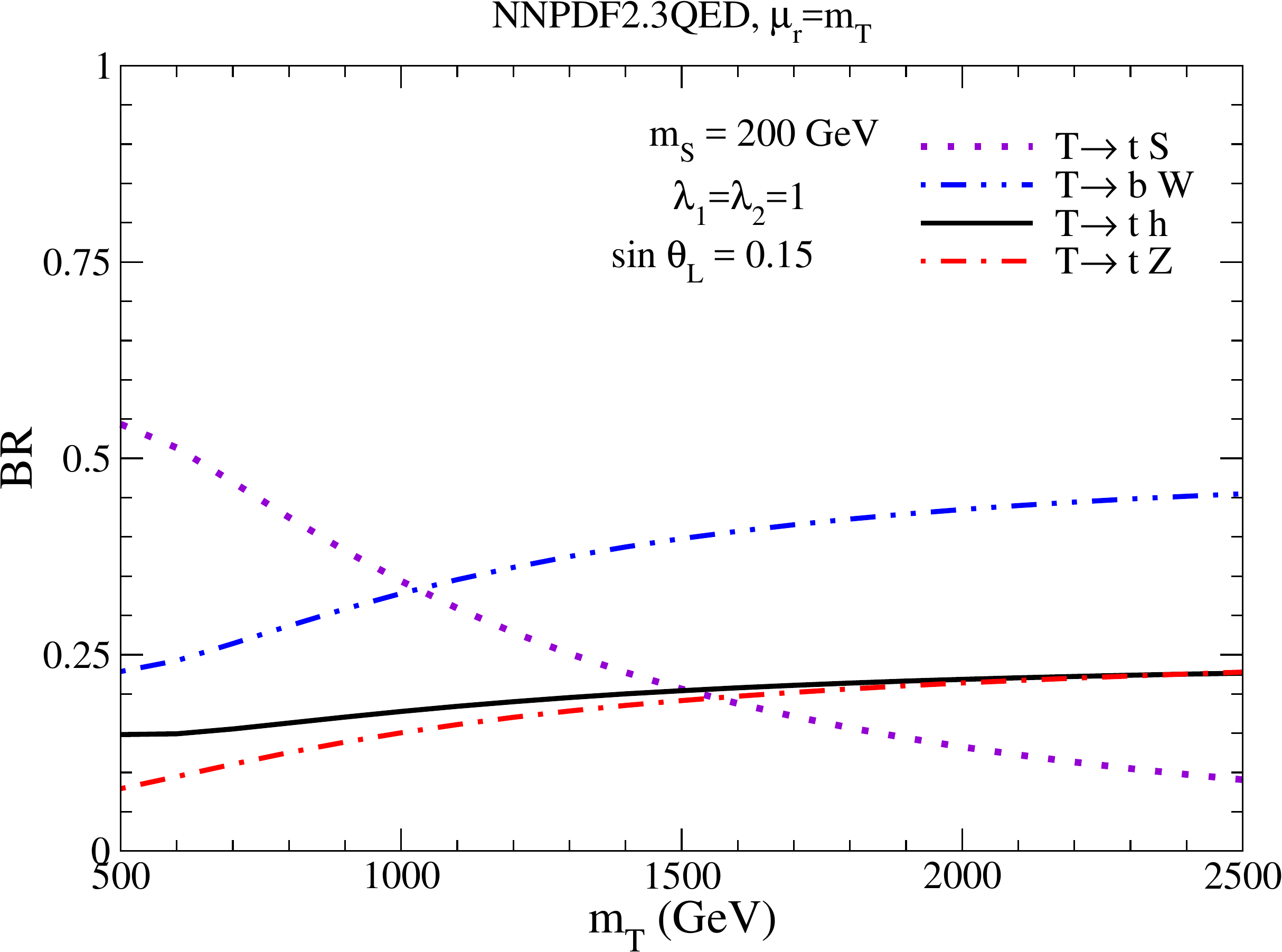}\label{fig:BR_mtp_ms200GeV}}
\subfigure[]{\includegraphics[width=0.45\textwidth,clip]{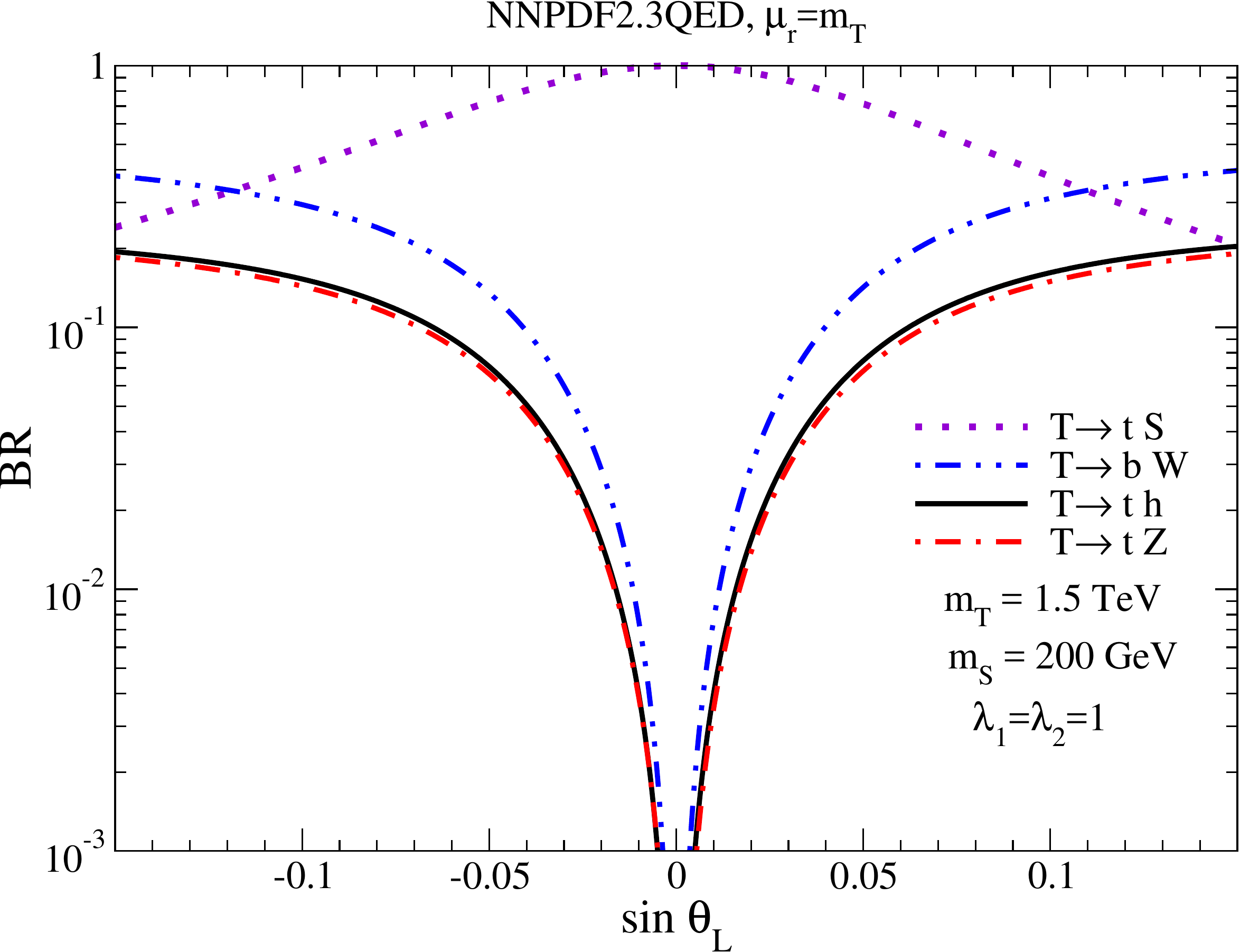}\label{fig:BR_sth_ms200GeV}}
\end{center}
\caption{\label{fig:BR_ms200GeV} Branching ratios of the top partner as a function of (a) top partner mass and (b) mixing angle $\sin\theta_L$ for (violet dotted) $T\rightarrow tS$, (blue dash-dot-dot) $T\rightarrow bW$, (black solid) $T\rightarrow th$, and (red dash-dot) $T\rightarrow tZ$.  For (a) the remaining parameters are set at $\lambda_{1,2}=1$, $m_S=200$~GeV, and $\sin\theta_L=0.15$; while for (b) we have $m_T=1.5$~TeV, $m_S=200$~GeV, and $\lambda_{1,2}=1$.}
\end{figure}

We first consider $m_S+m_t<m_T$, where $T\rightarrow tS$ is available.  Figure~\ref{fig:BR_ms200GeV} illustrates how the branching ratios of $T$ depend on the top partner mass and mixing angle.  Although not shown, we also calculated the branching ratios of $T\rightarrow t\gamma$ and $T\rightarrow tg$, but they are negligible in this regime.  As can be seen in Fig.~\ref{fig:BR_mtp_ms200GeV}, for smaller top partner masses the $T\rightarrow tS$ decay dominates while for larger masses the standard decays $T\rightarrow bW$, $T\rightarrow tZ$, and $T\rightarrow th$ dominate.  This can be understood by considering the partial widths in the $m_T\gg v,m_S$ limit and counting $\sin\theta_L\sim m_t/m_T\sim v/m_T$:
\begin{eqnarray}
\Gamma(T\rightarrow tS)&=&\frac{m_T}{32\pi}\left\{\left[\left(\lambda_{tT}^S\right)^2+\left(\lambda_{Tt}^S\right)^2\right](1+x_t^2-x_S^2)+4 x_t \lambda_{tT}^S\lambda_{Tt}^S\right\}\lambda^{1/2}(1,x_t^2,x_S^2)\nonumber\\
&\xrightarrow[m_T\gg v,m_S]{}&\frac{\lambda_1^2 m_T}{32\pi}\\
\Gamma(T\rightarrow th)&=&\frac{m_T}{32\pi}\left\{\left[\left(\lambda_{tT}^h\right)^2+\left(\lambda_{Tt}^h\right)^2\right](1+x_t^2-x_h^2)+4x_t\lambda_{tT}^h\lambda_{Tt}^h\right\}\lambda^{1/2}(1,x_t^2,x_h^2)\nonumber\\
&\xrightarrow[m_T\gg v]{}&\frac{m_T^3\sin^2\theta_L}{32\pi\,v^2} \label{eq:Tth}\\
\Gamma(T\rightarrow tZ)&=&\frac{g^2 m_T}{512\pi\,c_W^2}\sin^22\theta_L\left\{1+x_t^2-2\,x_Z^2+\frac{(1-x_t^2)^2}{x_Z^2}\right\}\lambda^{1/2}(1,x_t^2,x_Z^2)\nonumber\\
&\xrightarrow[m_T\gg v]{}&\frac{m_T^3\sin^2\theta_L}{32 \pi\,v^2} \label{eq:TtZ}\\
\Gamma(T\rightarrow bW)&=&\frac{g^2 m_T}{64\pi}\sin^2\theta_L\left\{1+x_b^2-2\,x_W^2+\frac{(1-x_b^2)^2}{x_W^2}\right\}\lambda^{1/2}(1,x_b^2,x_W^2)\nonumber\\
&\xrightarrow[m_T\gg v]{}&\frac{m_T^3\sin^2\theta_L}{16\pi\,v^2},\label{eq:TbW}
\end{eqnarray}   
where $x_i=m_i/m_T$ and $\lambda(x,y,z)=(x-y-z)^2-4\,y\,z$.  The decays into SM final states dominate since the partial widths scale as $m_T^3$ and the partial width $\Gamma(T\rightarrow tS)$ grows as $m_T$.   This can be understood via the Goldstone Equivalence Theorem and that the $W,Z,h$ couplings are proportional to mass for very heavy $m_T$.  In fact, the SM decays obey the expectation ${\rm BR}(T\rightarrow bW)\approx 2\,{\rm BR}(T\rightarrow tZ)\approx 2\,{\rm BR}(T\rightarrow th)\approx 50\%$\footnote{The exact pattern of branching ratios depends on the model and the quantum numbers of the top partner.  For a composite model in which the top partner predominantly decays into a top and scalar for the heavier top partners see Ref.~\cite{Serra:2015xfa}.  Also, see Ref.~\cite{Freitas:2017afm} for a discussion of the decay of level-2 KK fermions in a universal extra dimensional model, which do not obey the expected pattern from the Goldstone Equivalence Theorem.  Other exotic decay patterns can be found in Refs.~\cite{Dobrescu:2016pda,Bizot:2018tds}.}.  Of course, allowing mixing between the scalar and Higgs boson will slightly complicate this scenario, since the two mass eigenstate scalars will be superpositions of the gauge singlet scalar and Higgs boson.  Since the scalar would then have a component of the Higgs doublet,  the parametric dependence of the widths is
\begin{eqnarray}
\Gamma(T\rightarrow tS)&\sim& \sin^2\theta\,\sin^2\theta_L \frac{m_T^3}{v^2}+\cos^2\theta\,\lambda_1^2m_T\quad{\rm and}\nonumber\\
\Gamma(T\rightarrow th)&\sim& \cos^2\theta\sin^2\theta_L\frac{m_T^3}{v^2}+\sin^2\theta\,\lambda_1^2m_T,\label{eq:nonzeroscal}
\end{eqnarray}
 where $\theta$ is the scalar mixing angle and $m_T\gg v,m_S$.  Then $\Gamma(T\rightarrow tS)$ has a component that grows as $m_T^3$, but is suppressed by the scalar mixing angle.  
For simplicity, we are focusing on the scenario where the scalar mixing angle is zero, although, as is clear from Eq.~(\ref{eq:nonzeroscal}), the precise phenomenology will change for non-zero scalar mixing~\cite{Dolan:2016eki}.  However, while the branching ratios of the top partner can change, there are no new decay channels for the top partner in the non-zero scalar mixing scenario.  Hence, we still capture the major phenomenological aspects of this model.

Precisely when the SM final states dominate will also depend on the coupling constants $\lambda_{1,2}$ and mixing angle $\sin\theta_L$.  In Fig.~\ref{fig:BR_sth_ms200GeV} we show the dependence of the top partner branching ratios on $\sin\theta_L$ for $m_S=200$~GeV and $m_T=1.5$~TeV.  For larger mixing angles $|\sin\theta_L|\gtrsim 0.1-0.12$, the decay into bottom quark and $W$ dominates.  However, as expected for $\sin\theta_L\sim 0$ the branching ratio of $T\rightarrow tS$ is very nearly 100\% since the other tree level decay modes vanish.  

\begin{figure}[t]
\begin{center}
\begin{tabular}{cccc}
\hspace{0.0cm} \includegraphics[width=.35 \textwidth]{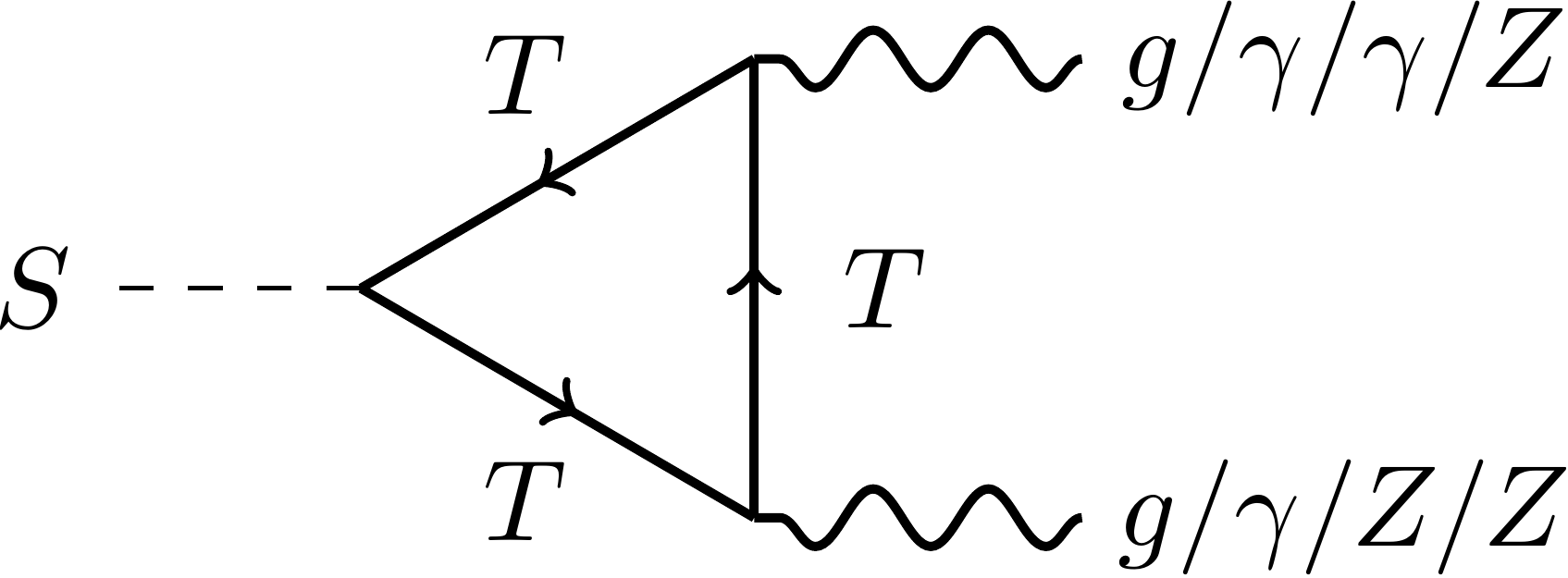} 
\end{tabular}
\end{center}
\caption{The scalar $S$ decays into $g g$, $\gamma \gamma$, $\gamma Z$ and $Z Z$ final states through $T$ loop in the zero-mixing limit ($\sin \theta_L \rightarrow 0$).}
\label{fig:Sdecay}
\end{figure}

When $T\rightarrow tS$ dominates, search strategies will strongly depend on the decay of the scalar.  If $S$ is allowed to have non-negligible mixing with the Higgs boson, the scalar will decay like a heavy Higgs boson.  That is, we would expect $S\rightarrow WW$, $S\rightarrow ZZ$, $S\rightarrow tt$, and $S\rightarrow hh$ to be tree level and dominate when they are allowed~\cite{Dolan:2016eki}.  If $\lambda_2=0$, as-well-as $a_1=b_1=b_3=0$ in Eq.~(\ref{eq:pot}), it is possible to apply a $Z_2$ symmetry on the top partner and scalar, $T\rightarrow -T$ and $S\rightarrow -S$, while the SM fields are even $SM\rightarrow SM$.   The only available decay mode is then $T\rightarrow tS$ and the scalar $S$ is a possible dark matter candidate.  Top partners are then pair produced and the signal is $T\bar{T}\rightarrow t\bar{t}+\slashed{E}_T$~\cite{Chala:2018qdf}.  The scenario we consider has no $Z_2$ symmetry and sets the scalar-Higgs mixing to zero.  Then the only decay channels available to the scalar $S$ are through loops of top quarks and top partners.  Depending on the precise mass of the scalar, the decays $S\rightarrow WW$, $S\rightarrow ZZ$, $S\rightarrow \gamma\gamma$, $S\rightarrow Z\gamma$, $S\rightarrow hh$, and $S\rightarrow gg$ will be possible.   The $S\rightarrow hh$ and $S\rightarrow WW$ decay rates are mixing angle suppressed since all contributing diagrams are dependent on $\lambda_{Tt}^h$, $\lambda_{tT}^h$, $\lambda_{TT}^h$, $\lambda_{tt}^S$, or $W-T-b$ in Eqs.~(\ref{eq:htt}-\ref{eq:gauge}).  Hence, in the $\sin\theta_L=0$ limit, the scalar $S$ decays to neutral gauge bosons, as shown in Fig.~\ref{fig:Sdecay}, and the branching ratios are determined by the gauge couplings.  Then $S\rightarrow gg$ and $T\rightarrow tS\rightarrow tgg$ are by far the dominate decay modes.

\subsubsection{$m_S>m_T-m_t$ and Long Lived Top Partners}
\label{sec:dechighms} 

\begin{figure}[tb]
\begin{center}
\subfigure[]{\includegraphics[width=0.45\textwidth,clip]{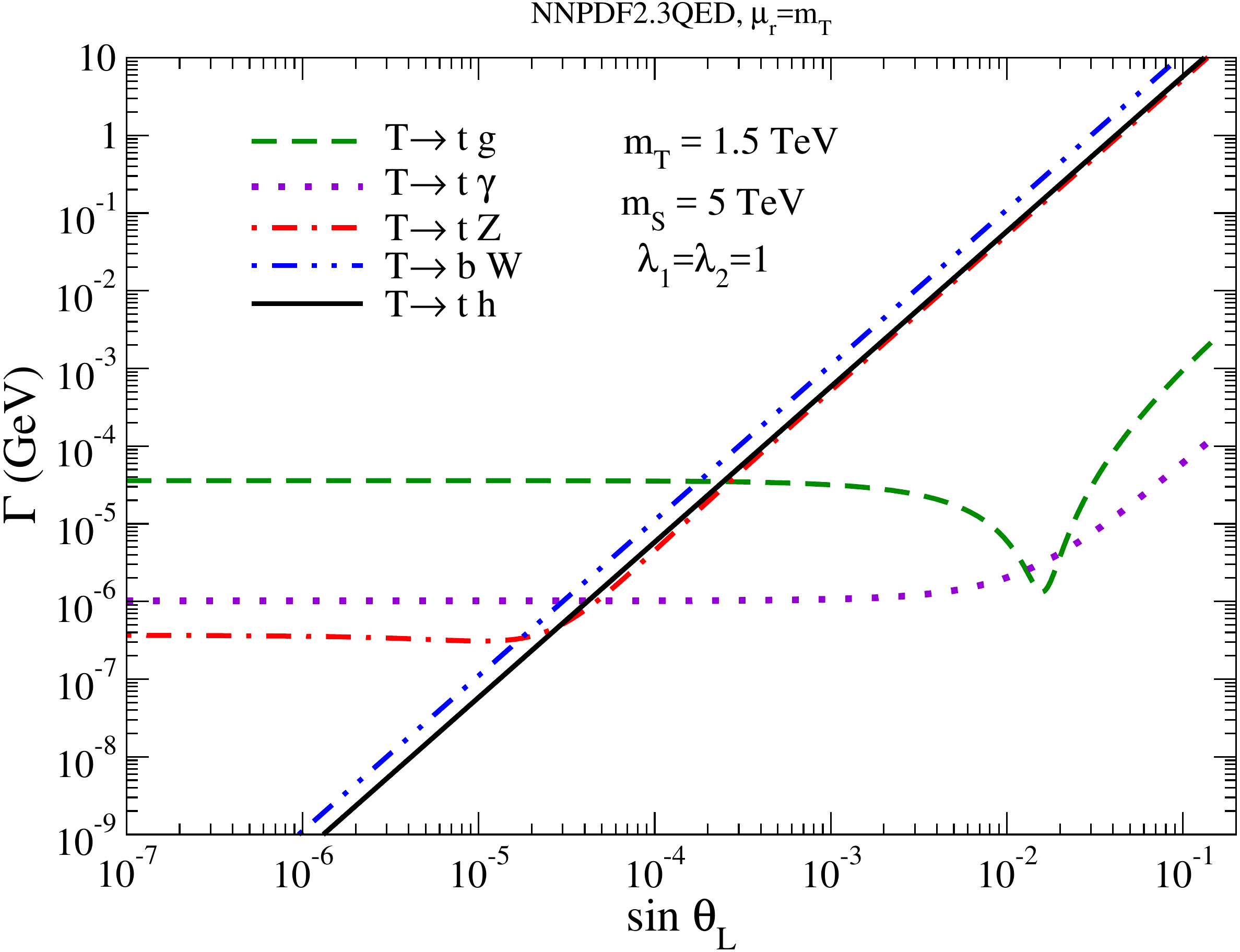}\label{fig:width_sth_ms5TeV}}
\subfigure[]{\includegraphics[width=0.45\textwidth,clip]{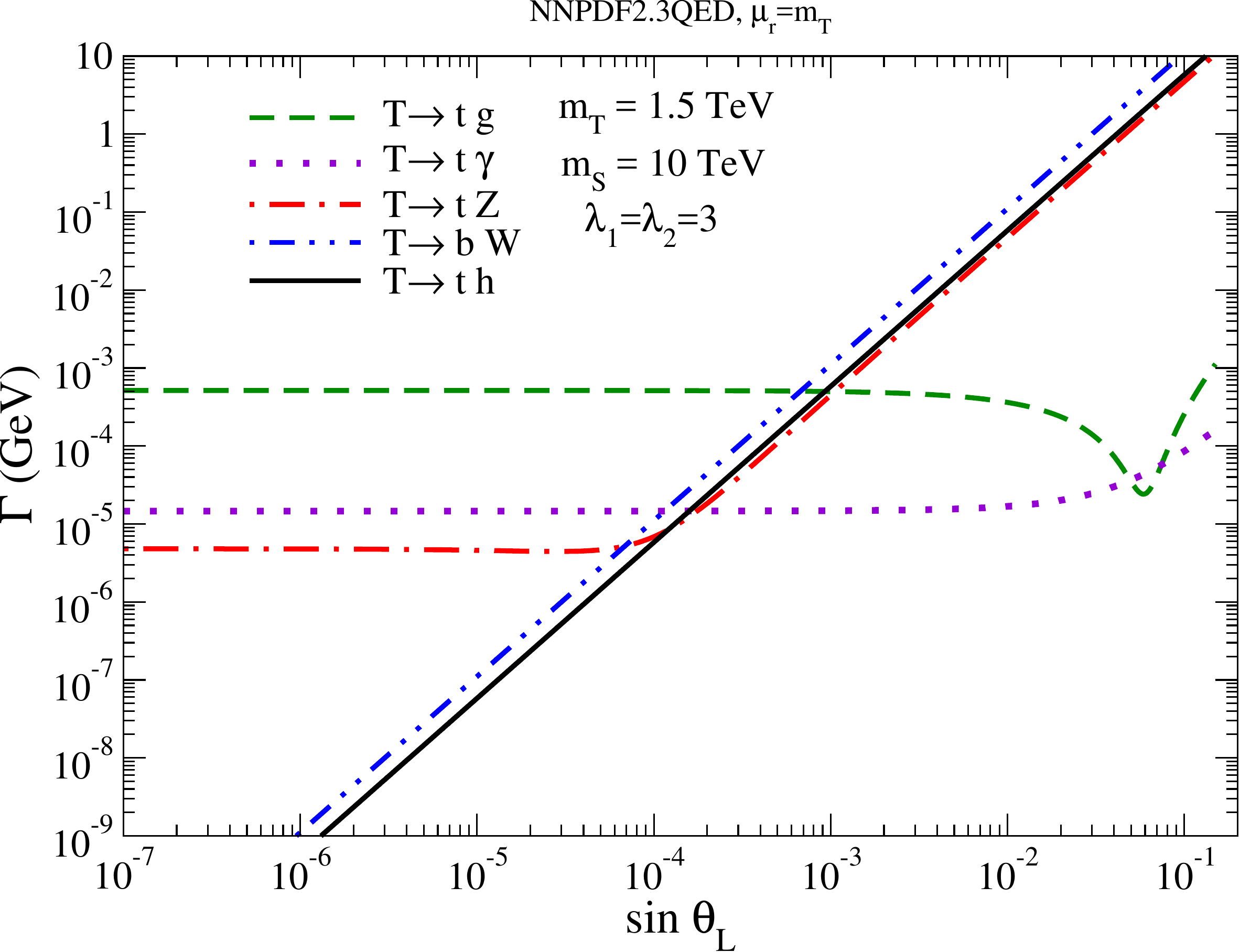}\label{fig:width_sth_ms10TeV}}
\subfigure[]{\includegraphics[width=0.45\textwidth,clip]{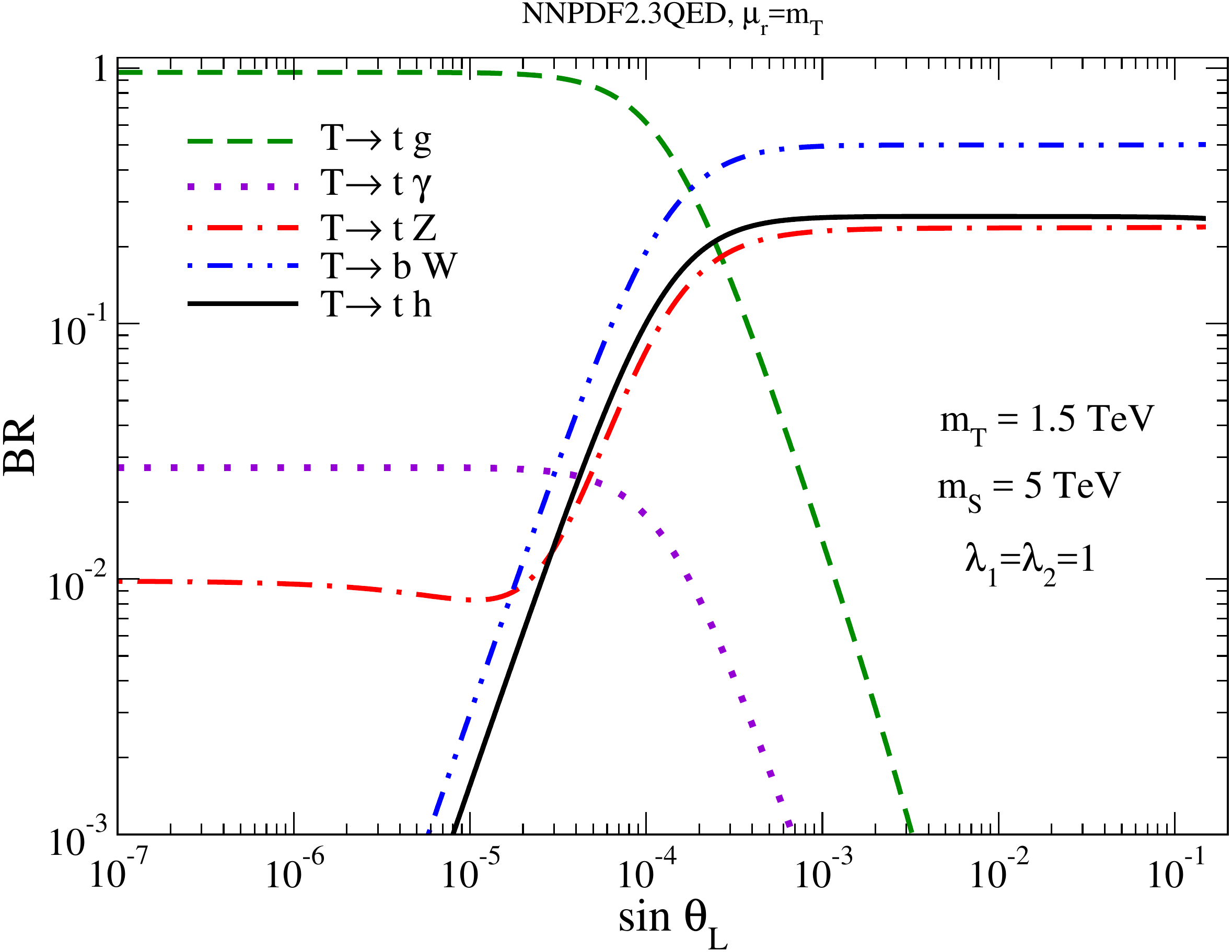}\label{fig:BR_sth_ms5TeV}}
\subfigure[]{\includegraphics[width=0.45\textwidth,clip]{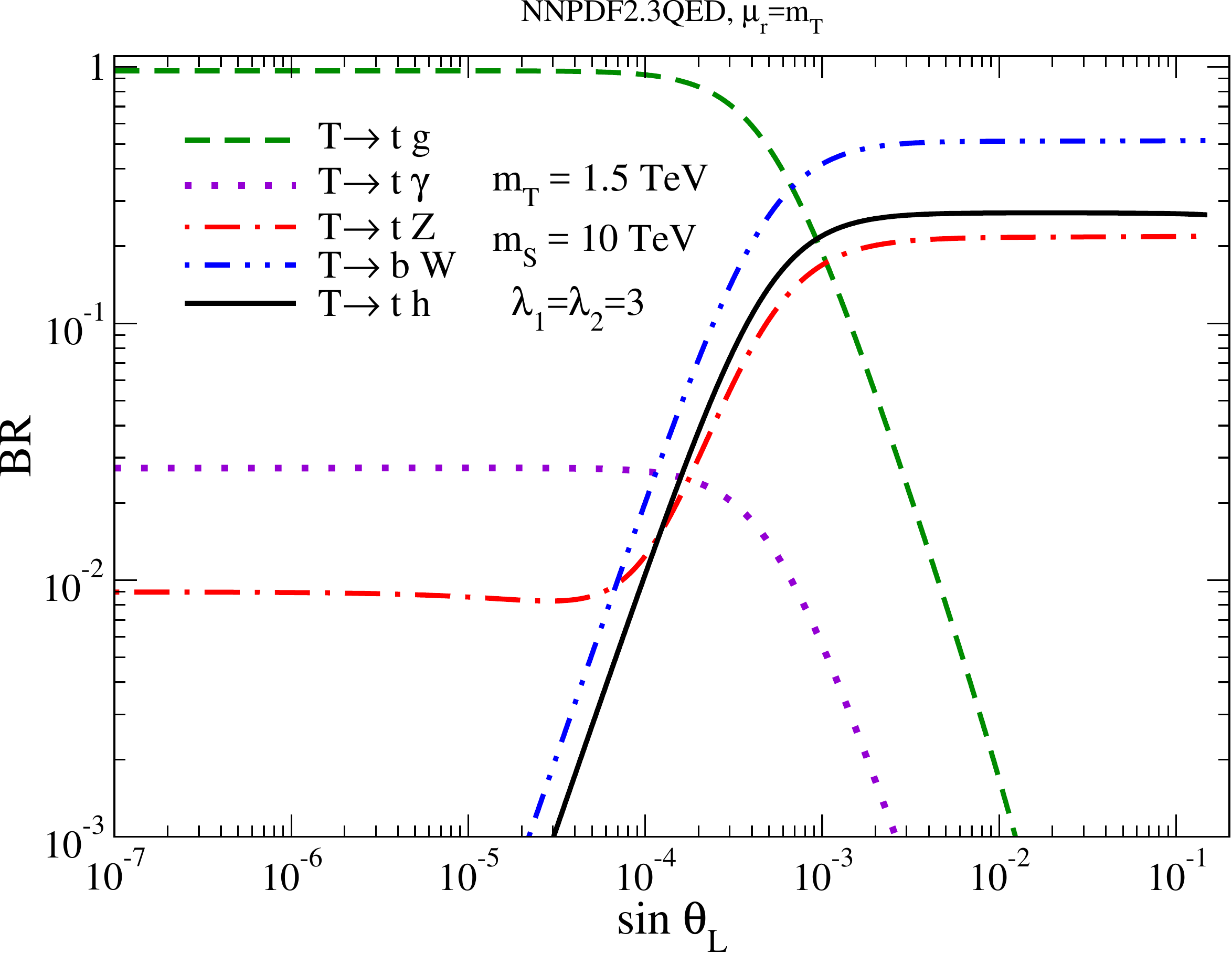}\label{fig:BR_sth_ms10TeV}}
\end{center}
\caption{\label{fig:decheavyS1} (a,b) Partial widths and (c,d) branching ratios of the top partner for $m_S>m_T-m_t$ for (green dashed) $T\rightarrow tg$, (violet dotted) $T\rightarrow t\gamma$, (red dash-dot) $T\rightarrow t Z$, (blue dash-dot-dot) $T\rightarrow bW$, and (black solid) $T\rightarrow th$.  These are shown for top partner mass $m_T=1.5$ TeV.  The remaining parameters are (a,c) $m_S=5$~TeV, $\lambda_{1,2}=1$ and (b,d) $m_S=10$~TeV, $\lambda_{1,2}=3$.  The renormalization scale is set to the top partner mass.}
\end{figure}

In Fig.~\ref{fig:decheavyS1} we show the (a,b) total widths and (c,d) branching ratios as a function of mixing angle $\sin\theta_L$ for scalar masses larger than top partner mass.  The top partner mass is $m_T=1.5$~TeV, the scalar masses and couplings are (a,c) $m_S=5$~TeV, $\lambda_{1,2}=1$ and (b,d) $m_S=10$~TeV, $\lambda_{1,2}=3$.  For mixing angle $\sin\theta_L\gtrsim 10^{-4}-10^{-3}$, the tree level decays dominate and the partial widths are independent of the scalar mass and couplings.  For $\sin\theta_L\lesssim 10^{-3}-10^{-4}$ the loop level decay $T\rightarrow tg$ is the main mode.  To determine the relative importance of the loop contributions it is useful to look at the $T\rightarrow tZ$ decay channel, since it is the only one for which we include both loop and tree level contributions.   At $\sin\theta_L\sim 10^{-5}-10^{-4}$ there is a clear transition between a $\Gamma(T\rightarrow tZ)\sim \sin^2\theta_L$ dependence expected at tree level and a width $\Gamma(T\rightarrow tZ)$ that is relatively independent of $\sin\theta_L$.  This is the passage between tree level and loop level dominance in $T\rightarrow tZ$.

The dependence of $\Gamma(T\rightarrow tg),\Gamma(T\rightarrow t\gamma),$ and $\Gamma(T\rightarrow tZ)$ on the model parameters at small angles can be understood by noting that for $\sin\theta_L\approx0$, $m_T\gg v$, and $m_S\gg m_T$, the EFT is Eq.~(\ref{eq:EFT}) is valid.  In this EFT, the partial widths are 
\begin{eqnarray}
\Gamma_{\rm EFT}(T\rightarrow tg)&\approx&\frac{\alpha_s\,C_F\,\lambda_1^2\lambda_2^2}{576\,\pi^4}\frac{m_T^5}{m_S^4}\left(1+\frac{3}{4}\log\frac{m_T^2}{m_S^2}\right)^2\label{eq:Ttg}\\
\Gamma_{\rm EFT}(T\rightarrow t\gamma)&\approx&\frac{\alpha\,\lambda_1^2\lambda_2^2}{1296\,\pi^4}\frac{m_T^5}{m_S^4}\left(1+\frac{3}{4}\log\frac{m_T^2}{m_S^2}\right)^2\label{eq:Ttgam}\\
\Gamma_{\rm EFT}(T\rightarrow tZ)&\approx&\frac{\alpha\,\lambda_1^2\lambda_2^2\, s_W^2}{1296\,\pi^4\, c_W^2}\frac{m_T^5}{m_S^4}\left(1+\frac{3}{4}\log\frac{m_T^2}{m_S^2}\right)^2.\label{eq:TtZloop}
\end{eqnarray}
Hence, the partial widths are independent of $\sin\theta_L$ and all have the same parametric dependence on the top partner mass, scalar mass, and couplings $\lambda_{1,2}$.

The branching ratios of the top partner in the $m_S>m_T-m_t$ regime are shown in Figs.~\ref{fig:BR_sth_ms5TeV} and~\ref{fig:BR_sth_ms10TeV}.  Although the values of the partial widths depend on the precise model parameters, the branching ratios are largely independent of model parameters for larger $\sin\theta_L\gtrsim 10^{-3}$ or small $\sin\theta_L\lesssim 10^{-5}$.   The behavior of the branching ratios for $\sin\theta_L\sim 10^{-5}-10^{-3}$ depends on the relative dominance of the tree level and loop level contribution, and hence the model parameters, as discussed above.  For mixing angles $\sin\theta_L\gtrsim 10^{-3}$, the tree level decays in to SM EW bosons $T\rightarrow bW$, $T\rightarrow tZ$ and $T\rightarrow th$ dominate and they obey the expected relation ${\rm BR}(T\rightarrow bW)\approx2\,{\rm BR}(T\rightarrow tZ)\approx 2\,{\rm BR}(T\rightarrow th)\approx 50\%$.  This can be understood by noting that in the heavy top partner regime, these partial widths only depend on $\sin\theta_L$ and $m_T$ and this dependence cancels in the ratios of the widths  in Eqs.~(\ref{eq:Tth}-\ref{eq:TbW}).  

For $\sin\theta_L\lesssim 10^{-4}$ the decay $T\rightarrow tg$ dominates, while for $\sin\theta_L\lesssim 10^{-5}$ all loop level decays dominate and the branching ratios are approximately independent of the model parameters.  For the EFT, since the partial widths in Eqs.~(\ref{eq:Ttg}-\ref{eq:TtZloop}) have the same parametric dependence, the branching ratios are independent of couplings $\lambda_{1,2}$ and masses $m_T,m_S$.  Hence, the branching ratios are largely determined by the gauge coupling constants and weak mixing angle.  Then the decay $T\rightarrow tg$ is by far the dominate mode due to the strong coupling constant.  
There are additional corrections from the Higgs vev to Eq.~(\ref{eq:EFT}) arising from neglected dimension-6 operators of the form
\begin{eqnarray}
\overline{Q_L}\sigma^{\mu\nu}\widetilde{\Phi} T_R B_{\mu\nu},\quad \overline{Q_L}\sigma^{\mu\nu}\widetilde{\Phi} T^A T_R G_{\mu\nu}^A.\label{eq:EFT1}
\end{eqnarray}
This can explain the $\mathcal{O}(10\%)$ differences between the branching ratios at $m_S=5$~TeV and $10$~TeV, as observed in Figs.~\ref{fig:BR_sth_ms5TeV} and~\ref{fig:BR_sth_ms10TeV}.
\begin{figure}[tb]
\begin{center}
\subfigure[]{\includegraphics[width=0.49\textwidth,clip]{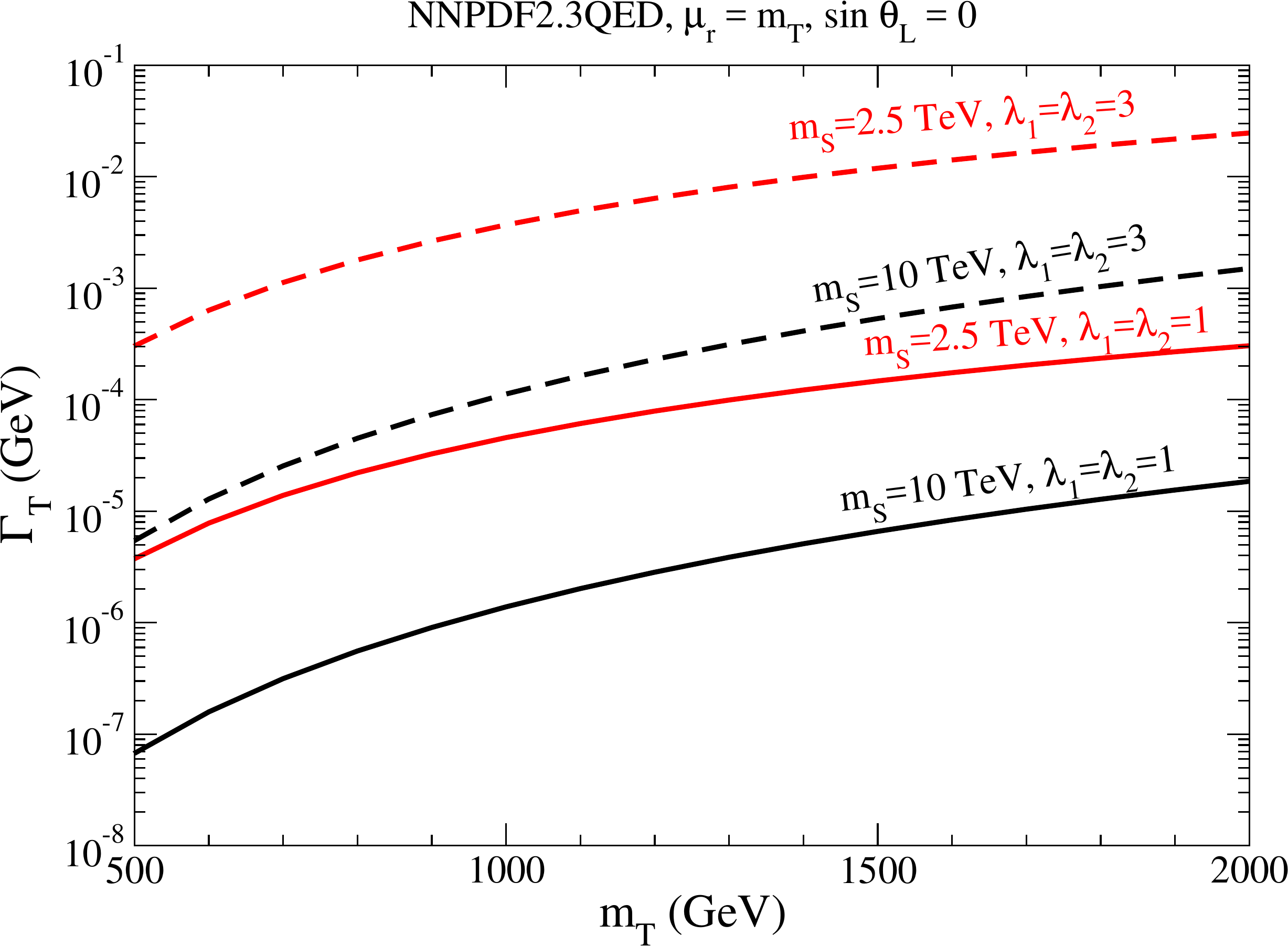}\label{fig:total_width_heavyMS}}
\subfigure[]{\includegraphics[width=0.49\textwidth,clip]{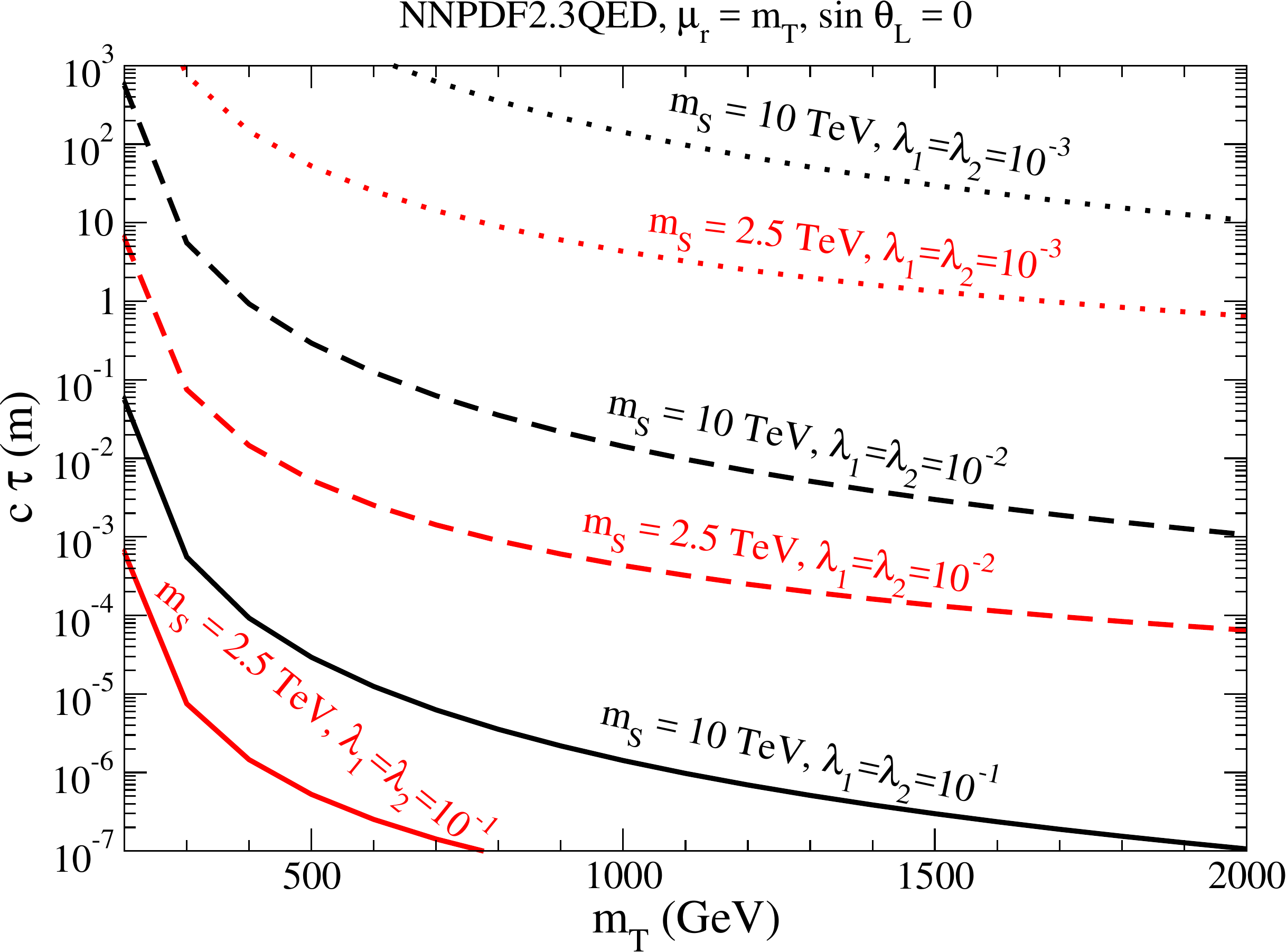}\label{fig:dec_length}}
\end{center}
\caption{\label{fig:decheavyS2} (a) Total width of top partner and (b) decay length of top partner for $\sin\theta_L=0$.  In both (a,b) the scalar mass is (red) $m_S=2.5$~TeV and (black) $m_S=10$~TeV.  In (a) the coupling constants are (dashed) $\lambda_{1,2}=3$ and (solid) $\lambda_{1,2}=1$.  In (b) the coupling constants are (dotted) $\lambda_{1,2}=10^{-3}$, (dashed) $\lambda_{1,2}=10^{-2}$, and (solid) $\lambda_{1,2}=10^{-1}$. The renormalization scale is set to the top partner mass.}
\end{figure}

For heavy scalars $m_S>m_T-m_t$ and zero mixing angle $\sin\theta_L=0$, Fig.~\ref{fig:decheavyS2} shows (a) the total width and (b) the decay length of the top partner for various parameter points.  If the decay width of a colored particle is less than $\Lambda_{QCD}\sim 100$~MeV~\cite{Patrignani:2016xqp}, we expect the particle to hadronize and bind with light quarks before it decays.  As can be clearly seen, when the loop level decays of the top partner are dominant, we have the total width $\Gamma_T<\Lambda_{QCD}$ for the vast majority of parameter space.  Hence, the top partner almost always hadronizes before it decays.  See for example Ref.~\cite{Buchkremer:2012dn} for a discussion of the phenomenology of top partner hadrons.  

At threshold it may be possible for pair produced top partners to bind and form exotic heavy quarkonia, $\eta_T=\langle T\overline{T}\rangle$.  This will be possible if the decay widths of $T$ and $\eta_T$ are less than the binding energy, $E_b$, of $\eta_T$. If this condition is not satisfied, the lifetime of $\eta_T$ will be less than the characteristic orbital time of the constituents and $\eta_T$ will not be a resonance.  Assuming that the binding force is essentially Coulombic, this condition is~\cite{Bigi:1986jk,Kats:2012ym}:
\begin{eqnarray}
\Gamma_T,\Gamma_{\eta_T}\lesssim |E_b|= \frac{C_F^2}{4} \alpha_s^2(m_T) m_T= 4~{\rm GeV}\left(\frac{\alpha_s(m_T)}{\alpha_S(1~{\rm TeV})}\right)^2\frac{m_T}{1~\rm TeV},~\label{eq:quarkonia}
\end{eqnarray} 
where $\Gamma_{\eta_T}$ is the $\eta_T$ decay width.  The precise decay pattern of the exotic quarkonia depend on the model parameters.  In addition to top partner decays, $\eta_T$ has decays into other SM final states.  The dominant mode is $\eta_T\rightarrow gg$~\cite{Buchkremer:2012dn,Kats:2012ym} with partial width $\Gamma(\eta_T\rightarrow gg)\sim 1-10$~MeV~\cite{Barger:1987xg,Kats:2012ym}.  Hence, if $\Gamma_T\ll \Gamma(\eta_T\rightarrow gg)$ the condition to form quarkonia in Eq.~(\ref{eq:quarkonia}) is always satisfied.  Additionally, we have ${\rm BR}(\eta_T\rightarrow gg)\approx 1$ and can employ a typical search for exotic quarkonia~\cite{Barger:1987xg,Kats:2012ym,Buchkremer:2012dn,Kuhn:1993cp}.  However, if $|E_b|\gtrsim \Gamma_T\gtrsim \Gamma(\eta_T\rightarrow gg)$, the top partner decays are expected to dominate the $\eta_T$ decays.  The top partners will decay according to the branching ratios in Fig.~\ref{fig:decheavyS1}.  For the parameter ranges in Fig.~\ref{fig:total_width_heavyMS}, the condition in Eq.~(\ref{eq:quarkonia}) is always satisfied.

As can be seen in Fig.~\ref{fig:dec_length}, for not too small couplings, the decay lengths of the top partners can be significant on the scales of collider experiments.  This leads to many exotic phenomena such as displaced vertices~\cite{Strassler:2006im,Graham:2012th,Liu:2015bma}, stopped particles~\cite{Drees:1990yw,Arvanitaki:2005nq,Graham:2011ah},  and long lived particles~\cite{Fairbairn:2006gg}.  The different decay lengths can be categorized as

\begin{itemize}
\item \bf{Prompt decays}:  Prompt decays have impact parameters $\lesssim 500~\mu$m~\cite{CMS:2014wda}.  For $m_S=2.5$~TeV, the top partner decays are prompt if $m_T\gtrsim$ $200-1000$~GeV and $10^{-1}\gtrsim\lambda_{1,2}\gtrsim 10^{-2}$.  For $m_S=10$~TeV, the decays are prompt if $m_T\gtrsim 300$~GeV and $\lambda_{1,2}\gtrsim 10^{-1}$.
\item \bf{Displaced vertices}:  If a particle's decay length is in the range $\mathcal{O}(1~{\rm mm})-\mathcal{O}(1~{\rm m})$ it can be reconstructed as a displaced vertex offset from the primary vertex of the proton-proton interaction~\cite{Liu:2015bma,CMS:2014wda,Aad:2015rba,Khachatryan:2016unx,Sirunyan:2017jdo,Aaboud:2017iio,ATLAS-CONF-2017-026}.  The top partner has these decay lengths for the following parameter regions:
\begin{eqnarray}
\lambda_{1,2}&\sim&10^{-3},\quad m_S=2.5~{\rm TeV},\quad m_T\gtrsim 1.5~{\rm TeV}\\
\lambda_{1,2}&\sim&10^{-2},\quad m_S=2.5~{\rm TeV},\quad m_T\lesssim 800~{\rm GeV}\\
\lambda_{1,2}&\sim&10^{-2},\quad m_S=10~{\rm TeV},\quad 400~{\rm GeV}\lesssim m_T\lesssim 2~{\rm TeV}\\
\lambda_{1,2}&\sim&10^{-1},\quad m_S=10~{\rm TeV},\quad m_T\lesssim 300~{\rm GeV}.
\end{eqnarray}
\item \bf{``Stable'' particles}:  It is possible for charged and colored particles to be stable on collider scales~\cite{Fairbairn:2006gg}.  Searches typically look for either high energy deposits in the trackers, measure time of flight with the muon systems, or search for decays in the hadronic calorimeter~\cite{Aad:2015asa,ATLAS-CONF-2016-103,Chatrchyan:2013oca,Khachatryan:2016sfv,ATLAS:2014fka,Aad:2015qfa,Aaboud:2016uth}.  These searches are sensitive to decay lengths of $\mathcal{O}(1~{\rm m})-\mathcal{O}(10~{\rm m})$ or longer.  For both $m_S=2.5$~TeV and $10$~TeV, top partners have these decay lengths for $\lambda_{1,2}\lesssim 10^{-3}$ and $m_T\lesssim 2$~TeV.  For $m_S=10$~TeV,  top partners also have these decay lengths for $\lambda_{1,2}\sim 10^{-2}$ and $m_T\lesssim 400$~GeV.
\item \bf{Stopped particles}:  Long lived colored particles hadronize and interact with the detectors, losing energy through ionization~\cite{Drees:1990yw,Arvanitaki:2005nq}.   It is possible for all the energy to be lost and the particles to stop inside the hadronic calorimeter~\cite{Arvanitaki:2005nq,Graham:2011ah}. For $\mathcal{O}(1~{\rm TeV})$ colored particles, nearly 100$\%$ with speeds below $\beta\sim 0.25-0.3$ will stop~\cite{Arvanitaki:2005nq}. If the particle's lifetime is $\gtrsim~\mathcal{O}(100~{\rm ns})$, they can be searched for as decays inside the hadronic calorimeter that are out of time with the bunch crossing~\cite{Graham:2011ah,Sirunyan:2017sbs,Khachatryan:2015jha,Aad:2013gva}.  For particles to stop in the calorimeter, they must be long lived on collider time scales.  Hence, much the same parameter space that gives ``stable'' particles gives stopped particles.
\end{itemize}

\subsubsection{Summary}
\label{sec:decsumm} 
\begin{figure}[tb]
\begin{center}
\subfigure[]{\includegraphics[width=0.49\textwidth,clip]{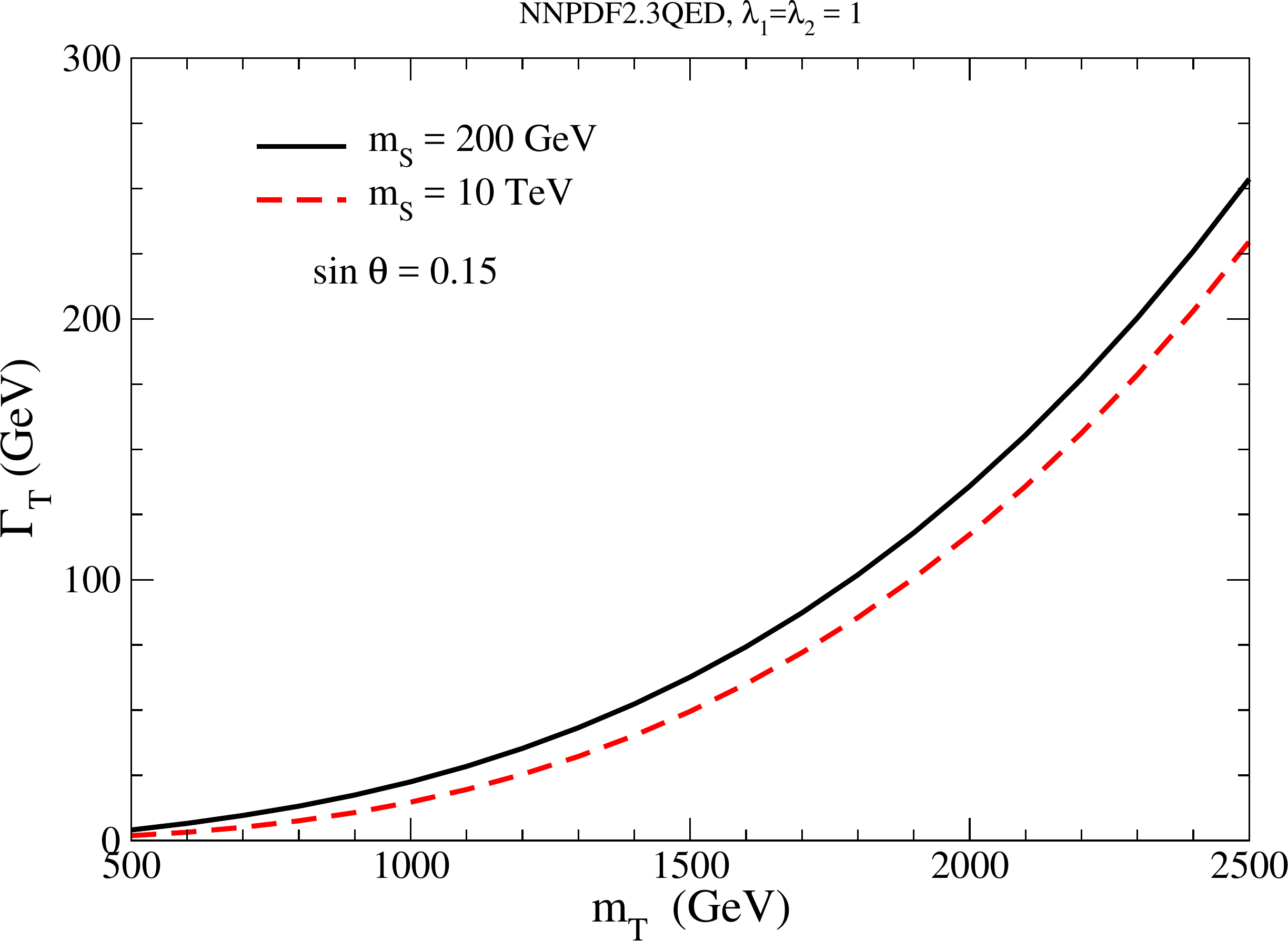}\label{fig:total_width}}
\subfigure[]{\includegraphics[width=0.49\textwidth,clip]{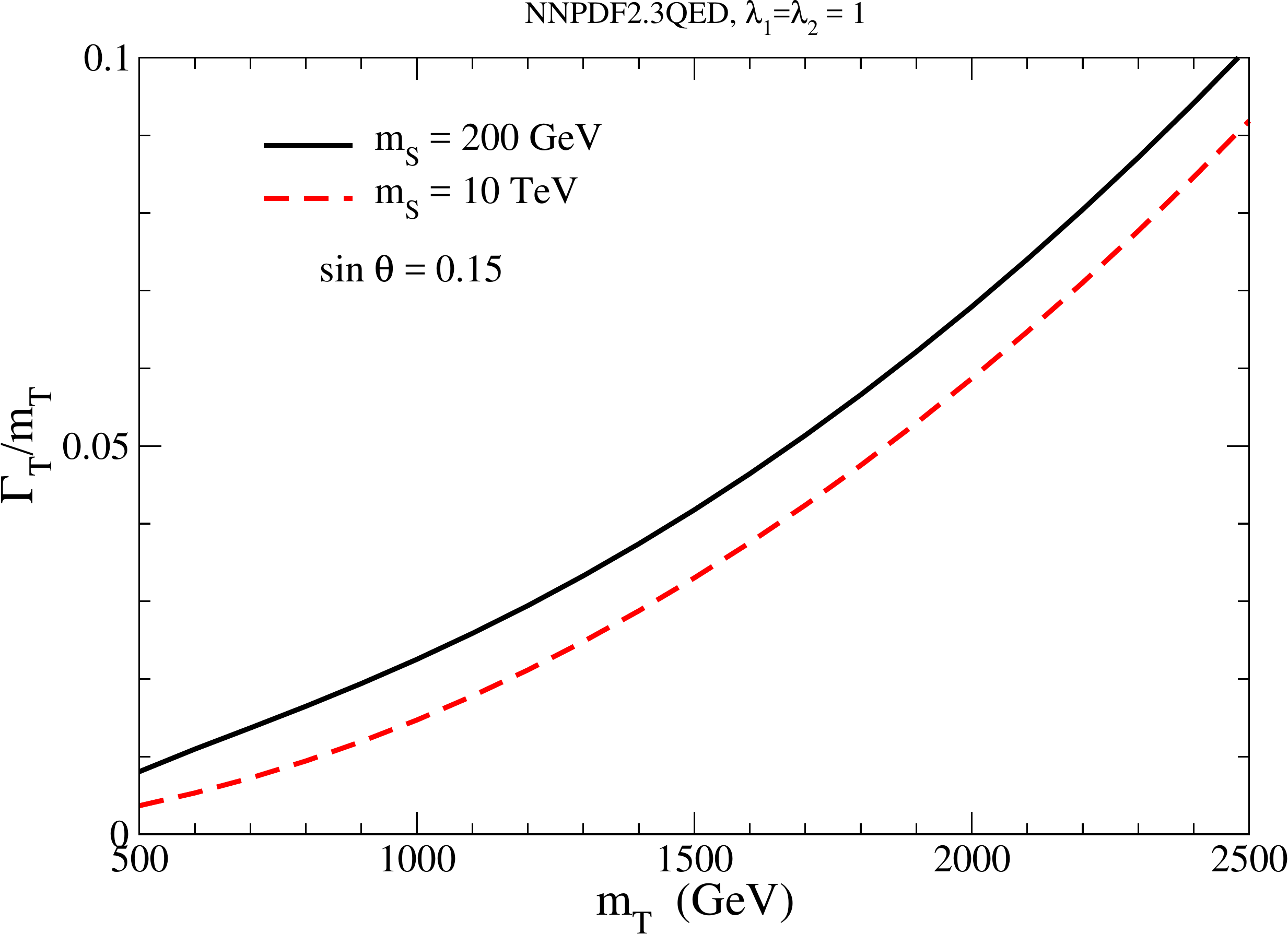}\label{fig:total_width_ratio}}
\end{center}
\caption{\label{fig:dec} (a) Total width of top partner and (b) width to mass ratio for $\lambda_{1,2}=1$, $\sin\theta_L=0.15$, and (black solid) $m_S=200$~GeV and (red dashed) $m_S=10$~TeV.}
\end{figure}

In Fig.~\ref{fig:dec} we show the (a) total width and (b) width to mass ratio for (black solid) $m_S=200$~GeV and (red dashed) $m_S=10$~TeV with $\sin\theta_L=0.15$.  For $m_S>m_T-m_t$, the $T\rightarrow St$ decay mode is no longer allowed. However, for non-negligible mixing angle, the tree level $T\rightarrow Wb$, $T\rightarrow tZ$, and $T\rightarrow th$ are still available and growing as $m_T^3$.  The result of decoupling $S$ is to suppress the total width by $\sim50\%$ for $m_T\sim500$~GeV and $\sim10\%$ for $m_T\sim 2.5$~TeV.  For both cases, although the width to Higgs and gauge bosons increases with $m_T^3$, the width to mass ratio never exceeds $10\%$ and can always be safely regarded as narrow.

\begin{table}[tb]
\begin{center}
\makebox[\textwidth][c]{
\begin{tabular}{|c||c|c|}\cline{2-3}\hline
&$\sin\theta_L\sim0$ &$\sin\theta_L\sim 0.1$\\\hline\hline
\multirow{2}{*}{$m_S<m_T-m_t$} & \multirow{2}{*}{${\rm BR}(T\rightarrow tS)\approx1$}& ${\rm BR}(T\rightarrow tS)\sim 0.1-0.5\quad{\rm BR}(T\rightarrow bW)\sim 0.25-0.5$\\
& &  ${\rm BR}(T\rightarrow th)\sim 0.1-0.25\quad{\rm BR}(T\rightarrow tZ)\sim 0.15-0.25$\\\hline
$m_S>m_T-m_t$ & ${\rm BR}(T\rightarrow tg)\approx 1$ & ${\rm BR}(T\rightarrow bW)\approx 2\,{\rm BR}(T\rightarrow tZ)\approx 2\,{\rm BR}(T\rightarrow th)\approx 0.5$\\\hline
\end{tabular}}
\end{center}
\caption{Branching ratios of dominant top partner decay modes for different mixing angles and mass regions.\label{tab:decmodes}}
\end{table}

\begin{table}[tb]
\begin{center}
\makebox[\textwidth][c]{
\begin{tabular}{|c|c|c|}\hline
$\sin\theta_L=0$& $m_S=2.5$~TeV & $m_S=10$ TeV\\\hline\hline
Prompt  &$\lambda_{1,2}\sim 10^{-1}-10^{-2}$, $m_T\gtrsim~0.2-1$~TeV & $\lambda_{1,2}\gtrsim 10^{-1}$, $m_T\gtrsim 300$~GeV\\\hline
\multirow{2}{*}{Displaced} & $\lambda_{1,2}\sim 10^{-3}$, $m_T\gtrsim 1.5$ TeV & $\lambda_{1,2}\sim 10^{-2}$, $400~{\rm GeV}\lesssim m_T\lesssim 2$~TeV\\
& or $\lambda_{1,2}\sim 10^{-2}$, $m_T\lesssim 800$~GeV   & or $\lambda_{1,2}\sim 10^{-1}$, $m_T\lesssim 300$~GeV\\\hline
``Stable'' & \multirow{2}{*}{$\lambda_{1,2}\lesssim 10^{-3}$, $m_T\lesssim 2$~TeV} & $\lambda_{1,2}\lesssim 10^{-3}$, $m_T\lesssim 2$~ TeV\\
/Stopped & & or $\lambda_{1,2}\sim 10^{-2}$, $m_T\lesssim 400$~GeV\\\hline
Hadronize & Perturbative $\lambda_{1,2}$ & Perturbative $\lambda_{1,2}$\\\hline
\end{tabular}}
\end{center}
\caption{For $\sin\theta_L=0$, Parameter spaces that give us prompt decays, displaced vertices, top partners that are stable on collider time scales, top partners that can stop in the calorimeters, and top partners that hadronize.\label{tab:displaced}}
\end{table}

We summarize our results for top partner decays in Tables~\ref{tab:decmodes} and~\ref{tab:displaced}.  The possible ranges of the dominant top partner decay modes for different mixing angle and scalar mass ranges are shown in Table~\ref{tab:decmodes}.  In Table~\ref{tab:displaced}, we give representative parameter regions that give various collider signatures of long lived top partners.

\section{Production and Decay of the Scalar}
\label{sec:proddecscalar}
\begin{figure}[tb]
\begin{center}
 \subfigure[]{\includegraphics[scale=0.75]{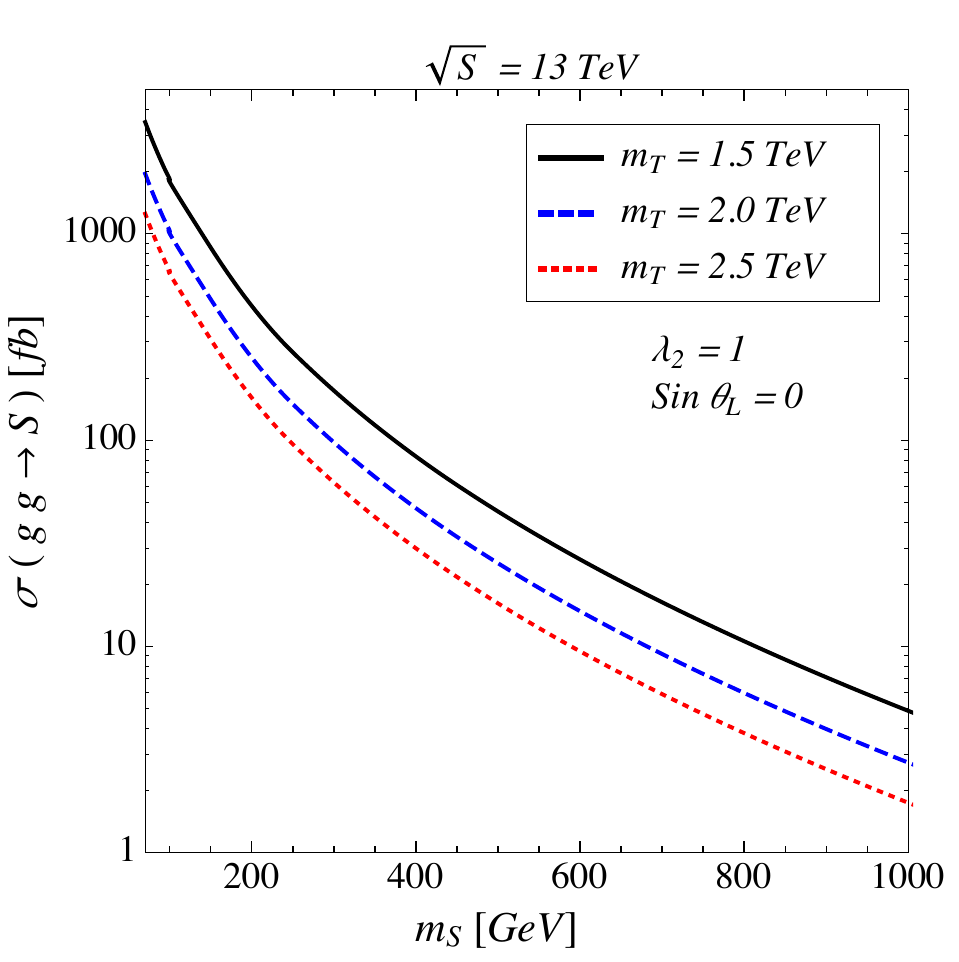}\label{fig:Sprod}} \hspace{0.0cm}
 \subfigure[]{\includegraphics[scale=0.75]{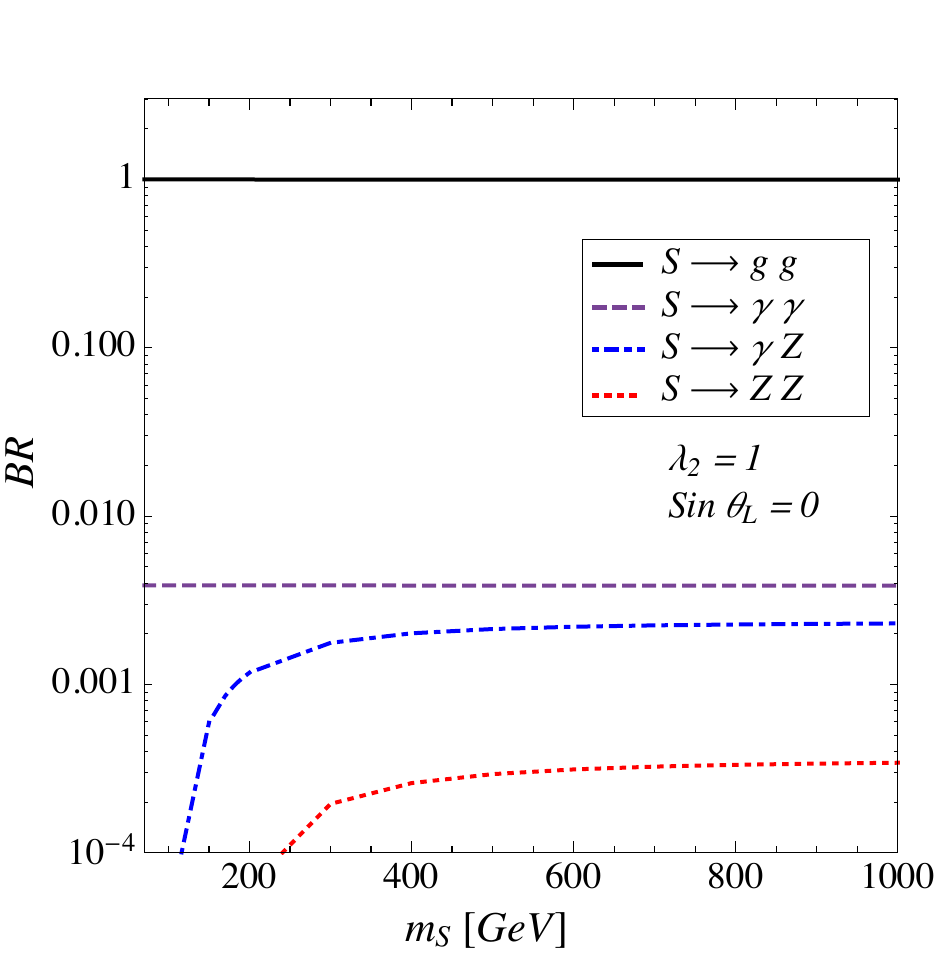} \label{fig:SBR}}
\caption{\label{fig:Srates} 
(a) The single $S$ production cross sections via gluon fusion at $\rm{N^3LO}$ accuracy in QCD at $\sqrt{S} = 13$~TeV for (black solid) $m_T=1.5$~TeV, (blue dash) $m_T=2$~TeV, and (red dot) $m_T=2.5$~TeV. (b) The branching ratios of $S$ decaying into (black solid) $g g$, (violet dash) $\gamma \gamma$, (blue dash-dot) $\gamma Z$, and (red dot) $Z Z$ final states. All plots are made in with zero top partner-top mixing ($\sin \theta_L \rightarrow 0$) with $\lambda_2=1$.} 
\end{center}
\end{figure}

We now discuss the production and decay of the scalar, $S$, in the model presented in Sec.~\ref{sec:model}.  We focus on the region of parameter space for which the scalar can be produced at the LHC with reasonable rates, i.e. $m_S\sim 100s$~GeV and $m_T>m_S$.   As mentioned in the previous section, the scalar can be produced via decays of the top partner.  The scalar can also be directly produced through gluon fusion mediated by top quark and top partner loops, similar to the loops in Fig.~\ref{fig:Sdecay}.  In Fig.~\ref{fig:Sprod}, we show the production cross sections for the scalar for various top partner masses and $\lambda_2=1$.  The scalar-Higgs and top partner-top mixing angles are set to zero.  In this limit, only the top partner loops contribute and for $m_T\gg m_S$ the cross section scales as $\sim \lambda^2_2 / m^2_{T}$.  Hence, the cross sections for different couplings and top partner masses can be easily obtained by rescaling these results.  The cross sections for scalar production are found by rescaling the N$^3$LO scalar gluon fusion production cross sections~\cite{Anastasiou:2016hlm}.  That is, we use the relevant Wilson coefficient for the $g-g-S$ contact interaction for $m_T\gg m_S$.

With $\sin\theta_L=0$ and no Higgs-scalar mixing, $S$ can decay into $g g$, $\gamma \gamma$, $\gamma Z$ and $Z Z$ through top partner loops, as shown in Fig.~\ref{fig:Sdecay}.  We show the branching ratios of the scalar $S$ in this limit in Fig.~\ref{fig:SBR}.   For $m_T\gg m_S$, all partial widths are proportional to $\lambda_2^2/m^2_T$.  Hence, the branching ratios are independent of the Yukawa coupling $\lambda_2$ and the top partner mass, and are determined by ratios of gauge couplings.  Due to the strong coupling, the dominant decay mode is into gluons with $\rm{BR}(S\rightarrow gg) \simeq 99 \%$.

\section{Experimental constraints}
\label{sec:const}
\begin{figure}[tb]
\begin{center}
\subfigure[]{\includegraphics[scale=0.7,clip]{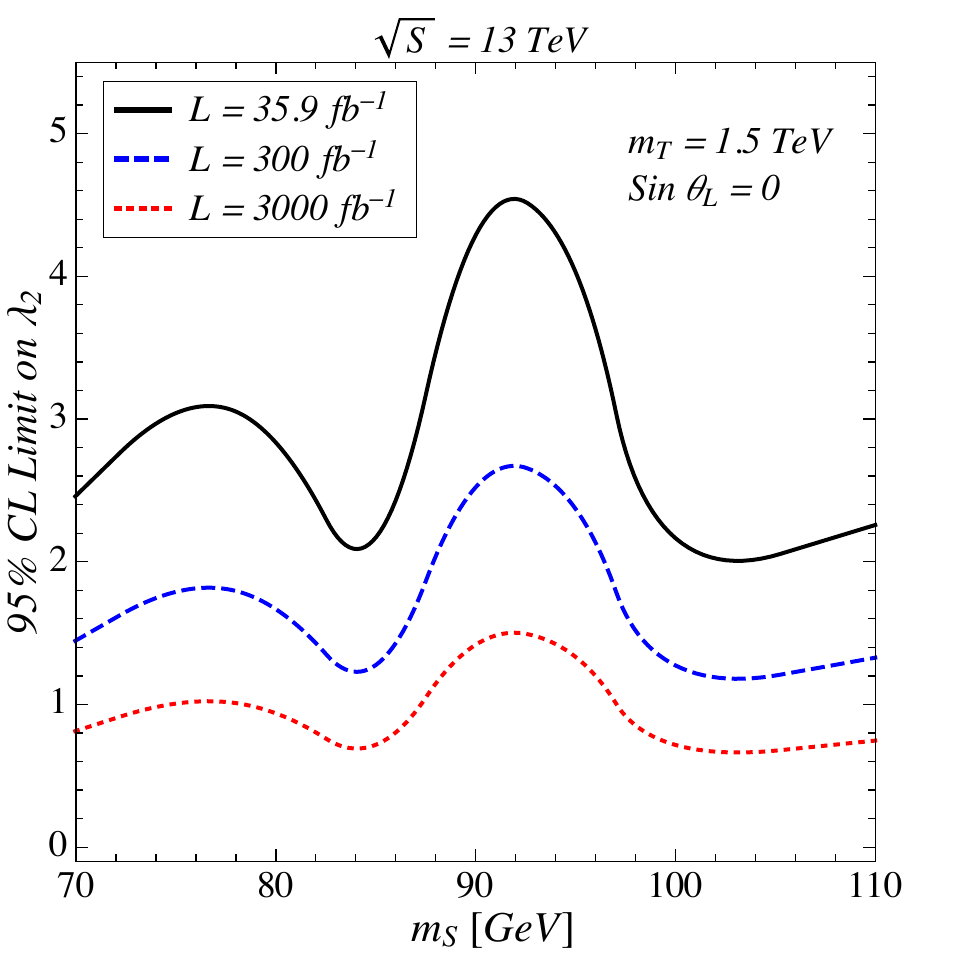}}
\subfigure[]{\includegraphics[scale=0.7,clip]{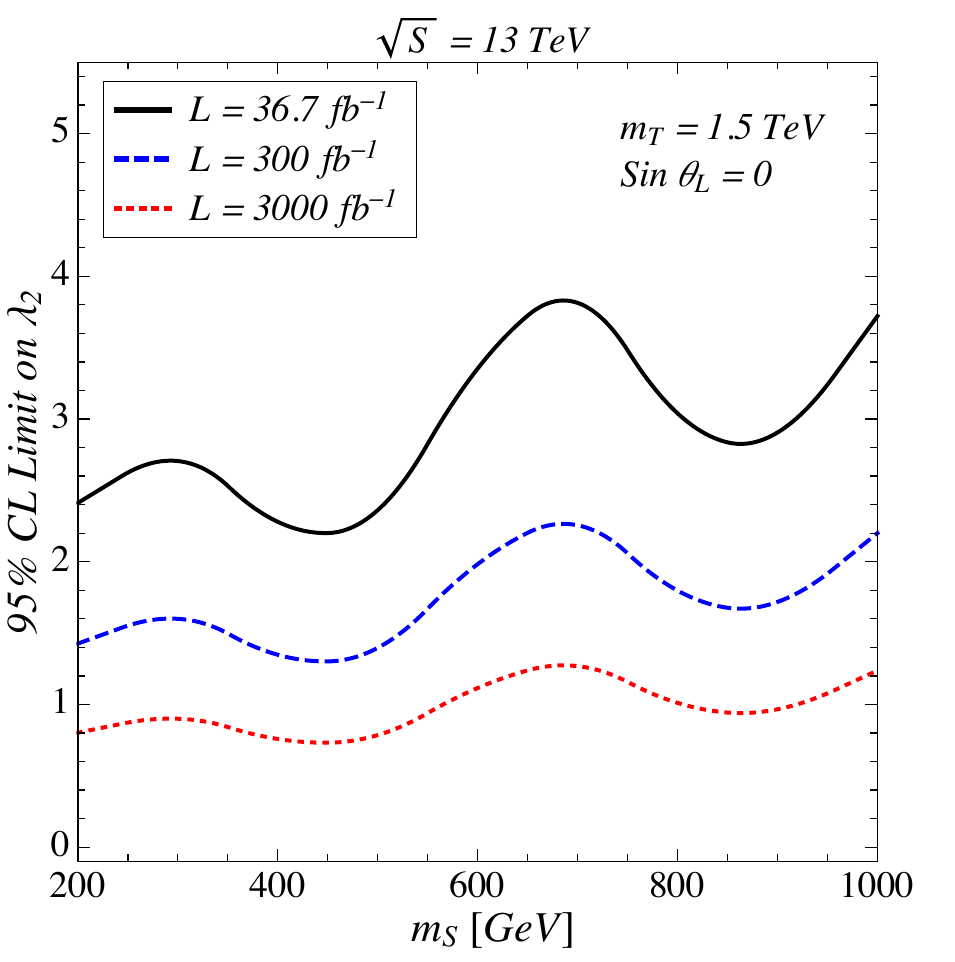}} \\
\subfigure[]{\includegraphics[scale=0.7,clip]{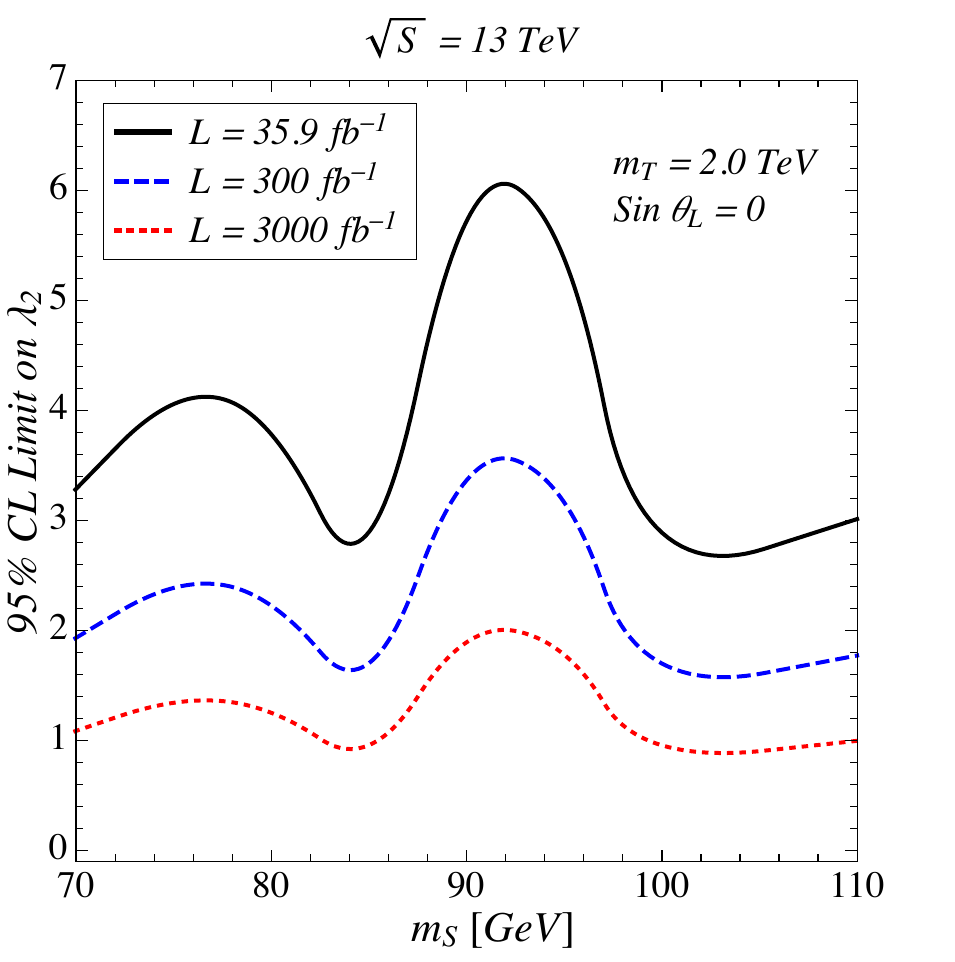}} 
\subfigure[]{\includegraphics[scale=0.7,clip]{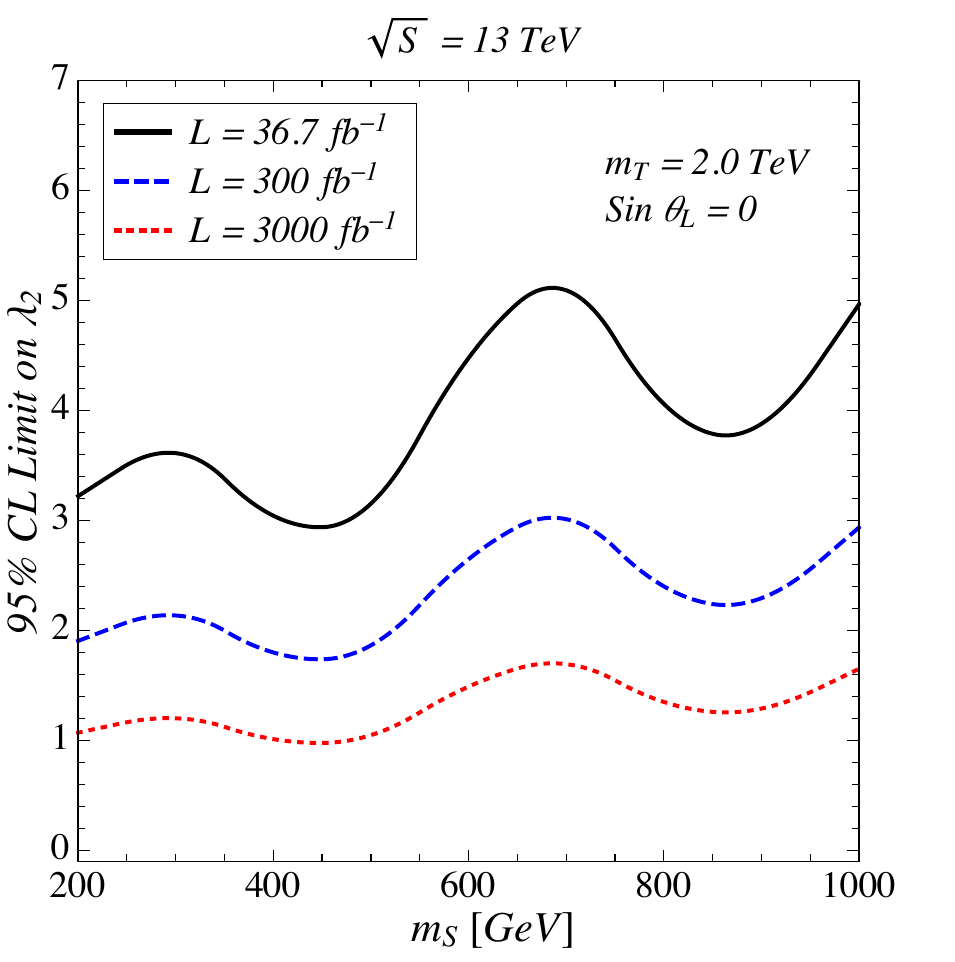}} 
\caption{\label{fig:Slimit}
The excluded regions of the parameter space in the (a,c) lower and (b,d) higher mass regions of $S$ for (a,b) $m_{T} = 1.5$ TeV and (c,d) $m_{T} = 2.0$ TeV. The regions above the black solid line are excluded diphoton searches~\cite{CMS-PAS-HIG-17-013,Aaboud:2017yyg} at the 13 TeV LHC. The blue dashed and red dotted curves are the projected exclusions with $300~ \rm{fb}^{-1}$ and $3000~ \rm{fb}^{-1}$ luminosities, respectively, via a naive rescaling based on the current limits.
} 
\end{center}
\end{figure}
Some of the most constraining limits on colored particles come from QCD pair production shown in Figs.~\ref{fig:ggTTs}-\ref{fig:qqTTs}.  The production is mediated by the strong force, and the rate is completely determined by the mass, spin, and color representation of the produced particles.  Hence, it is relatively model independent.  There have been many searches for top pair production, but their applicability depends on on the precise decay pattern of the top partner.  We summarize limits from pair production according to the mass and mixing categories in Table~\ref{tab:decmodes}:
\begin{itemize}
\item \bf{$\mathbf{m_S> m_T-m_t}$ and $\mathbf{\sin\theta_L\sim0.1}$}: The $T\rightarrow tS$ channel is forbidden, and the classic tree level decays $T\rightarrow th$, $T\rightarrow tZ$, and $T\rightarrow bW$ obey the expected relation ${\rm BR}(T\rightarrow bW)\approx 2\, {\rm BR}(T\rightarrow tZ)\approx 2\,{\rm BR}(T\rightarrow th)\approx 0.5$.  For this decay pattern recent studies of ATLAS~\cite{Aaboud:2017zfn, Aaboud:2017qpr} and CMS~\cite{Sirunyan:2017pks, Sirunyan:2017usq} excluded $m_T \lesssim 1.2  - 1.3$ TeV.
\item \bf{$\mathbf{m_S< m_T-m_t}$ and $\mathbf{\sin\theta_L\sim0.1}$}:  All tree level decays are available: $T\rightarrow tS,\,T\rightarrow th,\,T\rightarrow tZ,$ and $T\rightarrow bW$. The traditional searches for pair produced top partners $T\rightarrow th,\,T\rightarrow tZ$, and $T\rightarrow bW$~\cite{Aaboud:2017zfn, Aaboud:2017qpr,Sirunyan:2017pks, Sirunyan:2017usq} are then applicable. However, the branching ratios to $th$, $tZ$, and $bW$ do not obey the expected pattern ${\rm BR}(T\rightarrow bW)\approx 2\, {\rm BR}(T\rightarrow tZ)\approx 2\,{\rm BR}(T\rightarrow th)\approx 0.5$, as shown in Table~\ref{tab:decmodes} and Fig.~\ref{fig:BR_ms200GeV}.  Hence, the bounds are weakened.  To fill in the gaps, searches for $T\rightarrow tS$ will have to be performed~\cite{Dolan:2016eki}.  These will depend on the decay pattern of the scalar $S$, as discussed in Sections~\ref{sec:declowms} and \ref{sec:proddecscalar}.

\item \bf{$\mathbf{m_S> m_T-m_t}$ and $\mathbf{\sin\theta_L\sim0}$}:   All tree level decays are very suppressed, and the loop level decays are relevant: $T\rightarrow tg$, $T\rightarrow t\gamma$, and $T\rightarrow tZ$.  A recent CMS analysis~\cite{Sirunyan:2017yta} searched for pair-produced spin 3/2 vector-like excited quarks $T_{3/2}$ which exclusively decays as $T_{3/2} \rightarrow t g$.  The lower limit on the mass was found to be $ \sim 1.2$ TeV.  While ${\rm BR}(T\rightarrow tg)\sim 1$, the pair production rate of $T$ is different from $T_{3/2}$ since $T$ is spin 1/2. We recast the CMS search~\cite{Sirunyan:2017yta} to assess the current constraint on $T$ using NNLO pair production cross section~\cite{Sirunyan:2017usq, Czakon:2011xx, Czakon:2013goa, Czakon:2012pz, Czakon:2012zr, Cacciari:2011hy}.  The mass bound of this search is then $m_T\gtrsim 930$ GeV.

\item \bf{$\mathbf{m_S< m_T-m_t}$ and $\mathbf{\sin\theta_L\sim0}$}:  The decay channel $T\rightarrow tS$ dominates with branching ratio ${\rm BR}(T\rightarrow tS)\sim 1$.  This decay channel will require new search strategies~\cite{Dolan:2016eki}, which will depend on the decay pattern of the scalar $S$ and whether or not it mixes with the Higgs boson.  See Sections~\ref{sec:declowms} and ~\ref{sec:proddecscalar} for a discussion.
\end{itemize}
To be conservative, we will assume the strongest constraints from pair production and work in the regime $m_T\gtrsim 1.2-1.3$~TeV.

An alternative avenue to look for $T$ in the high mass region is the EW single production in association with jets or $W$~\cite{Willenbrock:1986cr,Han:2003wu,Han:2005ru}, as shown in Figs.~\ref{fig:Tjet}-\ref{fig:Tjet1}. Searches for the single production of $T$ in ATLAS~\cite{ATLAS-CONF-2016-072} and CMS~\cite{CMS-PAS-B2G-16-005, Sirunyan:2017ynj} have excluded  $m_T \lesssim 1 - 1.8$ TeV depending on the coupling strengths as well as branching ratios. For the $SU(2)_L$ singlet top partner model, this production mechanism vanishes as $\sin\theta_L\rightarrow 0$, and the constraints can be avoided.  

The most stringent constraints to the mixing between top partners and top quarks comes from EW precision measurements~\cite{He:2001tp,Chen:2017hak,Chen:2014xwa,Dawson:2012di, Aguilar-Saavedra:2013qpa}.  The oblique parameters constrain $|\sin\theta_L|\lesssim 0.16$ for $m_T=1$~TeV and $|\sin\theta_L|\lesssim 0.11$ for $m_T=2$~TeV~\cite{Chen:2017hak,Dawson:2012di, Aguilar-Saavedra:2013qpa}.  The collider bounds are considerably less constraining~\cite{ATLAS-CONF-2016-072}.

Recent scalar resonance searches at the LHC in the $g g$~\cite{CMS-PAS-EXO-16-056}, $\gamma \gamma$~\cite{CMS-PAS-HIG-17-013, Aaboud:2017yyg}, $\gamma Z$~\cite{Sirunyan:2017hsb} and $Z Z$~\cite{Aaboud:2017rel} channels can put significant constraints on the scalar mass and couplings. Despite the small branching ratio, the $S \rightarrow \gamma \gamma$ decay channel ($\rm{BR} \simeq 0.4 \%$) is the cleanest, setting the most stringent limit on $S$.  The experimental results are given for a low mass region $70~{\rm GeV}<m_S<110~{\rm GeV}$~\cite{CMS-PAS-HIG-17-013} and a high mass region $200~{\rm GeV}<m_S$~\cite{Aaboud:2017yyg}. Figure~\ref{fig:Slimit} demonstrates the excluded regions of the parameter space in the (a,c) lower and (b,d) higher $m_S$ regions, assuming for (a,b) $m_{T} = 1.5$ TeV and (c,d) $m_{T} = 2.0$ TeV.  Scalar-Higgs and top partner-top mixing angles have been set to zero.  The regions above these lines are excluded at the 13 TeV LHC.  We show results for the (black solid) current data, and projections to (blue dash) 300 fb$^{-1}$ and (red dot) 3 ab$^{-1}$.  We have assumed both systematic and statistical uncertainties scale as the square root of luminosity. The outlook for the projected limits at the high luminosity-LHC with 3 ab$^{-1}$ indicates that $\lambda_2$ is expected to be highly constrained $\lambda_2 \lesssim 1$ for the scalar $S$ mass of $ \sim 100 - 1000$ GeV. The bound can be relaxed as the top partner mass increases, since the cross section decreases as $1/m_T^2$.

\section{Signal Sensitivity at the High Luminosity-LHC}
\label{sec:results}

The loop-induced single $T$ production in association with a top quark, as shown in Fig.~\ref{fig:FeynLoop1}-\ref{fig:FeynLoopFin}, provides an unique event topology, offering useful handles to suppress the SM backgrounds. In this section, we present a detailed collider analysis for the high luminosity-LHC at $\sqrt{S} = 14$ TeV with 3 ab$^{-1}$ of data, and estimate the sensitivity reach in the final state
\bea
\label{eq:process} 
	p ~ p \rightarrow T ~ \overline{t}+t~\overline{T} \rightarrow S ~ t ~ \overline{t} \rightarrow g ~ g  ~ t ~ \overline{t} \, .
\eea
We focus on the $\sin\theta_L=0$ limit so that ${\rm BR}(T\rightarrow tS)\approx 1$ and ${\rm BR}(S\rightarrow gg)\approx1$.  Both $gg$- and $q \overline{q}$-initiated processes are taken into account in the analysis.\footnote{For $q \overline{q}$-initiated process, only the diagram with the s-channel gluon in Fig.~\ref{fig:FeynLoop1} is considered. We checked that contributions from the diagrams with the s-channel photon or $Z$ boson in Fig.~\ref{fig:FeynLoopZ} are negligible.} We focus on the semi-leptonic decay of the $t \overline{t}$ system in order to evade a contamination from the QCD multi-jet background. 

\subsection{Signal Generation}
\label{sec:signal}
To generate signal events described in Eq.~(\ref{eq:process}), we first implement the EFT in Eq.~(\ref{eq:EFT}) within the \texttt{MadGraph5\_aMC@NLO}~\cite{Alwall:2014hca} framework using \texttt{FeynRules}~\cite{Christensen:2008py,Alloul:2013bka}.  The vertices needed for $T$ and $S$ decays can be conveniently parametrized by the interaction Lagrangian\footnote{The kinematic distributions of final state particles can be sensitive to the chiral structure of the coupling $t-T-S$, since the polarization of the top quark propagates to daughter particles. Realizing sophisticated analysis to reflect all shapes of kinematic distributions is beyond the scope of our work. Here we will assume the relative size of the couplings is the same $\lambda^{S}_{t T} = \lambda^{S}_{T t}$.} 
\begin{eqnarray}
\lambda^{S}_{T t} S \overline{T}_L t_R + \lambda^{S}_{t T} S \overline{t}_L T_R \; +\; \rm{h.c.}
\end{eqnarray} 
and the effective operator 
\begin{eqnarray}
S G^{A}_{\mu \nu} G_{A}^{\mu \nu}.
\end{eqnarray}
We use the default \texttt{NNPDF2.3QED} parton distribution function~\cite{Ball:2013hta} with fixed factorization and renormalization scales set to $m_{T} + m_t$.  At generation level, we require all partons to pass cuts of 
\begin{eqnarray}
p_T > 30~{\rm GeV},\quad{\rm and}\quad ~| \eta |< 5,\label{eq:basepartons}
\end{eqnarray}
 while leptons are required to have 
\begin{eqnarray}
p_T^\ell > 30~{\rm GeV}\quad{\rm and}\quad ~| \eta^\ell |< 2.5,\label{eq:baseleptons}
\end{eqnarray}
where $p_T$ are transverse momentum, $\eta$ is rapidity, and $\ell$ indicates leptons. To acquire better statistics in dealing with the SM backgrounds, we demand 
\begin{eqnarray}
H_T > 700 \GeV,\label{eq:baseht}
\end{eqnarray} 
where $H_T$ denotes the scalar sum of the transverse momenta of all final state particles.

We will consider $m_S=110~\GeV$, $m_T=1.5$ TeV and $2$ TeV, and $\sin\theta_L=0$.  The $\sin\theta_L=0$ limit is particularly interesting in this model because the production and decay patterns of the top partner are different from the the traditional approaches, as discussed in Section~\ref{sec:proddec}.  We use such a small scalar mass so that the production cross section is maximized, as shown in Fig.~\ref{fig:prodmsc}.  However, the EFT in Eq.~(\ref{eq:EFT}) is not valid.  Thus, we reweight the matrix element of the EFT by the exact one-loop calculation on an event-by-event basis.  We also reweight the events according to the exact branching ratios of the decays $T  \rightarrow tS$ and $S  \rightarrow g g$.  Details of the $T$ production and decay calculation are given in Sections~\ref{sec:pro} and~\ref{sec:dec}, respectively, as-well-as the Appendices~\ref{Appe:renorm} and~\ref{Appe:vertrenorm}. Details of the scalar decay can be found in Section~\ref{sec:proddecscalar}. The reweighted events are showered and hadronized by \texttt{PYTHIA6}~\cite{Sjostrand:2006za} and clustered by the \texttt{FastJet}~\cite{Cacciari:2011ma} implementation of the anti-$k_T$ algorithm~\cite{Cacciari:2008gp} with a fixed cone size of $r = 0.4 \;(1.0)$ for a slim (fat) jet. We include simplistic detector effects based on the ATLAS detector performances~\cite{ATL-PHYS-PUB-2013-004}, and smear momenta and energies of reconstructed jets and leptons according to the value of their energies (see the details in Appendix~\ref{app:DR}).

\subsection{Background Generation}
\label{sec:BGD}

\begin{table}[tb]
\begin{center}
\scalebox{1.0}{
\begin{tabular}{|c|c|c|c|c|}
\hline
           Abbreviations           & Backgrounds                       & Matching       & $\sigma \cdot {\rm BR(fb)}$     \\ \hline	
  $t \overline{t} $                 &  $t \overline{t} + \rm{jets}$   &   4-flavor        &$  2.91 \times 10^{3} \; \rm fb $    \\ \hline
 \multirow{2}{*}{Single $t$} & $t W + \rm{jets}$                  &    5-flavor      &$  4.15 \times 10^{3} \; \rm fb $      \\
                                           & $tq+{\rm jets}$           &    4-flavor      & 	$ 77.2 \; \rm fb$                             \\ \hline
  $W$                                  &$W + \rm{jets}$                     &     5-flavor     &   $ 4.96  \times 10^{3} \; \rm fb $     \\ \hline  
  \multirow{2}{*}{$VV $}      & $W W + \rm{jets}$                &    4-flavor      &  $ 111 \; \rm fb $                                  \\ 
                                           &$W Z  + \rm{jets}$                 &    4-flavor      &  $ 43.5  \; \rm fb$                             \\

\hline		
\end{tabular}}
\end{center}
\caption{The summary of the SM backgrounds after generation level cuts Eqs.~(\ref{eq:basepartons}-\ref{eq:baseht}).  Matching refers to the either the 4-flavor or 5-flavor MLM matching.  $\sigma \cdot \rm BR$ denotes the production cross section (fb) times branching ratios including the top, $W$, and $Z$ decays.}
\label{tab:TotalBackG}
\end{table}

The SM backgrounds are generated by \texttt{MadGraph5\_aMC@NLO} at leading order accuracy in QCD at $\sqrt{S} = 14$ TeV with the \texttt{NNPDF2.3QED} parton distribution function~\cite{Ball:2013hta}. All events are subject to the cuts in Eqs.~(\ref{eq:basepartons}-\ref{eq:baseht}). We use the default variable renormalization and factorization scales.  The MLM-matching~\cite{Mangano:2006rw} scheme is used. The matching scales are chosen to be {\tt xqcut} = 30 GeV and {\tt Qcut} = 30 GeV for all backgrounds.

The most significant (irreducible) background is semi-leptonic $t \overline{t} + \rm{jets}$ matched up to two additional jets. The relevant EW  produced single-top backgrounds are $tW$ and $tq$, where $q$ is a light or $b$-quark.  The $tW$-channel is generated with up to three additional jets and one $W$ decays leptonically while the other decays hadronically. The $tq$ channel is generated with up to two additional jets and we only consider a top quark which decays leptonically.  Another relevant background includes $W + \rm{jets}$ with up to four additional jets and we only include a leptonically decaying $W$.  Much smaller backgrounds include $W W + \rm{jets}$ with up to three additional jets where one $W$ decays leptonically and the other hadronically. Finally, $W Z + \rm{jets}$ sample is generated with up to three additional jets where the $W$ is forced to decay leptonically and the $Z$ hadronically.  Although $WW$ and $WZ$ are small compared to the other backgrounds, they are still large compared to the signal.  A detailed summary of the backgrounds, the matching schemes, and their cross section after generation level cuts in Eqs.~(\ref{eq:basepartons}-\ref{eq:baseht}) is presented in Table.~\ref{tab:TotalBackG}.  It should be noted that $tW$ is the dominant contribution to single top, whereas Fig.~\ref{fig:prod_mtp} would seem to indicate that $tq$ should be dominant.  However, while EW $tq$ is dominant before cuts, the $H_T$ cut in Eq.~(\ref{eq:baseht}) greatly reduces $tq$ and $tW$ becomes the leading contribution.

All background events are fed into \texttt{PYTHIA6}~\cite{Sjostrand:2006za} for parton showering and hadronization, and then clustered by the \texttt{FastJet}~\cite{Cacciari:2011ma} implementation of the anti-$k_T$ algorithm \cite{Cacciari:2008gp}.  We use two cone sizes of $r = 0.4$ and $1.0$ for slim and fat jets, respectively. Momenta and energies of reconstructed jets and leptons are smeared in the same way of the signal event to reflect semi-realistic detector resolution effects.

\subsection{Signal Selection and Sensitivity}
\label{sec:SigSelect}

\begin{figure}[tb]
\begin{center}
\subfigure[]{\includegraphics[width=0.49\textwidth,clip]{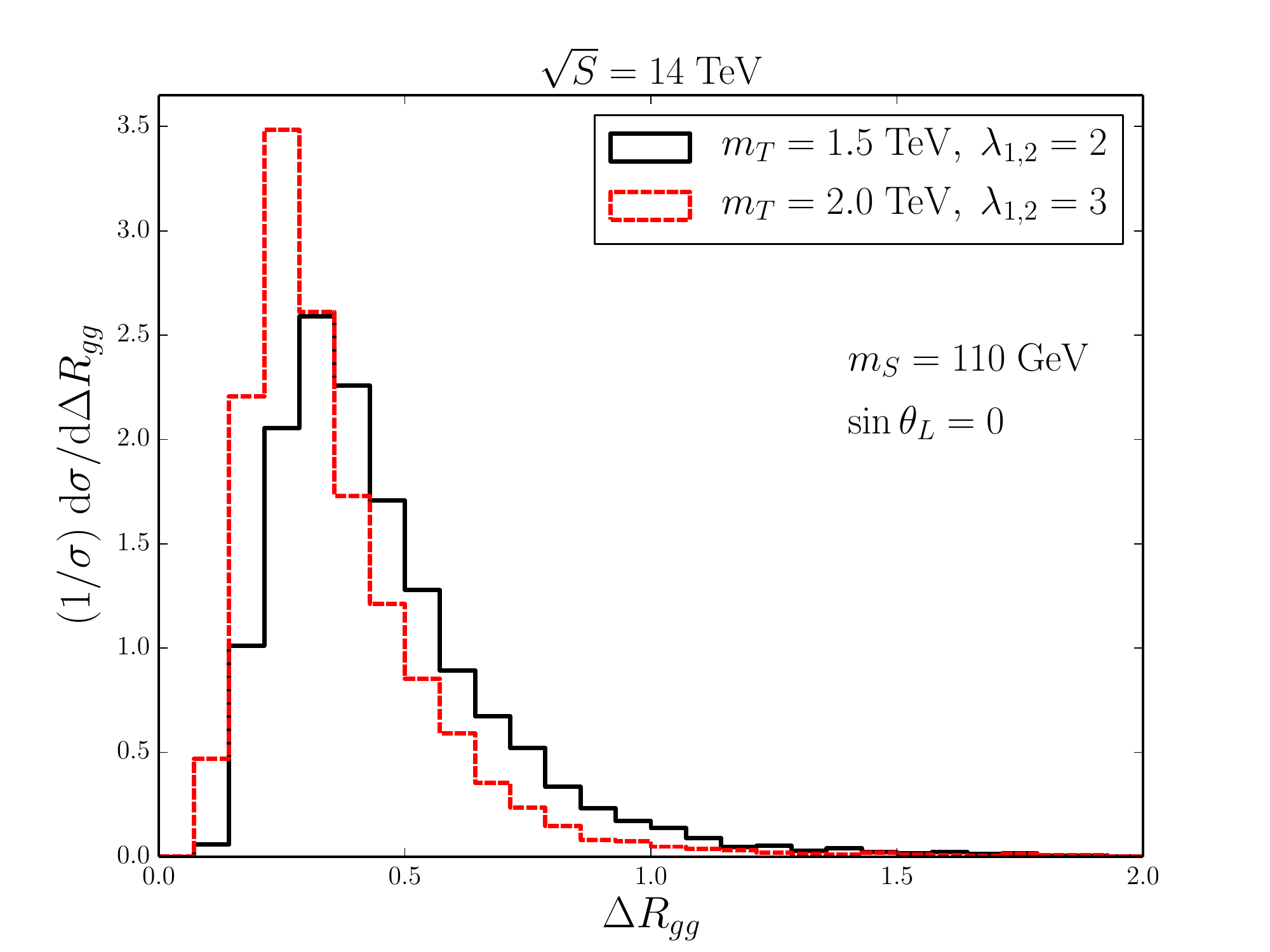}\label{fig:sdr}}
\subfigure[]{\includegraphics[width=0.49\textwidth,clip]{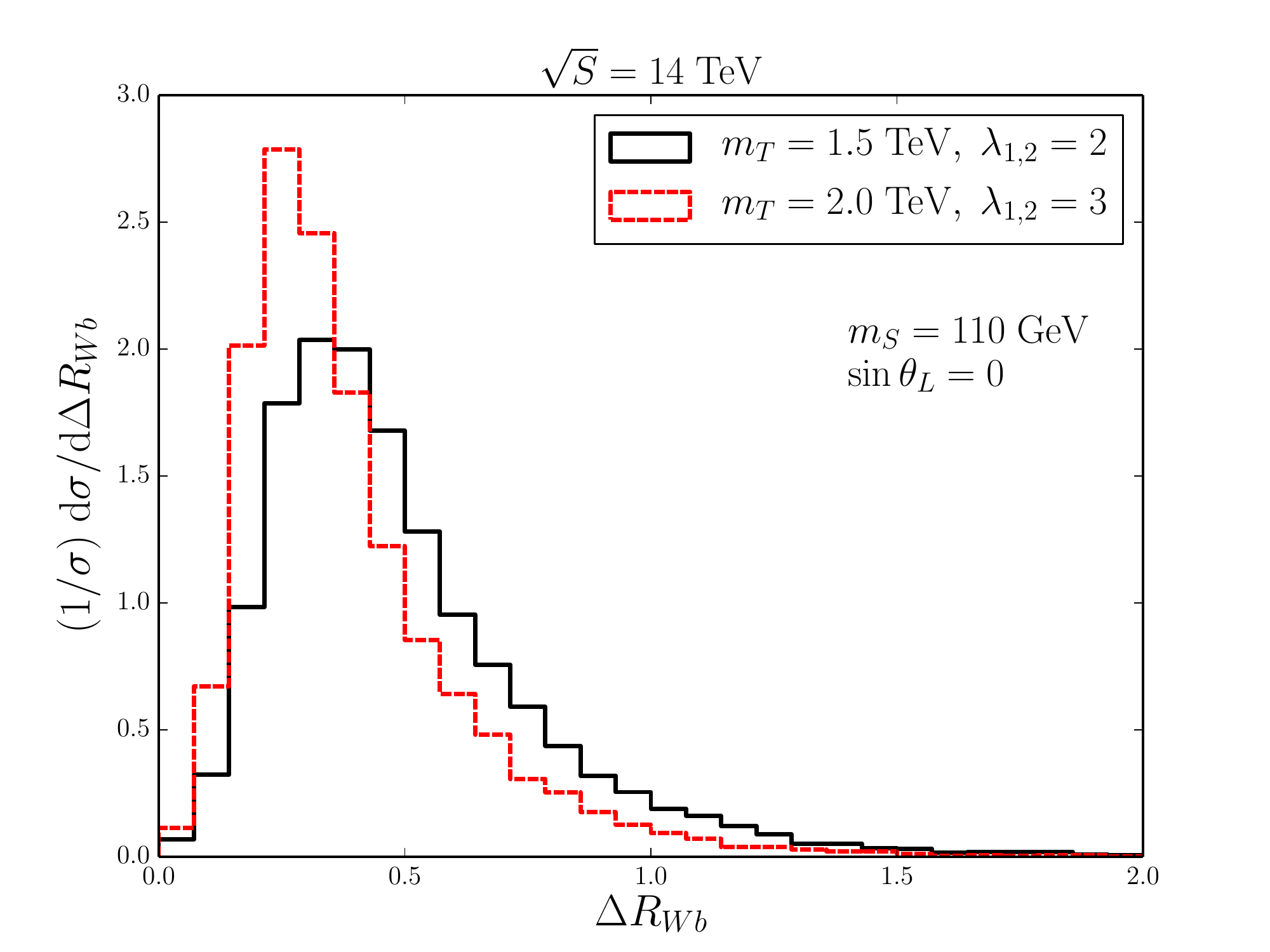}\label{fig:tdr}}\\
\subfigure[]{\includegraphics[width=0.49\textwidth,clip]{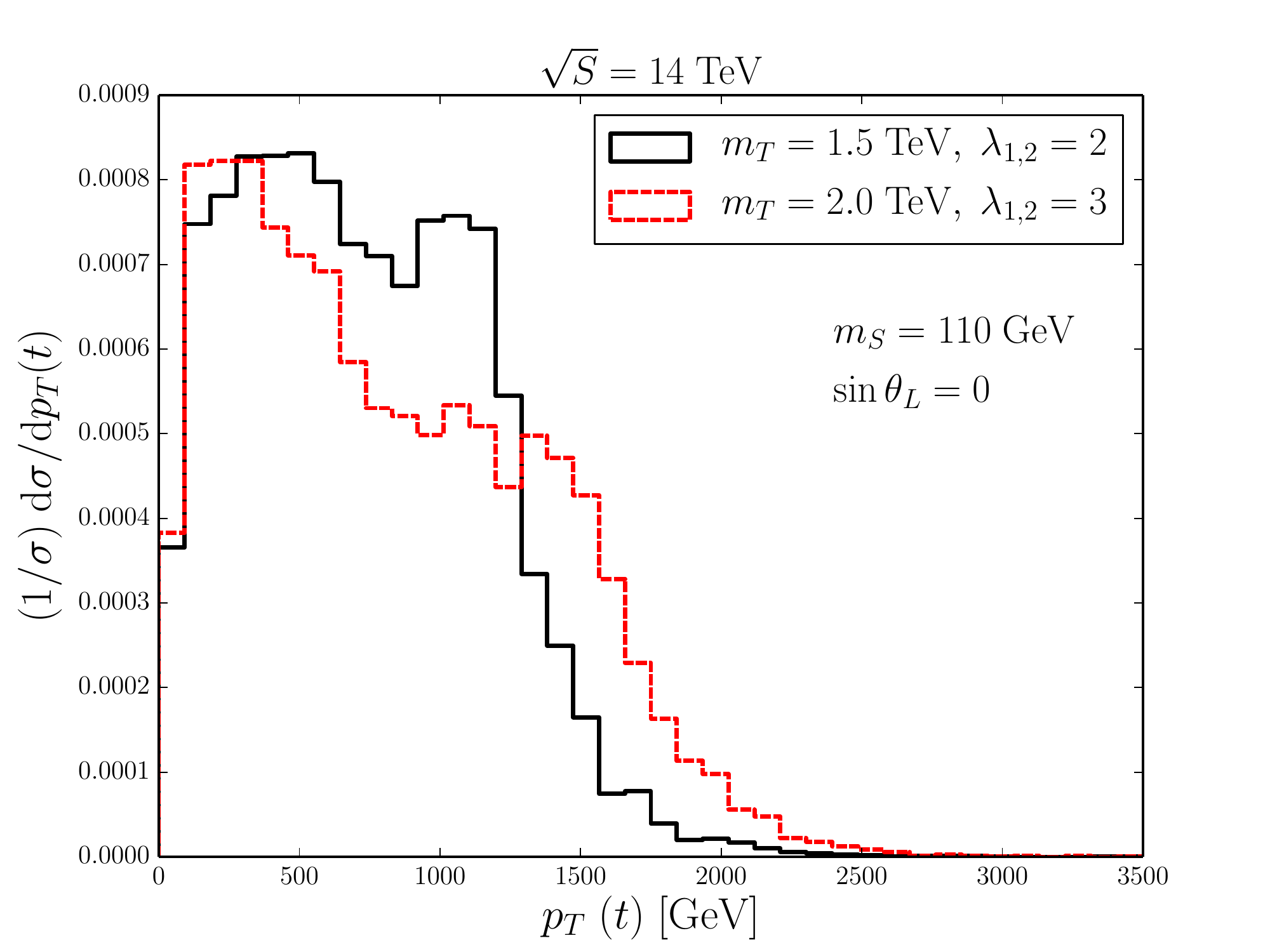}\label{fig:PartPT}}
\subfigure[]{\includegraphics[width=0.49\textwidth,clip]{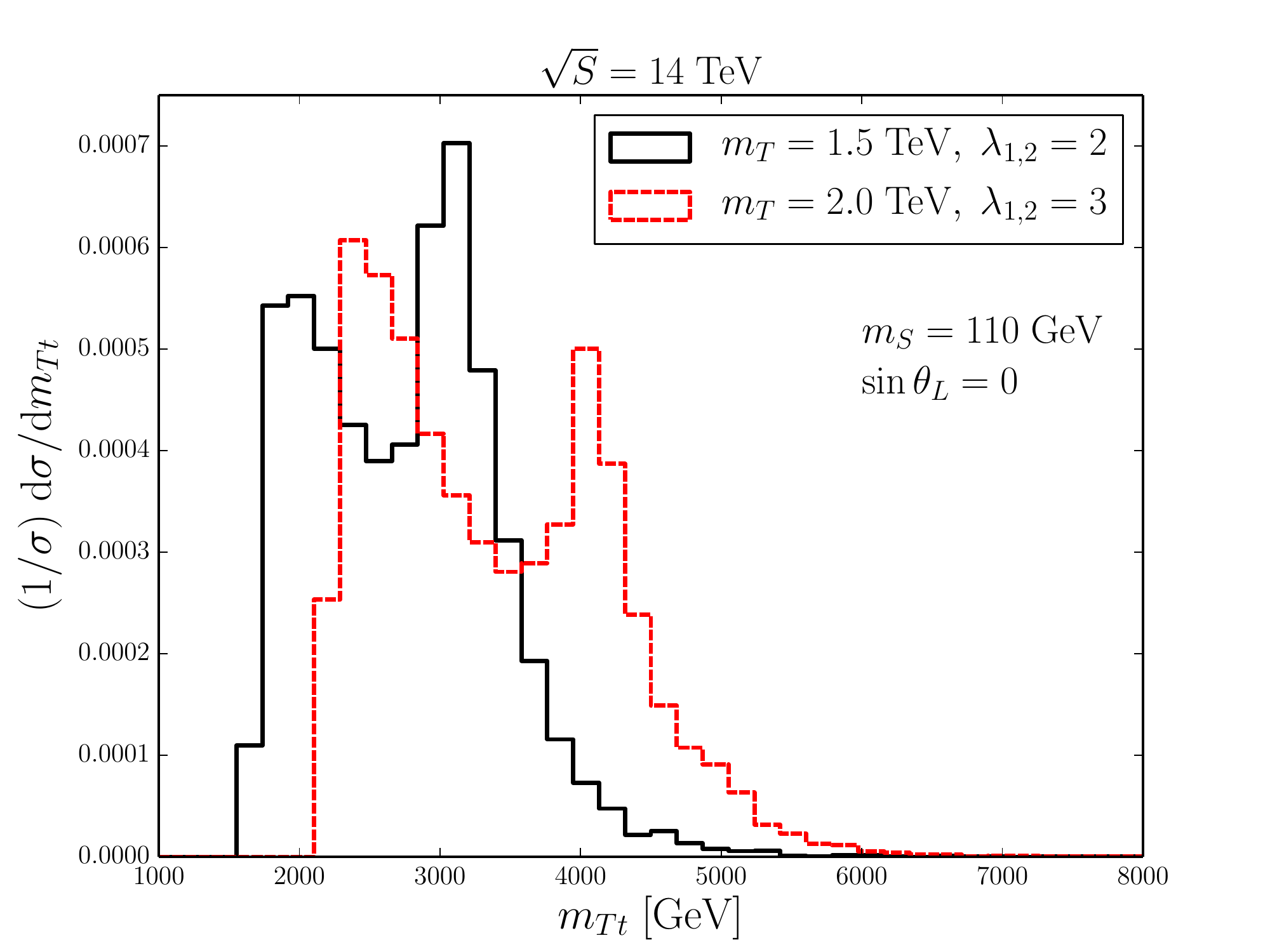}\label{fig:PartMTt}}
\end{center}
\caption{(a) $\Delta R_{gg}$ distribution of the scalar decay products $S\rightarrow gg$.  (b) $\Delta R_{Wb}$ distributions between the top quark decay products $t\rightarrow bW$ originating from the top partner decay $T\rightarrow tS$.  (c) Transverse momentum distribution of the top quark produced in association with the top partner $pp\rightarrow Tt$.  (d) Invariant mass of the top partner and top quark that are produced together $pp\rightarrow Tt$.   Distributions are at parton level and for both (black) $m_T=1.5$~TeV and (red) $m_T=2$~TeV.  The other model parameters are set to $m_{S} = 110~\GeV$ and  $\sin \theta_L = 0$.  The coupling constants are $\lambda_{1,2}=2$ for $m_T=1.5$~TeV and $\lambda_{1,2}=3$ for $m_T=2$~TeV}\label{fig:PartonPlots}
\end{figure}

Since we work in the parameter region that $m_T\gg m_S,m_t$, the top quark and $S$ arising from the heavy $T$ decay are kinematically boosted with high $p_T$.  Hence their decay products are highly collimated. To illustrate this, in Fig.~\ref{fig:sdr} we show $\Delta R_{gg}$ between the two gluons from the $S$ decay, and in Fig.~\ref{fig:tdr} we show $\Delta R_{Wb}$ between the $b$-quark and $W$ from the top quark decay originating from the $T$.  These plots are at partonic level before showering, hadronization, or detector effects have been considered.  The angular separation $\Delta R_{ij}$ is defined as
\begin{eqnarray}
\Delta R_{ij} = \sqrt{(\Delta\phi_{ij})^2+(\Delta \eta_{ij})^2},
\end{eqnarray}
where $\Delta\phi_{ij} = \phi_i-\phi_j$ is the difference of the azimuthal angles of particles $i,j$, and $\Delta\eta_{ij}=\eta_i-\eta_j$ is the difference of the rapidities of the particle $i,j$.  As can be seen, the distributions of $\Delta R_{gg}$ and $\Delta R_{Wb}$ peak at $\Delta R_{gg}\sim\Delta R_{Wb}\sim0.2-0.4$.  

The other top quark produced together with $T$ can also acquire a sizable $p_T$, as shown in Figs.~\ref{fig:PartPT}.   This can be understood via Fig.~\ref{fig:PartMTt}, where we show the top partner-top invariant mass $m_{Tt}$ distribution at partonic level.  In the $\sin\theta_L=0$ limit, only loops containing top partners contribute to $pp\rightarrow Tt$.  When $m_{Tt}\sim 2m_T$, the internal top partners can go on-shell, giving rise to the peaks in the $m_{Tt}$ distributions.  These peaks are quite pronounced.  Hence, there is a relatively strong Jacobian peak at $p_T\sim m_T$, causing the shoulder features in Fig.~\ref{fig:PartPT}.\footnote{We note that the peaks at $m_{Tt}\sim 2 m_T$ and $p_T\sim m_T$ are considerably more pronounced for $q\bar{q}$ initial states than they are for $gg$ initial states.  In fact, the $p_T$ spectrum of the top partner in the $gg$ initial states is smoothly falling from threshold, and the $p_T$ spectrum in the $q\bar{q}$ initial states grows until $p_T\sim m_T$ where it peaks and then falls off.  See also similar discussions presented in Ref.~\cite{Chway:2015lzg, Dawson:2015oha}.}

Since both tops and scalar are all boosted, we require that after showering, hadronization, and detector effects are accounted for that events contain at least one $r= 1.0$ fat jet with 
\begin{eqnarray}
p_T^j > 400 \GeV\quad{\rm and}\quad |\eta^j| < 2.5.\label{eq:basefatjet}
\end{eqnarray}
  The variable $r$ describes the cone-size of the anti-$k_T$ clustering algorithm~\cite{Cacciari:2008gp}, as described in Sections~\ref{sec:signal} and~\ref{sec:BGD}.  Additionally, our signal consists of one leptonically decaying top $t\rightarrow b\ell\nu$.  Hence, we require that our events have missing transverse energy
\begin{eqnarray}
\slashed{E}_T > 20 \GeV\label{eq:ETmiss},
\end{eqnarray}
at least one $r = 0.4$ slim jet with 
\begin{eqnarray}
p_T^j > 30 \GeV\quad{\rm and}\quad|\eta^j| < 2.5,\label{eq:basethinjet}
\end{eqnarray}
and exactly one  isolated lepton passing the cuts in Eq.~(\ref{eq:baseleptons}) and
\begin{eqnarray}
\mathit{mini-iso} > 0.7.\label{eq:isolepton}
\end{eqnarray}
The \textit{mini-iso}~\cite{Rehermann:2010vq} observable is defined as $p_T$ of a lepton divided by the total scalar sum of all charged particles' transverse energy (including the lepton) with $p_T > 1 \GeV$ in the cone of radius $\Delta R = 10 \GeV / p_T^\ell$.

Since both tops are highly boosted the signal contains a fat jet originating from a top quark.  Additionally, we have a fat jet originating from the decay of the scalar.  Both these fat jets will have unique internal substructures due to the daughter particles.  Such events are rare in the SM, and therefore serve as good handles to disentangle the SM backgrounds from our signal events. We use the \texttt{TemplateTagger v.1.0}~\cite{Backovic:2012jk} implementation of the Template Overlap Method (TOM)~\cite{Almeida:2010pa,  Backovic:2013bga} to tag massive boosted objects\footnote{For alternatives to the TOM see Ref.~\cite{Kasieczka:2017nvn} and references therein.}.  The TOM is based on an overlap $Ov_i^a$, where $a$ is a parent particle and $i$ is the number of daughter particles inside a fat jet.  The closer $Ov_i^a$ is to one, the more likely that a fat jet originated from the particle $a$.  This method is flexible enough to tag any type of heavy object and is weakly susceptible to pileup contamination~\cite{Backovic:2013bga}. A multi-dimensional TOM analysis~\cite{Backovic:2014ega, Backovic:2015bca} extends its capability to further unravel multiple boosted objects with different internal substructures, and significantly improves a net tagging efficiency of the hadronically-decaying top and scalar $S(\rightarrow g g)$ jets in the same event.   For a precise definition see Refs.~\cite{Almeida:2010pa,  Backovic:2013bga,Backovic:2012jk}.  For a $r=1.0$ fat jet to be tagged as the hadronic top, we demand a three-pronged top template overlap score 
\begin{eqnarray}
Ov_3^t > 0.6.\label{eq:fattop}
\end{eqnarray}
 We define a fat jet to be an $S$-candidate if it passes a two-pronged $S$ template overlap score and is not tagged as a top-fat jet:
\begin{eqnarray}
Ov_2^S > 0.5\quad{\rm and}\quad Ov_3^t < 0.6.\label{eq:fatscalar}
\end{eqnarray}

\begin{figure*}
  \centering
  \subfigure[]{\includegraphics[scale=0.36]{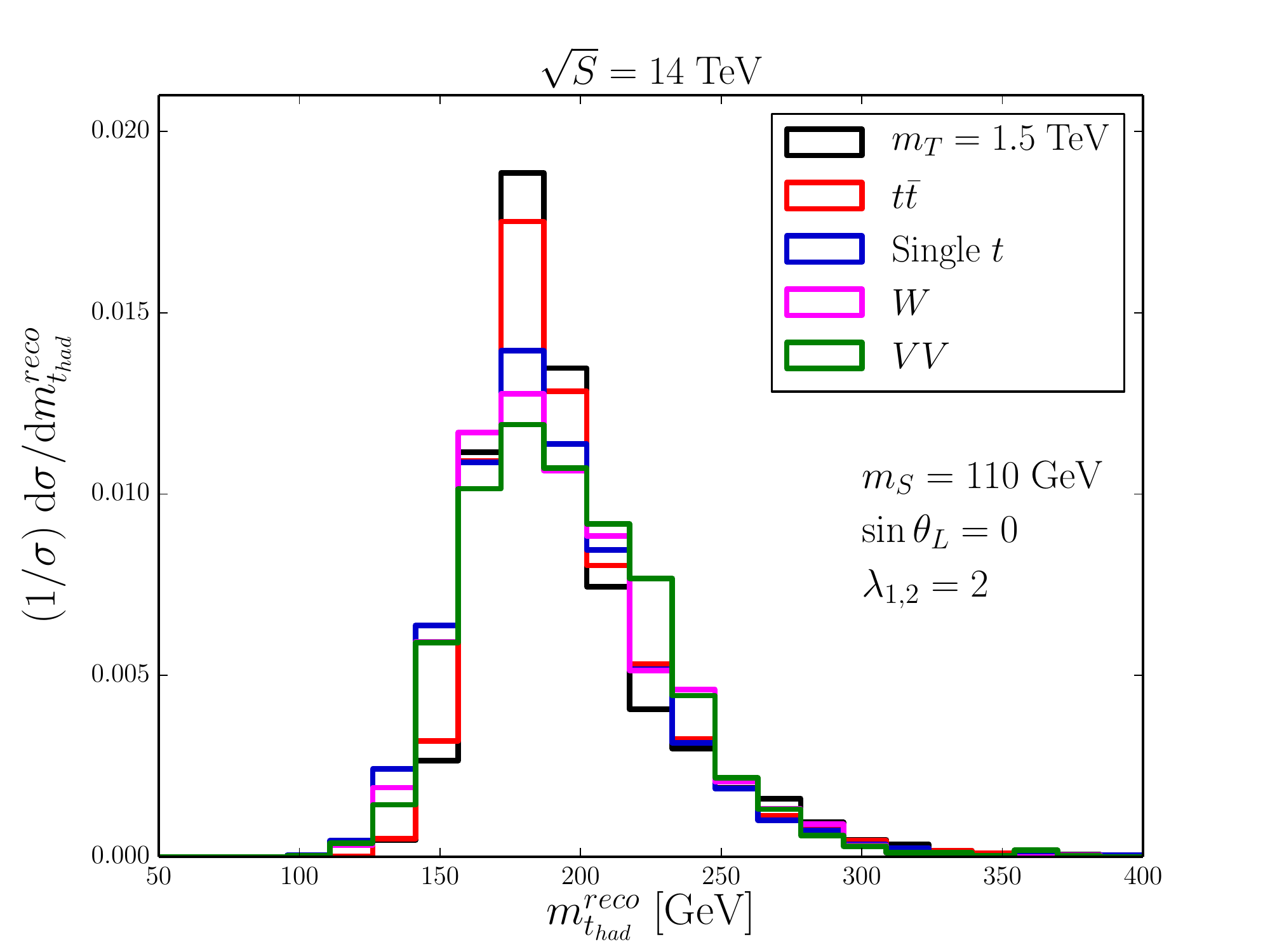}\label{fig:mthad}}
  \subfigure[]{\includegraphics[scale=0.36]{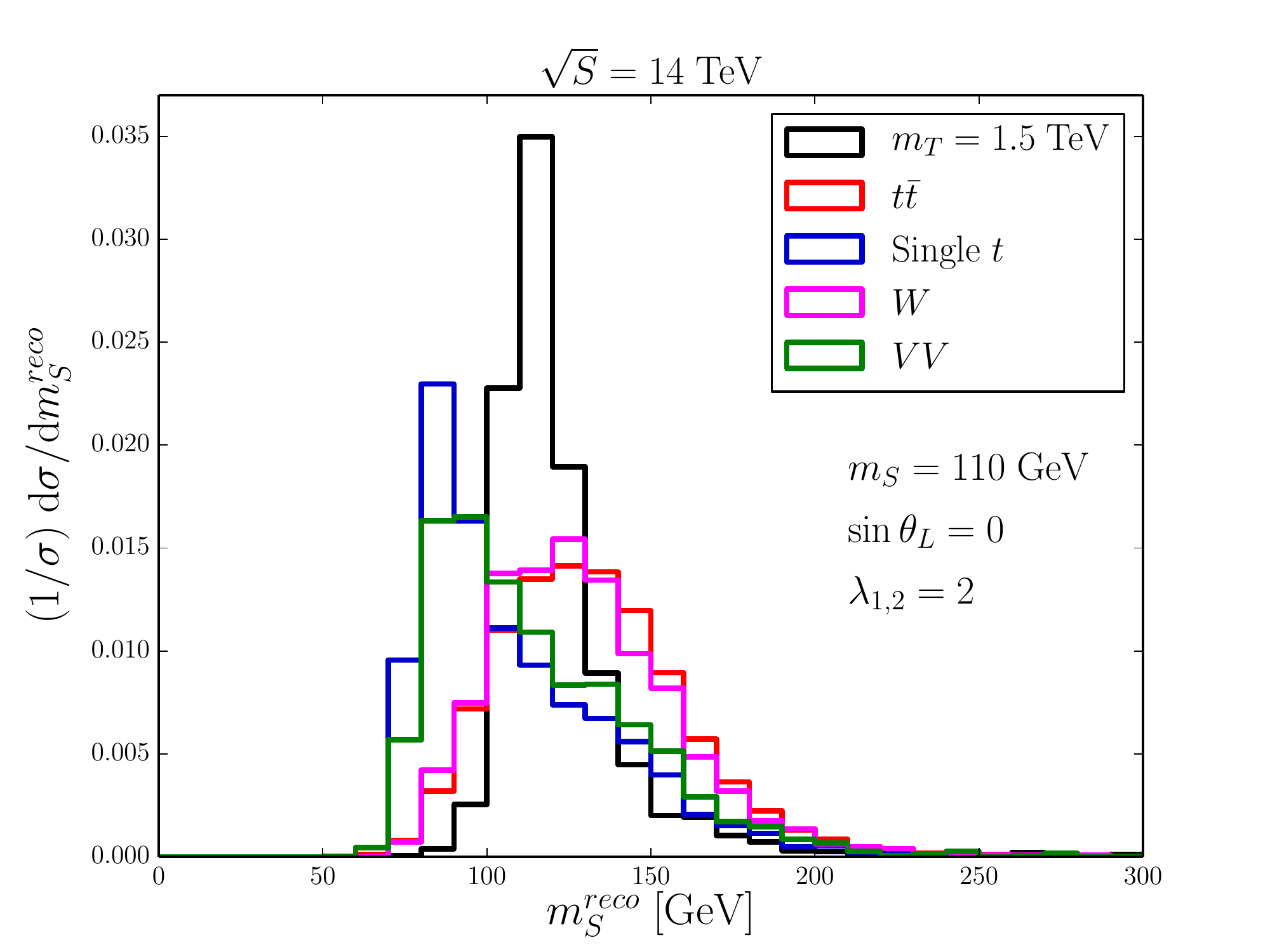}\label{fig:ms}} \\
  \subfigure[]{\includegraphics[scale=0.36]{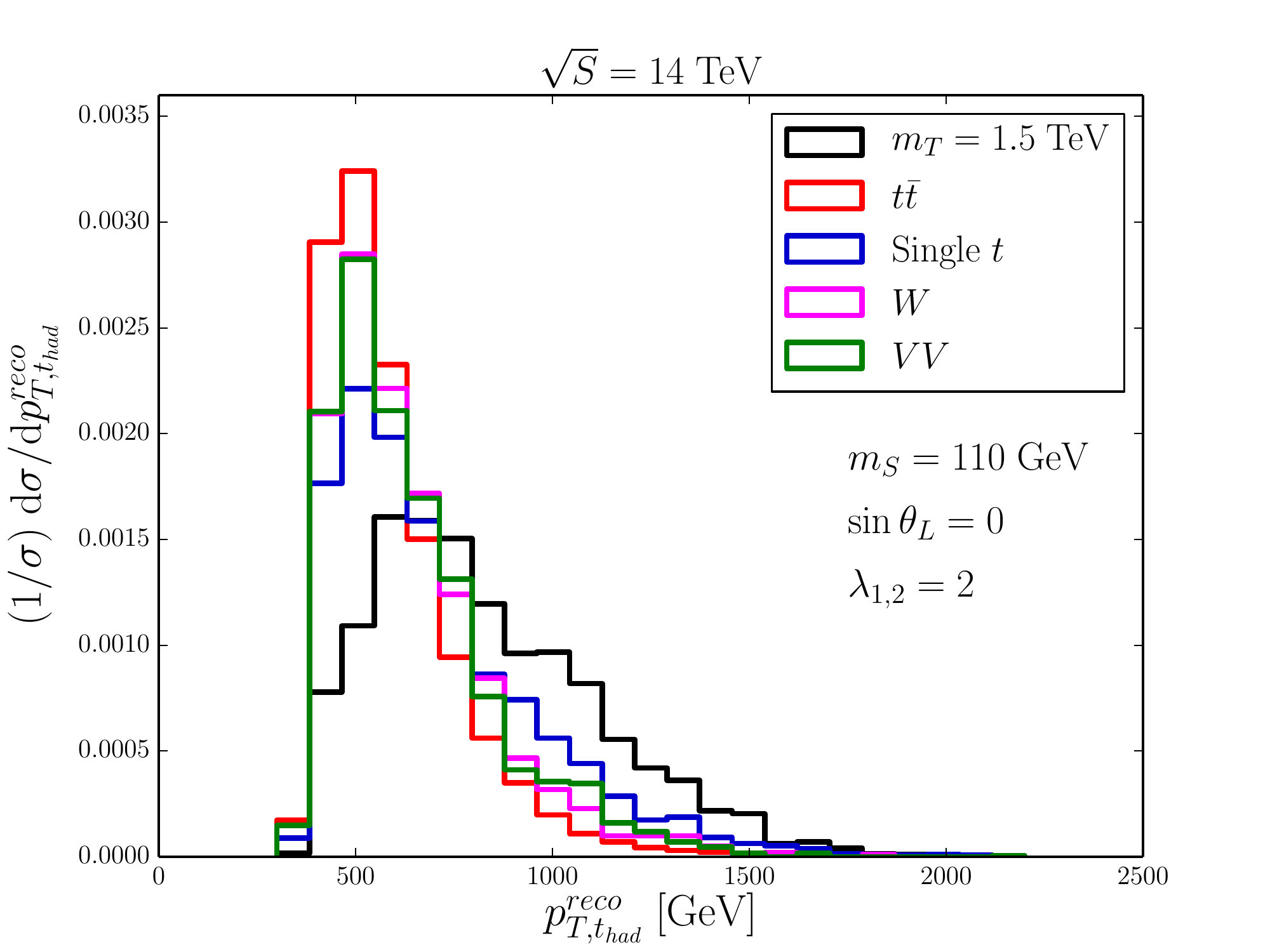}\label{fig:ptthad}}
  \subfigure[]{\includegraphics[scale=0.36]{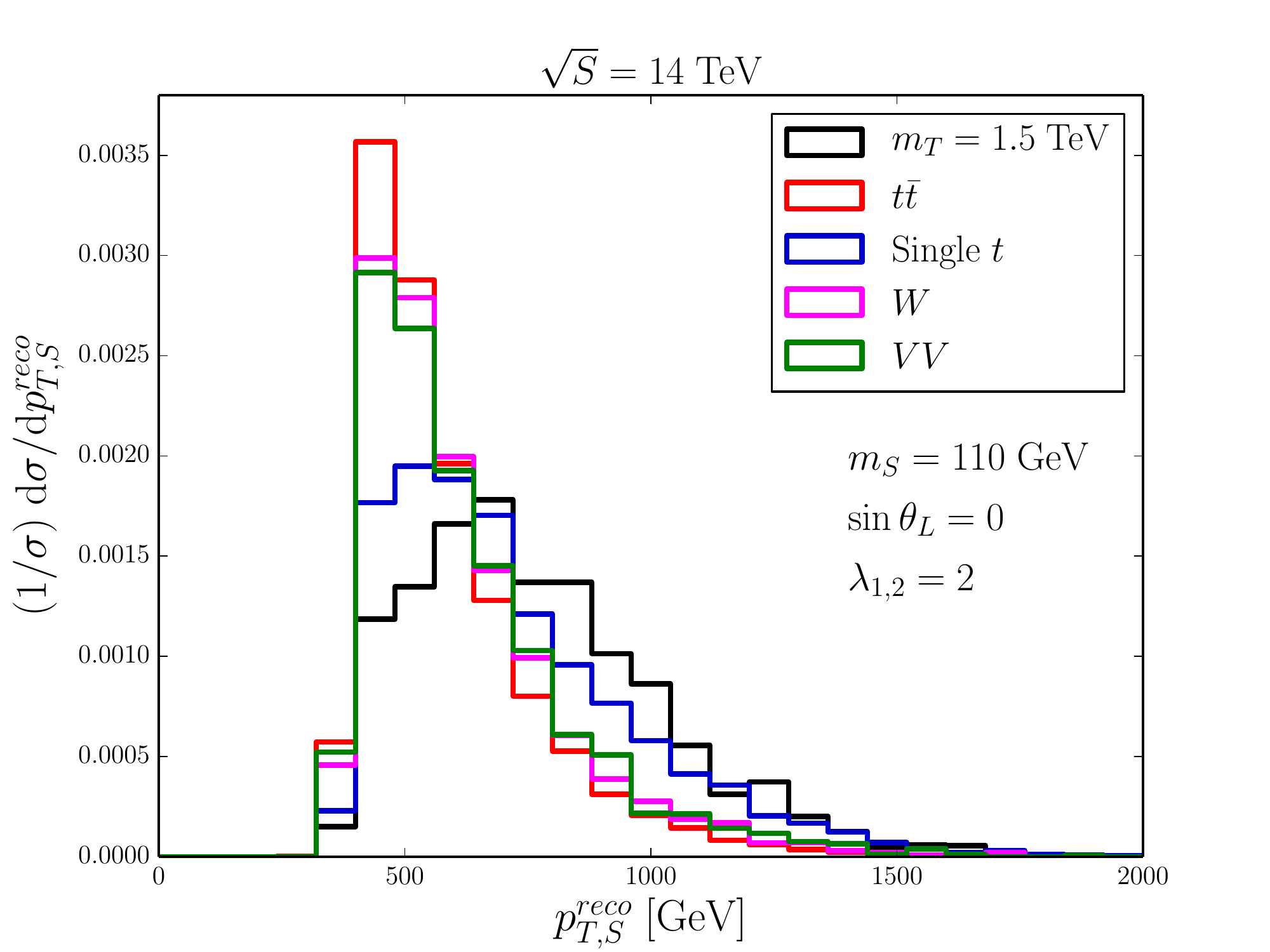}\label{fig:ptS}} 
  \caption{Reconstructed invariant mass distributions of the (a) top-tagged fat jet and (b) scalar-tagged fat jet for $m_{T} = 1.5\TeV$. The corresponding $p_T$ distributions are shown in (c) and (d) together with background distributions. All plots are generated based on events after showering, hadronization, and detector effects.  Our model parameters are $m_{S} = 110~\GeV$, $\lambda_{1,2} = 2$ and  $\sin \theta_L = 0$.  Basic cuts in Eqs.~(\ref{eq:basefatjet}-\ref{eq:isolepton}) have been applied.}\label{fig:PTplots15}
\end{figure*}

\begin{figure*}
  \centering
  \subfigure[]{\includegraphics[scale=0.36]{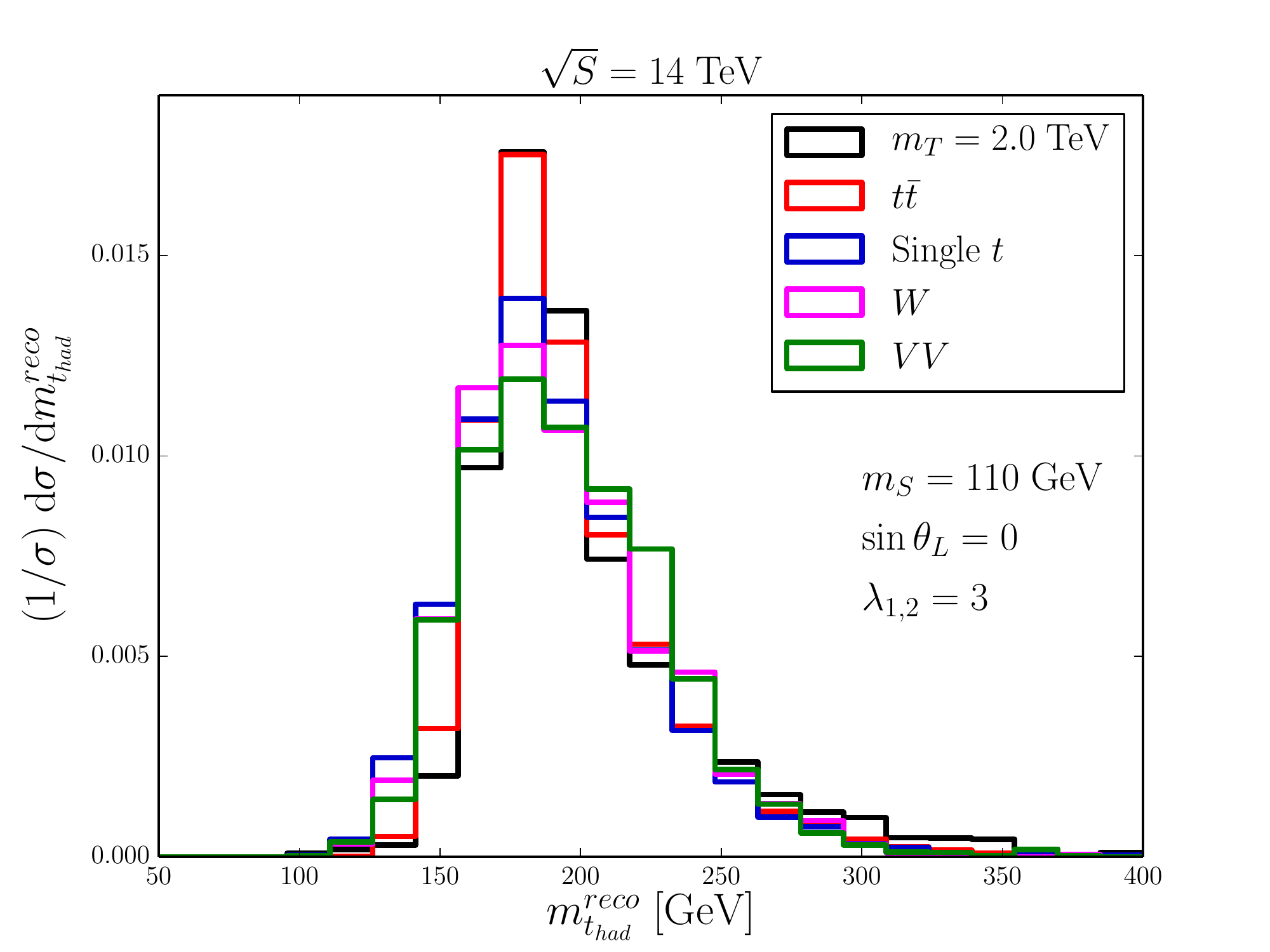}}
  \subfigure[]{\includegraphics[scale=0.36]{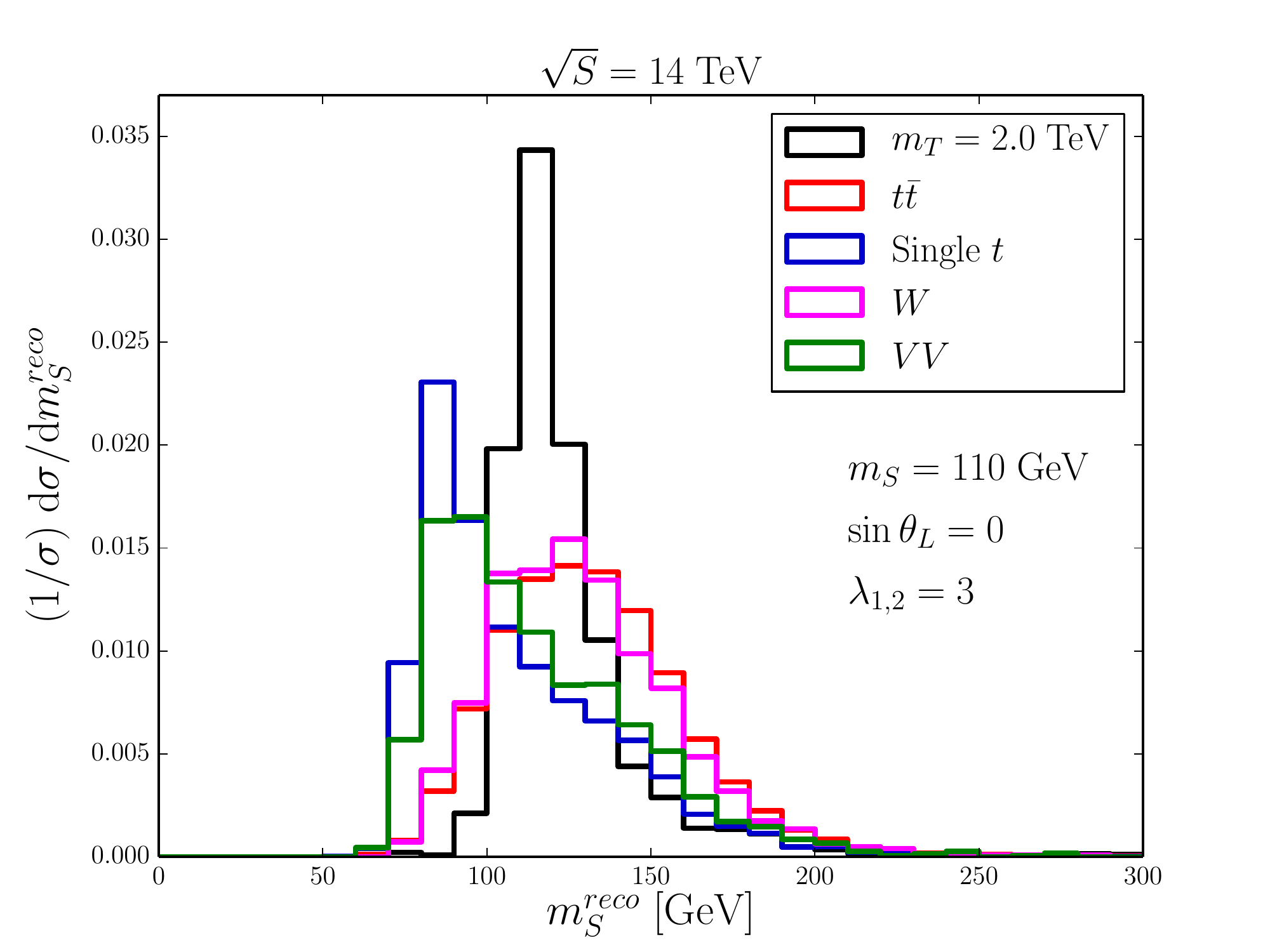}}\\
  \subfigure[]{\includegraphics[scale=0.36]{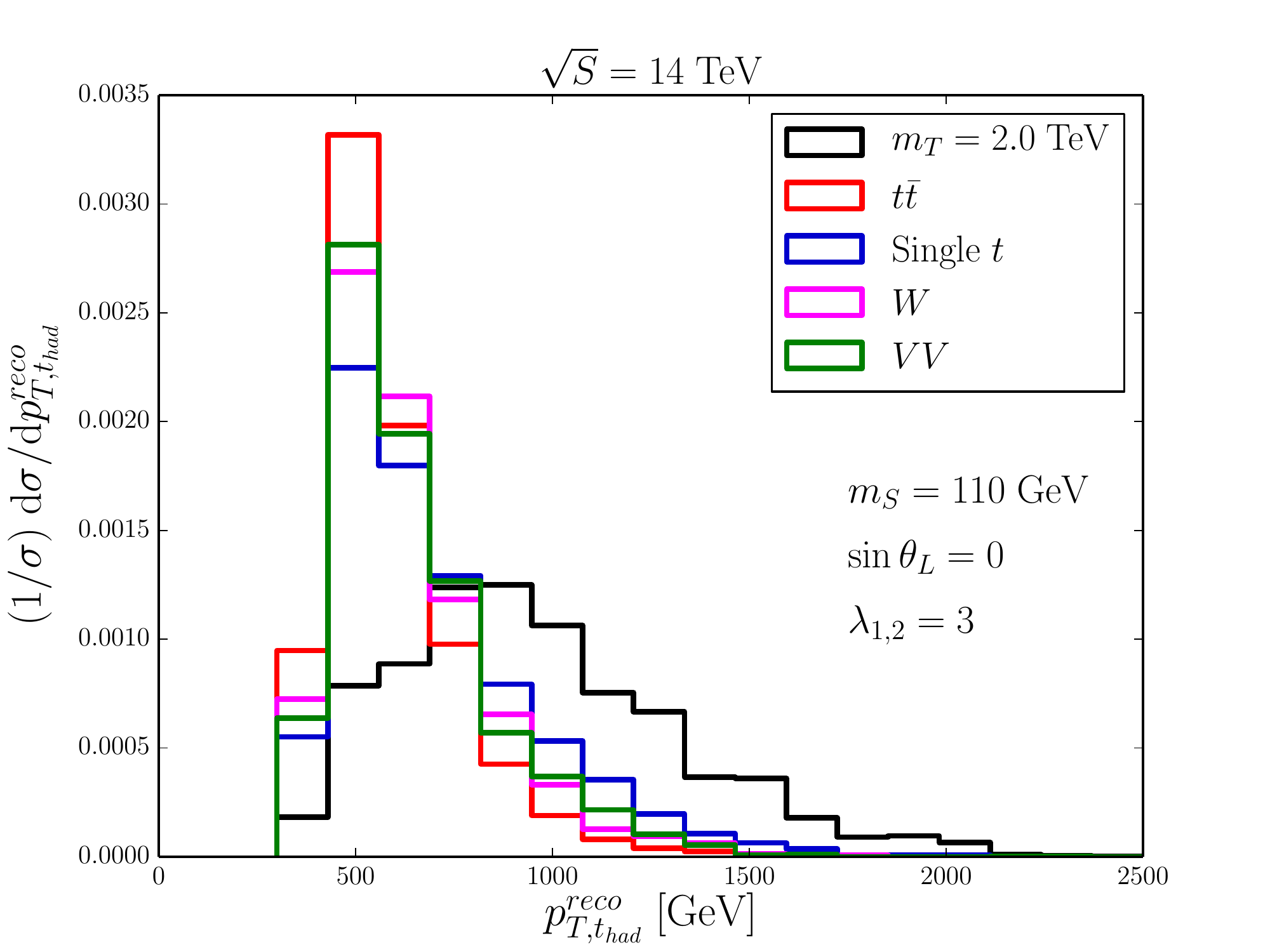}}
  \subfigure[]{\includegraphics[scale=0.36]{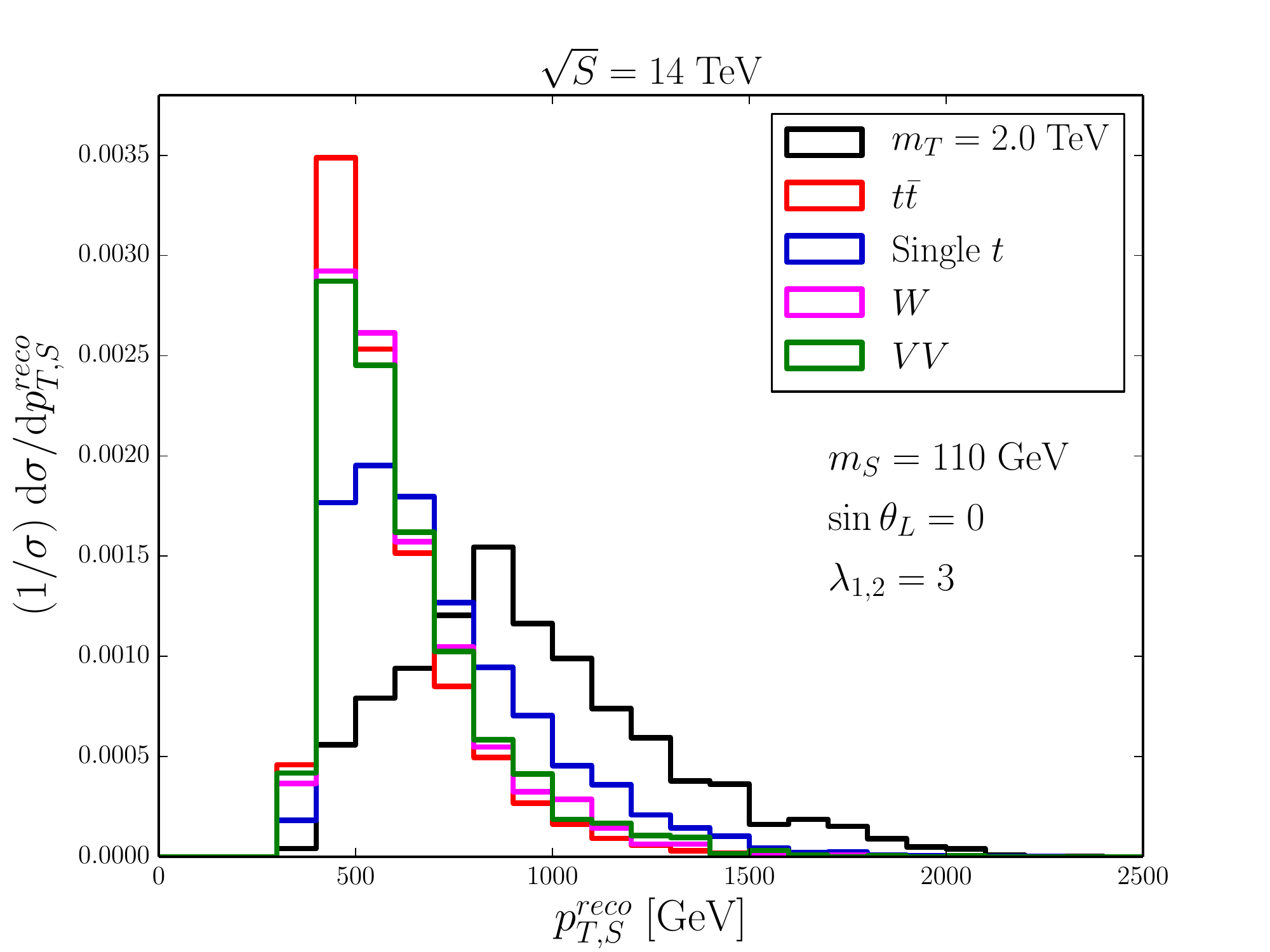} }
  \caption{Reconstructed invariant mass distributions of the (a) top-tagged fat jet and (b) scalar-tagged fat jet  for $m_{T} = 2\TeV$. The corresponding $p_T$ distributions are shown in (c) and (d) together with background distributions. All plots are generated based on events after showering, hadronization, and detector effects.  Our model parameters are $m_{S} = 110~\GeV$, $\lambda_{1,2} = 3$ and  $\sin \theta_L = 0$.  Basic cuts in Eqs.~(\ref{eq:basefatjet}-\ref{eq:isolepton}) have been applied.}\label{fig:PTplots20}
\end{figure*}

Figures~\ref{fig:mthad} and~\ref{fig:ms} show the resulting reconstructed invariant mass distributions of the top-tagged fat jet, $m_{t_{had}}^{reco}$, and the scalar-tagged fat jet, $m_S^{reco}$, respectively,  for $m_{T} = 1.5\TeV$ for both signal and background.  Both the signal and $t\bar{t}$ background $m_{t_{had}}^{reco}$ distributions are highly peaked at the top mass $m_{t}=173$~GeV, while the single top and vector boson backgrounds are not quite as peaked.  However, the reconstructed scalar mass $m_S^{reco}$ provides more separation from background.  For the signal, the $m_S^{reco}$ distribution is highly peaked at the scalar mass $m_S=110$~GeV, while the background is not.  Hence, for $m_T=1.5$~TeV we apply the cuts
\begin{eqnarray}
145~{\rm GeV}< &m_{t_{had}}^{reco}&<250~{\rm GeV}\quad{\rm and}\label{eq:mthadreco15}\\
 105~{\rm GeV}<&m_S^{reco}&<122~{\rm GeV}\label{eq:msreco}.
\end{eqnarray}     
Figures~\ref{fig:ptthad} and~\ref{fig:ptS} show the transverse momentum distribution of the top-tagged fat jet and scalar-tagged fat jet, respectively.  As can be clearly seen, the signal is harder than the background.  For $m_T=1.5$~TeV we place the further cut on the reconstructed scalar transverse momentum
\begin{eqnarray}
p^{reco}_{T,S}>540~{\rm GeV}.\label{eq:scalarpt15}
\end{eqnarray}
Finally, since the top-tagged fat jet should contain a $b$-quark, at least one $b$-tagged slim $r=0.4$ jet should be found inside the top-tagged fat jet.\footnote{In our semi-realistic approach for the $b$-jet identification, $r=0.4$ jets are classified into three categories where our heavy-flavor tagging algorithm iterates over all jets that are matched to $b$-hadrons or $c$-hadrons. If a $b$-hadron ($c$-hadron) is found inside, it is classified as a $b$-jet ($c$-jet). The remaining unmatched jets are called light-jets. Each jet candidate is further multiplied by a tag-rate~\cite{ATL-PHYS-PUB-2016-026}, where we apply a flat $b$-tag rate of $\epsilon_{b \rightarrow b} = 0.7$ and a mis-tag rate that a $c$-jet (light-jet) is misidentified as a $b$-jet of $\epsilon_{c \rightarrow b} = 0.2$ ($\epsilon_{j \rightarrow b} = 0.01)$. For a $r= 1.0$ fat jet to be $b$-tagged, on the other hand, we require that a $b$-tagged $r = 0.4$ jet is found inside a fat jet. To take into account the case where more than one $b$-jet might land inside a fat jet, we reweight a $b$-tagging efficiency depending on a $b$-tagging scheme described in Ref.~\cite{Backovic:2015bca}.}  We require that exactly one top-tagged fat jet passes the cut in Eq.~(\ref{eq:mthadreco15}) and has a $b$-tagged slim jet inside,
 and exactly one scalar-tagged fat jet passes the cuts in Eqs.~(\ref{eq:msreco}) and (\ref{eq:scalarpt15}):
\begin{eqnarray}
N^{1.5}_{t_{had}}=1\quad{\rm and}\quad N^{1.5}_S=1,\,{\rm respectively.}\label{eq:NTS15}
\end{eqnarray}

Table~\ref{tab:Cutflow1} is a cut-flow table showing the cumulative effects of cuts on signal and background rates.    Relative to the basic cuts in Eqs.(\ref{eq:basefatjet}-\ref{eq:isolepton}), under the requirement that $N_{t_{had}}^{1.5}=N_S^{1.5}=1$, the signal efficiency is $5.8\%$, while the major backgrounds  $t \overline{t}$ and single $t$ have efficiencies of $0.085\%$ and $0.057\%$, respectively.  The $W$ and $VV$ backgrounds are cut down to $0.0036\%$ and $0.0028\%$, respectively, greatly diminishing the overall size of backgrounds.

For $m_T=2$~TeV, the reconstructed invariant mass and transverse momentum distributions for the top-tagged and scalar-tagged fat jets are shown in Fig.~\ref{fig:PTplots20}.   The observations for $m_T=2$~TeV are largely the same as for $m_T=1.5$ TeV, except the $p_T$ spectrum of the top-tagged and scalar-tagged fat jets are harder for the signal.  Hence, for the targeted $m_{T} = 2.0\TeV$ analysis, we slightly tighten the $m_{t_{had}}^{reco}$ mass window
\begin{eqnarray}
155~{\rm GeV} < m_{t_{\rm{had}}}^{reco} < 250~{\rm GeV}.~\label{eq:mthadreco2}
\end{eqnarray} 
For the scalar-tagged fat jet we use the same mass window as Eq.~(\ref{eq:msreco}), but harden the transverse momentum cut:
\begin{eqnarray}
p_{T,S}^{reco} > 560~{\rm GeV}\label{eq:scalarpt2}
\end{eqnarray}
  However, as the $T$ mass scale increases, we confront the challenge that the signal cross section steeply decreases, weakening our significance.  To retain more signal events, we do not require that a $b$-tagged slim jet be found inside the top-tagged fat jet.
Hence, for $m_T=2$ TeV we require exactly one top-tagged fat jet that passes the cut in Eq.~(\ref{eq:mthadreco2}) and without the $b$-tagging requirement, and exactly one scalar-tagged fat jet that passes the cuts in Eqs.~(\ref{eq:msreco}) and~(\ref{eq:scalarpt2}):
\begin{eqnarray}
N^{2.0}_{t_{had}}=1\quad{\rm and}\quad N^{2.0}_S=1,\,{\rm respectively.}\label{eq:NTS2}
\end{eqnarray}
As can be seen in Table~\ref{tab:Cutflow1}, due to the relaxation of the $b$-tagging requirement, all efficiencies for background and signal are larger as compared to the $m_T=1.5$~TeV case.  However, the backgrounds are still efficiently suppressed, especially the backgrounds that do not contain top quarks.

To further separate signal from background, it is useful to fully reconstruct the event.  However, this means reconstructing the leptonically decaying top, $t_{lep}$, and the missing neutrino momentum.  First, to help reconstruct the top quark, we require that at least one of the slim jets passing the cuts in Eq.~(\ref{eq:basethinjet}) is also tagged as a $b$-jet and  meets the endpoint criteria 
\begin{eqnarray}
m_{b\ell} < 153.2 + \Gamma \GeV\label{eq:endpoint}
\end{eqnarray}
 where $m_{b\ell}$ is the invariant mass of the $b$-tagged slim jet and isolated lepton, and $\Gamma$ is a headroom to take into account effects of parton showering and hadronization. We choose ${\Gamma = 20 \GeV}$ to keep signal events up to $\sim 90 \%$. We then reconstruct the momentum of the missing neutrino following the prescription in Ref.~\cite{Barger:2006hm, Gopalakrishna:2010xm}. The total transverse momentum of the system is zero, so the transverse momentum of neutrino is just the missing transverse momentum.  However, the longitudinal component of the neutrino momentum is still unknown and cannot be determined via momentum conservation since the longitudinal momentum of the initial state is unknown at hadron colliders.  We will use the on-shell mass constraints that the invariant mass of the neutrino is $p_\nu^2=0$ and the invariant mass of the isolated lepton and neutrino satisfy $m^2_{\ell \nu} = m^2_{W}$.  Since these are quadratic equations, there are two possible solutions for the neutrino longitudinal momentum
\bea
\label{eq:Longi} 
	p^\nu_{L} = \frac{1}{2 ~ (p^\ell_{T})^2  } \Bigg( A ~ p^\ell_{L} \pm |\vec{p}_\ell| ~ \sqrt{ A^2 - 4 \,\left(p^\ell_{T}\right)^2\,  \slashed{E}^2_T   } \Bigg) \;,\label{eq:longneut}
\eea
where $ A =  m^2_W + 2 \vec{p}^{\; \ell}_{T} \cdot \vec{ \slashed{E}}_{T} $, $p^\ell_L$ is the lepton longitudinal momentum, $\vec{p}_\ell$ is the lepton's three-momentum, $\vec{p}_T^{\; \ell}$ is the lepton's transverse momentum vector, and $\vec{\slashed{E}}_T$ is the missing transverse energy vector.  To break the two fold-ambiguity of Eq.~(\ref{eq:longneut}) and to determine which $b$-jet originates from the leptonically decay top, we use the top quark mass constraint.  We select the $b$-jet and $p_L^\nu$ pair that minimizes the quantity
\begin{eqnarray}
| m^2_{b \ell \nu} - m^2_t |
\end{eqnarray}
where $m_{b\ell\nu}$ is the invariant mass of a $b$-jet, lepton, and neutrino system.   The resulting $b$-jet and neutrino momentum are used to reconstruct the leptonically decaying top, $t_{lep}$.  Once we have reconstructed the leptonically decay top, we require that it has the correct mass and has fairly high $p_T$:
\begin{eqnarray}
150~{\rm GeV}<&m^{reco}_{t_{lep}}&<210~{\rm GeV},\quad p_{T,t_{lep}}^{reco}>500~{\rm GeV},\quad{\rm for}\,m_T=1.5~{\rm TeV},\,{\rm and}\label{eq:mtlepreco15}\\
150~{\rm GeV}<&m^{reco}_{t_{lep}}&<220~{\rm GeV},\quad p_{T,t_{lep}}^{reco}>680~{\rm GeV},\quad{\rm for}\,m_T=2~{\rm TeV}.\label{eq:mtlepreco2}
\end{eqnarray}
As we can see from the fourth rows of Table~\ref{tab:Cutflow1}, as compared to the $N_{t_{had}}$ and $N_S$ cuts, after $t_{lep}$ reconstruction the vector boson backgrounds are reduced by $2-5$ orders of magnitude, the single top background efficiency is $1-4\%$, and the $t\bar{t}$ efficiency is $1-7\%$.  The signal efficiency is $20-40\%$.

\begin{figure}[tb]
  \centering
  \subfigure[]{\includegraphics[scale=0.36]{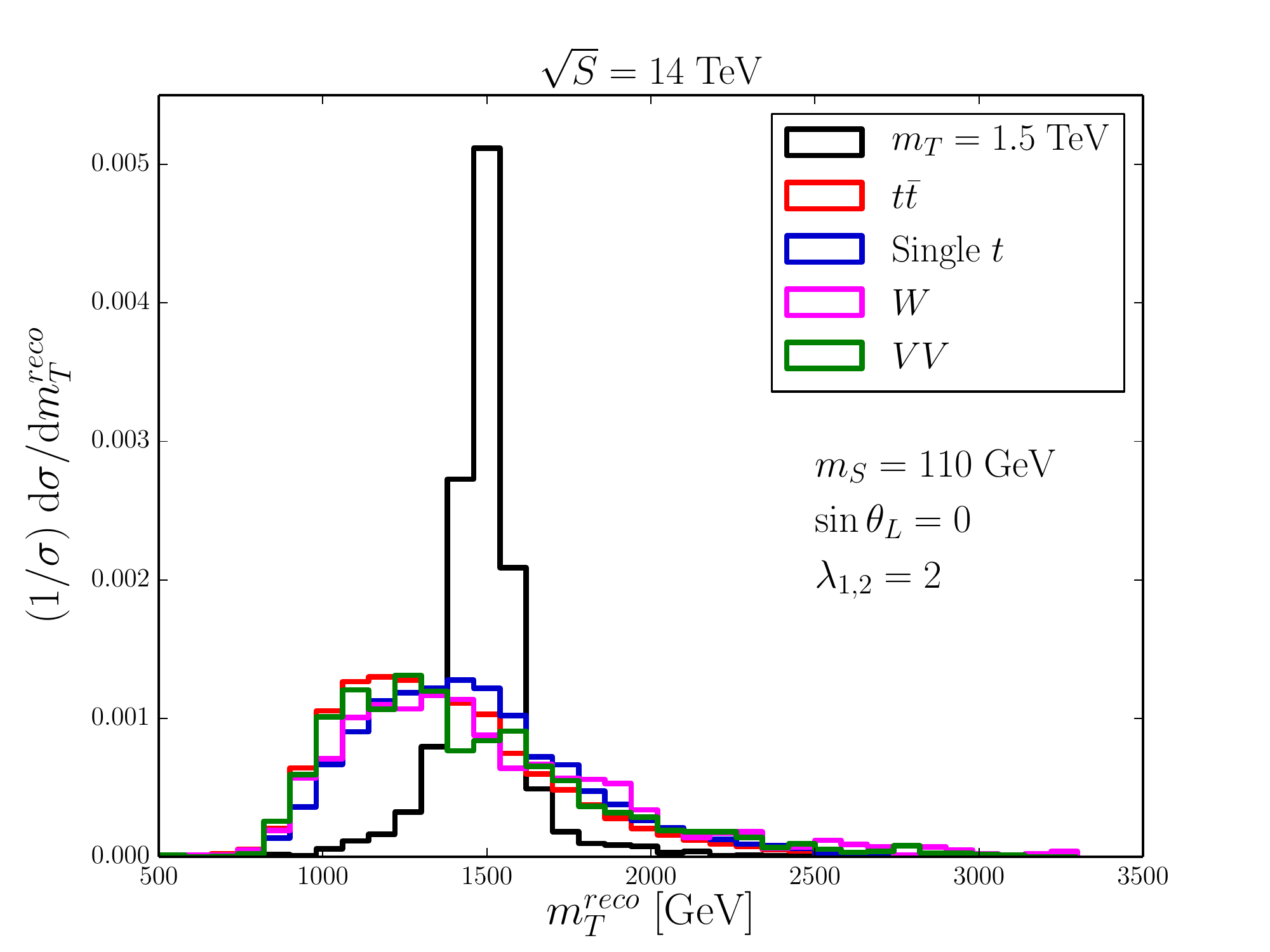}\label{fig:mtreco15}}
  \subfigure[]{\includegraphics[scale=0.36]{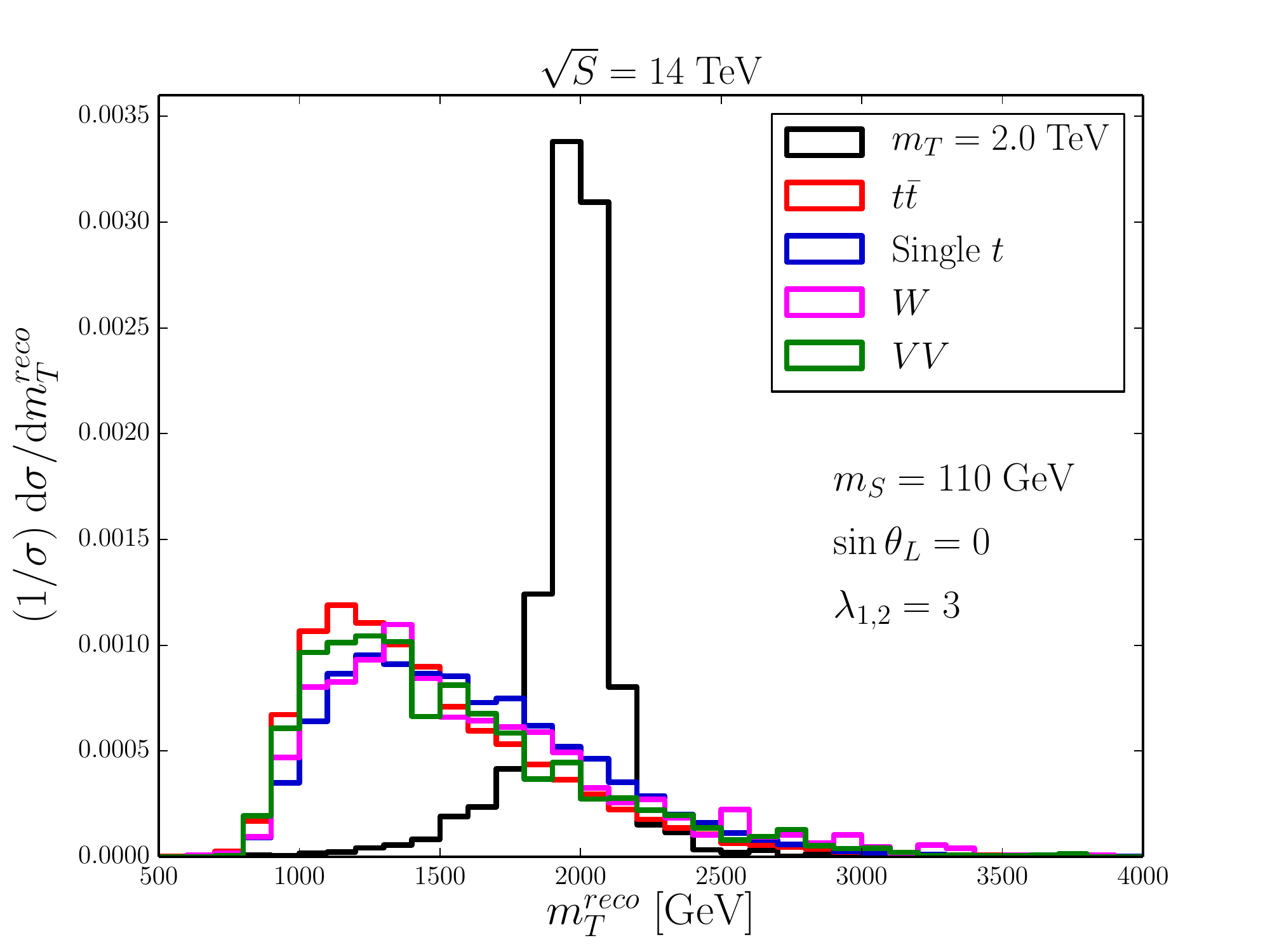}\label{fig:mtreco20}} \\
  \subfigure[]{\includegraphics[scale=0.36]{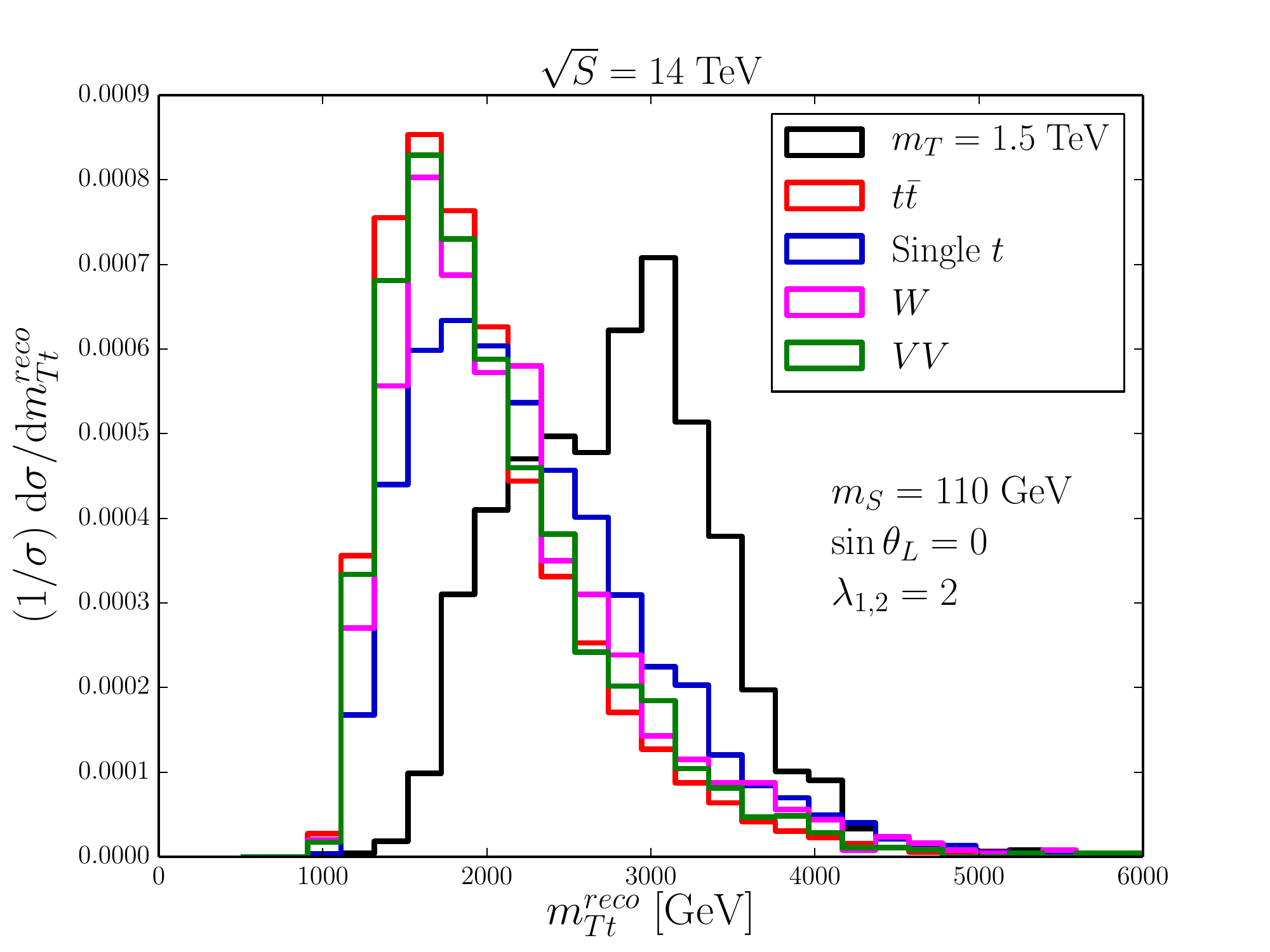}\label{fig:mttreco15} } 
  \subfigure[]{\includegraphics[scale=0.36]{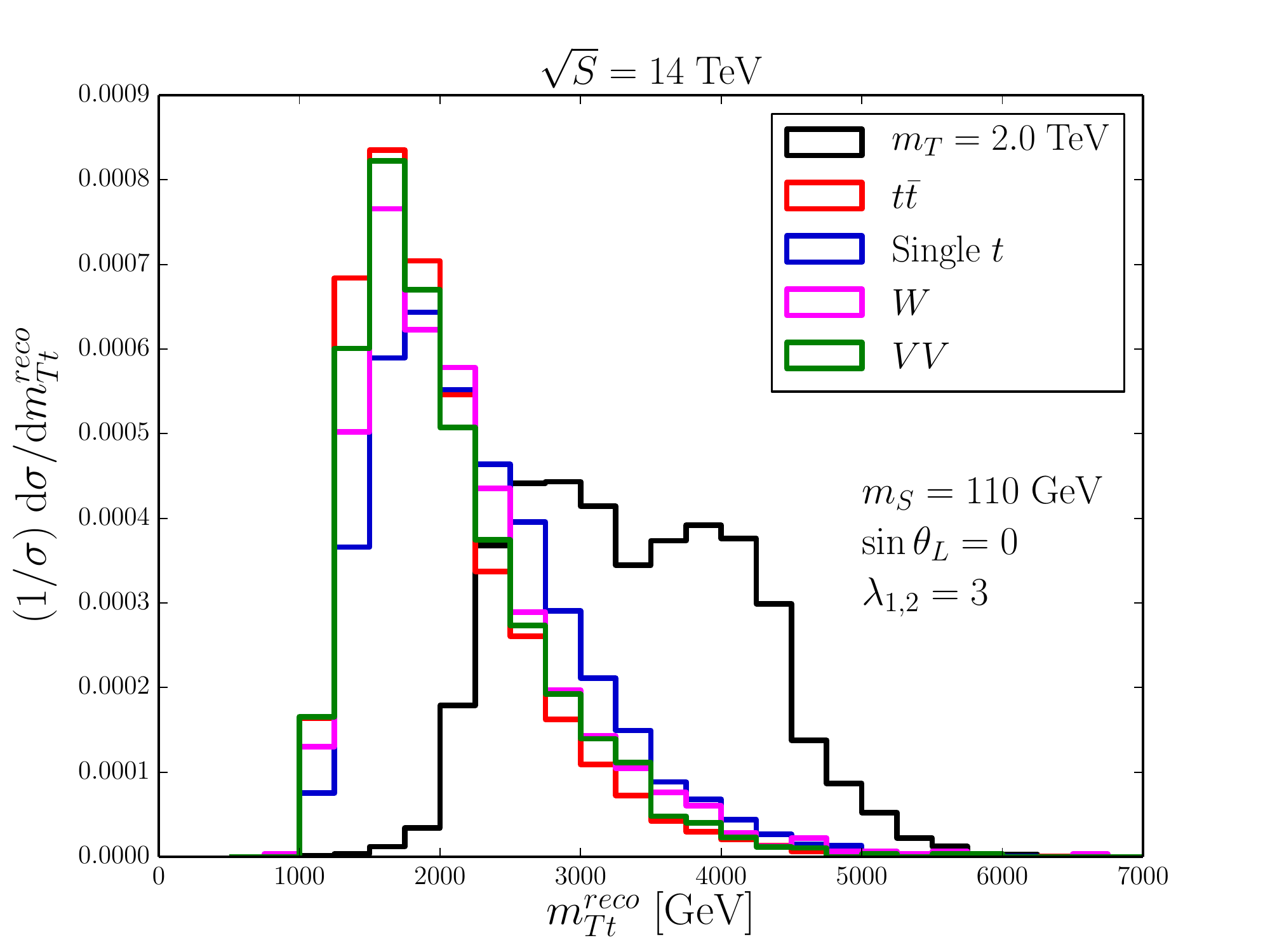}\label{fig:mttreco20} }
  \caption{The reconstructed $T$ invariant mass $m_T^{reco}$ distributions are shown for (a) $m_{T} =$ 1.5 TeV and (b) 2 TeV. The reconstructed system invariant mass $m_{T t}^{reco}$ distributions are shown for (c) $m_{T} =$ 1.5 TeV and (d) 2 TeV. All plots are generated based on events after showering, hadronization, and detector effects.  The other model parameters are set to $m_{S} = 110~\GeV$ and  $\sin \theta_L = 0$. The coupling constants are $\lambda_{1,2}=2$ for $m_T=1.5$~TeV and $\lambda_{1,2}=3$ for $m_T=2$~TeV.  Basic cuts in Eqs.~(\ref{eq:basefatjet}-\ref{eq:isolepton}) have been applied.} \label{tab:Mplots}
\end{figure}

Although the background is greatly reduced, to suppress it further relative to signal we will use the reconstructed top partner mass $m_T^{reco}$ and the total invariant mass of the reconstructed system.  While the top quarks and scalar are fully reconstructed, it is not clear yet which top quark originated from the top partner decay.  We select the pair $\{ S , t_{i}  \}$, where $i=had,lep$ denotes either the hadronic or leptonic top, that best reconstructs $T$ by minimizing the mass asymmetry variable 
\bea
\label{eq:MassAsy} 
	\Delta_m  = \Bigg| \frac{  m_{T} - m^{reco}_{S  t_{i}} }{ m_{T} + m^{reco}_{S  t_{i}} } \Bigg|~\;,
\eea
for each mass point $m_{T} = 1.5$ and $2~\TeV$, where $m^{reco}_{S  t_{i}}$ stands for the invariant mass of the pair $\{ S , t_{i}  \}$. The resulting reconstructed top partner invariant mass, $m_T^{reco}$, distributions are shown in Figs.~\ref{fig:mtreco15} and~\ref{fig:mtreco20} for $m_T=1.5$~TeV and $m_T=2$~TeV, respectively.  They clearly peak at the truth level top partner invariant mass.  Hence, we apply the cuts
\begin{eqnarray}
1400~{\rm GeV}<&m_T^{reco}&<1550~{\rm GeV}\quad{\rm for}\,m_T=1.5~{\rm TeV}\,{\rm and}\label{eq:mtreco15}\\
1860~{\rm GeV}<&m_T^{reco}&<2100~{\rm GeV}\quad{\rm for}\,m_T=2~{\rm TeV}.\label{eq:mtreco2}
\end{eqnarray}

One of the compelling features of the loop-induced single production channel is that the reconstructed system invariant mass $m^{reco}_{T t}$ distribution retains the peak-like structures at high invariant mass.  We show this in Figs.~\ref{fig:mttreco15} and~\ref{fig:mttreco20} for $m_T=1.5$~TeV and $2$~TeV, respectively.  Since the backgrounds are peaked at much lower invariant mass, they can be further suppressed with the cuts
\begin{eqnarray}
2865~{\rm GeV}&<&m_{Tt}^{reco}\quad{\rm for}\,m_T=1.5~{\rm TeV}\,{\rm and},\label{eq:mttreco15}\\
3000~{\rm GeV}&<&m_{Tt}^{reco}\quad{\rm for}\,m_T=2~{\rm TeV}.\label{eq:mttreco2}
\end{eqnarray} 

\begin{figure}[tb]
  \centering
  \subfigure[]{\includegraphics[scale=0.36]{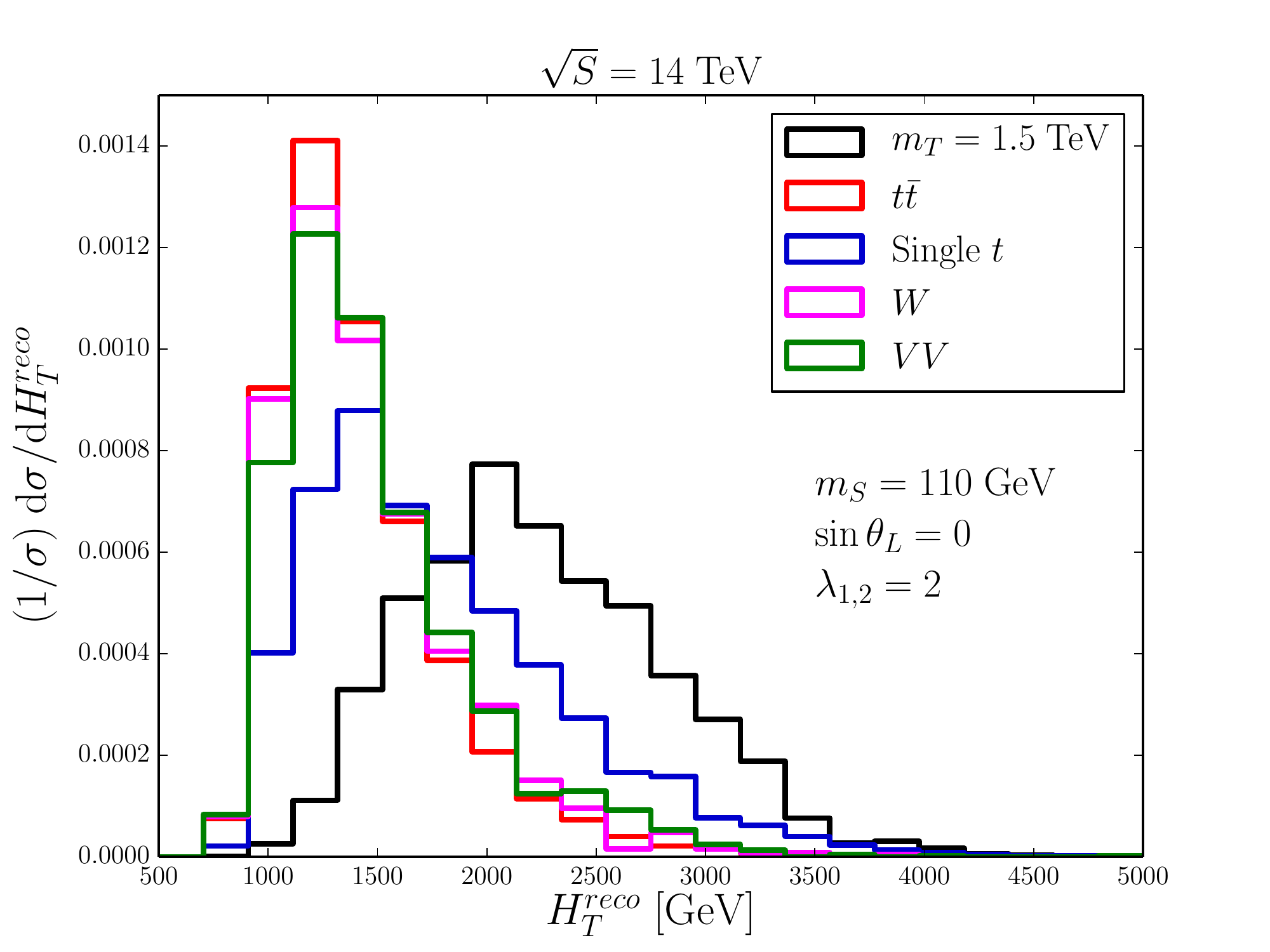}\label{fig:htreco15}}
 \subfigure[]{ \includegraphics[scale=0.35]{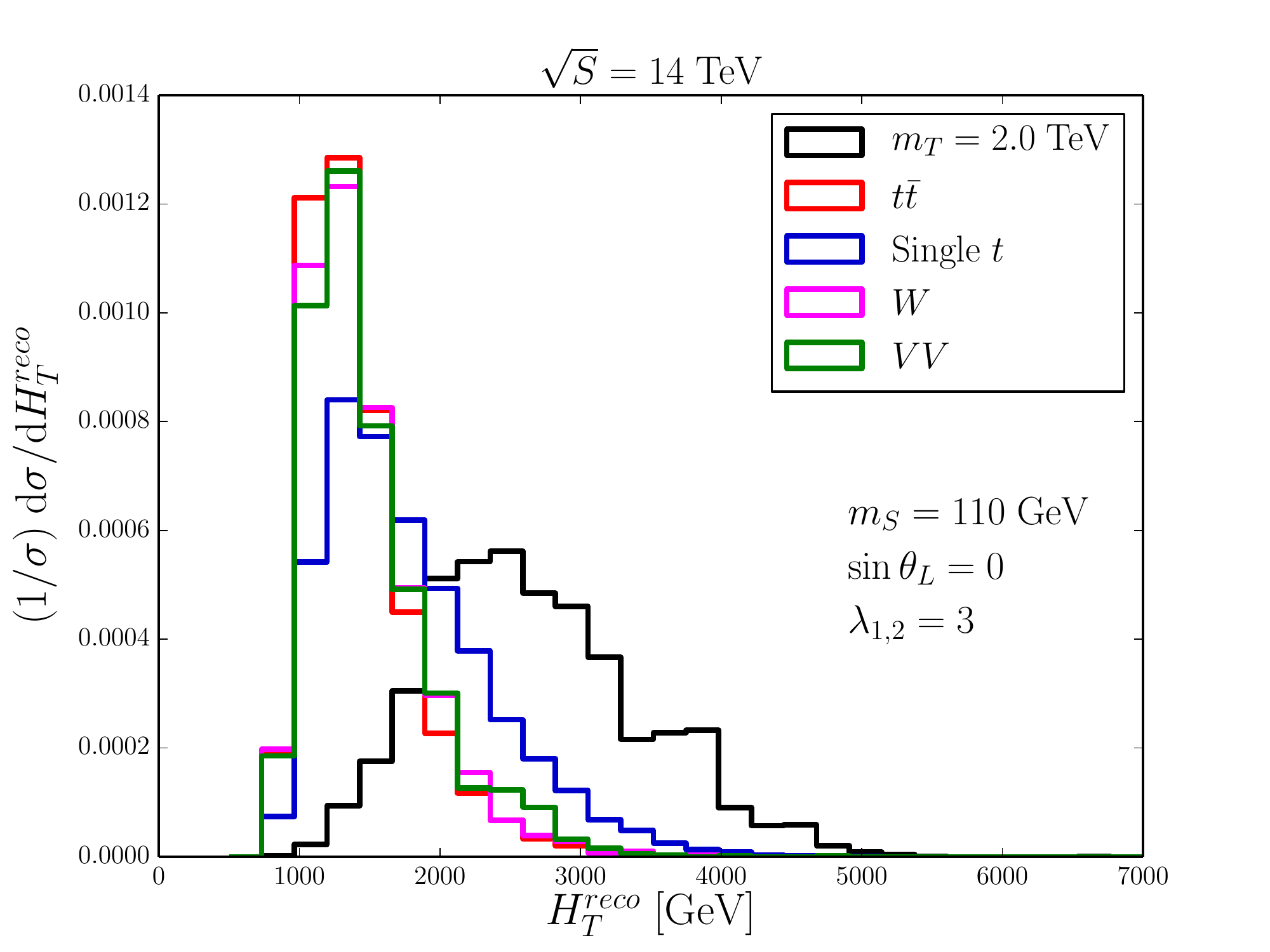}\label{fig:htreco20}}\\
  \subfigure[]{\includegraphics[scale=0.36]{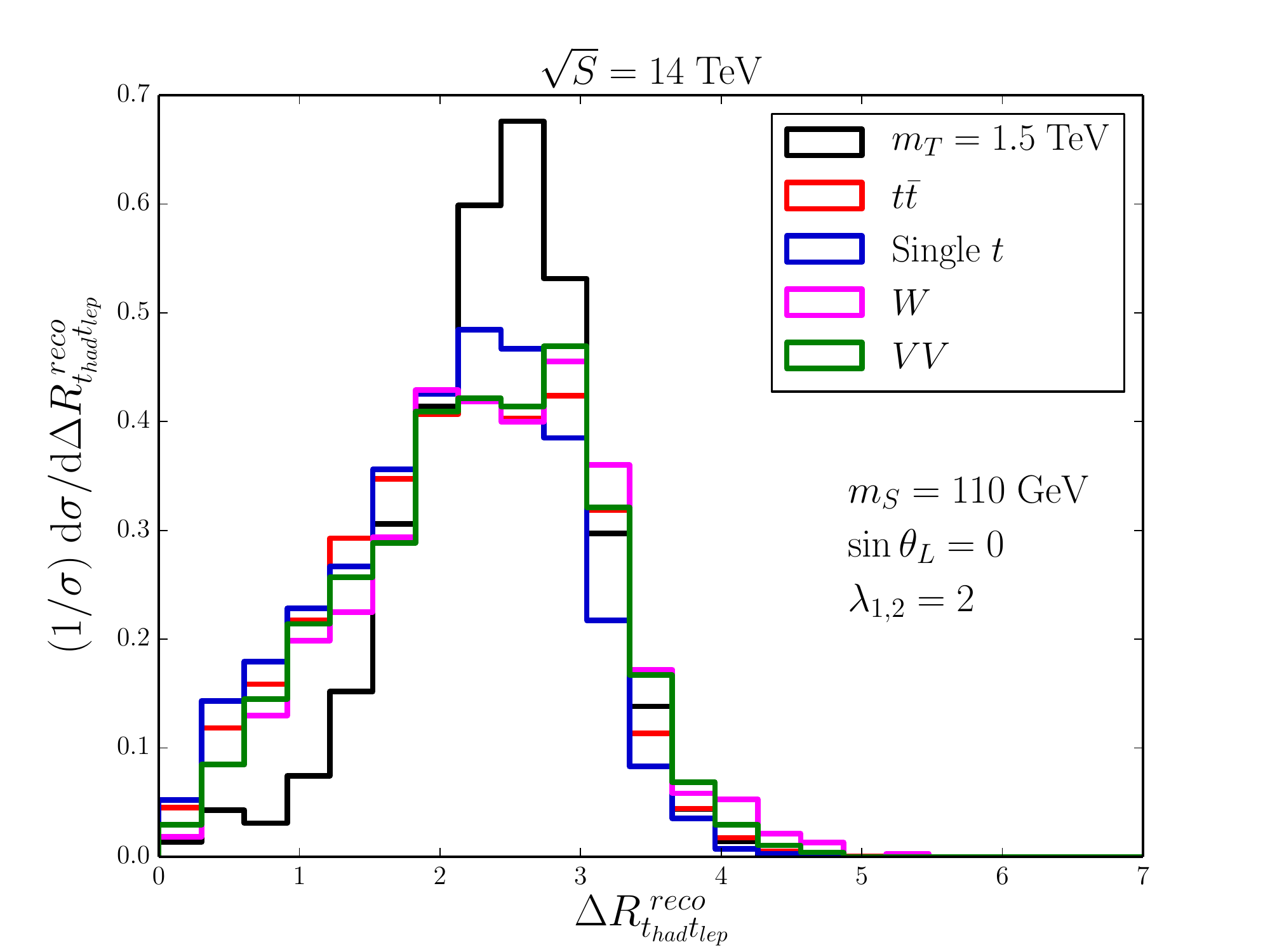}\label{fig:thtl15} }
 \subfigure[]{ \includegraphics[scale=0.35]{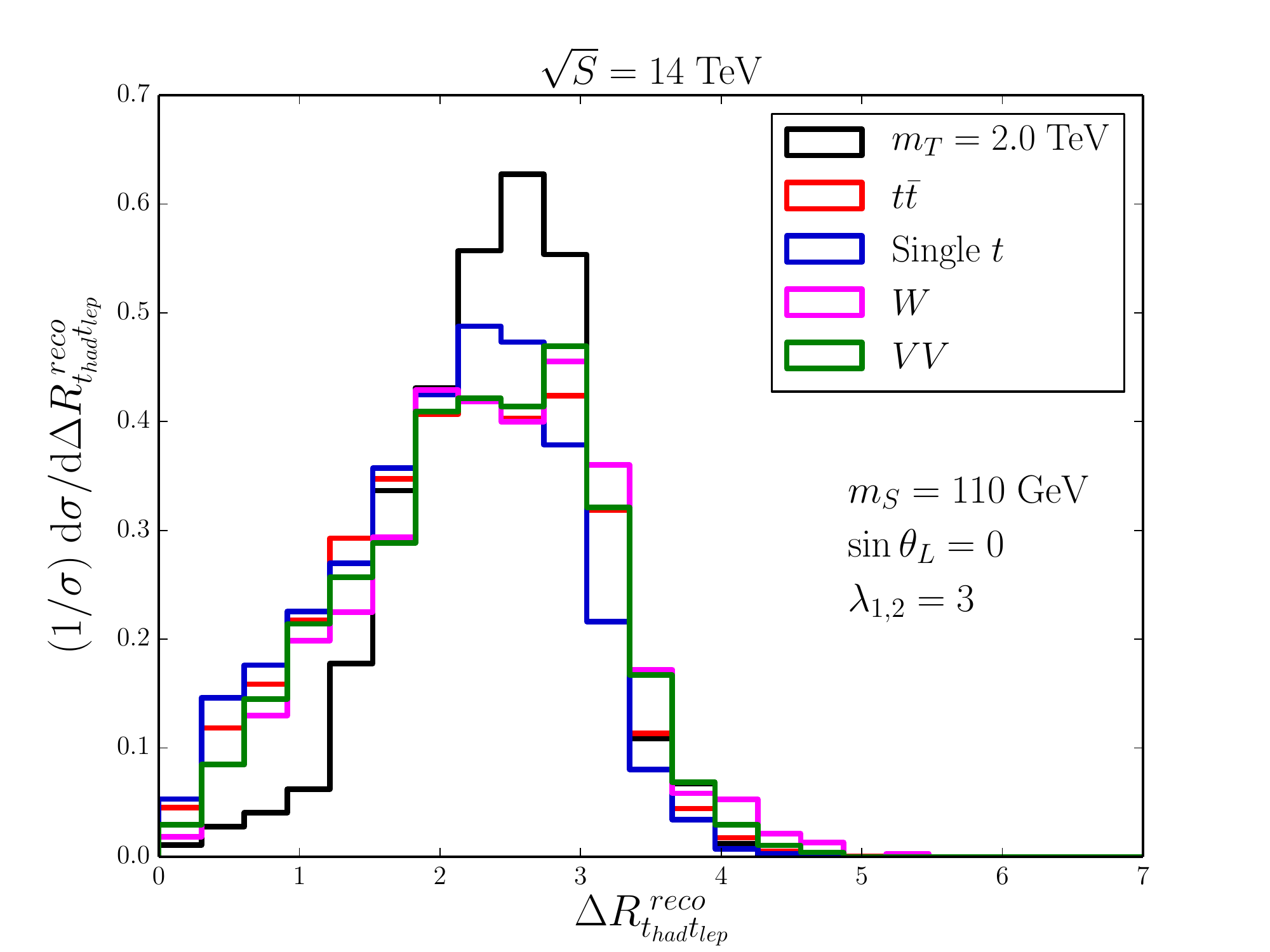}\label{fig:thtl20} }
  \caption{(a,b) $H_T^{reco}$, Eq.~(\ref{eq:HTreco}), distributions are shown for (a) $m_T=1.5$~TeV and (b) $m_T=2$~TeV.   (c,d) Distributions of the angular separation between the two reconstructed top quark, $\Delta R^{reco}_{t_{had}t_{lep}}$ are shown for (c) $m_T=1.5$~TeV and (d) $m_T=2$~TeV. All plots are generated based on events after showering, hadronization, and detector effects.  The other model parameters are set to $m_{S} = 110~\GeV$ and  $\sin \theta_L = 0$. The coupling constants are $\lambda_{1,2}=2$ for $m_T=1.5$~TeV and $\lambda_{1,2}=3$ for $m_T=2$~TeV.  Basic cuts in Eqs.~(\ref{eq:basefatjet}-\ref{eq:isolepton}) have been applied.}\label{fig:HTplots15}
\end{figure}

\begin{figure*}[tb]
  \centering

 \subfigure[]{ \includegraphics[scale=0.36]{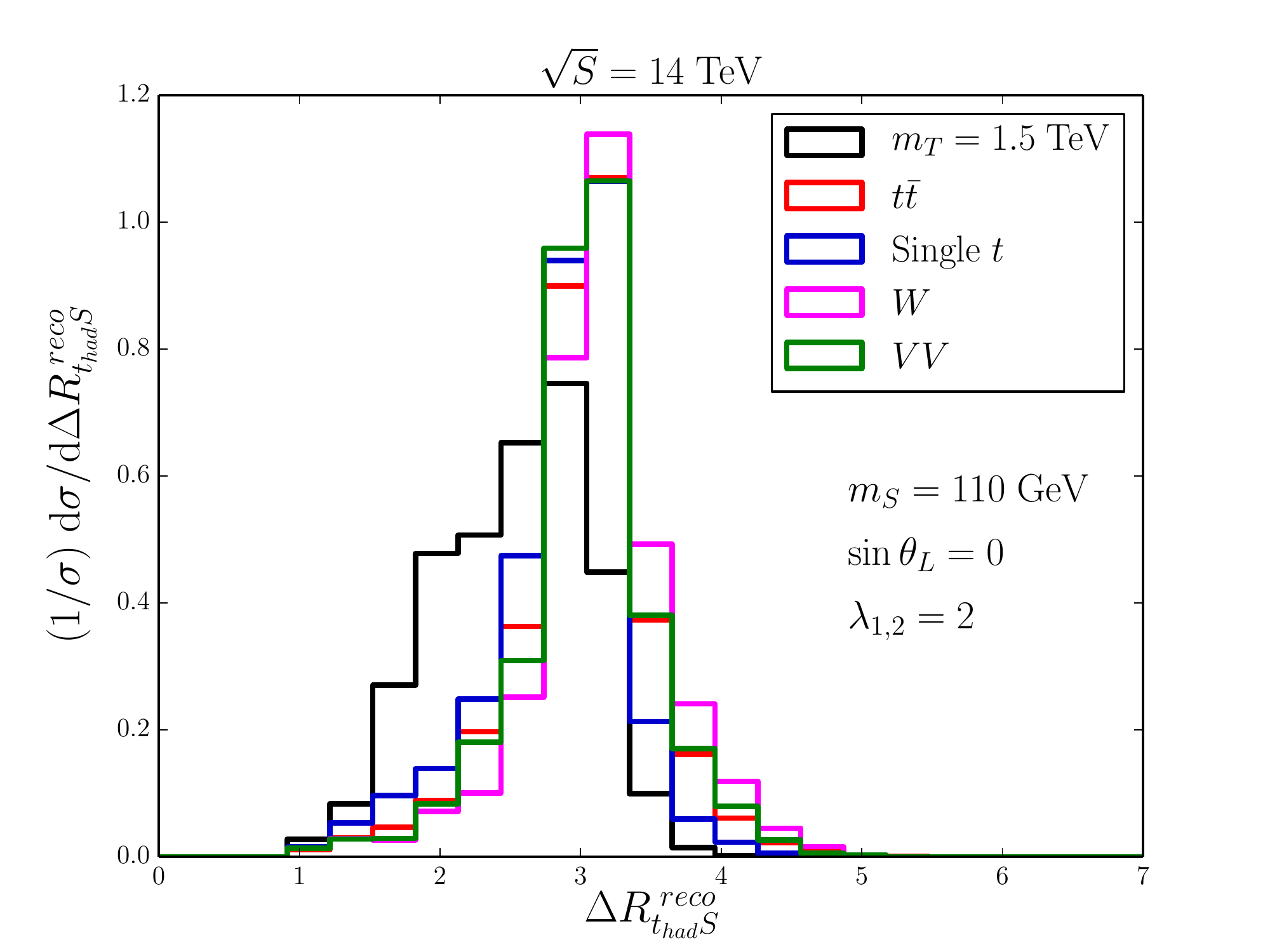}}
 \subfigure[]{ \includegraphics[scale=0.35]{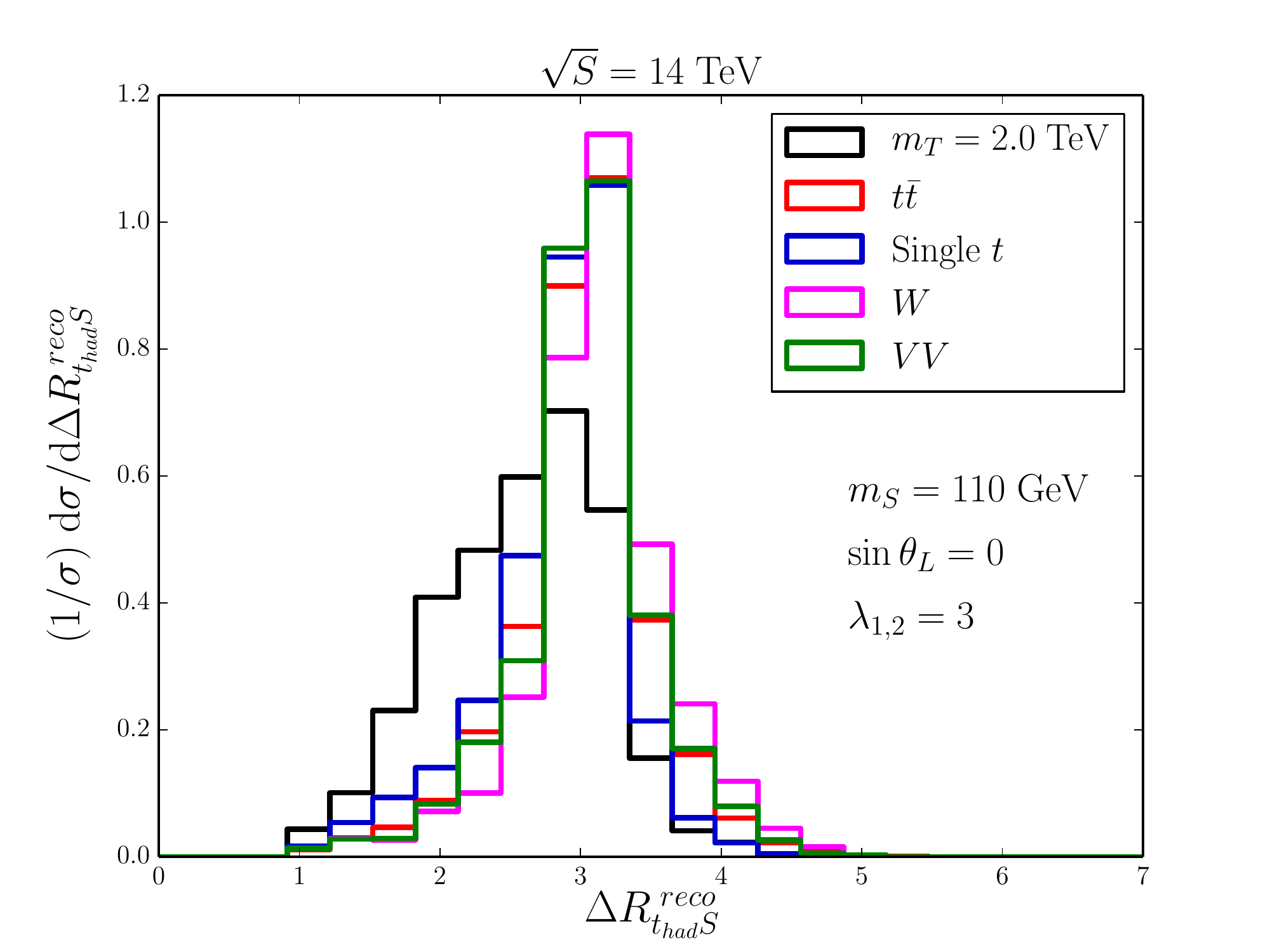}}\\
 \subfigure[]{ \includegraphics[scale=0.36]{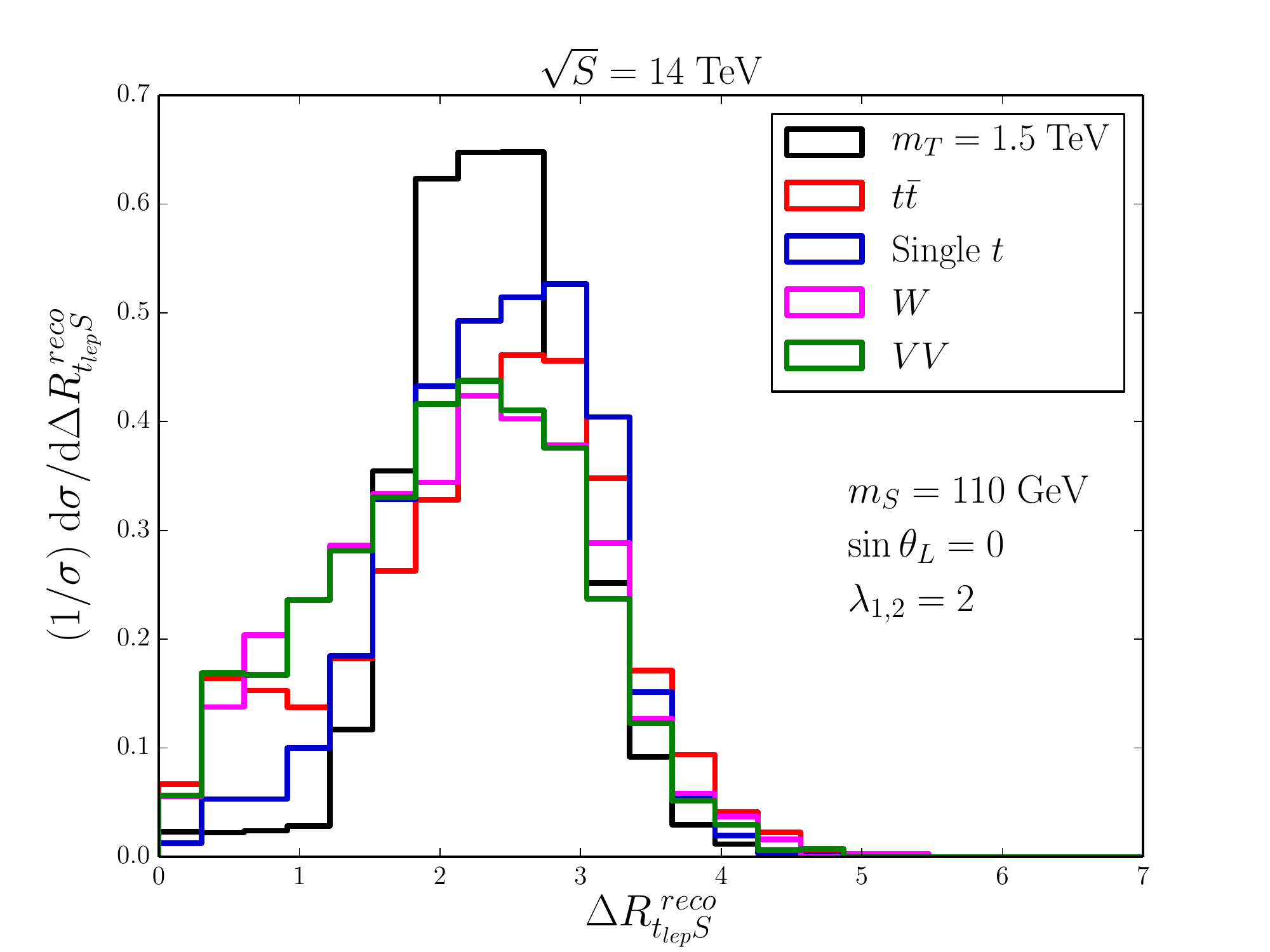}}
 \subfigure[]{ \includegraphics[scale=0.35]{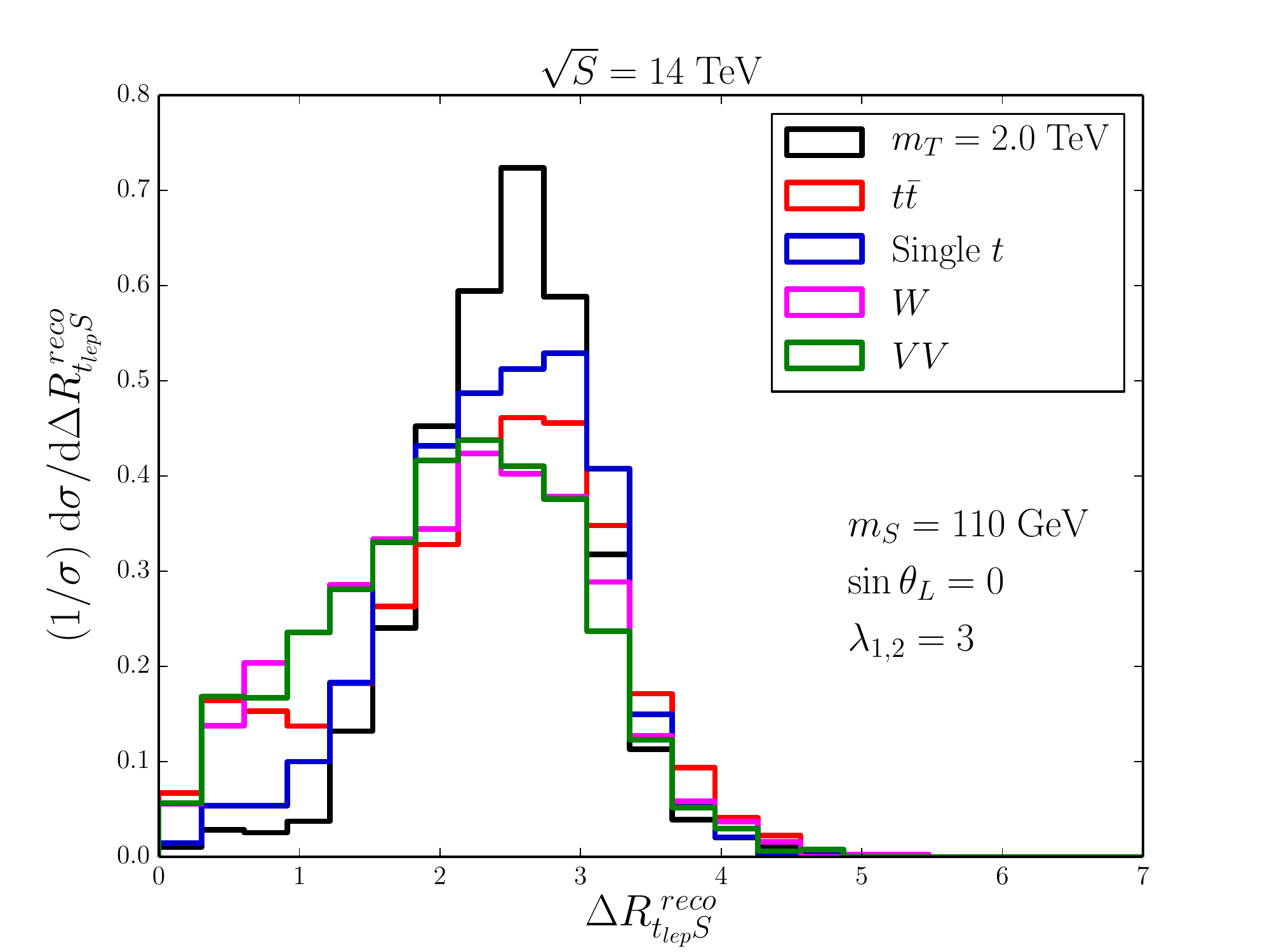}}
  \caption{Distributions of the angular separation between (a,b) the reconstructed $S$ and hadronically decaying top, $\Delta R^{reco}_{t_{had}S}$, and (c,d) the reconstructed $S$ and leptonically decaying top, $\Delta R^{reco}_{t_{lep}S}$, for (a,c) $m_T=1.5$~TeV and (b,d) $m_T=2$~TeV. All plots are generated based on events after showering, hadronization, and detector effects.  The other model parameters are set to $m_{S} = 110~\GeV$ and  $\sin \theta_L = 0$. The coupling constants are $\lambda_{1,2}=2$ for $m_T=1.5$~TeV and $\lambda_{1,2}=3$ for $m_T=2$~TeV.  Basic cuts in Eqs.~(\ref{eq:basefatjet}-\ref{eq:isolepton}) have been applied.}\label{fig:DRplots}
\end{figure*}

The effects of the the cuts in Eqs.~(\ref{eq:mtreco15}-\ref{eq:mttreco2}) on signal and background are shown in the fifth and sixth rows of Table~\ref{tab:Cutflow1}.  After these cuts the background and signal rates are comparable.    However, one final set of cuts is made to increase the significance of the signal.  We introduce a variable $H^{reco}_T$ defined as the scalar sum of the transverse momenta of the reconstructed top quarks and scalar
\bea
\label{eq:HTreco} 
	H^{reco}_T  =  p_{T,t_{had}}^{reco} + p_{T,t_{lep}}^{reco} + p_{T,S}^{reco} \;.
\eea
As demonstrated in Figs.~\ref{fig:htreco15} and~\ref{fig:htreco20} for $m_{T} =$ 1.5 TeV and 2 TeV, respectively, the signal is much harder than the background.

 Additionally, individual angular distance variables between the reconstructed tops and scalar $\Delta R^{reco}_{t_{{had}} t_{{lep}} }$, $\Delta R^{reco}_{t_{{had}} S}$ and $\Delta R^{reco}_{t_{{lep}} S}$ deliver additional handles in shaping and controlling the signal region.  This is shown in Figs.~\ref{fig:thtl15},~\ref{fig:thtl20}, and~\ref{fig:DRplots}, where background and signal clearly populate different regions of phase space.  Based on these observations, we apply a final set of cuts
\begin{eqnarray}
&&\left.\begin{aligned}
&2050~{\rm GeV} < H_T^{reco}\\
&\Delta R^{reco}_{t_{had}S}<3.41\\
&1.63<\Delta R_{t_{lep}S}^{reco}
    \end{aligned}\right\}
\quad {\rm for}\, m_T=1.5~{\rm TeV}\,{\rm and}\label{eq:fin15}\\
&&\left.\begin{aligned}
&2050~{\rm GeV} < H_T^{reco}\\
&1.79<\Delta R^{reco}_{t_{had}t_{lep}}\\
&1.58<\Delta R^{reco}_{t_{had}S}<3.6\\
&1.6<\Delta R_{t_{lep}S}^{reco}<3.1
    \end{aligned}\right\}
\quad {\rm for}\, m_T=2~{\rm TeV}.\label{eq:fin2}
\end{eqnarray}
As shown in the last row of the tables in Table~\ref{tab:Cutflow1}, these final cuts decrease the background cross section to below the signal rate, with good signal efficiency.

 To quantify the observability of our signal at the LHC, we compute a significance ($\sigma$) using the likelihood-ratio method~\cite{Cowan:2010js}
\beq
  \sigma \equiv
    \sqrt{-2\,\ln\bigg(\frac{L(B | Sig\!+\!B)}{L( Sig\!+\!B| Sig\!+\!B)}\bigg)}
  \;\;\;\;\; \text{with}\;\;\;
  L(x |n) =  \frac{x^{n}}{n !} e^{-x} \,,
\label{Eq:significance} \eeq
where $Sig$ and $B$ are the expected number of signal and background events, respectively. All significances in Table~\ref{tab:Cutflow1} are calculated for given luminosity of $3$ ab$^{-1}$ and given in the last row.  While the cuts from the basic to the reconstructed invariant mass $m_{Tt}^{reco}$ in Eqs.~(\ref{eq:basefatjet}-\ref{eq:mttreco2}) decrease background rates until they are comparable to signal, it is the final cuts in Eqs.~(\ref{eq:fin15}) and (\ref{eq:fin2}) that significantly increase the significance.  The final signal significance turns out to be $5.0$ for the benchmark parameter point $m_{T} = 1.5$ TeV, $\lambda_{1,2} = 2$, $m_{S} = 110~\GeV$ and $\sin \theta_L = 0$ assuming a luminosity of $3$ ab$^{-1}$. Although we can achieve the high significance, only $\sim 1.7$ signal events are expected.  While this may be enough to set constraints on the model, it is not enough for discovery. The sensitivity for heavier $T$ mass scales become weaker, where the final signal significance turns out to be $1.9$ for the benchmark parameter point $m_{T} = 2.0$ TeV, $\lambda_{1,2} = 3$, $m_{S} = 110~\GeV$ and $\sin \theta_L = 0$ with the same amount of the luminosity.  However, due to relaxation of $b$-tagging inside the hadronically decay top quark fat jet, we actually expect 2.4 signal events, more than $m_T=1.5$~TeV.

\begin{table*}[tb]
\begin{center}

\makebox[\textwidth][c]{
\begin{tabular}{|c|c|c|c|c|c|c|}
\hline
$m_{T} = 1.5$ TeV, $\lambda_{1,2} = 2$                         & Signal [fb]                         & $t \overline{t}$ [fb]                & Single $t$ [fb]            & $W$ [fb] 
                                                                                        & $VV$ [fb]                          & $\sigma$        \\ 
\hline \hline
Basic cuts, Eqs.~(\ref{eq:basefatjet}-\ref{eq:isolepton})                         & $0.055$                            &  $1.3 \times 10^{3}$   & $2.8 \times 10^{3}$ & $2.7 \times 10^{3}$
                                                                                        & $88$                                 & 0.036       \\  \hline
$N^{1.5}_{t_{had}}=N^{1.5}_S=1$, Eqs.~(\ref{eq:mthadreco15}-\ref{eq:NTS15})         & \multirow{1}{*}{$3.2 \times 10^{-3}$}          &  \multirow{1}{*}{1.11}                            & \multirow{1}{*}{1.6}                            & \multirow{1}{*}{0.098}                      & \multirow{1}{*}{$2.5 \times 10^{-3}$}         & \multirow{1}{*}{0.11}       \\\hline
Reconstructed $t_{lep}$, Eq.~(\ref{eq:mtlepreco15})                            & $1.2 \times 10^{-3}$          &  0.073                         & 0.070                       & $4.7 \times 10^{-4}$
                                                                                        & $\ll \mathcal{O}(10^{-5})$
& 0.17       \\  \hline
$1400~{\rm GeV} < m^{reco}_{T} < 1550$ GeV,                                         & \multirow{2}{*}{$9.2 \times 10^{-4}$}           &  \multirow{2}{*}{0.015}           & \multirow{2}{*}{$9.4 \times 10^{-3}$}   &\multirow{2}{*}{$\ll \mathcal{O}(10^{-5})$}  & \multirow{2}{*}{$\ll \mathcal{O}(10^{-5})$}
& \multirow{2}{*}{0.32}       \\  
Eq.~(\ref{eq:mtreco15})                                        &           &                             &    &                                                                                         &   &        \\\hline
$2865 \GeV < m_{Tt}^{reco}$, Eq.~(\ref{eq:mttreco15})       & $6.3 \times 10^{-4}$          & $1.5 \times 10^{-3}$ & $7.2 \times 10^{-5}$  & $ \ll \mathcal{O}(10^{-5})$
                                                                                        & $ \ll \mathcal{O}(10^{-5})$
& 0.81      \\  \hline
$2050 \GeV < H^{\rm reco}_T$   & \multirow{3}{*}{$5.8 \times 10^{-4}$}  & \multirow{3}{*}{$ \ll \mathcal{O}(10^{-5})$}
& \multirow{3}{*}{$ \ll \mathcal{O}(10^{-5})$}  &  \multirow{3}{*}{$ \ll \mathcal{O}(10^{-5})$}
                                                                                        & \multirow{3}{*}{$ \ll \mathcal{O}(10^{-5})$}
& \multirow{3}{*}{5.0}       \\  
$\Delta R^{reco}_{t_{{had}} S } < 3.41$, Eq.~(\ref{eq:fin15})       &                                          &                                                        &                             &  
                                                                                       &                                          &                                    \\  
$1.63 < \Delta R^{reco}_{t_{{lep}} S}$                                 &                                          &                                                        &                             &  
                                                                                       &                                          &                                    \\  \hline
\end{tabular}}

\vspace{20pt}

%
\makebox[\textwidth][c]{
\begin{tabular}{|c|c|c|c|c|c|c|}
\hline
$m_{T} = 2.0$ TeV, $\lambda_{1,2} = 3$                        & Signal [fb]                       & $t \overline{t}$ [fb]                & Single $t$ [fb]            & $W$ [fb] 
                                                                                        & $VV$ [fb]                         & $\sigma$        \\ 
\hline \hline
Basic cuts, Eqs.~(\ref{eq:basefatjet}-\ref{eq:isolepton})            & $0.040$                        &  $1.3\times10^{-3}$
& $2.8\times 10^{-3}$ 
& $2.7\times 10^{-3}$ 
                                                                                        & $88$
                           
& 0.027       \\  \hline
$N_{t_{\rm had} }^{2.0}= N_S^{2.0} = 1$, Eqs.(\ref{eq:msreco},\ref{eq:mthadreco2}-\ref{eq:NTS2})                                 & $5.4 \times 10^{-3}$       &  3.0                            & 14                            & 4.3
                                                                                        & 0.21

                               & 0.089       \\  \hline
Reconstructed $t_{lep}$, Eq.~(\ref{eq:mtlepreco2})                              & $1.2 \times 10^{-3}$      &  0.043                            & 0.096                           & $8.7 \times 10^{-5}$
                                                                                        & $6.8 \times 10^{-6}$      & 0.17       \\  \hline
$1860~{\rm GeV} < m^{reco}_{T} < 2100$ GeV,                                          & \multirow{2}{*}{$1.1 \times 10^{-3}$}      &  \multirow{2}{*}{0.010}                            & \multirow{2}{*}{$6.8 \times 10^{-3}$}     &\multirow{2}{*}{$\ll \mathcal{O}(10^{-5})$}

                                                                                        & \multirow{2}{*}{$6.8 \times 10^{-6}$}      & \multirow{2}{*}{0.44}       \\  
Eq.~(\ref{eq:mtreco2}) & & & & & & \\\hline
$3000 \GeV < m^{reco}_{Tt}$, Eq.~(\ref{eq:mttreco2})     & $9.4 \times 10^{-4}$      & $6.9 \times 10^{-3}$      & $6.8 \times 10^{-3}$    & $ \ll \mathcal{O}(10^{-5})$

                                                                                        & $3.4 \times 10^{-6}$      & 0.43      \\  \hline
$2050 \GeV < H^{\rm reco}_T$   & \multirow{4}{*}{$8.1 \times 10^{-4}$}  & \multirow{4}{*}{$3.2 \times 10^{-4}$}  & \multirow{4}{*}{$ \ll \mathcal{O}(10^{-5})$}

&  \multirow{4}{*}{$ \ll \mathcal{O}(10^{-5})$}

                                                                                        & \multirow{4}{*}{$ \ll \mathcal{O}(10^{-5})$}
 
& \multirow{4}{*}{1.9}       \\  
 $1.79 < \Delta R^{reco}_{t_{{had}} t_{{lep}} } $, Eq.~(\ref{eq:fin2})              &                                           &                                     &                                     &  
                                                                                       &                                           &                                    \\  
$1.58 < \Delta R^{reco}_{t_{{had}} S} < 3.6 $                      &                                           &                                      &                                     &  
                                                                                       &                                           &                                    \\  
$ 1.6 < \Delta R^{reco}_{t_{{lep}} S } < 3.1$                         &                                          &                                     &                             &  
                                                                                        &                                          &                                    \\  \hline
\end{tabular}}

\end{center}
\caption{Cumulative cut-flow tables showing the SM background and signal cross sections at two benchmark parameter points (top) $m_{T} = 1.5$ TeV and $\lambda_{1,2} = 2$ and (bottom) $m_{T} = 2.0$ TeV and $\lambda_{1,2} = 3$ where we fixed other parameters to $m_{S} = 110~\GeV$, $\sin \theta_L = 0$. Significances ($\sigma$) are calculated based on the likelihood-ratio method defined in Eq.(\ref{Eq:significance}) for given luminosity of $3$ ab$^{-1}$.  A summary of the backgrounds can be found in Table~\ref{tab:TotalBackG}.}
\label{tab:Cutflow1}
\end{table*}

\begin{figure}[t]
\begin{center}
\subfigure[]{\includegraphics[width=0.45\textwidth,clip]{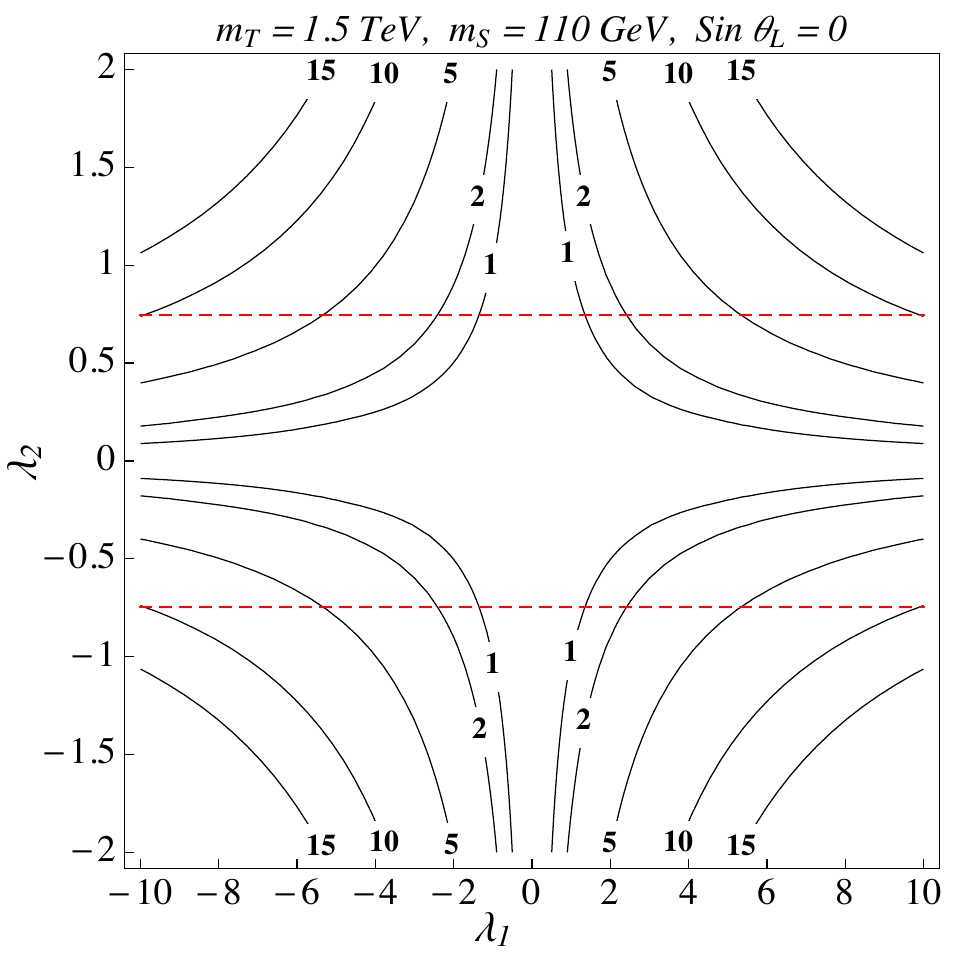}}
\subfigure[]{\includegraphics[width=0.45\textwidth,clip]{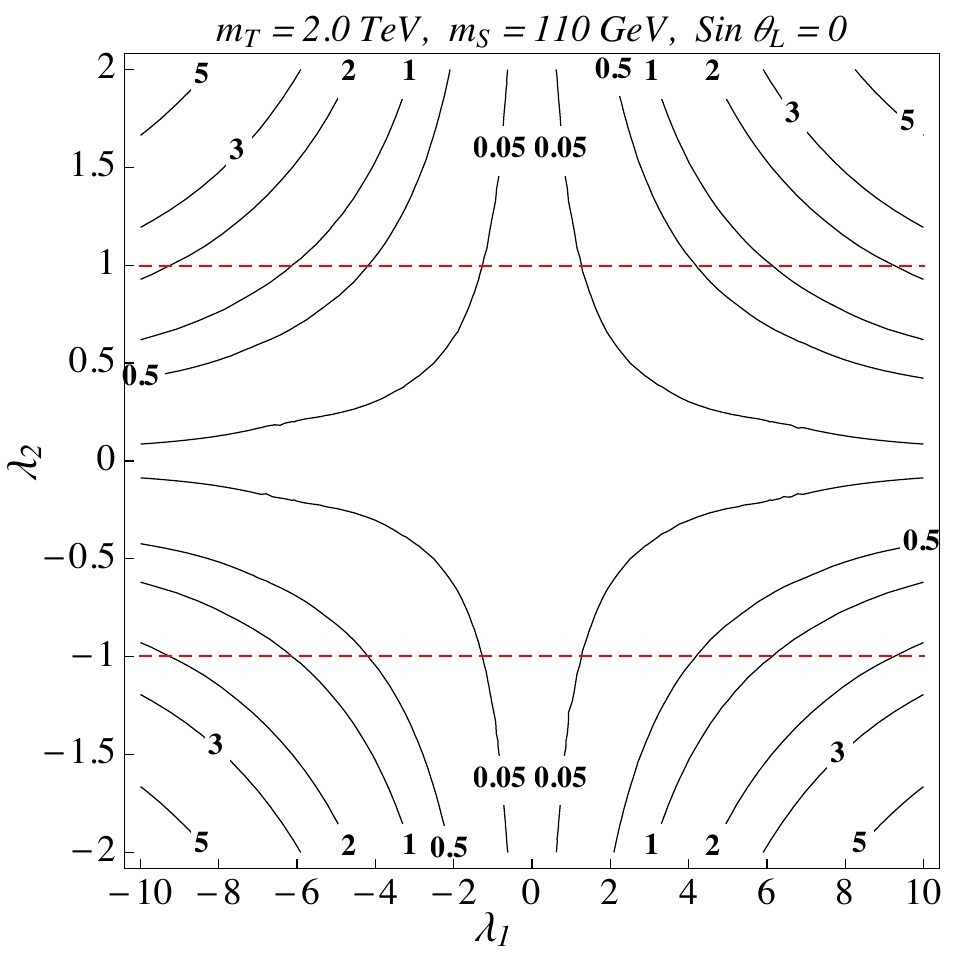}}
\end{center}
\caption{Contours of constant significance, Eq.~(\ref{Eq:significance}), for two benchmark $T$ mass scales (a) $m_{T} = 1.5$ TeV and (b) $m_{T} = 2.0$ TeV assuming a luminosity of 3~ab$^{-1}$.  We set other parameters to $m_{S} = 110~\GeV$ and $\sin \theta_L = 0$. The dashed red lines illustrate the bounds coming from the diphoton resonance searches as presented in Fig.~\ref{fig:Slimit}.}\label{fig:Exc}
\end{figure}

From these results, we can project sensitivities for many coupling constants.  For $\sin\theta_L=0$, the production cross section is proportional to $\lambda_1^2\lambda_2^2$, Eq.~(\ref{eq:propto}).  Additionally, the branching ratio of $T\rightarrow tS$ is essentially one.  Hence, we can simply scale the signal cross sections in Table~\ref{tab:Cutflow1} to determine significances for different coupling constants.  In Fig.~\ref{fig:Exc}, we summarize the final significance contours for two benchmark $T$ masses (a) $m_{T} = 1.5$ TeV and (b) $m_{T} = 2.0$ TeV, $m_S=110$~GeV, and $\sin\theta_L=0$.  These are for 3~ab$^{-1}$ of data. The solid black lines are contours of constant significance. The dashed red lines illustrate the bounds coming from the diphoton resonance searches as presented in Fig.~\ref{fig:Slimit}. At $95\%$ confidence level, using this channel the LHC will be able to exclude
\begin{eqnarray}
\sqrt{|\lambda_1\lambda_2|}&\gtrsim& 1.35\quad{\rm for}~m_T=1.5~{\rm TeV}\quad{\rm and}\\
\sqrt{|\lambda_1\lambda_2|}&\gtrsim& 3.04\quad{\rm for}~m_T=2~{\rm TeV}.
\end{eqnarray}
  Hence, the search for $Tt$ explores new parameter spaces in this model and is an important channel to consider.

\section{Conclusions}
\label{sec:conc}
We have studied a simple extension of the SM with a $SU(2)_L$ singlet fermionic top partner and gauge singlet scalar.  These top partners are ubiquitous in composite completions of the SM, and are needed to help make the Higgs natural.  Additionally, singlet scalars are present in many SM extensions and can provide a useful laboratory to categorize new physics signatures at the LHC.  While there have been many studies and searches for top partners, this model presents a unique phenomenology with many interesting characteristics.  At tree level, if the new scalar is light enough, the top partner has a new decay channel $T\rightarrow tS$ that can have a large branching ratio and will require new search strategies at the LHC~\cite{Dolan:2016eki}.  In particular, if the mixing angle between the top partner and top quark vanishes $\sin\theta_L=0$, then ${\rm BR}(T\rightarrow tS)\approx 1$, when it is kinematically allowed, as discussed in Section~\ref{sec:declowms}.  However, the precise decay channel of the scalar $S$ will depend on its mixing with the Higgs boson.  If that mixing is non-negligible, we can expect $S$ to decay much like a heavy Higgs, with the additional $S\rightarrow hh$ decay channel.  If the scalar-Higgs mixing is zero, $S$ will predominantly decay to gluon $S\rightarrow gg$ through top-partner loops.

Of particular interest to us, this model introduces many new loop induced production and decay modes of the top partner.  It is possible to produce the top partner in association with the top quark ($Tt$) through loops as shown in Fig.~\ref{fig:Prod_Feyn2}.  For the singlet top partner, the typical production mode is $T\overline{T}$ or single top partner production in association with a jet or $W$.  These single top partner production modes depend on the $T-b-W$ or $T-t-Z$ couplings, which are suppressed by the top partner-top mixing angle (Eq.~(\ref{eq:gauge})).  In the limit that this mixing angle goes to zero, these production modes vanish.  However, the loop induced diagrams for $Tt$ production persist.  As the LHC quickly saturates the phase space needed to pair produce the top partner, the $Tt$ channel will become increasingly important.  In fact, we found that for reasonable coupling constants, the $Tt$ production rate can overcome the $T\overline{T}$ production rate for top partner masses of $m_T\gtrsim 1.5$~TeV, as discussed in Section~\ref{sec:pro}.  Our results for top partner production are summarized in Table~\ref{tab:prodsum}.

Loop induced decays can also be quite interesting.  For non-negligible top partner-top mixing, the traditional decay modes $T\rightarrow tZ$, $T\rightarrow th$, and $T\rightarrow bW$ dominate.  However, similar to single top partner production, these decay modes vanish as top partner-top mixing vanishes.  In this limit, the scalar can mediate loop-induced the decay channels $T\rightarrow tg$, $T\rightarrow t\gamma$, and $T\rightarrow tZ$, through the loops shown in Figs.~\ref{fig:Ttg}-\ref{fig:TtZLoop}.  These loops do not vanish is the small mixing limit.  When $\sin\theta_L=0$ and $m_S>m_T$, these decays dominate.  Since the loops are all of a similar form, the branching ratios are determined by the gauge couplings and $T\rightarrow tg$ is the main decay mode.  While these decay channels have been searched for~\cite{Sirunyan:2017yta}, in this model they are loop induced and the top partner can be quite long lived, as discussed in Section~\ref{sec:dechighms}.  In fact, for most of the parameter range the top partner hadronizes before it decays.  For not too small couplings, it is possible to search for displaced vertices, ``stable'' particles, and stopped particles.  Our results for top partner decays are summarized in Tables~\ref{tab:decmodes} and \ref{tab:displaced}.

Whether $T\rightarrow tS$ dominates, $T\rightarrow tg$ dominates, or the top partner hadronizes and is long lived, new search strategies are needed at the LHC to fully probe the parameter space of this model.  To this end we have performed a collider study focusing on the exotic production mode $pp\rightarrow T\overline{t}+t\overline{T}$.  We focused on the small scalar mass case, in order to maximize the production rate, as shown in Fig.~\ref{fig:prodmsc}.  We also focused on $\sin\theta_L=0$, so that other single top partner modes decouple and the exotic $T\rightarrow tS\rightarrow tgg$  decay mode dominates.  This mode provided many boosted particles, allowing us to get a good handle on the signal.  This is a new production mode that provides an exotic signature at the LHC.  With 3 ab$^{-1}$ of data, we found that this production and decay mode can probe much of the parameter space inaccessible to other processes, as shown in Fig.~\ref{fig:Exc}.

As the LHC continues to gain data and new physics continues to remain elusive, it becomes imperative that we leave no rock unturned.  This means we must go beyond the simplest simplified models and search for new signals.  The model presented in this paper provides many new signatures of top partners that have not yet been searched for.  These included promptly decaying top partners with new decay channels, long live top partners with exotic decay channels, and new production channels for single top partner production.  In much of the parameter space, these signatures are available with reasonable masses and coupling constants.

\section*{Acknowledgments}
The authors would like to thank KC Kong for many helpful discussions and Zhen Liu for reading a preliminary draft and providing useful feedback.  IML is grateful to Sally Dawson for useful discussions and to the Mainz Institute for Theoretical Physics for its hospitality and its partial support during the completion of this work.  JHK is grateful to Tae Hyun Jung for valuable help and discussions. We also thank the HTCaaS group of the Korea Institute of Science and Technology Information (KISTI) for providing the necessary computing resources. This work is supported in part by United States Department of Energy grant number DE-SC0017988 by the University of Kansas General Research Fund allocation 2302091.  The data to reproduce the plots has been uploaded with the arXiv submission or is available upon request.

\appendix

\section{Wavefunction and Mass Renormalization of Top Partner}
\label{Appe:renorm}

We renormalize the bare Lagrangian of the top sector based on the on-shell wave function renormalization scheme~\cite{Espriu:2002xv,Kniehl:2009nz, Kniehl:2009kk, Kniehl:2006rc}, largely following the method of Ref.~\cite{Espriu:2002xv}. We start with the bare kinetic and mass terms of the top quark and top partner after electroweak symmetry breaking and mass diagonalization:
\begin{eqnarray}
\mathcal{L}^0_{kin,mass}&=&\overline{t}^0_{L}i\slashed{\partial}t^0_{L}+\overline{t}^0_{R}i\slashed{\partial}t^0_{R}+\overline{T}^0_{L}i\slashed{\partial}T^0_{L}+\overline{T}^0_{R}i\slashed{\partial}T^0_{R}\nonumber\\
&&-m_t^0(\overline{t}_L^0t_R^0+{\rm h.c.})-m_T^0(\overline{T}_L^0T_R^0+{\rm h.c.}),
\end{eqnarray}
where the superscript $0$ indicates bare quantities.  We allow for different wave-function renormalization constants for left- and right-handed fields, as well as for $\psi$ and $\overline{\psi}$:
 \bea
\label{eq:Topwfr} 
\begin{pmatrix} t^0_{\tau} \\ T^0_{\tau} \end{pmatrix}&=&\begin{pmatrix}\sqrt{Z_{tt}^\tau} & \sqrt{Z_{tT}^\tau}\\ \sqrt{Z_{Tt}^\tau} & \sqrt{Z_{TT}^\tau}\end{pmatrix}\begin{pmatrix} t_\tau \\ T_\tau\end{pmatrix} \simeq \begin{pmatrix} 1 + \frac{1}{2} \delta Z^{\tau}_{tt} & \frac{1}{2} \delta Z^{\tau}_{tT} \\ \frac{1}{2} \delta Z^{\tau}_{Tt}  & 1 + \frac{1}{2} \delta Z^{\tau}_{TT}   \end{pmatrix} \begin{pmatrix} t_{\tau} \\ T_{\tau} \end{pmatrix} \\ \nonumber
\begin{pmatrix} \overline{t}^0_{\tau} \\ \overline{T}^0_{\tau} \end{pmatrix}&=&\begin{pmatrix}\sqrt{\overline{Z}_{tt}^\tau} & \sqrt{\overline{Z}_{tT}^\tau}\\ \sqrt{\overline{Z}_{Tt}^\tau} & \sqrt{\overline{Z}_{TT}^\tau}\end{pmatrix}\begin{pmatrix} \overline{t}_\tau \\ \overline{T}_\tau\end{pmatrix} \simeq \begin{pmatrix} 1 + \frac{1}{2} \delta \overline{Z}^{\tau}_{tt} & \frac{1}{2} \delta \overline{Z}^{\tau}_{tT} \\ \frac{1}{2} \delta \overline{Z}^{\tau}_{Tt}  & 1 + \frac{1}{2} \delta \overline{Z}^{\tau}_{TT}   \end{pmatrix} \begin{pmatrix} \overline{t}_{\tau} \\ \overline{T}_{\tau} \end{pmatrix}
\eea
where $\tau=L,R$, $Z^\tau_{ij}$ and $\overline{Z}^\tau_{ij}$ are renormalization constants, $\delta Z^\tau_{ij}$ and $\delta \overline{Z}^\tau_{ij}$ are counterterms (CTs), and fields without the $0$ subscript are the physical, renormalized fields.  We renormalize the masses via
\begin{eqnarray}
m_t^0=m_t+\delta m_t\quad{\rm and}\quad m_T^0=m_T+\delta m_T.\label{eq:massrenorm}
\end{eqnarray}

To determine the wavefunction and mass CTs, we start with the two-point Feynman rules for the CTs at one-loop
\bea\nonumber\\\nonumber
\hspace*{-1cm} \includegraphics[scale=0.22]{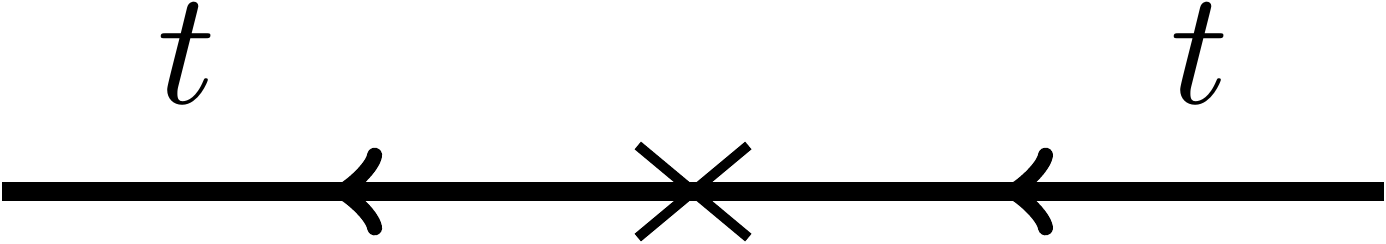} \; &=& \frac{i}{2}  \Bigg[ \big( \slashed{p} - m_t \big) \big( \delta Z^L_{tt} P_L + \delta Z^R_{tt} P_R \big) + \big( \delta \overline{Z}^L_{tt} P_R + \delta \overline{Z}^R_{tt} P_L \big) \big(\slashed{p} - m_t \big) \; - 2 \delta m_t  \Bigg]  \\ \nonumber
\hspace*{-1cm} \includegraphics[scale=0.22]{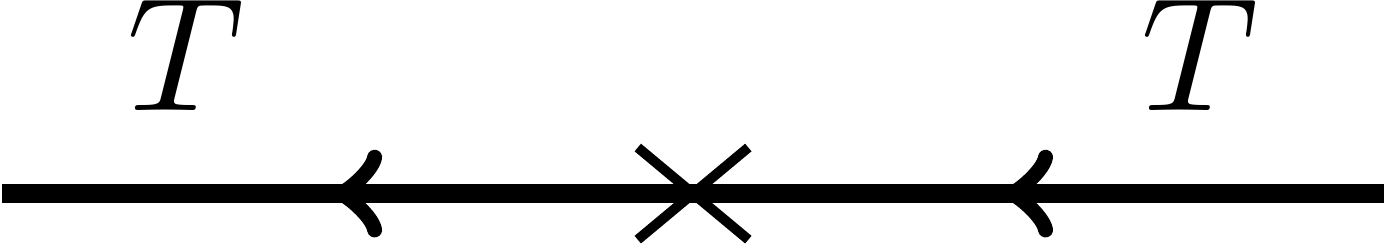} \; &=& \frac{i}{2}  \Bigg[  \big( \slashed{p} - m_T \big) \big( \delta Z^L_{TT} P_L + \delta Z^R_{TT} P_R \big) + \big( \delta \overline{Z}^L_{TT} P_R + \delta \overline{Z}^R_{TT} P_L \big) \big( \slashed{p} - m_T \big) - \; 2 \delta m_T  \Bigg]  \\ \nonumber
\hspace*{-1cm} \includegraphics[scale=0.22]{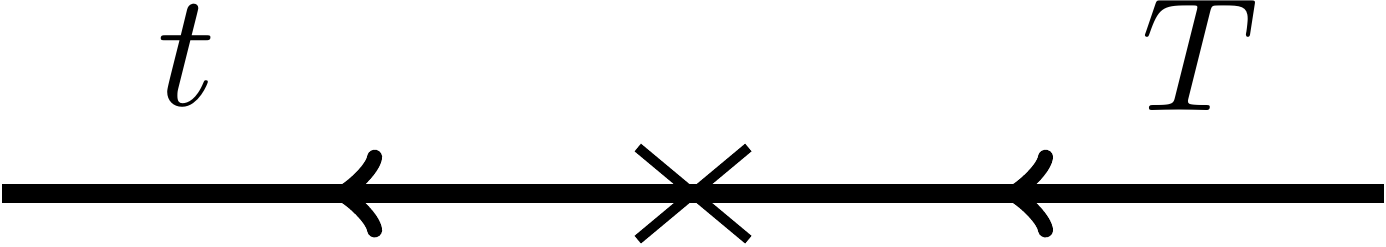} \; &=& \frac{i}{2}  \Bigg[ \big( \slashed{p} - m_t \big) \big( \delta Z^L_{tT} P_L + \delta Z^R_{tT} P_R \big) + \big( \delta \overline{Z}^L_{Tt} P_R + \delta \overline{Z}^R_{Tt} P_L \big) \big( \slashed{p} - m_T \big)  \Bigg]  \\ 
\hspace*{-1cm} \includegraphics[scale=0.22]{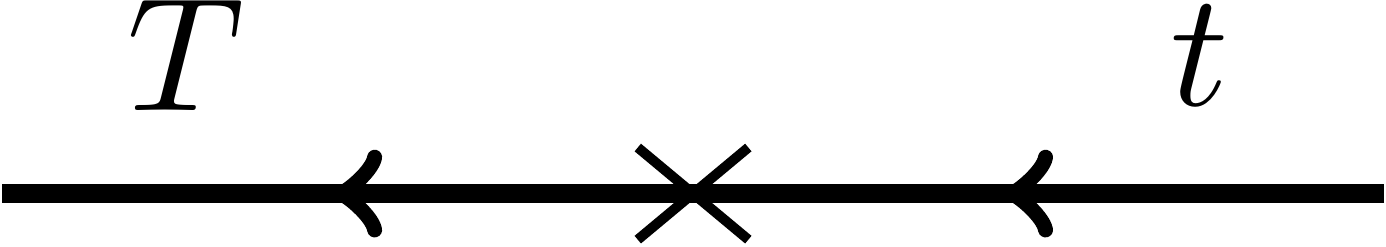} \; &=& \frac{i}{2}  \Bigg[  \big( \slashed{p} - m_T \big) \big( \delta Z^L_{Tt} P_L + \delta Z^R_{Tt} P_R \big) + \big( \delta \overline{Z}^L_{tT} P_R + \delta \overline{Z}^R_{tT} P_L \big) \big(\slashed{p} - m_t \big)  \Bigg],  \label{eq:FeynRules}
\eea
where the momentum $p$ is moving to the left with particle flow.  To calculate renormalization constants, we consider a propagator which mixes different families through radiative corrections
\bea
\label{eq:Propa} 
i S^{-1}_{i j} (p) = (\slashed{p} - m_i) \delta_{i j} - \hat{\Sigma}_{i j} (p)
\eea
where $\hat{\Sigma}_{i j} (\slashed{p})$ is a renormalized self-energy decomposed into all possible Dirac structures
\bea
\label{eq:SelfEnergy3} \nonumber
\hat{\Sigma}_{i j} (\slashed{p}) &=& \slashed{p} P_R  \bigg( \Sigma^{\gamma R}_{i j} (p^2) - \frac{1}{2} \delta \overline{Z}^{R}_{j i} - \frac{1}{2} \delta Z^{R}_{i j} \bigg)
+ \slashed{p} P_L  \bigg( \Sigma^{\gamma L}_{i j} (p^2) - \frac{1}{2} \delta \overline{Z}^{L}_{j i} - \frac{1}{2} \delta Z^{L}_{i j} \bigg)   \\   \nonumber
&+& P_R \bigg( \Sigma^{R}_{i j} (p^2) + \frac{1}{2} \big( \delta \overline{Z}^{L}_{j i} m_j + \delta Z^{R}_{i j}m_i \big) \; + \delta m_{i} \delta_{i j}  \bigg)   \\ 
&+& P_L \bigg( \Sigma^{L}_{i j} (p^2) + \frac{1}{2} \big( \delta \overline{Z}^{R}_{j i} m_j + \delta Z^{L}_{i j} m_i \big) \; + \delta m_{i} \delta_{i j}  \bigg),
\eea
and $\Sigma_{ij}(\slashed{p})$ is the one-loop one-particle irreducible unrenormalized two point function:
\begin{eqnarray}
\includegraphics[scale=0.5]{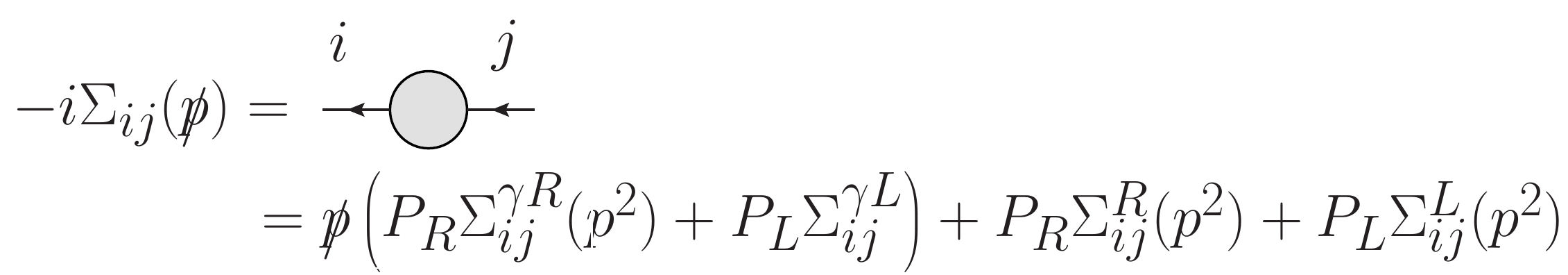}
\end{eqnarray}
Off diagonal wave function renormalization constants can be obtained by using the renormalization conditions that $i-j$ mixing vanishes when either $i$ or $j$ are on-shell:
\bea
\label{eq:OnShell2} \nonumber
\bar{u}_i (p) \widetilde{Re}  \bigg(   \hat{\Sigma}_{i j} (\slashed{p}) \bigg)    \bigg \arrowvert_{p^2 \; \rightarrow \; m^2_i} = 0   \\ 
\widetilde{\rm Re}  \bigg( \hat{\Sigma}_{i j} (\slashed{p}) \bigg) u_j (p)  \bigg \arrowvert_{p^2 \; \rightarrow \; m^2_j} = 0,  
\eea
where $\widetilde{Re}$ indicates that the real and complex pieces of the coupling constants are retained, but the absorptive pieces of the loop integrals are dropped~\cite{Denner:1991kt}.
The off-diagonal wave-function renormalization constants are then~\cite{Espriu:2002xv}
\bea
\label{eq:Zs_2} \nonumber\\\nonumber
\hspace*{-0.5cm}  \delta \overline{Z}^{L}_{j i} &=& \frac{2}{m^2_i - m^2_j} \widetilde{Re}  \Bigg( \bigg[ m_i m_j \Sigma^{\gamma R}_{i j} (m^2_i)  + m^2_i \Sigma^{\gamma L}_{i j} (m^2_i) 
 + m_j \Sigma^{R}_{i j} (m^2_i) + m_i \Sigma^{L}_{i j} (m^2_i)       \bigg] \Bigg) \\   \nonumber
\hspace*{-0.5cm}  \delta \overline{Z}^{R}_{j i} &=& \frac{2}{m^2_i - m^2_j} \widetilde{Re}  \Bigg( \bigg[ m_i m_j \Sigma^{\gamma L}_{i j} (m^2_i) + m^2_i \Sigma^{\gamma R}_{i j} (m^2_i) 
 + m_i \Sigma^{R}_{i j} (m^2_i) + m_j \Sigma^{L}_{i j} (m^2_i)        \bigg]  \Bigg) \\  \nonumber
\hspace*{-0.5cm} \delta Z^{L}_{i j} &=& \frac{2}{m^2_j - m^2_i} \widetilde{Re}  \Bigg( \bigg[ m_i m_j \Sigma^{\gamma R}_{i j} (m^2_j) + m^2_j \Sigma^{\gamma L}_{i j} (m^2_j) 
 + m_j \Sigma^{R}_{i j} (m^2_j) + m_i \Sigma^{L}_{i j} (m^2_j)      \bigg]  \Bigg) \\   
\hspace*{-0.5cm} \delta Z^{R}_{i j} &=& \frac{2}{m^2_j - m^2_i} \widetilde{Re}  \Bigg(  \bigg[ m_i m_j \Sigma^{\gamma L}_{i j} (m^2_j) + m^2_j \Sigma^{\gamma R}_{i j} (m^2_j) 
 + m_j \Sigma^{L}_{i j} (m^2_j) + m_i \Sigma^{R}_{i j} (m^2_j)       \bigg]   \Bigg).
\end{eqnarray}

Now we turn to the diagonal entries of the propagator Eq.(\ref{eq:Propa}).  We impose three conditions~\cite{Espriu:2002xv}, two of which are the normal pole and residue constraints.  These conditions are imposed after explicitly inverting $S^{-1}_{ii}$.
\begin{enumerate}
\item The numerator of $S_{i i}$ should not be chiral when the particle is on-shell $p^2=m^2_i$.
\item The propagator $S_{ii}$ should have a pole at $p^2=m^2_i$.
\item When on-shell, the propagator should have unit residue:
\begin{eqnarray}
\lim_{p^2\rightarrow m_i^2} (\slashed{p}-m_i)(S_{ii})=i
\end{eqnarray} 
\end{enumerate}
See Ref.~~\cite{Espriu:2002xv} for details of the calculation.  For completeness, we summarize their results here:
\begin{eqnarray}
\delta m_i&=&-\frac{1}{2}\widetilde{Re}\left[m_i \left(\Sigma^{\gamma L}_{ii}(m_i^2)+\Sigma^{\gamma R}_{ii}(m_i^2)\right)+\Sigma^L_{ii}(m_i^2)+\Sigma^R_{ii}(m_i^2)\right]\nonumber\\
\delta\overline{Z}^L_{ii}&=&\widetilde{Re}\left[\Sigma^{\gamma L}_{ii}(m_i^2)-X-\frac{\alpha_i}{2}+D\right]\nonumber\\
\delta\overline{Z}^R_{ii}&=&\widetilde{Re}\left[\Sigma^{\gamma R}_{ii}(m_i^2)+X-\frac{\alpha_i}{2}+D\right]\nonumber\\
\delta{Z}^L_{ii}&=&\widetilde{Re}\left[\Sigma^{\gamma L}_{ii}(m_i^2)+X+\frac{\alpha_i}{2}+D\right]\nonumber\\
\delta{Z}^R_{ii}&=&\widetilde{Re}\left[\Sigma^{\gamma R}_{ii}(m_i^2)-X+\frac{\alpha_i}{2}+D\right],
\end{eqnarray}
where
\begin{eqnarray}
D&=&m_i^2\left(\Sigma_{ii}^{\gamma L '}(m_i^2)+\Sigma_{ii}^{\gamma R'}(m_i^2)\right)+m_i\left(\Sigma^{L'}_{ii}(m_i^2)+\Sigma^{R'}_{ii}(m_i^2)\right)\nonumber\\
X&=&\frac{1}{2m_i}\left(\Sigma^R_{ii}(m_i^2)-\Sigma_{ii}^L(m_i^2)\right),
\end{eqnarray}
and the primes indicate derivative with respect to the argument $p^2$. The $\alpha_i$ are arbitrary constants that reflect that there are not enough renormalization conditions to fully determine the wavefunction and mass CTs. We will choose $\alpha_i=0$.

\subsection{Off-diagonal Mass Counterterms}
\label{Appe:OffDia} 
When renormalizing, it is possible to have off-diagonal mass CTs as well as the diagonal CTs in Eq.~(\ref{eq:massrenorm}).  Some literature includes the off-diagonal CTs~\cite{Kniehl:2009nz, Kniehl:2009kk, Kniehl:2006rc}, while others do not~\cite{Espriu:2002xv}.  The two approaches are equivalent, and it is a choice whether or not to include them.  This is because the off-diagonal renormalization conditions in Eq.~(\ref{eq:OnShell2}) are insufficient to uniquely solve for both the off-diagonal wave-functions CTs and the off-diagonal mass CTs.

We start by adding off-diagonal mass CTs, and will assume all mass terms are real.  Tildes indicate fields in the non-zero mass CT scheme.  After mass renormalization, but before wave function renormalization, the mass terms are
\begin{eqnarray}
-\mathcal{L}_{mass}&=&(m_T+\delta m_T)\overline{\widetilde{T}^0}\widetilde{T}^0+(m_t+\delta m_t)\overline{\widetilde{t}^0}\widetilde{t}^0+\delta m^L_{tT}\overline{\widetilde{t}^0_R}\widetilde{T}^0_L+\delta m^L_{Tt}\overline{\widetilde{T}^0_R}\widetilde{t}^0_L\nonumber\\
&&+\delta m^R_{tT}\overline{\widetilde{t}^0_L}\widetilde{T}^0_R+\delta m^R_{Tt}\overline{\widetilde{T}^0_L}\widetilde{t}^0_R.\label{eq:massCT}
\end{eqnarray}
The hermiticity of the mass terms requires that
\begin{eqnarray}
\delta m_{tT}^L=\delta m_{Tt}^R\quad{\rm and}\quad \delta m_{Tt}^L=\delta m_{tT}^R.
\end{eqnarray}
These mass terms can be diagonalized via the usual bi-unitary transformation
\begin{eqnarray}
\begin{pmatrix} \widetilde{t}^0_\tau \\ \widetilde{T}^0_\tau\end{pmatrix}=U_\tau \begin{pmatrix} t^0_\tau \\ T^0_\tau \end{pmatrix},
\end{eqnarray}
where $\tau=L,R$.  Writing $U_\tau\approx 1+i h_\tau$, where $h_\tau$ is Hermitian, we find at one-loop order
\bea
\label{eq:DiaSqua6} 
i h_{L,i j} &=& \frac{ - ( m_i \; \delta m^L_{i j} + \delta m^{R}_{i j} \; m_j ) }{m^2_i - m^2_j}  \;\;\;\;\;\; (\text{when}\;\;\; i \neq j )\\\nonumber
i h_{R,i j} &=& \frac{ -( m_i \; \delta m^R_{i j} + \delta m^{L}_{i j} \; m_j ) }{m^2_i - m^2_j}  \;\;\;\;\;\; (\text{when}\;\;\; i \neq j )  \\ \nonumber
i h_{R,i i} &=& ih_{L,ii}= 0,
\eea
where we have chosen $h_{R,ii}=h_{L,ii}=0$ since they are unconstrained by the diagonalization condition.

After diagonalization, the mass terms becomes
\begin{eqnarray}
-\mathcal{L}_{mass}&=&(m_T+\delta m_T)\overline{T^0}T^0+(m_t+\delta m_t)\overline{t^0}t^0.
\end{eqnarray}
This is precisely the form that we would have in Section~\ref{Appe:renorm}.  Hence, the fields without tildes correspond to the field in Sec.~\ref{Appe:renorm}, as the notation indicates.  With this identification, it is possible to to relate the counterterm matrices:
\begin{eqnarray}
\begin{pmatrix} \widetilde{t}^0_\tau \\ \widetilde{T}^0_\tau\end{pmatrix}&\approx&(1+i h_\tau) \begin{pmatrix} t^0_\tau \\ T^0_\tau \end{pmatrix}\approx \left(1+i h_\tau+\frac{1}{2}\delta Z^\tau\right)\begin{pmatrix} t_\tau \\ T_\tau \end{pmatrix}\nonumber\\
&\approx&\left(1+\frac{1}{2}\delta \widetilde{Z}^\tau\right)\begin{pmatrix} t_\tau \\ T_\tau \end{pmatrix},\label{eq:diagrenorm}
\end{eqnarray} 
where $\delta Z^\tau$ is the wavefunction CT matrix in Eq.~(\ref{eq:Topwfr}) and $\delta \widetilde{Z}^\tau$ is an equivalent wavefunction CT matrix with non-zero mass CTs.  We have used the fact that after full renormalization the schemes with and without the off-diagonal counterterms have to produce the same renormalized physical fields.   That is, whether we diagonalize the mass CT matrix then perform wave-function renormalization or perform wave-function renormalization and find a scheme to determine the off-diagonal mass CTs, the final renormalized fields should be the same.  Hence, on the right-hand-side of Eq.~(\ref{eq:diagrenorm}), the final renormalized fields are the same.  

We can then read off the relationship between the wave-function CTs with nonzero or zero mass CTs:
\begin{eqnarray}
\delta \widetilde{Z}^\tau=\delta Z^\tau+2\,i\,h_\tau.\label{eq:wvfcntrans}
\end{eqnarray}
Similarly, the relationship for the renormalization of the barred fields is
\begin{eqnarray}
\delta \overline{\widetilde{Z}}{}^\tau=\delta \overline{Z}^\tau-2\,i\,h^*_\tau.\label{eq:barwvfcntrans}
\end{eqnarray}
Hence, any scheme to choose the off-diagonal mass CTs is equivalent at one-loop order and Eqs.~(\ref{eq:wvfcntrans},\ref{eq:barwvfcntrans}) together with the matrix elements in Eq.~(\ref{eq:DiaSqua6}) give the transformation between the different schemes.  As previously mentioned, the ambiguity arises because the off-diagonal renormalization conditions in Eq.~(\ref{eq:OnShell2}) are insufficient to solve for both the off-diagonal wave-functions CTs and the off-diagonal mass CTs.  So we chose $\delta m^\tau_{ij}=0$ for simplicity.

\section{Vertex Counterterms and Mixing Angle Renormalization}
\label{Appe:vertrenorm}
We now turn to renormalization of the interactions between the top partner and top quark.  The only interactions that we consider at one-loop and are $T-t-g$, $T-t-\gamma$, and $T-t-Z$.  These have the added complication that flavor changing interactions need to be renormalized, including quark mixing~\cite{Deshpande:1981zq,Denner:1990yz,Espriu:2002xv,Kniehl:2009nz, Kniehl:2009kk, Kniehl:2006rc}. 

Since there are no tree-level interactions between $T-t-g$ and $T-t-\gamma$, the vertex counterterms originate from wavefunction renormalization.  For $T-t-g$ the, the counterterms in Sec.~\ref{Appe:renorm} are sufficient.  For the $T-t-\gamma$ interaction, the $Z-\gamma$ wavefunction renormalization must also be considered.  Following~\cite{Jegerlehner:1991dq}, the counterterms are
\begin{eqnarray}
A^0_\mu&=&\sqrt{Z_\gamma} A_\mu-\frac{1}{2}\left(\delta Z_{\gamma Z}+\Delta_0\right)Z_\mu\approx\left(1+\frac{1}{2}\delta Z_\gamma\right)A_\mu-\frac{1}{2}\left(\delta Z_{\gamma Z}+\Delta_0\right)Z_\mu\\
Z^0_\mu&=&\frac{1}{2}\Delta_0 A_\mu+\sqrt{Z_Z}Z_\mu\approx \frac{1}{2}\Delta_0 A_\mu+\left(1+\frac{1}{2}\delta Z_Z\right)Z_\mu,
\end{eqnarray}
where, again, the superscript $0$ indicates unrenormalized quantities.
To find $\Delta_0$ and $\delta Z_{Z\gamma}$, construct the renormalized two-point function
\begin{eqnarray}
\includegraphics[height=0.13\textwidth,clip]{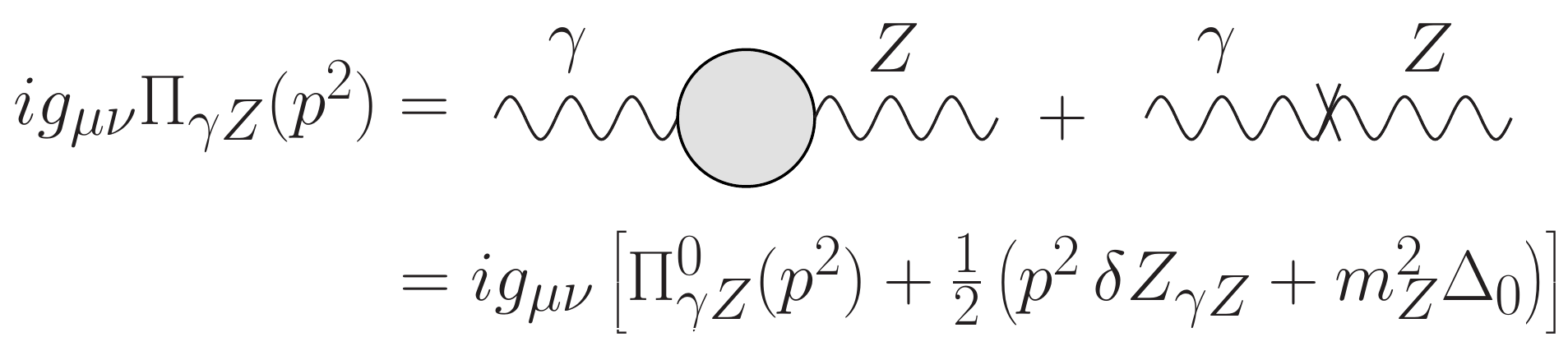},\label{eq:gamZ2pt}
\end{eqnarray}
where $\Pi^0_{\gamma Z}$ is the unrenormalized two-point loop functions.  Demanding that on-shell the mixing goes to zero
\begin{eqnarray}
\Pi_{\gamma Z}(0)=\Pi_{\gamma Z}(m_Z^2)=0,
\end{eqnarray}
the result is
\begin{eqnarray}
\Delta_0=-\frac{2\,\Pi_{\gamma Z}^0(0)}{m_Z^2}\quad{\rm and}\quad \delta Z_{\gamma Z}=-\frac{2\,\widetilde{Re}\left[\Pi^0_{\gamma Z}(m_Z^2)-\Pi^0_{\gamma Z}(0)\right]}{m_Z^2}.
\end{eqnarray}

For $T-t-Z$ we need mixing angle and coupling constant renormalization as-well-as wave-function CTs.  The wave-function renormalization $\delta Z_Z$ can be determined by the usual requirements that the $Z$-propagator has a pole at $p^2=m_Z^2$ and that it has unit residue:
\begin{eqnarray}
\delta Z_Z=\widetilde{Re}\left[\frac{d\Pi^0_{ZZ}}{dp^2}\right]_{p^2=m_Z^2},
\end{eqnarray}
where $\Pi^0_{ZZ}(p^2)$ is an unrenormalized two-point function defined similarly to $\Pi_{\gamma Z}^0(p^2)$ in Eq.~\ref{eq:gamZ2pt}.  The coupling constant and mixing angle CTs are defined as
\begin{eqnarray}
e^0&=&e\left(1+\delta e\right)\\
s_W^0&=&s_W\left(1+\delta s_W\right)\\
c_W^0&=&c_W\left(1-\delta s_W \frac{s_W^2}{c_W^2}\right)\\
\frac{e^0}{s_W^0c_W^0}&=&\frac{e}{s_W c_W}\left(1+\delta g_Z\right)\\
\delta g_Z&=&1+\delta e+\delta s_W\left(t_W^2-1\right)\\
\theta_L^0&=&\theta_L\left(1+\delta \theta_L\right),
\end{eqnarray}
where $t_W=s_W/c_W$.  We refer the reader to Ref.~\cite{Jegerlehner:1991dq} for details on calculating $\delta e$ and $\delta s_W$.

The relevant vertex counterterms are then 
\begin{eqnarray}
&\includegraphics[height=0.11\textwidth,clip]{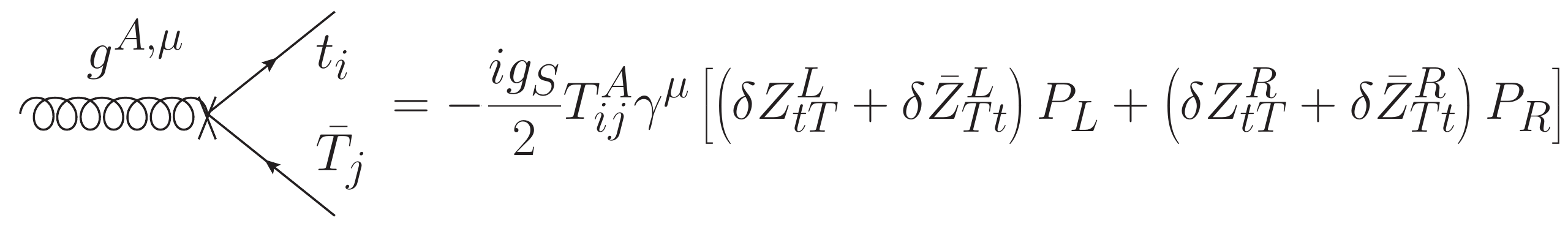}&\\
&\includegraphics[height=0.11\textwidth,clip]{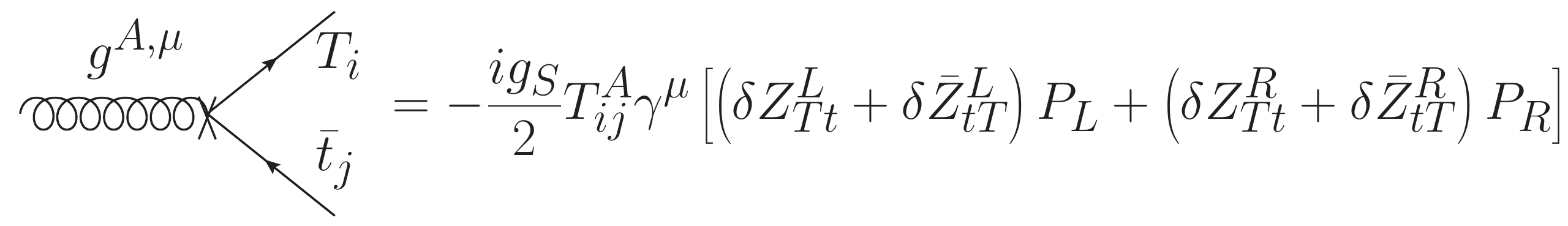}&\\
&\includegraphics[height=0.11\textwidth,clip]{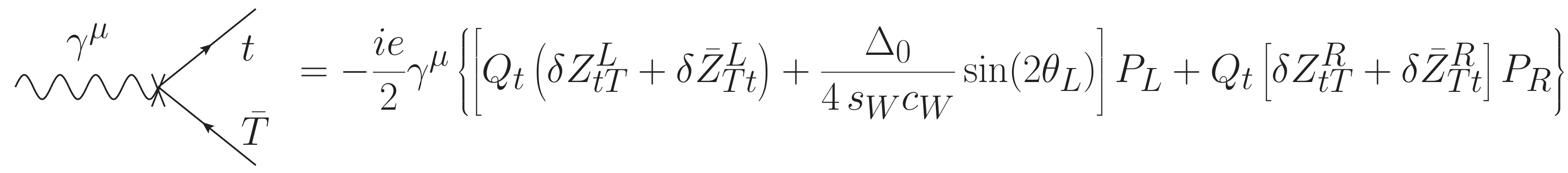}&\\
&\includegraphics[height=0.11\textwidth,clip]{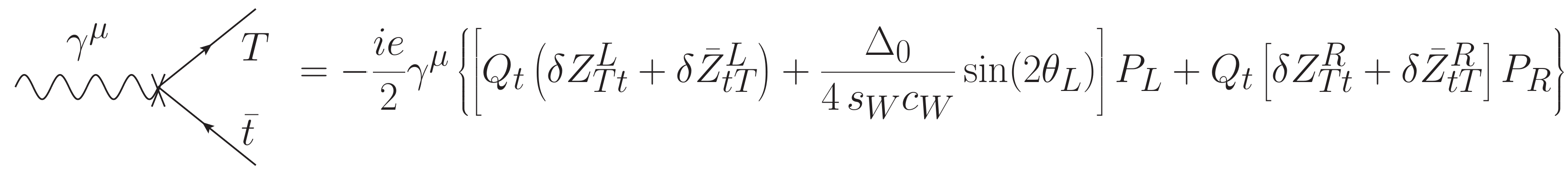}&\\
&\includegraphics[height=0.22\textwidth,clip]{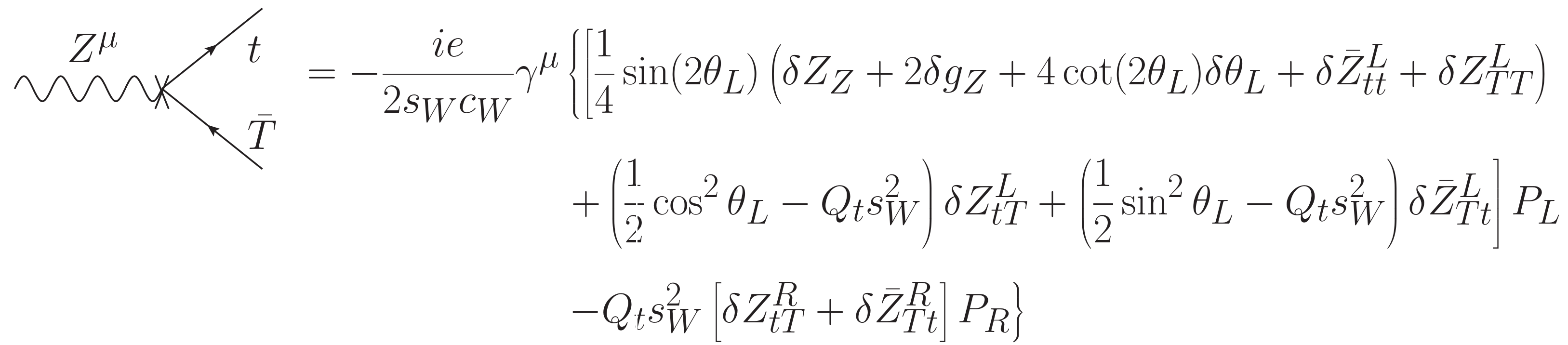}&\\
&\includegraphics[height=0.22\textwidth,clip]{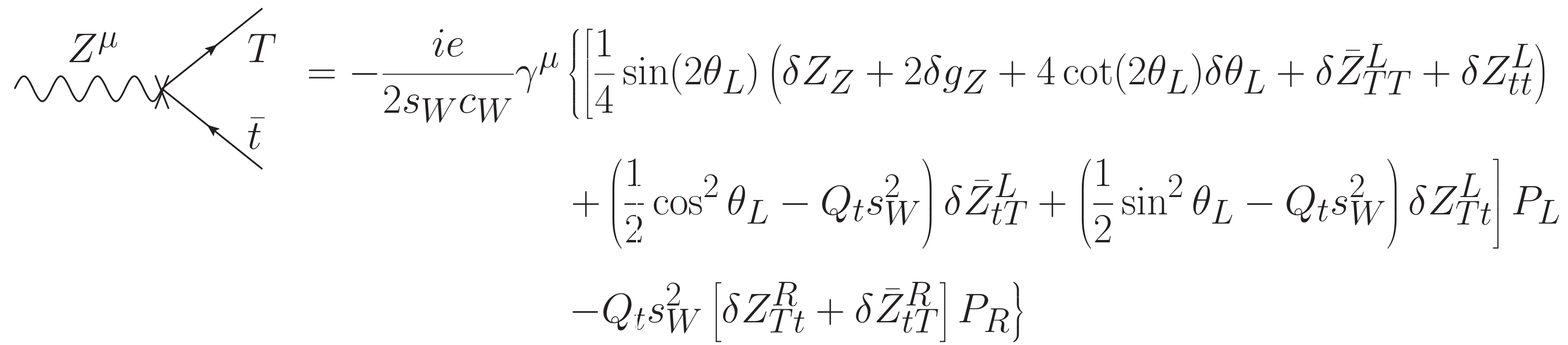}&
\end{eqnarray}
The final piece needed in the mixing angle CT, $\delta \theta_L$.  We focus on renormalizing $T\rightarrow tZ$.  First, define
\begin{eqnarray}
\frac{1}{2}\sin(2\theta_L)\delta g_L&\equiv&\frac{1}{4}\sin(2\theta_L)\left(\delta Z_Z+2\delta g_Z+4\cot(2\theta_L)\delta \theta_L+\delta \overline{Z}_{tt}^L+\delta Z_{TT}^L\right)\nonumber\\
&&+\left(\frac{1}{2}\cos^2\theta_L-Q_t s^2_W\right)\delta {Z}_{tT}^L+\left(\frac{1}{2}\sin^2\theta_L-Q_t s^2_W\right)\delta \overline{Z}_{Tt}^L.
\end{eqnarray}
We then calculate $T\rightarrow tZ$ and determine $\delta g_L$ in the $\overline{\rm MS}$ scheme.  At one loop for $T\rightarrow tZ$ diagrams with the scalar $S$, Higgs, Goldstones, $W$, and $Z$ are included.  In this way, all corrections from Yukawa couplings are included in a gauge invariant way.  Diagrams with gluons are not included, since they are corrections to the tree level $T\rightarrow tZ$, and so vanish as the tree level $T\rightarrow tZ$ vanishes.  In addition, gluons have a separate gauge parameter from the EW sector and are not needed for gauge invariance.  We have verified that in the limit $m_t,m_Z,\sin\theta_L\rightarrow 0$ limit that the Lorentz structure of the EFT in Eq.~(\ref{eq:EFT}) is recovered.

\section{Parameterization of Detector Resolution Effects}
\label{app:DR}

We include detector effects based on the ATLAS detector performances~\cite{ATL-PHYS-PUB-2013-004}. The jet energy resolution is parametrized by noise ($N$), stochastic ($S$), and constant ($C$) terms
\begin{align}
\frac{\sigma}{E} =  \sqrt{
\bigg( \frac{N}{E} \bigg)^2 +\bigg( \frac{S}{\sqrt{E}}\bigg)^2  +C^2~,
}
\end{align}
where in our analysis we use $N=5.3$, $S=0.74$ and $C=0.05$ for jets; and $N=0.3$, $S=0.1$, and $C=0.01$ for electrons.

The muon energy resolution is derived by the Inner Detector (ID) and Muon Spectrometer (MS) resolution functions
\begin{align}
\sigma = \frac{ \sigma_{\text{ID}}~ \sigma_{\text{MS} }  } { \sqrt{ \sigma^2_{\text{ID}} + \sigma^2_{\text{MS}   } } }~,
\end{align}
where
\bea
\label{eq:MuonSmear}
\sigma_{\text{ID}} &=& E~\sqrt{ a^2_1 + ( a_2 ~ E )^2   } \\
\sigma_{\text{MS}} &=& E~\sqrt{ \bigg( \frac{b_0}{E} \bigg)^2 + b^2_1 + (b_2~E)^2   }~\;.
\eea
We use $a_1 = 0.023035$, $a_2 = 0.000347$, $b_0 = 0.12$, $b_1 = 0.03278$ and $b_2 = 0.00014$ in our study.

\bibliographystyle{JHEP}
\bibliography{lit}

\providecommand{\href}[2]{#2}\begingroup\raggedright\begin{thebibliography}{100}

\bibitem{Agashe:2004rs}
K.~Agashe, R.~Contino and A.~Pomarol, \emph{{The Minimal composite Higgs
  model}}, \href{https://doi.org/10.1016/j.nuclphysb.2005.04.035}{\emph{Nucl.
  Phys.} {\bfseries B719} (2005) 165--187},
  [\href{https://arxiv.org/abs/hep-ph/0412089}{{\ttfamily hep-ph/0412089}}].

\bibitem{Agashe:2005dk}
K.~Agashe and R.~Contino, \emph{{The Minimal composite Higgs model and
  electroweak precision tests}},
  \href{https://doi.org/10.1016/j.nuclphysb.2006.02.011}{\emph{Nucl. Phys.}
  {\bfseries B742} (2006) 59--85},
  [\href{https://arxiv.org/abs/hep-ph/0510164}{{\ttfamily hep-ph/0510164}}].

\bibitem{Agashe:2006at}
K.~Agashe, R.~Contino, L.~Da~Rold and A.~Pomarol, \emph{{A Custodial symmetry
  for $Zb \bar b$}},
  \href{https://doi.org/10.1016/j.physletb.2006.08.005}{\emph{Phys. Lett.}
  {\bfseries B641} (2006) 62--66},
  [\href{https://arxiv.org/abs/hep-ph/0605341}{{\ttfamily hep-ph/0605341}}].

\bibitem{Contino:2006qr}
R.~Contino, L.~Da~Rold and A.~Pomarol, \emph{{Light custodians in natural
  composite Higgs models}},
  \href{https://doi.org/10.1103/PhysRevD.75.055014}{\emph{Phys. Rev.}
  {\bfseries D75} (2007) 055014},
  [\href{https://arxiv.org/abs/hep-ph/0612048}{{\ttfamily hep-ph/0612048}}].

\bibitem{Giudice:2007fh}
G.~F. Giudice, C.~Grojean, A.~Pomarol and R.~Rattazzi, \emph{{The
  Strongly-Interacting Light Higgs}},
  \href{https://doi.org/10.1088/1126-6708/2007/06/045}{\emph{JHEP} {\bfseries
  06} (2007) 045}, [\href{https://arxiv.org/abs/hep-ph/0703164}{{\ttfamily
  hep-ph/0703164}}].

\bibitem{Azatov:2011qy}
A.~Azatov and J.~Galloway, \emph{{Light Custodians and Higgs Physics in
  Composite Models}},
  \href{https://doi.org/10.1103/PhysRevD.85.055013}{\emph{Phys. Rev.}
  {\bfseries D85} (2012) 055013},
  [\href{https://arxiv.org/abs/1110.5646}{{\ttfamily 1110.5646}}].

\bibitem{Serra:2015xfa}
J.~Serra, \emph{{Beyond the Minimal Top Partner Decay}},
  \href{https://doi.org/10.1007/JHEP09(2015)176}{\emph{JHEP} {\bfseries 09}
  (2015) 176}, [\href{https://arxiv.org/abs/1506.05110}{{\ttfamily
  1506.05110}}].

\bibitem{ArkaniHamed:2002qy}
N.~Arkani-Hamed, A.~G. Cohen, E.~Katz and A.~E. Nelson, \emph{{The Littlest
  Higgs}}, \href{https://doi.org/10.1088/1126-6708/2002/07/034}{\emph{JHEP}
  {\bfseries 07} (2002) 034},
  [\href{https://arxiv.org/abs/hep-ph/0206021}{{\ttfamily hep-ph/0206021}}].

\bibitem{ArkaniHamed:2002pa}
N.~Arkani-Hamed, A.~G. Cohen, T.~Gregoire and J.~G. Wacker,
  \emph{{Phenomenology of electroweak symmetry breaking from theory space}},
  \href{https://doi.org/10.1088/1126-6708/2002/08/020}{\emph{JHEP} {\bfseries
  08} (2002) 020}, [\href{https://arxiv.org/abs/hep-ph/0202089}{{\ttfamily
  hep-ph/0202089}}].

\bibitem{Low:2002ws}
I.~Low, W.~Skiba and D.~Tucker-Smith, \emph{{Little Higgses from an
  antisymmetric condensate}},
  \href{https://doi.org/10.1103/PhysRevD.66.072001}{\emph{Phys. Rev.}
  {\bfseries D66} (2002) 072001},
  [\href{https://arxiv.org/abs/hep-ph/0207243}{{\ttfamily hep-ph/0207243}}].

\bibitem{Chang:2003un}
S.~Chang and J.~G. Wacker, \emph{{Little Higgs and custodial SU(2)}},
  \href{https://doi.org/10.1103/PhysRevD.69.035002}{\emph{Phys. Rev.}
  {\bfseries D69} (2004) 035002},
  [\href{https://arxiv.org/abs/hep-ph/0303001}{{\ttfamily hep-ph/0303001}}].

\bibitem{Csaki:2003si}
C.~Csaki, J.~Hubisz, G.~D. Kribs, P.~Meade and J.~Terning, \emph{{Variations of
  little Higgs models and their electroweak constraints}},
  \href{https://doi.org/10.1103/PhysRevD.68.035009}{\emph{Phys. Rev.}
  {\bfseries D68} (2003) 035009},
  [\href{https://arxiv.org/abs/hep-ph/0303236}{{\ttfamily hep-ph/0303236}}].

\bibitem{Perelstein:2003wd}
M.~Perelstein, M.~E. Peskin and A.~Pierce, \emph{{Top quarks and electroweak
  symmetry breaking in little Higgs models}},
  \href{https://doi.org/10.1103/PhysRevD.69.075002}{\emph{Phys. Rev.}
  {\bfseries D69} (2004) 075002},
  [\href{https://arxiv.org/abs/hep-ph/0310039}{{\ttfamily hep-ph/0310039}}].

\bibitem{Chen:2003fm}
M.-C. Chen and S.~Dawson, \emph{{One loop radiative corrections to the rho
  parameter in the littlest Higgs model}},
  \href{https://doi.org/10.1103/PhysRevD.70.015003}{\emph{Phys. Rev.}
  {\bfseries D70} (2004) 015003},
  [\href{https://arxiv.org/abs/hep-ph/0311032}{{\ttfamily hep-ph/0311032}}].

\bibitem{Berger:2012ec}
J.~Berger, J.~Hubisz and M.~Perelstein, \emph{{A Fermionic Top Partner:
  Naturalness and the LHC}},
  \href{https://doi.org/10.1007/JHEP07(2012)016}{\emph{JHEP} {\bfseries 07}
  (2012) 016}, [\href{https://arxiv.org/abs/1205.0013}{{\ttfamily 1205.0013}}].

\bibitem{Willenbrock:1986cr}
S.~S.~D. Willenbrock and D.~A. Dicus, \emph{{Production of Heavy Quarks from W
  Gluon Fusion}}, \href{https://doi.org/10.1103/PhysRevD.34.155}{\emph{Phys.
  Rev.} {\bfseries D34} (1986) 155}.

\bibitem{Han:2003wu}
T.~Han, H.~E. Logan, B.~McElrath and L.-T. Wang, \emph{{Phenomenology of the
  little Higgs model}},
  \href{https://doi.org/10.1103/PhysRevD.67.095004}{\emph{Phys. Rev.}
  {\bfseries D67} (2003) 095004},
  [\href{https://arxiv.org/abs/hep-ph/0301040}{{\ttfamily hep-ph/0301040}}].

\bibitem{Han:2005ru}
T.~Han, H.~E. Logan and L.-T. Wang, \emph{{Smoking-gun signatures of little
  Higgs models}},
  \href{https://doi.org/10.1088/1126-6708/2006/01/099}{\emph{JHEP} {\bfseries
  01} (2006) 099}, [\href{https://arxiv.org/abs/hep-ph/0506313}{{\ttfamily
  hep-ph/0506313}}].

\bibitem{DeSimone:2012fs}
A.~De~Simone, O.~Matsedonskyi, R.~Rattazzi and A.~Wulzer, \emph{{A First Top
  Partner Hunter's Guide}},
  \href{https://doi.org/10.1007/JHEP04(2013)004}{\emph{JHEP} {\bfseries 04}
  (2013) 004}, [\href{https://arxiv.org/abs/1211.5663}{{\ttfamily 1211.5663}}].

\bibitem{Backovic:2015bca}
M.~Backovic, T.~Flacke, J.~H. Kim and S.~J. Lee, \emph{{Search Strategies for
  TeV Scale Fermionic Top Partners with Charge 2/3}},
  \href{https://doi.org/10.1007/JHEP04(2016)014}{\emph{JHEP} {\bfseries 04}
  (2016) 014}, [\href{https://arxiv.org/abs/1507.06568}{{\ttfamily
  1507.06568}}].

\bibitem{Liu:2016jho}
Y.-B. Liu, \emph{{Search for single production of the heavy vectorlike $T$
  quark with $T\to th$ and $h\to \gamma\gamma$ at the high-luminosity LHC}},
  \href{https://doi.org/10.1103/PhysRevD.95.035013}{\emph{Phys. Rev.}
  {\bfseries D95} (2017) 035013},
  [\href{https://arxiv.org/abs/1612.05851}{{\ttfamily 1612.05851}}].

\bibitem{Aguilar-Saavedra:2013qpa}
J.~A. Aguilar-Saavedra, R.~Benbrik, S.~Heinemeyer and M.~P{\'e}rez-Victoria,
  \emph{{Handbook of vectorlike quarks: Mixing and single production}},
  \href{https://doi.org/10.1103/PhysRevD.88.094010}{\emph{Phys. Rev.}
  {\bfseries D88} (2013) 094010},
  [\href{https://arxiv.org/abs/1306.0572}{{\ttfamily 1306.0572}}].

\bibitem{Ortiz:2014iza}
N.~Gutierrez~Ortiz, J.~Ferrando, D.~Kar and M.~Spannowsky,
  \emph{{Reconstructing singly produced top partners in decays to Wb}},
  \href{https://doi.org/10.1103/PhysRevD.90.075009}{\emph{Phys. Rev.}
  {\bfseries D90} (2014) 075009},
  [\href{https://arxiv.org/abs/1403.7490}{{\ttfamily 1403.7490}}].

\bibitem{Matsedonskyi:2014mna}
O.~Matsedonskyi, G.~Panico and A.~Wulzer, \emph{{On the Interpretation of Top
  Partners Searches}},
  \href{https://doi.org/10.1007/JHEP12(2014)097}{\emph{JHEP} {\bfseries 12}
  (2014) 097}, [\href{https://arxiv.org/abs/1409.0100}{{\ttfamily 1409.0100}}].

\bibitem{Liu:2015kmo}
N.~Liu, L.~Wu, B.~Yang and M.~Zhang, \emph{{Single top partner production in
  the Higgs to diphoton channel in the Littlest Higgs Model with $T$-parity}},
  \href{https://doi.org/10.1016/j.physletb.2015.12.066}{\emph{Phys. Lett.}
  {\bfseries B753} (2016) 664--669},
  [\href{https://arxiv.org/abs/1508.07116}{{\ttfamily 1508.07116}}].

\bibitem{Backovic:2015lfa}
M.~Backovi{\'c}, T.~Flacke, J.~H. Kim and S.~J. Lee, \emph{{Discovering heavy
  new physics in boosted $Z$ channels: $Z \to l^+l^-$ vs $Z\to \nu \bar \nu$}},
  \href{https://doi.org/10.1103/PhysRevD.92.011701}{\emph{Phys. Rev.}
  {\bfseries D92} (2015) 011701},
  [\href{https://arxiv.org/abs/1501.07456}{{\ttfamily 1501.07456}}].

\bibitem{Zhang:2017nsn}
Y.-J. Zhang, L.~Han and Y.-B. Liu, \emph{{Single production of the top partner
  in the $T\rightarrow tZ$ channel at the LHeC}},
  \href{https://doi.org/10.1016/j.physletb.2017.02.051}{\emph{Phys. Lett.}
  {\bfseries B768} (2017) 241--247}.

\bibitem{Liu:2017sdg}
Y.-B. Liu and Y.-Q. Li, \emph{{Search for single production of the vector-like
  top partner at the 14 TeV LHC}},
  \href{https://doi.org/10.1140/epjc/s10052-017-5228-4}{\emph{Eur. Phys. J.}
  {\bfseries C77} (2017) 654},
  [\href{https://arxiv.org/abs/1709.06427}{{\ttfamily 1709.06427}}].

\bibitem{Lavoura:1992np}
L.~Lavoura and J.~P. Silva, \emph{{The Oblique corrections from vector - like
  singlet and doublet quarks}},
  \href{https://doi.org/10.1103/PhysRevD.47.2046}{\emph{Phys. Rev.} {\bfseries
  D47} (1993) 2046--2057}.

\bibitem{Maekawa:1995ha}
N.~Maekawa, \emph{{Electroweak symmetry breaking by vector - like fermions'
  condensation with small S and T parameters}},
  \href{https://doi.org/10.1103/PhysRevD.52.1684}{\emph{Phys. Rev.} {\bfseries
  D52} (1995) 1684--1692}.

\bibitem{He:2001tp}
H.-J. He, N.~Polonsky and S.-f. Su, \emph{{Extra families, Higgs spectrum and
  oblique corrections}},
  \href{https://doi.org/10.1103/PhysRevD.64.053004}{\emph{Phys. Rev.}
  {\bfseries D64} (2001) 053004},
  [\href{https://arxiv.org/abs/hep-ph/0102144}{{\ttfamily hep-ph/0102144}}].

\bibitem{Dawson:2012di}
S.~Dawson and E.~Furlan, \emph{{A Higgs Conundrum with Vector Fermions}},
  \href{https://doi.org/10.1103/PhysRevD.86.015021}{\emph{Phys. Rev.}
  {\bfseries D86} (2012) 015021},
  [\href{https://arxiv.org/abs/1205.4733}{{\ttfamily 1205.4733}}].

\bibitem{Ellis:2014dza}
S.~A.~R. Ellis, R.~M. Godbole, S.~Gopalakrishna and J.~D. Wells, \emph{{Survey
  of vector-like fermion extensions of the Standard Model and their
  phenomenological implications}},
  \href{https://doi.org/10.1007/JHEP09(2014)130}{\emph{JHEP} {\bfseries 09}
  (2014) 130}, [\href{https://arxiv.org/abs/1404.4398}{{\ttfamily 1404.4398}}].

\bibitem{Chen:2017hak}
C.-Y. Chen, S.~Dawson and E.~Furlan, \emph{{Vectorlike fermions and Higgs
  effective field theory revisited}},
  \href{https://doi.org/10.1103/PhysRevD.96.015006}{\emph{Phys. Rev.}
  {\bfseries D96} (2017) 015006},
  [\href{https://arxiv.org/abs/1703.06134}{{\ttfamily 1703.06134}}].

\bibitem{Barger:2007im}
V.~Barger, P.~Langacker, M.~McCaskey, M.~J. Ramsey-Musolf and G.~Shaughnessy,
  \emph{{LHC Phenomenology of an Extended Standard Model with a Real Scalar
  Singlet}}, \href{https://doi.org/10.1103/PhysRevD.77.035005}{\emph{Phys.
  Rev.} {\bfseries D77} (2008) 035005},
  [\href{https://arxiv.org/abs/0706.4311}{{\ttfamily 0706.4311}}].

\bibitem{OConnell:2006rsp}
D.~O'Connell, M.~J. Ramsey-Musolf and M.~B. Wise, \emph{{Minimal Extension of
  the Standard Model Scalar Sector}},
  \href{https://doi.org/10.1103/PhysRevD.75.037701}{\emph{Phys. Rev.}
  {\bfseries D75} (2007) 037701},
  [\href{https://arxiv.org/abs/hep-ph/0611014}{{\ttfamily hep-ph/0611014}}].

\bibitem{Pruna:2013bma}
G.~M. Pruna and T.~Robens, \emph{{Higgs singlet extension parameter space in
  the light of the LHC discovery}},
  \href{https://doi.org/10.1103/PhysRevD.88.115012}{\emph{Phys. Rev.}
  {\bfseries D88} (2013) 115012},
  [\href{https://arxiv.org/abs/1303.1150}{{\ttfamily 1303.1150}}].

\bibitem{Chen:2014ask}
C.-Y. Chen, S.~Dawson and I.~M. Lewis, \emph{{Exploring resonant di-Higgs boson
  production in the Higgs singlet model}},
  \href{https://doi.org/10.1103/PhysRevD.91.035015}{\emph{Phys. Rev.}
  {\bfseries D91} (2015) 035015},
  [\href{https://arxiv.org/abs/1410.5488}{{\ttfamily 1410.5488}}].

\bibitem{Buttazzo:2015bka}
D.~Buttazzo, F.~Sala and A.~Tesi, \emph{{Singlet-like Higgs bosons at present
  and future colliders}},
  \href{https://doi.org/10.1007/JHEP11(2015)158}{\emph{JHEP} {\bfseries 11}
  (2015) 158}, [\href{https://arxiv.org/abs/1505.05488}{{\ttfamily
  1505.05488}}].

\bibitem{Robens:2015gla}
T.~Robens and T.~Stefaniak, \emph{{Status of the Higgs Singlet Extension of the
  Standard Model after LHC Run 1}},
  \href{https://doi.org/10.1140/epjc/s10052-015-3323-y}{\emph{Eur. Phys. J.}
  {\bfseries C75} (2015) 104},
  [\href{https://arxiv.org/abs/1501.02234}{{\ttfamily 1501.02234}}].

\bibitem{Dawson:2015haa}
S.~Dawson and I.~M. Lewis, \emph{{NLO corrections to double Higgs boson
  production in the Higgs singlet model}},
  \href{https://doi.org/10.1103/PhysRevD.92.094023}{\emph{Phys. Rev.}
  {\bfseries D92} (2015) 094023},
  [\href{https://arxiv.org/abs/1508.05397}{{\ttfamily 1508.05397}}].

\bibitem{Costa:2015llh}
R.~Costa, M.~M{\"u}hlleitner, M.~O.~P. Sampaio and R.~Santos, \emph{{Singlet
  Extensions of the Standard Model at LHC Run 2: Benchmarks and Comparison with
  the NMSSM}}, \href{https://doi.org/10.1007/JHEP06(2016)034}{\emph{JHEP}
  {\bfseries 06} (2016) 034},
  [\href{https://arxiv.org/abs/1512.05355}{{\ttfamily 1512.05355}}].

\bibitem{Kanemura:2015fra}
S.~Kanemura, M.~Kikuchi and K.~Yagyu, \emph{{Radiative corrections to the Higgs
  boson couplings in the model with an additional real singlet scalar field}},
  \href{https://doi.org/10.1016/j.nuclphysb.2016.04.005}{\emph{Nucl. Phys.}
  {\bfseries B907} (2016) 286--322},
  [\href{https://arxiv.org/abs/1511.06211}{{\ttfamily 1511.06211}}].

\bibitem{Kanemura:2016lkz}
S.~Kanemura, M.~Kikuchi and K.~Yagyu, \emph{{One-loop corrections to the Higgs
  self-couplings in the singlet extension}},
  \href{https://doi.org/10.1016/j.nuclphysb.2017.02.004}{\emph{Nucl. Phys.}
  {\bfseries B917} (2017) 154--177},
  [\href{https://arxiv.org/abs/1608.01582}{{\ttfamily 1608.01582}}].

\bibitem{Robens:2016xkb}
T.~Robens and T.~Stefaniak, \emph{{LHC Benchmark Scenarios for the Real Higgs
  Singlet Extension of the Standard Model}},
  \href{https://doi.org/10.1140/epjc/s10052-016-4115-8}{\emph{Eur. Phys. J.}
  {\bfseries C76} (2016) 268},
  [\href{https://arxiv.org/abs/1601.07880}{{\ttfamily 1601.07880}}].

\bibitem{Lewis:2017dme}
I.~M. Lewis and M.~Sullivan, \emph{{Benchmarks for Double Higgs Production in
  the Singlet Extended Standard Model at the LHC}},
  \href{https://doi.org/10.1103/PhysRevD.96.035037}{\emph{Phys. Rev.}
  {\bfseries D96} (2017) 035037},
  [\href{https://arxiv.org/abs/1701.08774}{{\ttfamily 1701.08774}}].

\bibitem{Kanemura:2017gbi}
S.~Kanemura, M.~Kikuchi, K.~Sakurai and K.~Yagyu, \emph{{H-COUP: a program for
  one-loop corrected Higgs boson couplings in non-minimal Higgs sectors}},
  \href{https://arxiv.org/abs/1710.04603}{{\ttfamily 1710.04603}}.

\bibitem{Dawson:2017jja}
S.~Dawson and M.~Sullivan, \emph{{Enhanced di-Higgs boson production in the
  complex Higgs singlet model}},
  \href{https://doi.org/10.1103/PhysRevD.97.015022}{\emph{Phys. Rev.}
  {\bfseries D97} (2018) 015022},
  [\href{https://arxiv.org/abs/1711.06683}{{\ttfamily 1711.06683}}].

\bibitem{Fox:2011qc}
P.~J. Fox, D.~Tucker-Smith and N.~Weiner, \emph{{Higgs friends and counterfeits
  at hadron colliders}},
  \href{https://doi.org/10.1007/JHEP06(2011)127}{\emph{JHEP} {\bfseries 06}
  (2011) 127}, [\href{https://arxiv.org/abs/1104.5450}{{\ttfamily 1104.5450}}].

\bibitem{Ellis:2015oso}
J.~Ellis, S.~A.~R. Ellis, J.~Quevillon, V.~Sanz and T.~You, \emph{{On the
  Interpretation of a Possible $\sim 750$ GeV Particle Decaying into $\gamma
  \gamma$}}, \href{https://doi.org/10.1007/JHEP03(2016)176}{\emph{JHEP}
  {\bfseries 03} (2016) 176},
  [\href{https://arxiv.org/abs/1512.05327}{{\ttfamily 1512.05327}}].

\bibitem{McDermott:2015sck}
S.~D. McDermott, P.~Meade and H.~Ramani, \emph{{Singlet Scalar Resonances and
  the Diphoton Excess}},
  \href{https://doi.org/10.1016/j.physletb.2016.02.033}{\emph{Phys. Lett.}
  {\bfseries B755} (2016) 353--357},
  [\href{https://arxiv.org/abs/1512.05326}{{\ttfamily 1512.05326}}].

\bibitem{Falkowski:2015swt}
A.~Falkowski, O.~Slone and T.~Volansky, \emph{{Phenomenology of a 750 GeV
  Singlet}}, \href{https://doi.org/10.1007/JHEP02(2016)152}{\emph{JHEP}
  {\bfseries 02} (2016) 152},
  [\href{https://arxiv.org/abs/1512.05777}{{\ttfamily 1512.05777}}].

\bibitem{Anandakrishnan:2015yfa}
A.~Anandakrishnan, J.~H. Collins, M.~Farina, E.~Kuflik and M.~Perelstein,
  \emph{{Odd Top Partners at the LHC}},
  \href{https://doi.org/10.1103/PhysRevD.93.075009}{\emph{Phys. Rev.}
  {\bfseries D93} (2016) 075009},
  [\href{https://arxiv.org/abs/1506.05130}{{\ttfamily 1506.05130}}].

\bibitem{Gupta:2015zzs}
R.~S. Gupta, S.~J{\"a}ger, Y.~Kats, G.~Perez and E.~Stamou, \emph{{Interpreting
  a 750 GeV Diphoton Resonance}},
  \href{https://doi.org/10.1007/JHEP07(2016)145}{\emph{JHEP} {\bfseries 07}
  (2016) 145}, [\href{https://arxiv.org/abs/1512.05332}{{\ttfamily
  1512.05332}}].

\bibitem{Han:2015dlp}
H.~Han, S.~Wang and S.~Zheng, \emph{{Scalar Explanation of Diphoton Excess at
  LHC}}, \href{https://doi.org/10.1016/j.nuclphysb.2016.04.002}{\emph{Nucl.
  Phys.} {\bfseries B907} (2016) 180--186},
  [\href{https://arxiv.org/abs/1512.06562}{{\ttfamily 1512.06562}}].

\bibitem{Knapen:2015dap}
S.~Knapen, T.~Melia, M.~Papucci and K.~Zurek, \emph{{Rays of light from the
  LHC}}, \href{https://doi.org/10.1103/PhysRevD.93.075020}{\emph{Phys. Rev.}
  {\bfseries D93} (2016) 075020},
  [\href{https://arxiv.org/abs/1512.04928}{{\ttfamily 1512.04928}}].

\bibitem{Craig:2015lra}
N.~Craig, P.~Draper, C.~Kilic and S.~Thomas, \emph{{Shedding Light on Diphoton
  Resonances}}, \href{https://doi.org/10.1103/PhysRevD.93.115023}{\emph{Phys.
  Rev.} {\bfseries D93} (2016) 115023},
  [\href{https://arxiv.org/abs/1512.07733}{{\ttfamily 1512.07733}}].

\bibitem{Dolan:2016eki}
M.~J. Dolan, J.~L. Hewett, M.~Kr{\"a}mer and T.~G. Rizzo, \emph{{Simplified
  Models for Higgs Physics: Singlet Scalar and Vector-like Quark
  Phenomenology}}, \href{https://doi.org/10.1007/JHEP07(2016)039}{\emph{JHEP}
  {\bfseries 07} (2016) 039},
  [\href{https://arxiv.org/abs/1601.07208}{{\ttfamily 1601.07208}}].

\bibitem{Nakamura:2017irk}
K.~Nakamura, K.~Nishiwaki, K.-y. Oda, S.~C. Park and Y.~Yamamoto,
  \emph{{Di-higgs enhancement by neutral scalar as probe of new colored
  sector}}, \href{https://doi.org/10.1140/epjc/s10052-017-4835-4}{\emph{Eur.
  Phys. J.} {\bfseries C77} (2017) 273},
  [\href{https://arxiv.org/abs/1701.06137}{{\ttfamily 1701.06137}}].

\bibitem{Ham:2004cf}
S.~W. Ham, Y.~S. Jeong and S.~K. Oh, \emph{{Electroweak phase transition in an
  extension of the standard model with a real Higgs singlet}},
  \href{https://doi.org/10.1088/0954-3899/31/8/017}{\emph{J. Phys.} {\bfseries
  G31} (2005) 857--871},
  [\href{https://arxiv.org/abs/hep-ph/0411352}{{\ttfamily hep-ph/0411352}}].

\bibitem{Profumo:2007wc}
S.~Profumo, M.~J. Ramsey-Musolf and G.~Shaughnessy, \emph{{Singlet Higgs
  phenomenology and the electroweak phase transition}},
  \href{https://doi.org/10.1088/1126-6708/2007/08/010}{\emph{JHEP} {\bfseries
  08} (2007) 010}, [\href{https://arxiv.org/abs/0705.2425}{{\ttfamily
  0705.2425}}].

\bibitem{Espinosa:2011ax}
J.~R. Espinosa, T.~Konstandin and F.~Riva, \emph{{Strong Electroweak Phase
  Transitions in the Standard Model with a Singlet}},
  \href{https://doi.org/10.1016/j.nuclphysb.2011.09.010}{\emph{Nucl. Phys.}
  {\bfseries B854} (2012) 592--630},
  [\href{https://arxiv.org/abs/1107.5441}{{\ttfamily 1107.5441}}].

\bibitem{No:2013wsa}
J.~M. No and M.~Ramsey-Musolf, \emph{{Probing the Higgs Portal at the LHC
  Through Resonant di-Higgs Production}},
  \href{https://doi.org/10.1103/PhysRevD.89.095031}{\emph{Phys. Rev.}
  {\bfseries D89} (2014) 095031},
  [\href{https://arxiv.org/abs/1310.6035}{{\ttfamily 1310.6035}}].

\bibitem{Curtin:2014jma}
D.~Curtin, P.~Meade and C.-T. Yu, \emph{{Testing Electroweak Baryogenesis with
  Future Colliders}},
  \href{https://doi.org/10.1007/JHEP11(2014)127}{\emph{JHEP} {\bfseries 11}
  (2014) 127}, [\href{https://arxiv.org/abs/1409.0005}{{\ttfamily 1409.0005}}].

\bibitem{Huang:2015tdv}
P.~Huang, A.~Joglekar, B.~Li and C.~E.~M. Wagner, \emph{{Probing the
  Electroweak Phase Transition at the LHC}},
  \href{https://doi.org/10.1103/PhysRevD.93.055049}{\emph{Phys. Rev.}
  {\bfseries D93} (2016) 055049},
  [\href{https://arxiv.org/abs/1512.00068}{{\ttfamily 1512.00068}}].

\bibitem{Huang:2016cjm}
P.~Huang, A.~J. Long and L.-T. Wang, \emph{{Probing the Electroweak Phase
  Transition with Higgs Factories and Gravitational Waves}},
  \href{https://doi.org/10.1103/PhysRevD.94.075008}{\emph{Phys. Rev.}
  {\bfseries D94} (2016) 075008},
  [\href{https://arxiv.org/abs/1608.06619}{{\ttfamily 1608.06619}}].

\bibitem{Chen:2017qcz}
C.-Y. Chen, J.~Kozaczuk and I.~M. Lewis, \emph{{Non-resonant Collider
  Signatures of a Singlet-Driven Electroweak Phase Transition}},
  \href{https://doi.org/10.1007/JHEP08(2017)096}{\emph{JHEP} {\bfseries 08}
  (2017) 096}, [\href{https://arxiv.org/abs/1704.05844}{{\ttfamily
  1704.05844}}].

\bibitem{Fichet:2016xpw}
S.~Fichet, G.~von Gersdorff, E.~Pont{\'o}n and R.~Rosenfeld, \emph{{The Global
  Higgs as a Signal for Compositeness at the LHC}},
  \href{https://doi.org/10.1007/JHEP01(2017)012}{\emph{JHEP} {\bfseries 01}
  (2017) 012}, [\href{https://arxiv.org/abs/1608.01995}{{\ttfamily
  1608.01995}}].

\bibitem{Aaboud:2017zfn}
{\scshape ATLAS} collaboration, M.~Aaboud et~al., \emph{{Search for pair
  production of heavy vector-like quarks decaying to high-p$_{T}$ W bosons and
  b quarks in the lepton-plus-jets final state in pp collisions at $
  \sqrt{s}=13 $ TeV with the ATLAS detector}},
  \href{https://doi.org/10.1007/JHEP10(2017)141}{\emph{JHEP} {\bfseries 10}
  (2017) 141}, [\href{https://arxiv.org/abs/1707.03347}{{\ttfamily
  1707.03347}}].

\bibitem{Aaboud:2017qpr}
{\scshape ATLAS} collaboration, M.~Aaboud et~al., \emph{{Search for pair
  production of vector-like top quarks in events with one lepton, jets, and
  missing transverse momentum in $ \sqrt{s}=13 $ TeV $pp$ collisions with the
  ATLAS detector}}, \href{https://doi.org/10.1007/JHEP08(2017)052}{\emph{JHEP}
  {\bfseries 08} (2017) 052},
  [\href{https://arxiv.org/abs/1705.10751}{{\ttfamily 1705.10751}}].

\bibitem{Sirunyan:2017pks}
{\scshape CMS} collaboration, A.~M. Sirunyan et~al., \emph{{Search for pair
  production of vector-like quarks in the bW$\overline{\mathrm{b}}$W channel
  from proton-proton collisions at $\sqrt{s} =$ 13 TeV}},
  \href{https://arxiv.org/abs/1710.01539}{{\ttfamily 1710.01539}}.

\bibitem{Sirunyan:2017usq}
{\scshape CMS} collaboration, A.~M. Sirunyan et~al., \emph{{Search for pair
  production of vector-like T and B quarks in single-lepton final states using
  boosted jet substructure in proton-proton collisions at $\sqrt{s}=13$ TeV}},
  \href{https://doi.org/10.1007/JHEP11(2017)085}{\emph{JHEP} {\bfseries 11}
  (2017) 085}, [\href{https://arxiv.org/abs/1706.03408}{{\ttfamily
  1706.03408}}].

\bibitem{DeRujula:1983ak}
A.~De~Rujula, L.~Maiani and R.~Petronzio, \emph{{Search for Excited Quarks}},
  \href{https://doi.org/10.1016/0370-2693(84)90930-4}{\emph{Phys. Lett.}
  {\bfseries 140B} (1984) 253--258}.

\bibitem{Kuhn:1984rj}
J.~H. Kuhn and P.~M. Zerwas, \emph{{Excited Quarks and Leptons}},
  \href{https://doi.org/10.1016/0370-2693(84)90618-X}{\emph{Phys. Lett.}
  {\bfseries 147B} (1984) 189--196}.

\bibitem{Baur:1987ga}
U.~Baur, I.~Hinchliffe and D.~Zeppenfeld, \emph{{Excited Quark Production at
  Hadron Colliders}},
  \href{https://doi.org/10.1142/S0217751X87000661}{\emph{Int. J. Mod. Phys.}
  {\bfseries A2} (1987) 1285}.

\bibitem{Baur:1989kv}
U.~Baur, M.~Spira and P.~M. Zerwas, \emph{{Excited Quark and Lepton Production
  at Hadron Colliders}},
  \href{https://doi.org/10.1103/PhysRevD.42.815}{\emph{Phys. Rev.} {\bfseries
  D42} (1990) 815--824}.

\bibitem{Han:2010rf}
T.~Han, I.~Lewis and Z.~Liu, \emph{{Colored Resonant Signals at the LHC:
  Largest Rate and Simplest Topology}},
  \href{https://doi.org/10.1007/JHEP12(2010)085}{\emph{JHEP} {\bfseries 12}
  (2010) 085}, [\href{https://arxiv.org/abs/1010.4309}{{\ttfamily 1010.4309}}].

\bibitem{Sirunyan:2017yta}
{\scshape CMS} collaboration, A.~M. Sirunyan et~al., \emph{{Search for pair
  production of excited top quarks in the lepton + jets final state}},
  \href{https://doi.org/10.1016/j.physletb.2018.01.049}{\emph{Phys. Lett.}
  {\bfseries B778} (2018) 349},
  [\href{https://arxiv.org/abs/1711.10949}{{\ttfamily 1711.10949}}].

\bibitem{Aad:2015zhl}
{\scshape ATLAS, CMS} collaboration, G.~Aad et~al., \emph{{Combined Measurement
  of the Higgs Boson Mass in $pp$ Collisions at $\sqrt{s}=7$ and 8 TeV with the
  ATLAS and CMS Experiments}},
  \href{https://doi.org/10.1103/PhysRevLett.114.191803}{\emph{Phys. Rev. Lett.}
  {\bfseries 114} (2015) 191803},
  [\href{https://arxiv.org/abs/1503.07589}{{\ttfamily 1503.07589}}].

\bibitem{ATLAS-CONF-2017-046}
{\scshape ATLAS Collaboration} collaboration, \emph{{Measurement of the Higgs
  boson mass in the $H\rightarrow ZZ^*\rightarrow 4\ell$ and
  $H\rightarrow\gamma\gamma$ channels with $\sqrt{s}$=13TeV $pp$ collisions
  using the ATLAS detector}},  Tech. Rep. ATLAS-CONF-2017-046, CERN, Geneva,
  Jul, 2017.

\bibitem{Sirunyan:2017exp}
{\scshape CMS} collaboration, A.~M. Sirunyan et~al., \emph{{Measurements of
  properties of the Higgs boson decaying into the four-lepton final state in pp
  collisions at $ \sqrt{s}=13 $ TeV}},
  \href{https://doi.org/10.1007/JHEP11(2017)047}{\emph{JHEP} {\bfseries 11}
  (2017) 047}, [\href{https://arxiv.org/abs/1706.09936}{{\ttfamily
  1706.09936}}].

\bibitem{Chatrchyan:2012xdj}
{\scshape CMS} collaboration, S.~Chatrchyan et~al., \emph{{Observation of a new
  boson at a mass of 125 GeV with the CMS experiment at the LHC}},
  \href{https://doi.org/10.1016/j.physletb.2012.08.021}{\emph{Phys. Lett.}
  {\bfseries B716} (2012) 30--61},
  [\href{https://arxiv.org/abs/1207.7235}{{\ttfamily 1207.7235}}].

\bibitem{Aad:2012tfa}
{\scshape ATLAS} collaboration, G.~Aad et~al., \emph{{Observation of a new
  particle in the search for the Standard Model Higgs boson with the ATLAS
  detector at the LHC}},
  \href{https://doi.org/10.1016/j.physletb.2012.08.020}{\emph{Phys. Lett.}
  {\bfseries B716} (2012) 1--29},
  [\href{https://arxiv.org/abs/1207.7214}{{\ttfamily 1207.7214}}].

\bibitem{Dawson:2012mk}
S.~Dawson, E.~Furlan and I.~Lewis, \emph{{Unravelling an extended quark sector
  through multiple Higgs production?}},
  \href{https://doi.org/10.1103/PhysRevD.87.014007}{\emph{Phys. Rev.}
  {\bfseries D87} (2013) 014007},
  [\href{https://arxiv.org/abs/1210.6663}{{\ttfamily 1210.6663}}].

\bibitem{Patrignani:2016xqp}
{\scshape Particle Data Group} collaboration, C.~Patrignani et~al.,
  \emph{{Review of Particle Physics}},
  \href{https://doi.org/10.1088/1674-1137/40/10/100001}{\emph{Chin. Phys.}
  {\bfseries C40} (2016) 100001}.

\bibitem{Hahn:2000kx}
T.~Hahn, \emph{{Generating Feynman diagrams and amplitudes with FeynArts 3}},
  \href{https://doi.org/10.1016/S0010-4655(01)00290-9}{\emph{Comput. Phys.
  Commun.} {\bfseries 140} (2001) 418--431},
  [\href{https://arxiv.org/abs/hep-ph/0012260}{{\ttfamily hep-ph/0012260}}].

\bibitem{Christensen:2008py}
N.~D. Christensen and C.~Duhr, \emph{{FeynRules - Feynman rules made easy}},
  \href{https://doi.org/10.1016/j.cpc.2009.02.018}{\emph{Comput. Phys. Commun.}
  {\bfseries 180} (2009) 1614--1641},
  [\href{https://arxiv.org/abs/0806.4194}{{\ttfamily 0806.4194}}].

\bibitem{Alloul:2013bka}
A.~Alloul, N.~D. Christensen, C.~Degrande, C.~Duhr and B.~Fuks,
  \emph{{FeynRules 2.0 - A complete toolbox for tree-level phenomenology}},
  \href{https://doi.org/10.1016/j.cpc.2014.04.012}{\emph{Comput. Phys. Commun.}
  {\bfseries 185} (2014) 2250--2300},
  [\href{https://arxiv.org/abs/1310.1921}{{\ttfamily 1310.1921}}].

\bibitem{Hahn:1998yk}
T.~Hahn and M.~Perez-Victoria, \emph{{Automatized one loop calculations in
  four-dimensions and D-dimensions}},
  \href{https://doi.org/10.1016/S0010-4655(98)00173-8}{\emph{Comput. Phys.
  Commun.} {\bfseries 118} (1999) 153--165},
  [\href{https://arxiv.org/abs/hep-ph/9807565}{{\ttfamily hep-ph/9807565}}].

\bibitem{Ball:2013hta}
{\scshape NNPDF} collaboration, R.~D. Ball, V.~Bertone, S.~Carrazza,
  L.~Del~Debbio, S.~Forte, A.~Guffanti et~al., \emph{{Parton distributions with
  QED corrections}},
  \href{https://doi.org/10.1016/j.nuclphysb.2013.10.010}{\emph{Nucl. Phys.}
  {\bfseries B877} (2013) 290--320},
  [\href{https://arxiv.org/abs/1308.0598}{{\ttfamily 1308.0598}}].

\bibitem{Buckley:2014ana}
A.~Buckley, J.~Ferrando, S.~Lloyd, K.~Nordstr{\"o}m, B.~Page, M.~R{\"u}fenacht
  et~al., \emph{{LHAPDF6: parton density access in the LHC precision era}},
  \href{https://doi.org/10.1140/epjc/s10052-015-3318-8}{\emph{Eur. Phys. J.}
  {\bfseries C75} (2015) 132},
  [\href{https://arxiv.org/abs/1412.7420}{{\ttfamily 1412.7420}}].

\bibitem{Greco:2014aza}
D.~Greco and D.~Liu, \emph{{Hunting composite vector resonances at the LHC:
  naturalness facing data}},
  \href{https://doi.org/10.1007/JHEP12(2014)126}{\emph{JHEP} {\bfseries 12}
  (2014) 126}, [\href{https://arxiv.org/abs/1410.2883}{{\ttfamily 1410.2883}}].

\bibitem{Sirunyan:2017bfa}
{\scshape CMS} collaboration, A.~M. Sirunyan et~al., \emph{{Search for a heavy
  resonance decaying to a top quark and a vector-like top quark at $
  \sqrt{s}=13 $ TeV}},
  \href{https://doi.org/10.1007/JHEP09(2017)053}{\emph{JHEP} {\bfseries 09}
  (2017) 053}, [\href{https://arxiv.org/abs/1703.06352}{{\ttfamily
  1703.06352}}].

\bibitem{Dobrescu:2009vz}
B.~A. Dobrescu, K.~Kong and R.~Mahbubani, \emph{{Prospects for top-prime quark
  discovery at the Tevatron}},
  \href{https://doi.org/10.1088/1126-6708/2009/06/001}{\emph{JHEP} {\bfseries
  06} (2009) 001}, [\href{https://arxiv.org/abs/0902.0792}{{\ttfamily
  0902.0792}}].

\bibitem{Barcelo:2011wu}
R.~Barcelo, A.~Carmona, M.~Chala, M.~Masip and J.~Santiago, \emph{{Single
  Vectorlike Quark Production at the LHC}},
  \href{https://doi.org/10.1016/j.nuclphysb.2011.12.012}{\emph{Nucl. Phys.}
  {\bfseries B857} (2012) 172--184},
  [\href{https://arxiv.org/abs/1110.5914}{{\ttfamily 1110.5914}}].

\bibitem{Bini:2011zb}
C.~Bini, R.~Contino and N.~Vignaroli, \emph{{Heavy-light decay topologies as a
  new strategy to discover a heavy gluon}}, {\emph{JHEP} {\bfseries 01} (2012)
  }.

\bibitem{Freitas:2017afm}
A.~Freitas, K.~Kong and D.~Wiegand, \emph{{Radiative corrections to masses and
  couplings in Universal Extra Dimensions}},
  \href{https://arxiv.org/abs/1711.07526}{{\ttfamily 1711.07526}}.

\bibitem{Dobrescu:2016pda}
B.~A. Dobrescu and F.~Yu, \emph{{Exotic Signals of Vectorlike Quarks}},
  \href{https://arxiv.org/abs/1612.01909}{{\ttfamily 1612.01909}}.

\bibitem{Bizot:2018tds}
N.~Bizot, G.~Cacciapaglia and T.~Flacke, \emph{{Common exotic decays of top
  partners}},  \href{https://arxiv.org/abs/1803.00021}{{\ttfamily 1803.00021}}.

\bibitem{Chala:2018qdf}
M.~Chala, R.~Gr{\"o}ber and M.~Spannowsky, \emph{{Searches for vector-like
  quarks at future colliders and implications for composite Higgs models with
  dark matter}},  \href{https://arxiv.org/abs/1801.06537}{{\ttfamily
  1801.06537}}.

\bibitem{Buchkremer:2012dn}
M.~Buchkremer and A.~Schmidt, \emph{{Long-lived heavy quarks : a review}},
  \href{https://doi.org/10.1155/2013/690254}{\emph{Adv. High Energy Phys.}
  {\bfseries 2013} (2013) 690254},
  [\href{https://arxiv.org/abs/1210.6369}{{\ttfamily 1210.6369}}].

\bibitem{Bigi:1986jk}
I.~I.~Y. Bigi, Y.~L. Dokshitzer, V.~A. Khoze, J.~H. Kuhn and P.~M. Zerwas,
  \emph{{Production and Decay Properties of Ultraheavy Quarks}},
  \href{https://doi.org/10.1016/0370-2693(86)91275-X}{\emph{Phys. Lett.}
  {\bfseries B181} (1986) 157--163}.

\bibitem{Kats:2012ym}
Y.~Kats and M.~J. Strassler, \emph{{Probing Colored Particles with Photons,
  Leptons, and Jets}}, \href{https://doi.org/10.1007/JHEP11(2012)097,
  10.1007/JHEP07(2016)009}{\emph{JHEP} {\bfseries 11} (2012) 097},
  [\href{https://arxiv.org/abs/1204.1119}{{\ttfamily 1204.1119}}].

\bibitem{Barger:1987xg}
V.~D. Barger, E.~W.~N. Glover, K.~Hikasa, W.-Y. Keung, M.~G. Olsson, C.~J.
  Suchyta, III et~al., \emph{{Superheavy Quarkonium Production and Decays: A
  New Higgs Signal}}, \href{https://doi.org/10.1103/PhysRevD.35.3366,
  10.1103/PhysRevD.38.1632.2}{\emph{Phys. Rev.} {\bfseries D35} (1987) 3366}.

\bibitem{Kuhn:1993cp}
J.~H. Kuhn and E.~Mirkes, \emph{{Exotic bound state production at hadron
  colliders}}, \href{https://doi.org/10.1016/0370-2693(93)90573-Z}{\emph{Phys.
  Lett.} {\bfseries B311} (1993) 301--306},
  [\href{https://arxiv.org/abs/hep-ph/9305231}{{\ttfamily hep-ph/9305231}}].

\bibitem{Strassler:2006im}
M.~J. Strassler and K.~M. Zurek, \emph{{Echoes of a hidden valley at hadron
  colliders}},
  \href{https://doi.org/10.1016/j.physletb.2007.06.055}{\emph{Phys. Lett.}
  {\bfseries B651} (2007) 374--379},
  [\href{https://arxiv.org/abs/hep-ph/0604261}{{\ttfamily hep-ph/0604261}}].

\bibitem{Graham:2012th}
P.~W. Graham, D.~E. Kaplan, S.~Rajendran and P.~Saraswat, \emph{{Displaced
  Supersymmetry}}, \href{https://doi.org/10.1007/JHEP07(2012)149}{\emph{JHEP}
  {\bfseries 07} (2012) 149},
  [\href{https://arxiv.org/abs/1204.6038}{{\ttfamily 1204.6038}}].

\bibitem{Liu:2015bma}
Z.~Liu and B.~Tweedie, \emph{{The Fate of Long-Lived Superparticles with
  Hadronic Decays after LHC Run 1}},
  \href{https://doi.org/10.1007/JHEP06(2015)042}{\emph{JHEP} {\bfseries 06}
  (2015) 042}, [\href{https://arxiv.org/abs/1503.05923}{{\ttfamily
  1503.05923}}].

\bibitem{Drees:1990yw}
M.~Drees and X.~Tata, \emph{{Signals for heavy exotics at hadron colliders and
  supercolliders}},
  \href{https://doi.org/10.1016/0370-2693(90)90508-4}{\emph{Phys. Lett.}
  {\bfseries B252} (1990) 695--702}.

\bibitem{Arvanitaki:2005nq}
A.~Arvanitaki, S.~Dimopoulos, A.~Pierce, S.~Rajendran and J.~G. Wacker,
  \emph{{Stopping gluinos}},
  \href{https://doi.org/10.1103/PhysRevD.76.055007}{\emph{Phys. Rev.}
  {\bfseries D76} (2007) 055007},
  [\href{https://arxiv.org/abs/hep-ph/0506242}{{\ttfamily hep-ph/0506242}}].

\bibitem{Graham:2011ah}
P.~W. Graham, K.~Howe, S.~Rajendran and D.~Stolarski, \emph{{New Measurements
  with Stopped Particles at the LHC}},
  \href{https://doi.org/10.1103/PhysRevD.86.034020}{\emph{Phys. Rev.}
  {\bfseries D86} (2012) 034020},
  [\href{https://arxiv.org/abs/1111.4176}{{\ttfamily 1111.4176}}].

\bibitem{Fairbairn:2006gg}
M.~Fairbairn, A.~C. Kraan, D.~A. Milstead, T.~Sjostrand, P.~Z. Skands and
  T.~Sloan, \emph{{Stable massive particles at colliders}},
  \href{https://doi.org/10.1016/j.physrep.2006.10.002}{\emph{Phys. Rept.}
  {\bfseries 438} (2007) 1--63},
  [\href{https://arxiv.org/abs/hep-ph/0611040}{{\ttfamily hep-ph/0611040}}].

\bibitem{CMS:2014wda}
{\scshape CMS} collaboration, V.~Khachatryan et~al., \emph{{Search for
  Long-Lived Neutral Particles Decaying to Quark-Antiquark Pairs in
  Proton-Proton Collisions at $\sqrt{s} =$ 8 TeV}},
  \href{https://doi.org/10.1103/PhysRevD.91.012007}{\emph{Phys. Rev.}
  {\bfseries D91} (2015) 012007},
  [\href{https://arxiv.org/abs/1411.6530}{{\ttfamily 1411.6530}}].

\bibitem{Aad:2015rba}
{\scshape ATLAS} collaboration, G.~Aad et~al., \emph{{Search for massive,
  long-lived particles using multitrack displaced vertices or displaced lepton
  pairs in pp collisions at $\sqrt{s}$ = 8 TeV with the ATLAS detector}},
  \href{https://doi.org/10.1103/PhysRevD.92.072004}{\emph{Phys. Rev.}
  {\bfseries D92} (2015) 072004},
  [\href{https://arxiv.org/abs/1504.05162}{{\ttfamily 1504.05162}}].

\bibitem{Khachatryan:2016unx}
{\scshape CMS} collaboration, V.~Khachatryan et~al., \emph{{Search for R-parity
  violating supersymmetry with displaced vertices in proton-proton collisions
  at $\sqrt{s}$ = 8 TeV}},
  \href{https://doi.org/10.1103/PhysRevD.95.012009}{\emph{Phys. Rev.}
  {\bfseries D95} (2017) 012009},
  [\href{https://arxiv.org/abs/1610.05133}{{\ttfamily 1610.05133}}].

\bibitem{Sirunyan:2017jdo}
{\scshape CMS} collaboration, A.~M. Sirunyan et~al., \emph{{Search for new
  long-lived particles at $\sqrt{s} =$ 13 TeV}},
  \href{https://arxiv.org/abs/1711.09120}{{\ttfamily 1711.09120}}.

\bibitem{Aaboud:2017iio}
{\scshape ATLAS} collaboration, M.~Aaboud et~al., \emph{{Search for long-lived,
  massive particles in events with displaced vertices and missing transverse
  momentum in $\sqrt{s}$ = 13 TeV $pp$ collisions with the ATLAS detector}},
  \href{https://arxiv.org/abs/1710.04901}{{\ttfamily 1710.04901}}.

\bibitem{ATLAS-CONF-2017-026}
{\scshape ATLAS Collaboration} collaboration, \emph{{Search for long-lived,
  massive particles in events with displaced vertices and missing transverse
  momentum in 13 TeV $pp$ collisions with the ATLAS detector}},  Tech. Rep.
  ATLAS-CONF-2017-026, CERN, Geneva, Apr, 2017.

\bibitem{Aad:2015asa}
{\scshape ATLAS} collaboration, G.~Aad et~al., \emph{{Search for pair-produced
  long-lived neutral particles decaying in the ATLAS hadronic calorimeter in
  $pp$ collisions at $\sqrt{s}$ = 8 TeV}},
  \href{https://doi.org/10.1016/j.physletb.2015.02.015}{\emph{Phys. Lett.}
  {\bfseries B743} (2015) 15--34},
  [\href{https://arxiv.org/abs/1501.04020}{{\ttfamily 1501.04020}}].

\bibitem{ATLAS-CONF-2016-103}
{\scshape ATLAS Collaboration} collaboration, \emph{{Search for long-lived
  neutral particles decaying in the hadronic calorimeter of ATLAS at $\sqrt{s}
  = 13~\mathrm{TeV}$ in $3.2~\mathrm{fb^{-1}}$ of data}},  Tech. Rep.
  ATLAS-CONF-2016-103, CERN, Geneva, Sep, 2016.

\bibitem{Chatrchyan:2013oca}
{\scshape CMS} collaboration, S.~Chatrchyan et~al., \emph{{Searches for
  long-lived charged particles in pp collisions at $\sqrt{s}$=7 and 8 TeV}},
  \href{https://doi.org/10.1007/JHEP07(2013)122}{\emph{JHEP} {\bfseries 07}
  (2013) 122}, [\href{https://arxiv.org/abs/1305.0491}{{\ttfamily 1305.0491}}].

\bibitem{Khachatryan:2016sfv}
{\scshape CMS} collaboration, V.~Khachatryan et~al., \emph{{Search for
  long-lived charged particles in proton-proton collisions at $\sqrt s=$
  13  TeV}}, \href{https://doi.org/10.1103/PhysRevD.94.112004}{\emph{Phys.
  Rev.} {\bfseries D94} (2016) 112004},
  [\href{https://arxiv.org/abs/1609.08382}{{\ttfamily 1609.08382}}].

\bibitem{ATLAS:2014fka}
{\scshape ATLAS} collaboration, G.~Aad et~al., \emph{{Searches for heavy
  long-lived charged particles with the ATLAS detector in proton-proton
  collisions at $ \sqrt{s}=8 $ TeV}},
  \href{https://doi.org/10.1007/JHEP01(2015)068}{\emph{JHEP} {\bfseries 01}
  (2015) 068}, [\href{https://arxiv.org/abs/1411.6795}{{\ttfamily 1411.6795}}].

\bibitem{Aad:2015qfa}
{\scshape ATLAS} collaboration, G.~Aad et~al., \emph{{Search for metastable
  heavy charged particles with large ionisation energy loss in pp collisions at
  $\sqrt{s} = 8$ TeV using the ATLAS experiment}},
  \href{https://doi.org/10.1140/epjc/s10052-015-3609-0}{\emph{Eur. Phys. J.}
  {\bfseries C75} (2015) 407},
  [\href{https://arxiv.org/abs/1506.05332}{{\ttfamily 1506.05332}}].

\bibitem{Aaboud:2016uth}
{\scshape ATLAS} collaboration, M.~Aaboud et~al., \emph{{Search for heavy
  long-lived charged $R$-hadrons with the ATLAS detector in 3.2 fb$^{-1}$ of
  proton--proton collision data at $\sqrt{s} = 13$ TeV}},
  \href{https://doi.org/10.1016/j.physletb.2016.07.042}{\emph{Phys. Lett.}
  {\bfseries B760} (2016) 647--665},
  [\href{https://arxiv.org/abs/1606.05129}{{\ttfamily 1606.05129}}].

\bibitem{Sirunyan:2017sbs}
{\scshape CMS} collaboration, A.~M. Sirunyan et~al., \emph{{Search for decays
  of stopped exotic long-lived particles produced in proton-proton collisions
  at $\sqrt{s}=$ 13 TeV}},  \href{https://arxiv.org/abs/1801.00359}{{\ttfamily
  1801.00359}}.

\bibitem{Khachatryan:2015jha}
{\scshape CMS} collaboration, V.~Khachatryan et~al., \emph{{Search for Decays
  of Stopped Long-Lived Particles Produced in Proton-Proton Collisions at
  $\sqrt{s}= 8\,\text {TeV} $}},
  \href{https://doi.org/10.1140/epjc/s10052-015-3367-z}{\emph{Eur. Phys. J.}
  {\bfseries C75} (2015) 151},
  [\href{https://arxiv.org/abs/1501.05603}{{\ttfamily 1501.05603}}].

\bibitem{Aad:2013gva}
{\scshape ATLAS} collaboration, G.~Aad et~al., \emph{{Search for long-lived
  stopped R-hadrons decaying out-of-time with pp collisions using the ATLAS
  detector}}, \href{https://doi.org/10.1103/PhysRevD.88.112003}{\emph{Phys.
  Rev.} {\bfseries D88} (2013) 112003},
  [\href{https://arxiv.org/abs/1310.6584}{{\ttfamily 1310.6584}}].

\bibitem{Anastasiou:2016hlm}
C.~Anastasiou, C.~Duhr, F.~Dulat, E.~Furlan, T.~Gehrmann, F.~Herzog et~al.,
  \emph{{CP-even scalar boson production via gluon fusion at the LHC}},
  \href{https://doi.org/10.1007/JHEP09(2016)037}{\emph{JHEP} {\bfseries 09}
  (2016) 037}, [\href{https://arxiv.org/abs/1605.05761}{{\ttfamily
  1605.05761}}].

\bibitem{CMS-PAS-HIG-17-013}
{\scshape CMS Collaboration} collaboration, \emph{{Search for new resonances in
  the diphoton final state in the mass range between 70 and 110 GeV in pp
  collisions at $\sqrt{s}=$ 8 and 13 TeV}},  Tech. Rep. CMS-PAS-HIG-17-013,
  CERN, Geneva, 2017.

\bibitem{Aaboud:2017yyg}
{\scshape ATLAS} collaboration, M.~Aaboud et~al., \emph{{Search for new
  phenomena in high-mass diphoton final states using 37 fb$^{-1}$ of
  proton--proton collisions collected at $\sqrt{s}=13$ TeV with the ATLAS
  detector}}, \href{https://doi.org/10.1016/j.physletb.2017.10.039}{\emph{Phys.
  Lett.} {\bfseries B775} (2017) 105--125},
  [\href{https://arxiv.org/abs/1707.04147}{{\ttfamily 1707.04147}}].

\bibitem{Czakon:2011xx}
M.~Czakon and A.~Mitov, \emph{{Top++: A Program for the Calculation of the
  Top-Pair Cross-Section at Hadron Colliders}},
  \href{https://doi.org/10.1016/j.cpc.2014.06.021}{\emph{Comput. Phys. Commun.}
  {\bfseries 185} (2014) 2930},
  [\href{https://arxiv.org/abs/1112.5675}{{\ttfamily 1112.5675}}].

\bibitem{Czakon:2013goa}
M.~Czakon, P.~Fiedler and A.~Mitov, \emph{{Total Top-Quark Pair-Production
  Cross Section at Hadron Colliders Through $O(α\frac{4}{S})$}},
  \href{https://doi.org/10.1103/PhysRevLett.110.252004}{\emph{Phys. Rev. Lett.}
  {\bfseries 110} (2013) 252004},
  [\href{https://arxiv.org/abs/1303.6254}{{\ttfamily 1303.6254}}].

\bibitem{Czakon:2012pz}
M.~Czakon and A.~Mitov, \emph{{NNLO corrections to top pair production at
  hadron colliders: the quark-gluon reaction}},
  \href{https://doi.org/10.1007/JHEP01(2013)080}{\emph{JHEP} {\bfseries 01}
  (2013) 080}, [\href{https://arxiv.org/abs/1210.6832}{{\ttfamily 1210.6832}}].

\bibitem{Czakon:2012zr}
M.~Czakon and A.~Mitov, \emph{{NNLO corrections to top-pair production at
  hadron colliders: the all-fermionic scattering channels}},
  \href{https://doi.org/10.1007/JHEP12(2012)054}{\emph{JHEP} {\bfseries 12}
  (2012) 054}, [\href{https://arxiv.org/abs/1207.0236}{{\ttfamily 1207.0236}}].

\bibitem{Cacciari:2011hy}
M.~Cacciari, M.~Czakon, M.~Mangano, A.~Mitov and P.~Nason, \emph{{Top-pair
  production at hadron colliders with next-to-next-to-leading logarithmic
  soft-gluon resummation}},
  \href{https://doi.org/10.1016/j.physletb.2012.03.013}{\emph{Phys. Lett.}
  {\bfseries B710} (2012) 612--622},
  [\href{https://arxiv.org/abs/1111.5869}{{\ttfamily 1111.5869}}].

\bibitem{ATLAS-CONF-2016-072}
{\scshape ATLAS Collaboration} collaboration, \emph{{Search for single
  production of vector-like quarks decaying into $Wb$ in $pp$ collisions at
  $\sqrt{s} =$ 13 TeV with the ATLAS detector}},  Tech. Rep.
  ATLAS-CONF-2016-072, CERN, Geneva, Aug, 2016.

\bibitem{CMS-PAS-B2G-16-005}
{\scshape CMS Collaboration} collaboration, \emph{{Search for a vectorlike top
  partner produced through electroweak interaction and decaying to a top quark
  and a Higgs boson using boosted topologies in the all-hadronic final state}},
   Tech. Rep. CMS-PAS-B2G-16-005, CERN, Geneva, 2016.

\bibitem{Sirunyan:2017ynj}
{\scshape CMS} collaboration, A.~M. Sirunyan et~al., \emph{{Search for single
  production of a vector-like T quark decaying to a Z boson and a top quark in
  proton-proton collisions at sqrt(s) = 13 TeV}},
  \href{https://arxiv.org/abs/1708.01062}{{\ttfamily 1708.01062}}.

\bibitem{Chen:2014xwa}
C.-Y. Chen, S.~Dawson and I.~M. Lewis, \emph{{Top Partners and Higgs Boson
  Production}}, \href{https://doi.org/10.1103/PhysRevD.90.035016}{\emph{Phys.
  Rev.} {\bfseries D90} (2014) 035016},
  [\href{https://arxiv.org/abs/1406.3349}{{\ttfamily 1406.3349}}].

\bibitem{CMS-PAS-EXO-16-056}
{\scshape CMS Collaboration} collaboration, \emph{{Searches for dijet
  resonances in pp collisions at $\sqrt{s}=13~\mathrm{TeV}$ using data
  collected in 2016.}},  Tech. Rep. CMS-PAS-EXO-16-056, CERN, Geneva, 2017.

\bibitem{Sirunyan:2017hsb}
{\scshape CMS} collaboration, A.~M. Sirunyan et~al., \emph{{Search for
  Z$\gamma$ resonances using leptonic and hadronic final states in
  proton-proton collisions at $\sqrt{s}=$ 13 TeV}},
  \href{https://arxiv.org/abs/1712.03143}{{\ttfamily 1712.03143}}.

\bibitem{Aaboud:2017rel}
{\scshape ATLAS} collaboration, M.~Aaboud et~al., \emph{{Search for heavy $ZZ$
  resonances in the $\ell^+\ell^-\ell^+\ell^-$ and $\ell^+\ell^-\nu\bar\nu$
  final states using proton proton collisions at $\sqrt{s}= 13$ TeV with the
  ATLAS detector}},  \href{https://arxiv.org/abs/1712.06386}{{\ttfamily
  1712.06386}}.

\bibitem{Alwall:2014hca}
J.~Alwall, R.~Frederix, S.~Frixione, V.~Hirschi, F.~Maltoni, O.~Mattelaer
  et~al., \emph{{The automated computation of tree-level and next-to-leading
  order differential cross sections, and their matching to parton shower
  simulations}}, \href{https://doi.org/10.1007/JHEP07(2014)079}{\emph{JHEP}
  {\bfseries 07} (2014) 079},
  [\href{https://arxiv.org/abs/1405.0301}{{\ttfamily 1405.0301}}].

\bibitem{Sjostrand:2006za}
T.~Sjostrand, S.~Mrenna and P.~Z. Skands, \emph{{PYTHIA 6.4 Physics and
  Manual}}, \href{https://doi.org/10.1088/1126-6708/2006/05/026}{\emph{JHEP}
  {\bfseries 05} (2006) 026},
  [\href{https://arxiv.org/abs/hep-ph/0603175}{{\ttfamily hep-ph/0603175}}].

\bibitem{Cacciari:2011ma}
M.~Cacciari, G.~P. Salam and G.~Soyez, \emph{{FastJet User Manual}},
  \href{https://doi.org/10.1140/epjc/s10052-012-1896-2}{\emph{Eur. Phys. J.}
  {\bfseries C72} (2012) 1896},
  [\href{https://arxiv.org/abs/1111.6097}{{\ttfamily 1111.6097}}].

\bibitem{Cacciari:2008gp}
M.~Cacciari, G.~P. Salam and G.~Soyez, \emph{{The Anti-k(t) jet clustering
  algorithm}}, \href{https://doi.org/10.1088/1126-6708/2008/04/063}{\emph{JHEP}
  {\bfseries 04} (2008) 063},
  [\href{https://arxiv.org/abs/0802.1189}{{\ttfamily 0802.1189}}].

\bibitem{ATL-PHYS-PUB-2013-004}
\emph{{Performance assumptions for an upgraded ATLAS detector at a
  High-Luminosity LHC}},  Tech. Rep. ATL-PHYS-PUB-2013-004, CERN, Geneva, Mar,
  2013.

\bibitem{Mangano:2006rw}
M.~L. Mangano, M.~Moretti, F.~Piccinini and M.~Treccani, \emph{{Matching matrix
  elements and shower evolution for top-quark production in hadronic
  collisions}},
  \href{https://doi.org/10.1088/1126-6708/2007/01/013}{\emph{JHEP} {\bfseries
  01} (2007) 013}, [\href{https://arxiv.org/abs/hep-ph/0611129}{{\ttfamily
  hep-ph/0611129}}].

\bibitem{Chway:2015lzg}
D.~Chway, R.~Dermíšek, T.~H. Jung and H.~D. Kim, \emph{{Gluons to Diphotons
  via New Particles with Half the Signal’s Invariant Mass}},
  \href{https://doi.org/10.1103/PhysRevLett.117.061801}{\emph{Phys. Rev. Lett.}
  {\bfseries 117} (2016) 061801},
  [\href{https://arxiv.org/abs/1512.08221}{{\ttfamily 1512.08221}}].

\bibitem{Dawson:2015oha}
S.~Dawson, A.~Ismail and I.~Low, \emph{{What’s in the loop? The anatomy of
  double Higgs production}},
  \href{https://doi.org/10.1103/PhysRevD.91.115008}{\emph{Phys. Rev.}
  {\bfseries D91} (2015) 115008},
  [\href{https://arxiv.org/abs/1504.05596}{{\ttfamily 1504.05596}}].

\bibitem{Rehermann:2010vq}
K.~Rehermann and B.~Tweedie, \emph{{Efficient Identification of Boosted
  Semileptonic Top Quarks at the LHC}},
  \href{https://doi.org/10.1007/JHEP03(2011)059}{\emph{JHEP} {\bfseries 03}
  (2011) 059}, [\href{https://arxiv.org/abs/1007.2221}{{\ttfamily 1007.2221}}].

\bibitem{Backovic:2012jk}
M.~Backović and J.~Juknevich, \emph{{TemplateTagger v1.0.0: A Template
  Matching Tool for Jet Substructure}},
  \href{https://doi.org/10.1016/j.cpc.2013.12.018}{\emph{Comput. Phys. Commun.}
  {\bfseries 185} (2014) 1322--1338},
  [\href{https://arxiv.org/abs/1212.2978}{{\ttfamily 1212.2978}}].

\bibitem{Almeida:2010pa}
L.~G. Almeida, S.~J. Lee, G.~Perez, G.~Sterman and I.~Sung, \emph{{Template
  Overlap Method for Massive Jets}},
  \href{https://doi.org/10.1103/PhysRevD.82.054034}{\emph{Phys. Rev.}
  {\bfseries D82} (2010) 054034},
  [\href{https://arxiv.org/abs/1006.2035}{{\ttfamily 1006.2035}}].

\bibitem{Backovic:2013bga}
M.~Backovic, O.~Gabizon, J.~Juknevich, G.~Perez and Y.~Soreq, \emph{{Measuring
  boosted tops in semi-leptonic $t\bar t$ events for the standard model and
  beyond}}, \href{https://doi.org/10.1007/JHEP04(2014)176}{\emph{JHEP}
  {\bfseries 04} (2014) 176},
  [\href{https://arxiv.org/abs/1311.2962}{{\ttfamily 1311.2962}}].

\bibitem{Kasieczka:2017nvn}
G.~Kasieczka, T.~Plehn, M.~Russell and T.~Schell, \emph{{Deep-learning Top
  Taggers or The End of QCD?}},
  \href{https://doi.org/10.1007/JHEP05(2017)006}{\emph{JHEP} {\bfseries 05}
  (2017) 006}, [\href{https://arxiv.org/abs/1701.08784}{{\ttfamily
  1701.08784}}].

\bibitem{Backovic:2014ega}
M.~Backović, T.~Flacke, J.~H. Kim and S.~J. Lee, \emph{{Boosted Event
  Topologies from TeV Scale Light Quark Composite Partners}},
  \href{https://doi.org/10.1007/JHEP04(2015)082}{\emph{JHEP} {\bfseries 04}
  (2015) 082}, [\href{https://arxiv.org/abs/1410.8131}{{\ttfamily 1410.8131}}].

\bibitem{ATL-PHYS-PUB-2016-026}
{\scshape ATLAS Collaboration} collaboration, \emph{{Expected performance for
  an upgraded ATLAS detector at High-Luminosity LHC}},  Tech. Rep.
  ATL-PHYS-PUB-2016-026, CERN, Geneva, Oct, 2016.

\bibitem{Barger:2006hm}
V.~Barger, T.~Han and D.~G.~E. Walker, \emph{{Top Quark Pairs at High Invariant
  Mass: A Model-Independent Discriminator of New Physics at the LHC}},
  \href{https://doi.org/10.1103/PhysRevLett.100.031801}{\emph{Phys. Rev. Lett.}
  {\bfseries 100} (2008) 031801},
  [\href{https://arxiv.org/abs/hep-ph/0612016}{{\ttfamily hep-ph/0612016}}].

\bibitem{Gopalakrishna:2010xm}
S.~Gopalakrishna, T.~Han, I.~Lewis, Z.-g. Si and Y.-F. Zhou, \emph{{Chiral
  Couplings of W' and Top Quark Polarization at the LHC}},
  \href{https://doi.org/10.1103/PhysRevD.82.115020}{\emph{Phys. Rev.}
  {\bfseries D82} (2010) 115020},
  [\href{https://arxiv.org/abs/1008.3508}{{\ttfamily 1008.3508}}].

\bibitem{Cowan:2010js}
G.~Cowan, K.~Cranmer, E.~Gross and O.~Vitells, \emph{{Asymptotic formulae for
  likelihood-based tests of new physics}},
  \href{https://doi.org/10.1140/epjc/s10052-011-1554-0,
  10.1140/epjc/s10052-013-2501-z}{\emph{Eur. Phys. J.} {\bfseries C71} (2011)
  1554}, [\href{https://arxiv.org/abs/1007.1727}{{\ttfamily 1007.1727}}].

\bibitem{Espriu:2002xv}
D.~Espriu, J.~Manzano and P.~Talavera, \emph{{Flavor mixing, gauge invariance
  and wave function renormalization}},
  \href{https://doi.org/10.1103/PhysRevD.66.076002}{\emph{Phys. Rev.}
  {\bfseries D66} (2002) 076002},
  [\href{https://arxiv.org/abs/hep-ph/0204085}{{\ttfamily hep-ph/0204085}}].

\bibitem{Kniehl:2009nz}
B.~A. Kniehl and A.~Sirlin, \emph{{Novel formulations of CKM matrix
  renormalization}}, \href{https://doi.org/10.1063/1.3293809}{\emph{AIP Conf.
  Proc.} {\bfseries 1182} (2009) 327--330},
  [\href{https://arxiv.org/abs/0906.2670}{{\ttfamily 0906.2670}}].

\bibitem{Kniehl:2009kk}
B.~A. Kniehl and A.~Sirlin, \emph{{A Novel Formulation of
  Cabibbo-Kobayashi-Maskawa Matrix Renormalization}},
  \href{https://doi.org/10.1016/j.physletb.2009.02.024}{\emph{Phys. Lett.}
  {\bfseries B673} (2009) 208--210},
  [\href{https://arxiv.org/abs/0901.0114}{{\ttfamily 0901.0114}}].

\bibitem{Kniehl:2006rc}
B.~A. Kniehl and A.~Sirlin, \emph{{Simple On-Shell Renormalization Framework
  for the Cabibbo-Kobayashi-Maskawa Matrix}},
  \href{https://doi.org/10.1103/PhysRevD.74.116003}{\emph{Phys. Rev.}
  {\bfseries D74} (2006) 116003},
  [\href{https://arxiv.org/abs/hep-th/0612033}{{\ttfamily hep-th/0612033}}].

\bibitem{Denner:1991kt}
A.~Denner, \emph{{Techniques for calculation of electroweak radiative
  corrections at the one loop level and results for W physics at LEP-200}},
  \href{https://doi.org/10.1002/prop.2190410402}{\emph{Fortsch. Phys.}
  {\bfseries 41} (1993) 307--420},
  [\href{https://arxiv.org/abs/0709.1075}{{\ttfamily 0709.1075}}].

\bibitem{Deshpande:1981zq}
N.~G. Deshpande and G.~Eilam, \emph{{FLAVOR CHANGING ELECTROMAGNETIC
  TRANSITIONS}}, \href{https://doi.org/10.1103/PhysRevD.26.2463}{\emph{Phys.
  Rev.} {\bfseries D26} (1982) 2463}.

\bibitem{Denner:1990yz}
A.~Denner and T.~Sack, \emph{{Renormalization of the Quark Mixing Matrix}},
  \href{https://doi.org/10.1016/0550-3213(90)90557-T}{\emph{Nucl. Phys.}
  {\bfseries B347} (1990) 203--216}.

\bibitem{Jegerlehner:1991dq}
F.~Jegerlehner, \emph{{Renormalizing the standard model}}, {\emph{Conf. Proc.}
  {\bfseries C900603} (1990) 476--590}.

\end{thebibliography}\endgroup

\end{document}